\documentclass[a4paper]{book}
\usepackage[utf8x]{inputenc}
\usepackage{blindtext, rotating} 
\usepackage{amsmath} 
\usepackage{amssymb} 
\usepackage{dsfont} 
\usepackage{textcomp}  
\usepackage{mathtools} 
\usepackage{bm}
\usepackage{booktabs} 
\usepackage[
	colorlinks=true
	,breaklinks
	]{hyperref} 
\usepackage[hyphenbreaks]{breakurl} 
\usepackage{xcolor}
\definecolor{c1}{rgb}{0,0,1} 
\definecolor{c2}{rgb}{0,0.3,0.9} 
\definecolor{c3}{rgb}{0.3,0,0.9} 
\hypersetup{
    linkcolor={c1}, 
    citecolor={c2}, 
    urlcolor={c3} 
}
\usepackage[authoryear,square,numbers]{natbib} 
\usepackage[nottoc]{tocbibind} 
\usepackage{graphicx} 
\usepackage{longtable} 
\usepackage{bigstrut} 
\usepackage{enumerate} 
\usepackage{todonotes} 
\usepackage{makeidx} 
\makeindex

\usepackage[top=2cm, bottom=1.8cm,left=2.5cm,right=2.5cm]{geometry} 
\usepackage{fancyhdr} 

\begin{document}


\pagestyle{empty}
\title{Nonlinear dynamics and chaos in multidimensional disordered Hamiltonian systems}
\author{Bertin Many Manda}
\date{}
\maketitle
%
%
%
%
%
%


%

\begin{titlepage}

\newcommand{\HRule}{\rule{\linewidth}{0.5mm}} 

\center 
 

\textsc{\LARGE University of Cape Town}\\[1.5cm] 
\textsc{\Large Department of Mathematics and Applied Mathematics}\\[0.5cm] 
\textsc{\large Phd Dissertation}\\[0.5cm] 


\HRule \\[0.4cm]
{ \huge \bfseries  Nonlinear dynamics and chaos in multidimensional disordered Hamiltonian systems}\\[0.4cm] 
\HRule \\[1.5cm]
 

\begin{minipage}{0.4\textwidth}
\begin{flushleft} \large
\emph{Author:}\\
Bertin \textsc{Many Manda} 
\end{flushleft}
\end{minipage}
~
\begin{minipage}{0.4\textwidth}
\begin{flushright} \large
\emph{Supervisor:} \\
A/Prof. Charalampos \textsc{Skokos} 
\end{flushright}
\end{minipage}\\[4cm]



{\large \today}\\[3cm] 


\includegraphics[width=0.35\textwidth]{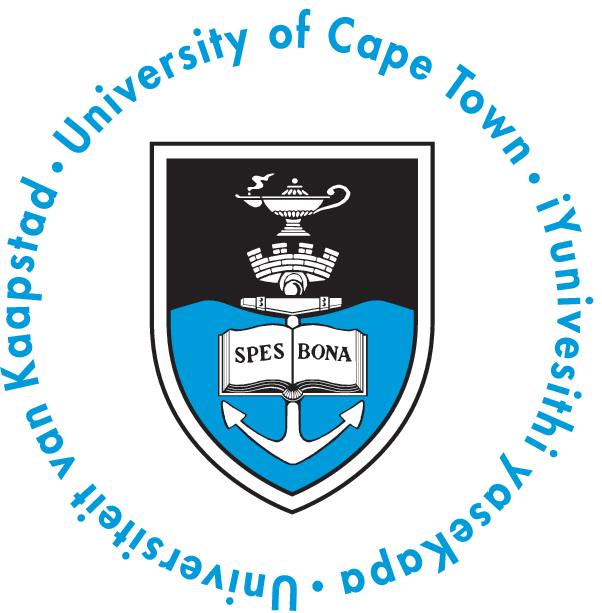}\\[1cm] 
 

\vfill 

\end{titlepage}

\pagestyle{plain}

\chapter*{\label{chap:abstract}Abstract}
\addcontentsline{toc}{chapter}{Abstract}
\pagestyle{fancy}
\fancyhf{}
\fancyhead[OC]{\leftmark}
\fancyhead[EC]{\rightmark}
\cfoot{\thepage}
In this thesis we study the chaotic behavior of multidimensional Hamiltonian systems in the presence of nonlinearity and disorder.
It is known that any localized initial excitation in a large enough linear disordered system spreads for a finite amount of time and then halts forever.
This phenomenon is called Anderson localization (AL).
What happens to AL when nonlinearity is introduced is an interesting question which has been considered in several studies over the past decades. 
Recent works focussing on two widely--applicable systems, namely the disordered Klein-Gordon (DKG) lattice of anharmonic oscillators and the disordered discrete nonlinear Schr\"odinger (DDNLS) equation, mainly in one spatial dimension suggest that nonlinearity eventually destroys AL.
This leads to an infinite diffusive spreading of initially localized wave packets whose extent (measured for instance through the wave packet's second moment $m_2$) grows in time $t$ as $t^{\alpha_m}$ with $0< \alpha_m < 1$.
However, the characteristics and the asymptotic fate of such evolutions still remain an issue of intense debate due to their computational difficulty, especially in systems of more than one spatial dimension.
Two different spreading regimes, the so-called weak and strong chaos regimes, have been theoretically predicted and numerically identified.
As the spreading of initially localized wave packets is a non-equilibrium thermalization process related to the ergodic and chaotic properties of the system, in our work we investigate the properties of chaos studying the behavior of observables related to the system's tangent dynamics. 
In particular, we consider the DDNLS model of one (1D) and two (2D) spatial dimensions and develop robust, efficient and fast numerical integration schemes for the long-time evolution of the phase space and tangent dynamics of these systems. 
Implementing these integrators, we perform extensive numerical simulations for various sets of parameter values.
We present, to the best of our knowledge for the first time, detailed computations of the time evolution of the system's maximum Lyapunov exponent (MLE--$\Lambda$) i.e.~the most commonly used chaos indicator, and the related deviation vector distribution (DVD). 
We find that although the systems' MLE decreases in time following a power law $t^{\alpha_\Lambda}$ with $\alpha_\Lambda <0$ for both the weak and strong chaos cases, no crossover to the behavior $\Lambda \propto t^{-1}$ (which is indicative of regular motion) is observed.
By investigating a large number of weak and strong chaos cases, we determine the different $\alpha_\Lambda$ values for the 1D and 2D systems.
In addition, the analysis of the DVDs reveals the existence of random fluctuations of chaotic hotspots with increasing amplitudes inside the excited part of the wave packet, which assist in homogenizing chaos and contribute to the thermalization of more lattice sites.  
Furthermore, we show the existence of a dimension-free relation between the wave packet spreading and its degree of chaoticity between the 1D and 2D  DDNLS systems. 
The generality of our findings is confirmed, as similar behaviors to the ones observed for the DDNLS systems are also present in the case of DKG models.

\chapter*{\label{chap:declaration}Plagiarism declaration}
\addcontentsline{toc}{chapter}{Plagiarism declaration}
\pagestyle{fancy}
\fancyhf{}
\fancyhead[OC]{\leftmark}
\fancyhead[EC]{\rightmark}
\cfoot{\thepage}

I, the undersigned, hereby declare that the work contained in this thesis is my original work, and that any work done by others or by myself previously has been acknowledged and referenced accordingly.

\includegraphics[height=4cm]{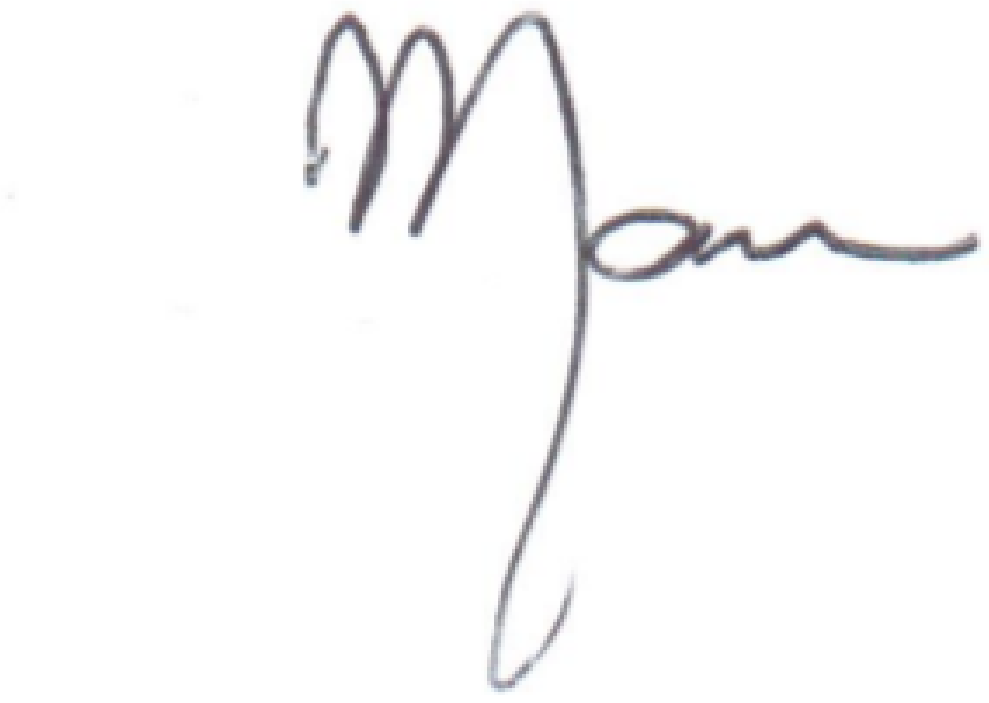} 
\vspace{0.5cm}
\hrule
\vspace{0.5cm}
Bertin \textsc{MANY MANDA}, 19 October 2020.
\chapter*{\label{chapter:acknowledgements}Acknowledgements}
\addcontentsline{toc}{chapter}{Acknowledgments}
\pagestyle{fancy}
\fancyhf{}
\fancyhead[OC]{\leftmark}
\fancyhead[EC]{\rightmark}
\cfoot{\thepage}

The completion of my doctoral project was only made possible due to the outstanding support and guidance of my supervisor A/Prof. Haris Skokos.
The several months he spent proof-reading this thesis is a glance into how engaged and supportive he has been during these last $4$ years.
 
I am grateful to my colleagues Bob Senyange, Malcolm Hillebrand and Henok Moges for proof-reading this work and for the innumerable fruitful discussions I had with them.
We spent hours arguing on analytical and numerical methods, which contributed enormously to improving several aspects of my doctoral work.

During my PhD work, I had the opportunity to interact with many other people whose feedback increased my understanding of nonlinear lattice dynamics and deeply shaped the content of this work.
More specifically I mention Prof.~Anastasios~Bountis (Nazarbayev University, Kazakhstan), Dr.~George~Kalosakas (University of Patras, Greece), Dr.~Carlo~Danieli (Max Planck Institute for the Physics of Complex Systems, Germany), Dr.~Thudiyangal~Mithun (University of Massachusetts, USA), Dr.~Konstantinos~Kaloudis (Nazarbayev University, Kazakhstan), Dr.~Thomas~Oikonomou (Nazarbayev University, Kazakhstan) and Mr.~Malcolm~Hillebrand (University of Cape Town, South Africa) with whom I had the privilege to co-author research papers.
In addition, the insights I obtained from discussions with Prof.~Sergej~Flach (Institute for Basic Science, Korea), Prof.~Panayotis~Kevrekidis (University of Massachusetts, USA) and Dr.~Yagmur~Kati (Institute for Basic Science, Korea) are also encompassed in this work.
Furthermore, I thank Prof.~‪Giorgos~Tsironis (University of Crete, Greece) who as an organizer financially assisted me to participate in the International Conference `Nonlinear Localization in Lattices' which was held in Spetses, Greece in June 2018 and where I met most of the researchers I mentioned above.
 
Finally, I would like to thank my whole family. 
My brothers and sisters for their encouragement and my mother Marie Bernadette Biloa Onana who made travelling to South Africa possible through great financial and personal sacrifices.
 
This work was funded by the National Research Foundation of South Africa under the Grant Number: $98974$.
I also received additional support from the University of Cape Town.
I thank the Centre for High Performance Computing of South Africa and the University of Cape Town's ICTS High Performance Computing Centre for providing the computational resources and the personal assistance necessary for the successful completion of this work.

\chapter*{\label{chap:publication}Publications and conference presentations}
\addcontentsline{toc}{chapter}{Publications and conference presentations}
\pagestyle{fancy}
\fancyhf{}
\fancyhead[OC]{\leftmark}
\fancyhead[EC]{\rightmark}
\cfoot{\thepage}

\subsection*{Publications related to this thesis:}
\begin{enumerate}
    \item[P1.] B.~Senyange, {\bf B.~Many~Manda}, and Ch.~Skokos, {\it Characteristics of chaos evolution in one-dimensional disordered nonlinear lattices,} Physical Review E, 2018, 98, 052229. 
            DOI:~\url{doi.org/10.1103/PhysRevE.98.052229}.
    \item[P2.] C. Danieli, {\bf B.~Many~Manda}, T.~Mithun, and Ch.~Skokos, {\it Computational efficiency of numerical integration methods for the tangent dynamics of many-body Hamiltonian systems in one and two spatial dimensions}, Mathematics in Engineering, 2019, 1, 447. 
            DOI:~\url{doi.org/10.3934/mine.2019.3.447}.
    \item[P3.] {\bf B.~Many~Manda}, B.~Senyange, and Ch.~Skokos, {\it Chaotic wave-packet spreading in two-dimensional disordered nonlinear lattices}, Physical Review E, 2020, 101, 032206. 
            DOI:~\url{doi.org/10.1103/PhysRevE.101.032206}.
\end{enumerate}

\subsection*{Other publications:}
\begin{enumerate}
    \item[P4.] M.~Hillebrand, {\bf B.~Many~Manda}, G.~Kalosakas, E.~Gerlach, and Ch.~Skokos, {\it Chaotic dynamics of graphene and graphene nanoribbons}, Chaos: An Interdisciplinary Journal of Nonlinear Science, 2020, 30, 063150. 
            DOI:~\url{doi.org/10.1063/5.0007761}.
    \item[P5.] B.~Senyange, J.-J.~Plessis, {\bf B.~Many~Manda}, and Ch.~Skokos, {\it Properties of Normal Modes in a Modified Disordered Klein-Gordon Lattice: From Disorder to Order}, Nonlinear Phenomena in Complex Systems: An Interdisciplinary Journal, 2020, 23, 165. 
            DOI:~\url{doi.org/10.33581/1561-4085-2020-23-2-165-171}.
    \item[P6.] A.~Bountis, K.~Kaloudis, Th.~Oikonomou, {\bf B.~Many~Manda}, and Ch.~Skokos, {\it Stability properties of 1-dimensional Hamiltonian lattices with non-analytic potentials}, International Journal of Bifurcation and Chaos, 2020, 30, 2030047.
            DOI:~\url{doi.org/10.1142/S0218127420300475}. 
    \item[P7.] T.~Mithun, A.~Maluckov, {\bf B.~Many~Manda}, Ch.~Skokos, A.~Bishop, A.~Saxena, A.~Khare, and P.~G.~Kevrekidis, {\it Thermalization in the one-dimensional Salerno model lattice}, Submitted to Physical Review E, 2020.
            Preprint version:~\url{physics.class-ph/2012.03652}.
\end{enumerate}

\subsection*{Conference presentations:}
\begin{enumerate}
    \item Poster presentation: {\it Dynamical behavior of a 2D planar model of graphene}.
        The International Conference on Nonlinear Localization in Lattices, Spetses, Greece, June 2018.
    \item Poster presentation: {\it Chaotic wave packet spreading in nonlinear disordered media}.
        The UCT \& SU's Science Postgraduate Symposium, Stellenbosch, South Africa, September 2019. 
    \item Oral presentation: {\it Chaotic wave packet spreading in the one- and two-dimensional disordered nonlinear Schr\"odinger equations}.
        The $62^{nd}$ Annual Congress of the South African Mathematical Society, Cape Town, South Africa, December 2019.
    
\end{enumerate}

\tableofcontents

\chapter{\label{chap:general introduction}General introduction}

\section{\label{sec:dynamics_disordered_systems}Dynamics of disordered systems}

Disordered dynamical systems intend to model the heterogeneity of nature.
They are obtained by breaking the spatial translational symmetry (isotropy) of Hamiltonian models by attributing to one or many of their parameters random values to each single degree of freedom.
In principle, the random numbers are drawn from any chosen probability distribution and could directly affect the eigen-energies (see e.g.~\citep{anderson1958absence,gredeskul1992propagation,kramer1993localization,shepelyansky1993delocalization,kopidakis2008absence,pikovsky2008destruction}), the on-site interactions (see e.g.~\citep{kopidakis2008absence,skokos2009delocalization,skokos2010spreading,laptyeva2010crossover,bodyfelt2011nonlinear}), the inter-particle interactions (see e.g.~\citep{prosen1992energy,dauxois1993entropy,barre2001lyapunov,roy2012spreading,martinez2016superdiffusive,hillebrand2019heterogeneity,ngapasare2019chaos,pikovsky2020scaling}) and the particle's masses (see e.g.~\citep{achilleos2016energy,achilleos2018chaos,ngapasare2019chaos,hillebrand2020chaotic}) among others.
The study of disordered systems has a long standing history, starting in the mid $20^\textrm{th}$ century with the explosion of modern technologies such as semiconductors.
Studying a linear electron tight-binding approximation of the latter material, P.~W.~Anderson~\citep{anderson1958absence,kramer1993localization} discovered that initially localized electron matter waves remain in that state for an infinitely long time.
This phenomenon is now referred as {\it Anderson localization} (AL) and has been experimentally shown to occur in a vast number of physical systems (see e.g.~\citep{wiersma1997localization,chabanov2000statistical,runge2003manifestations,genack2005signatures,storzer2006observation,billy2008direct,hu2008localization,kondov2011three,sheinfux2017observation}).

The question of what happens to AL when nonlinearity is introduced has attracted extensive work in theory and simulations~\citep{shepelyansky1993delocalization,shapiro2007expansion,kopidakis2008absence,pikovsky2008destruction,flach2009universal,skokos2009delocalization,garcia2009delocalization,veksler2009spreading,mulansky2010spreading,laptyeva2010crossover,skokos2010spreading,krimer2010statistics,flach2010spreading,johansson2010kam,basko2011weak,bodyfelt2011nonlinear,bodyfelt2011wave,aubry2011kam,ivanchenko2011anderson,molina2012optical,vermersch2012interacting,michaely2012effective,basko2012local,mulansky2012scaling,laptyeva2012subdiffusion,milovanov2012localization,lucioni2013modeling,vermersch2013spectral,mulansky2013energy,skokos2013nonequilibrium,laptyeva2013nonlinearb,tieleman2014chaoticity,ivanchenko2014quantum,antonopoulos2014complex,ermann2014destruction,laptyeva2014nonlinear,basko2014kinetic,flach2015nonlinear,achilleos2016energy,flach2016spreading,martinez2016superdiffusive,achilleos2018chaos,antonopoulos2017analyzing,iomin2017subdiffusion,sato2019anomalous,vakulchyk2019wave,rivas2020seltrapping}, as well as in experiments~\citep{schwartz2007transport,roati2008anderson,lahini2008anderson,lucioni2011observation}.
Two typical disordered Hamiltonian models are at the center of these studies, namely the disordered Klein-Gordon (DKG) chain of nonlinear oscillators and the disordered discrete nonlinear Schr\"odinger equation (DDNLS).
These systems were extensively studied in several of the works mentioned above.
There it was shown that for large enough nonlinearity and disorder strengths, AL is destroyed.
This leads to different dynamical subdiffusive spreading regimes which are characterized by a growing second moment of the wave packet, according to the power law $t^{a_m}$ with $0<a_m < 1$.
In particular, the {\it weak} chaos with $a_m = 1/(1 + 2D)$ ($D$ being the lattice's dimensionality)~\citep{flach2009universal,skokos2009delocalization,flach2010spreading} and {\it strong chaos} dynamical regimes where $a_m = 1/(1 + D)$ can occur~\citep{laptyeva2010crossover,flach2010spreading,bodyfelt2011nonlinear}.

The remarkable success of the theoretical framework explaining the diffusion process comes from the fact that nonlinearity couples normal modes, which interact with each other and allow the transfer of energy~\citep{gredeskul1992propagation,shepelyansky1993delocalization,flach2009universal,flach2010spreading}.
This transfer happens in a chaotic manner whose evolution allows the randomization and thermalization of the wave packet~\citep{flach2010spreading,basko2011weak,skokos2013nonequilibrium,senyange2018characteristics,manda2020chaotic}. 
Nevertheless, whether the system can sustain the underlying spreading for all times remains a controversial issue.
Indeed a possible slowing down of the wave packet spreading is suggested by~\citep{mulansky2011scaling,pikovsky2011scaling} as well as an eventual halt of it by~\citep{johansson2010kam,aubry2011kam}, but the vast majority of the numerical simulations do not support such behavior~\citep{shepelyansky1993delocalization,pikovsky2008destruction,flach2009universal,skokos2009delocalization,laptyeva2010crossover,skokos2010spreading,bodyfelt2011nonlinear,laptyeva2012subdiffusion,skokos2013nonequilibrium,kati2020density}.

A notable distinction between all these possible outcomes is related to the tangent dynamics of the system which supports its chaotic behavior.
Indeed, indicators of chaos often appears to be very sensitive to changes in the system and therefore could act as effective dynamical order parameters in several complex models~\citep{pikovsky2016lyapunov} and in particular in multidimensional Hamiltonian systems (see for e.g.~\citep{barre2001lyapunov,hillebrand2019heterogeneity}).  
Therefore, understanding the characteristics of chaos in this context appears to be a fundamental, open question. 
Several works have been performed trying to fill this void.
The connection between chaos and nonlinear wave spreading is investigated in~\citep{mulansky2011scaling} for small sized systems after their regularization, while an interesting chaotic behavior has been observed in the spreading of wave packets in~\citep{tieleman2014chaoticity}.
In addition,~\citep{michaely2012effective} analyzed the temporal and spectral properties of the effective nonlinear strengths within the wave packet and identified it as noisy.
Furthermore,~\citep{vermersch2013spectral} computed the spectral entropy, which is related to chaotic and ergodic properties of systems and measured the {\it temporal} evolution of chaos, but only for relatively small times.
Much closer to what we have studied in this thesis, the {\it asymptotic temporal} evolution of chaos was studied for the one-dimensional (1D) DKG system in the weak chaos regime by~\citep{skokos2013nonequilibrium}.
There, it was depicted a constantly decaying chaos strength during the wave packet's evolution, and was visualized the fluctuations of localized chaotic hotspots mentioned in previous studies (see e.g.~\citep{basko2011weak,basko2012local,michaely2012effective}) as necessary for the wave packet thermalization.

\section{\label{sec:subject_of_thesis}Subject of the thesis}
One aim of this thesis is to show that the chaotic processes which were observed in~\citep{skokos2013nonequilibrium} for the weak chaos regime in the 1D DKG system, which possesses a single integral of motion, also appear in the 1D DDNLS model which admits an extra integral of motion compared to the 1D DKG system, thus generalizing in some sense the results obtained in that paper.
In addition, we will take that study a step further, investigating also the dynamics in the strong chaos case; as well as providing a `complete' quantification of the system's chaoticity through the calculation of observables in the system's tangent space, such as the maximum Lyapunov exponent (MLE)~\citep{skokos2010lyapunov,pikovsky2016lyapunov} or the deviation vector distribution (DVD) statistics~\citep{dauxois1997modulational,barre2001lyapunov,skokos2013nonequilibrium}.
These two quantities are complementary: While the MLE estimates the mean chaos strength of the entire system, the DVD looks at local properties of chaos giving us information about which lattice sites contribute the most to the overall chaotic dynamics.
A basic novelty of our work is that we do not only focus on the 1D version of the system, but we also expand our study to the two-dimensional (2D) version of the system whose subdiffusive spreading has barely been investigated~\citep{garcia2009delocalization}.  

In order to reach our goals, we perform large scale numerical experiments in both the 1D and 2D DDNLS lattices, as well as semi-analytical calculations in some cases to interpret our findings.
We note that the equations of motion of these DDNLS models are particularly difficult to integrate from a computational point of view.
This is demonstrated by the fact that results for the DKG systems, whose numerical integration is a much easier task, are typically presented for final times which are one or two orders of magnitude larger than the DDNLS systems in both one (see e.g.~\citep{senyange2018characteristics}) and two spatial dimensions (see e.g.~\citep{manda2020chaotic}).
Another important point we want to emphasize is the lack of efficient numerical integration techniques for the evolution of the tangent dynamics of the DDNLS systems in one and two spatial dimensions on top of the difficulty of numerically evolving the equations of motion of these systems.
Therefore, we dedicate an important part of this work to creating novel numerical methods for the fast and accurate long time integration of the equations of motion and the variational equations for both the 1D and 2D models of the DDNLS system using the so-called {\it tangent map method}~\citep{skokos2010numerical,gerlach2011dynamical,gerlach2012efficient}.  
The utilization of the tangent map method takes advantage of symplectic integrators (SIs) which are specifically designed for the numerical evolution of trajectories of Hamiltonian systems.
The implementation of the techniques we develop in this work allow us to reach computational performances comparable to the ones seen in previous studies, where only the integration of the equations of motion was done despite adding the significant computational load of the variational equations.
Consequently this approach helps us to extract robust asymptotic behaviors of the wave packet chaotic dynamics (quantified through the evolution of the MLE for instance).

As we will see later in this thesis, systems exhibiting AL offer a perfect testbed for the analysis of chaotic processes in nonlinear dynamical systems as the destruction of AL is a purely nonlinear phenomenon~\citep{shepelyansky1993delocalization,garcia2009delocalization,flach2009universal}.
It is remarkable that the interplay between disorder and nonlinearity results in the localization of chaos, a behavior which will be discussed in some detail in our work.
We will also explain how the combined action of the non vanishing chaotic strength along with the random fluctuations of localized chaos spots are characteristics of the chaotic dynamics of wave packet spreading in disordered media.
Consequently, as long as they persist the propagation of wave packets never stops, even if the system is getting closer to the integrable limit.
This result is general as it was also observed for the DKG model in one~\citep{skokos2013nonequilibrium,senyange2018characteristics,senyange2020chaotic} and two~\citep{manda2020chaotic,senyange2020chaotic} spatial dimensions.

The thesis is organized as follows.
In Chap.~\ref{chap:spreading}, we discuss the notion of AL, emphasizing its manifestation in the 1D disordered discrete linear Schr\"odinger (DDLS) model, and review the theory of spreading of nonlinear wave packets.
In Chap.~\ref{chapter:num_integration}, we introduce the numerical quantities we calculate in our study and describe in detail the numerical techniques we developed for the integration of the equations of motion and the variational equations of the DDNLS models.
In Chap.~\ref{chap:chaotic_one_dimensional}, we present our numerical results for the chaotic behavior of the 1D DDNLS model, while Chap.~\ref{chap:chaotic_two_dimensional} is dedicated to a similar analysis but for the 2D DDNLS model.
Finally, in Chap.~\ref{chapter:conclusion and future outlook} we conclude our work, summarize our findings and present a few ideas for future investigations.

\chapter{\label{chap:spreading}Localization and spreading of wave packets in disordered lattices}
\pagestyle{fancy}
\fancyhf{}
\fancyhead[OC]{\leftmark}
\fancyhead[EC]{\rightmark}
\cfoot{\thepage}


\section*{\label{sec:intro}Introduction}

In this chapter, we discuss the different regimes of wave packets' dynamics we encounter in random disordered lattices.
The word `disorder' mentioned here could refer to impurities, vacancies, dislocations, distortions or defects which occur in idealistic {\it perfect periodic lattices}.
A periodic lattice is invariant under space translation, which means that interactions are the same at every location within the lattice, and thus constitutes an idealistic representation of physical systems in nature.

The first part of this chapter is dedicated to a rather concise review of the extended work done on the analysis of models obtained through linear approximations of disordered systems.
In this context we show how initially localized wave packets remain in that state forever. 
This exceptionally important result was initially obtained by Anderson in his seminal paper~\citep{anderson1958absence} where he tried to understand the {\it metal insulator transition}.
Nowadays it is well known that this so-called Anderson localization (AL) phenomenon appears in several other fields, making it an important problem to study in modern physics (see e.g.~\citep{izrailev1990simple,lahini2008anderson,lahini2009observation,krimer2009delocalization,krimer2010statistics,savin2010suppression}).
We can depart from linear approximations to obtain more realistic models by adding nonlinear couplings between system's components, through the inclusion of higher order terms from Taylor series expansions of the interaction potential functions.
For example, high intensity photon beams propagating through optical waveguides create a nonlinear response of the medium, which in turn affects the wave propagation~\citep{schwartz2007transport,lahini2008anderson}.
Thus, the second part of this chapter focuses on the effects of nonlinearity on wave localization in nonlinear disordered systems.
In that part, we review what we believe are the most important aspects of this behavior.

The chapter is organized as follows.
In Sec.~\ref{sec:anderson_localization} we present a fundamental linear disorder model in order to define AL.
In Sec.~\ref{sec:destruction_and_loc} we present the nonlinear models we are going to investigate, while in Sec.~\ref{sec:dynamical_regimes_chirikov} we elaborate using (semi-) analytical arguments the different dynamical behaviors of the nonlinear wave packet in disordered lattices.
Section~\ref{sec:statistical_physics_spreading} is similar to Sec.~\ref{sec:dynamical_regimes_chirikov}, but we base our analysis on statistical physics arguments.
In Sec.~\ref{sec:spreading_mechanism} we explain the mechanism of effective wave packet propagation in the different dynamical regimes, while in Sec.~\ref{sec:incommensureate_wave_packet}, we discuss the implications of the existence of initial extended wave packets which turn out to be of practical importance for numerical simulations of two-dimensional systems.
Finally, in Sec.~\ref{sec:conclusion_anderson_localization_spreading} we summarize the content of the chapter. 

\section{\label{sec:anderson_localization}Anderson localization}

The classical theory of electron transport in conductive metals predicts that electron delocalization is due to nonelastic multiple scatterings of the electron, which collides with atomic nuclei which are several times heavier, leading to the well know result of Ohm's law with the Drude conductivity $\sigma_E \propto \tau$. 
Here $\tau$ is the average time elapsed between two collisions and $\sigma_E$ is the conductivity within the metal.
Therefore, $v \tau$ is the electron mean-free-path (with $v$ being the average electron's velocity), which characterizes the mean distance between two successive scatterings of the electron.
An interesting question which arises is what happens if disorder (in the form of a random potential around each nuclei generated from an arbitrary source) is introduced and tuned up higher and higher within the system?
Will the conductivity be reduced due to the increasingly smaller electron mean-free-path? 
Or, will this process result to something more peculiar?

\begin{figure}[!htb]
    \centering
    \includegraphics[width=0.6\textwidth]{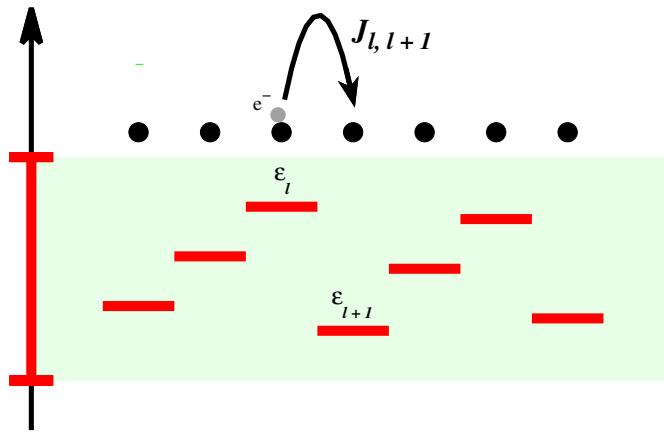}
    \caption{Schematic representation of the electron tight-binding model. 
            The electron [grey dot] is attached to atom's nucleus [black dots] which are considered to be fixed. 
            Each atomic site possesses random on-site electron binding energies [red bars] selected from a uniform distribution on the interval $[-W/2, W/2]$ of measure $W$ [red vertical segment in the left].
            The electron at site $l$, can tunnel either to site $l - 1$ or site $l + 1$ with equal transmission probabilities $\mathcal{J}$ often referred as hopping [black arrow on the top of the electron].
            Thus, the elements of $\mathcal{J}$ of the form $\mathcal{J}_{l, l-1}$ and $\mathcal{J}_{l, l+1}$ are equal and taken to unity (and zero otherwise).
            }
    \label{fig:anderson_model}
\end{figure}

Anderson decided to tackle this problem in the mean-field approximation of the electron tight binding model whose Hamiltonian reads~\citep{anderson1958absence,kramer1993localization}
\begin{equation}
    \mathcal{H}_L= \sum _{l=1}^{N} \epsilon _l \lvert \psi _l \rvert ^2 - \mathcal{J}_{l, l + 1} \left(\psi _{l+1} \psi _l^\star + \psi _{l} \psi _{l + 1}^\star\right).
    \label{eq:hamilton_anderson_model}
\end{equation}
Here $\psi _l \in \mathbb{C}$ is the wave function, which describes the state of the electron at site $l$ and $\psi _l^\star$ is its complex conjugate. 
The off-diagonal matrix elements $\mathcal{J}_{l, m}$, $1\leq l, m \leq N$, with $N$ denoting the size of the lattice, describes the strength of the hopping between sites $l$ and $m$. 
In addition, $\epsilon _l$ is a random parameter with units of the energy/frequency (here the system is rescaled such that $\hbar = 1$).
Its value at each site is drawn from a uniform probability distribution on the interval $[-W/2, W/2]$ with density $\mathcal{P}_{W} = 1/W$. 
Consequently, $W$ is the disorder strength as it defines the size of the interval from which the random values $\epsilon _l$ are drawn. 
In Eq.~\eqref{eq:hamilton_anderson_model} we consider that the electron can only diffuse to its nearest neighbors as also shown in Fig.~\ref{fig:anderson_model}.
In addition, we also assume that the probabilities for the electron to tunnel to the left or right site are equal and we set all the non-zero elements of the matrix $\mathcal{J}$ to $\mathcal{J}_{l, m} = 1$.

In Eq.~\eqref{eq:hamilton_anderson_model} the $N$ pairs $\left(\psi_l, i\psi _l ^\star\right)$ represent a set of canonical generalized coordinates of the system's phase space.
The equations of motion of the Hamilton function~\eqref{eq:hamilton_anderson_model} are then obtained through 
\begin{equation}
    \frac{d\psi _l}{dt} = \frac{\partial \mathcal{H}_L}{\partial \left(i \psi _l ^\star\right)},
    \label{eq:hamilton_eq_anderson_model}
\end{equation}
leading to the DDLS equation  
\begin{equation}
    i \dot{\psi} _l = \epsilon _l \psi _l - \left( \psi _{l - 1} + \psi _{l + 1} \right),
    \label{eq:eq_mot_anderson}
\end{equation}
where the dot refers to the derivative with respect to time $t$.

We note that spatial localization means that the probability $\lvert \psi _l \rvert^2$ of the presence of the electron at site $l$ exponentially decays, so that the electron does not tunnel out of the lattice~\citep{kramer1993localization,krimer2010statistics}.
Over the years several methods have been developed in order to analytically and numerically observe such localization~\citep{kramer1993localization}.
For instance, a crude approach consists of putting a lump of matter (preferably located at a single site) into the system and let it evolve for a very long time.
If localization occurs, then at the end of the evolution we should expect a spreading of the matter to few subsequent lattice sites and a halting of this process.
In this case, the number of sites on which the amount of matter is not exponentially small is a characteristic length of the system that we refer as {\it localization length} or {\it localization volume}.
We shall emphasize the difference between these two appellations in the later paragraphs.

The normal modes (NMs) of a linear system represent an orthonormal basis of the system's phase space. 
Therefore, any system's wave packet can be written as a weighted sum over these modes and the wave packet localization implies that the modes are also localized. 
Following this observation, one can therefore focus in analyzing the characteristic of the NMs of the DDLS system~\eqref{eq:eq_mot_anderson}. 

\begin{figure}[!h]
    \centering 
    \includegraphics[width=0.475\textwidth, height=0.4\linewidth]{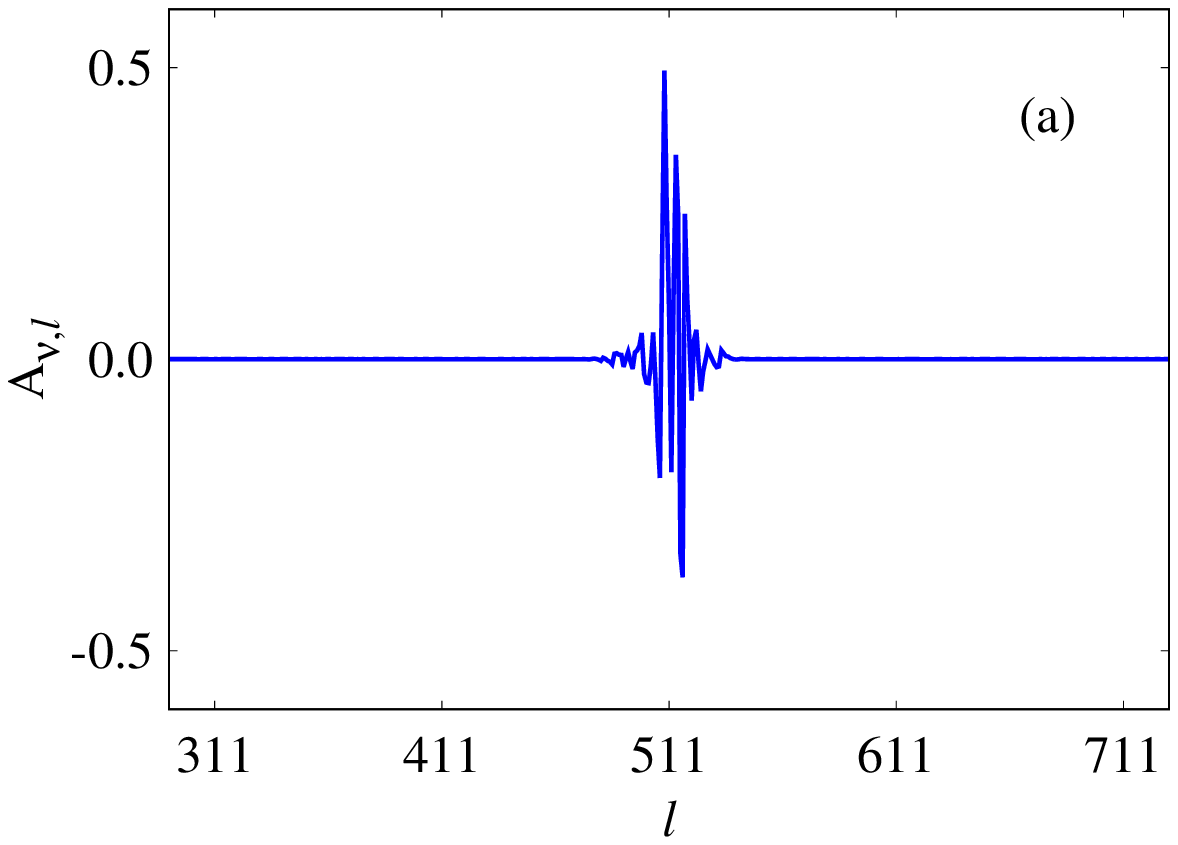}
    \includegraphics[width=0.475\textwidth, height=0.4\linewidth]{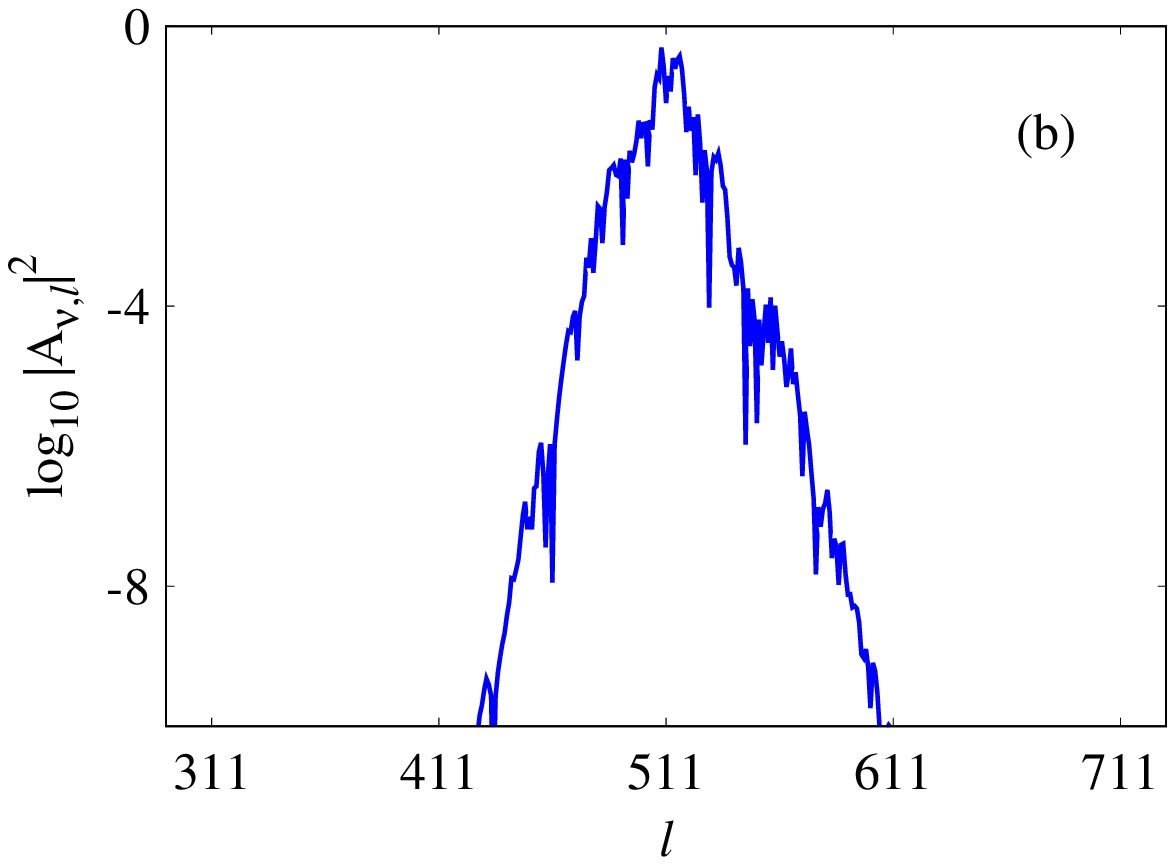}
    \caption{A representative NM of Hamiltonian~\eqref{eq:hamilton_anderson_model} with $N = 1021$ sites located around the center of the lattice (mode number $\nu = 357$). 
    Amplitudes (a) $A_{\nu, l}$ and (b) $\lvert A_{\nu, l}\rvert^2$.
    Panel (b) is in semi-logarithmic scale.
    The system disordered strength is $W=4$.
    }
    \label{fig:normal_modes_localization_anderson_model}
\end{figure}

Let us outline the steps of such procedure.
The solution of linear wave equations consists of plane waves of the form
\begin{equation}
\psi _l = A_l e^{-i E t},
\label{eq:stationary_states_anderson}
\end{equation}
where $A_l$ is the amplitude of the wave at site $l$ and $E$ its frequency.
Substituting Eq.~\eqref{eq:stationary_states_anderson} into Eq.~\eqref{eq:eq_mot_anderson} we obtain the equivalent eigenvalue problem
\begin{equation}
E A_l = \epsilon _l A_l - A_{l - 1} - A_{l + 1},
\label{eq:eigenvalue-vectors_eq}
\end{equation}
which reduces in a matrix form to 
\begin{equation}
    E \bm{I} \bm{\Psi} = \bm{\tilde{A}} \bm{\Psi},
    \label{eq:eigenvalue-vector_eq_1}
\end{equation}
where $\bm{\Psi} = \left(A_1, A_2, \ldots, A_N\right)$, 
\begin{equation}
    \bm{\tilde{A}} = 
    \begin{pmatrix}
        \epsilon _ 1 & -1 & 0 & \ldots & 0 & 0 \\
        -1 & \epsilon _2 & -1 & \ldots & 0 & 0 \\
        \vdots & \vdots & \vdots & \ddots & \vdots & \vdots \\
        0 & 0 & 0 & \ldots & \epsilon _{N - 1} & -1  \\
        0 & 0 & 0 & \ldots & -1 & \epsilon _{N} \\
    \end{pmatrix} 
    \mbox{ and }
    \bm{I} = 
    \begin{pmatrix}
        1 & 0 & 0 & \ldots & 0 & 0 \\
        0 & 1 & 0 & \ldots & 0 & 0 \\
        \vdots & \vdots & \vdots & \ddots & \vdots & \vdots \\
        0 & 0 & 0 & \ldots & 1 & 0  \\
        0 & 0 & 0 & \ldots & 0 & 1 \\
    \end{pmatrix}.
    \label{eq:eigenvalue-vector_eq_2}
\end{equation}
Solving the eigenvalue problem~\eqref{eq:eigenvalue-vector_eq_1} is equivalent to diagonalizing the matrix $\bm{\tilde{A}}$. 
We then obtain the normalized eigenstates $\bm{\Psi}_{\nu} = \left(A_{\nu, 1}, A_{\nu, 2}, \ldots, A_{\nu, N}\right)$ with $\lVert \bm{\Psi}_\nu\rVert^2 = \sum _{l} A_{\nu, l}^2 = 1$ associated with the eigenfrequencies $E_\nu$\footnote{We note that we set $\hbar = 1$ so that the energy $E_\nu$ is numerically identical to the frequency $\nu$, as they are related through the equation $E_\nu = \hbar \nu$.}. 
An example of a computed NM $\bm{\Psi}_\nu$ through this process of numerical diagonalization of $\bm{\tilde{A}}$ (see Eqs.~\eqref{eq:eigenvalue-vector_eq_1} and~\eqref{eq:eigenvalue-vector_eq_2}) is shown in Fig.~\ref{fig:normal_modes_localization_anderson_model}.
The parameters of system~\eqref{eq:hamilton_anderson_model} are set to $W = 4$ and $N = 1021$.
The computed NMs are sorted by increasing order of their centers' spatial position which is computed as $\overline{l}_\nu = \sum_l l \lvert A_{\nu, l} \rvert^2$.
In Fig.~\ref{fig:normal_modes_localization_anderson_model}, we plot the profile of the mode whose center is closest to the lattice's middle (the mode's number in the ordering in $\nu = 357$), associated with the eigenfrequency $E_{\nu} \approx -0.916$.
The spatial exponential localization of the mode is evident from the results of Fig.~\ref{fig:normal_modes_localization_anderson_model}(b).

The NMs of a system exhibiting AL are characterized by an asymptotic exponential spatial decay from their center (here referred as $\overline{l}_\nu$)~\citep{kramer1993localization} of the form 
\begin{equation}
    \bm{\Psi}_{\nu} \propto \zeta_\nu ^{-1/2} e^{-(l - \overline{l}_\nu)/\zeta _\nu},
    \label{eq:asymtotic_behavior_normal_mode}
\end{equation}
where $\zeta _\nu$ is the so-called localization length of mode $\nu$.
The $\zeta_\nu$ values have to be finite and bounded from above for all the frequencies $\nu$ \footnote{In contrary, an extended mode exhibits an unbounded localization length  i.e. $\zeta _\nu \rightarrow \infty$. 
In this situation, AL cannot survive as the energy will be carried by the extended mode through the whole system~\eqref{eq:hamilton_anderson_model}.}.

In Eq.~\eqref{eq:asymtotic_behavior_normal_mode} the key quantity is the localization length $\zeta _\nu$ as it is directly related to the mode's profile.
In~\citep{kramer1993localization}, several numerical and analytical methods were presented for the estimation of $\zeta_\nu$ (e.g. random matrix theory, Green's functions, transfer matrix methods {\it etc.}). 
Using these methods, the authors showed that in the limit of weak disorder the localization length can be expressed as
\begin{equation}
    \zeta (E_\nu) = \zeta _\nu = \frac{24\left(4 - E_\nu^2\right)}{W^2}.
    \label{eq:loc_vol_1d_anderson_small_w}
\end{equation}
From Eq.~\eqref{eq:loc_vol_1d_anderson_small_w}, we see that the localization length $\zeta _\nu$ takes its largest value at the center of the spectrum, i.e.~when $E_\nu = 0$, becoming 
\begin{equation}
    \zeta (0) = \zeta _0 = 96/W^2.    
\end{equation}
On the other hand, in the case of strong disorder, the largest value of the localization length becomes~\citep{laptyeva2013nonlinear}
\begin{equation}
    \zeta _0 = \frac{1}{\ln \left( \frac{W}{2} \right)}.
    \label{eq:loc_vol_1d_anderson_big_w}
\end{equation}

In order to numerically probe the validity of Eqs.~\eqref{eq:loc_vol_1d_anderson_small_w} and~\eqref{eq:loc_vol_1d_anderson_big_w}, we compute the statistical properties of NMs.
The spatial extent of the NMs of the 1D DDLS model~\eqref{eq:hamilton_anderson_model} has been studied through extensive numerical computations in~\citep{krimer2010statistics}.
In that paper, the spatial volume occupied by a mode $\nu$ is considered as the number of lattice sites where the amplitude $A_{\nu, l}$ is not exponentially small and quantified by a quantity which was named localization volume $V_\nu$~\citep{krimer2010statistics,laptyeva2012subdiffusion}.
This quantify was computed as~\citep{krimer2010statistics}  
\begin{equation}
    V_\nu = \sqrt{12m_2^{(\nu)} } + 1, 
    \label{eq:loc_vol_m2}
\end{equation}
where $m_2 ^{(\nu)} = \sum _{l} (l - \overline{l}_\nu)^2 \lvert A_{\nu, l} \rvert ^2$ is the mode's second moment with $\overline{l}_\nu = \sum _{l} l \lvert A_{\nu, l} \rvert ^2$ being the mode's center.
In addition, one could also consider as an estimation of the mode's localization volume the number of sites which contribute the most to the NM spatial distribution, something which can be quantified by the mode's participation number~\citep{krimer2010statistics}
\begin{equation}
    P_{\nu} = \frac{1}{\sum _{l} \lvert A _{\nu, l}\rvert ^4}.
    \label{eq:loc_vol_p}
\end{equation}
Thus, averaging these quantities for the NMs belonging to the spectra of several disorder realizations, we compute the average localization volume $\overline{V}$ of the NMs in the 1D DDLS system~\eqref{eq:hamilton_anderson_model}.
It is worth mentioning that the numerical analysis often focusses on the central part of the frequency spectrum where $\zeta \propto \zeta _0$.
Consequently, for that frequency region, there exists a direct proportionality between the localization volume $\overline{V}$ and localization length $\zeta _0$~\citep{krimer2010statistics,senyange2020properties} such that 
\begin{equation}
    \overline{V} \approx \frac{a_\zeta}{W^2},
    \label{eq:localization_volume_length_chap_anderson}
\end{equation}
where $a_\zeta$ is a real constant.
The value of $a_\zeta$ of Eq.~\eqref{eq:localization_volume_length_chap_anderson} oscillates within the interval $[330, 360]$ (see e.g.~\citep{krimer2010statistics,flach2010spreading,flach2016spreading,kati2020density}).
\begin{figure}[!htb]
    \centering 
    \includegraphics[width=0.45\textwidth, height=0.4\linewidth, trim={1.15cm, 0.6cm, 0cm, 0cm}, clip]{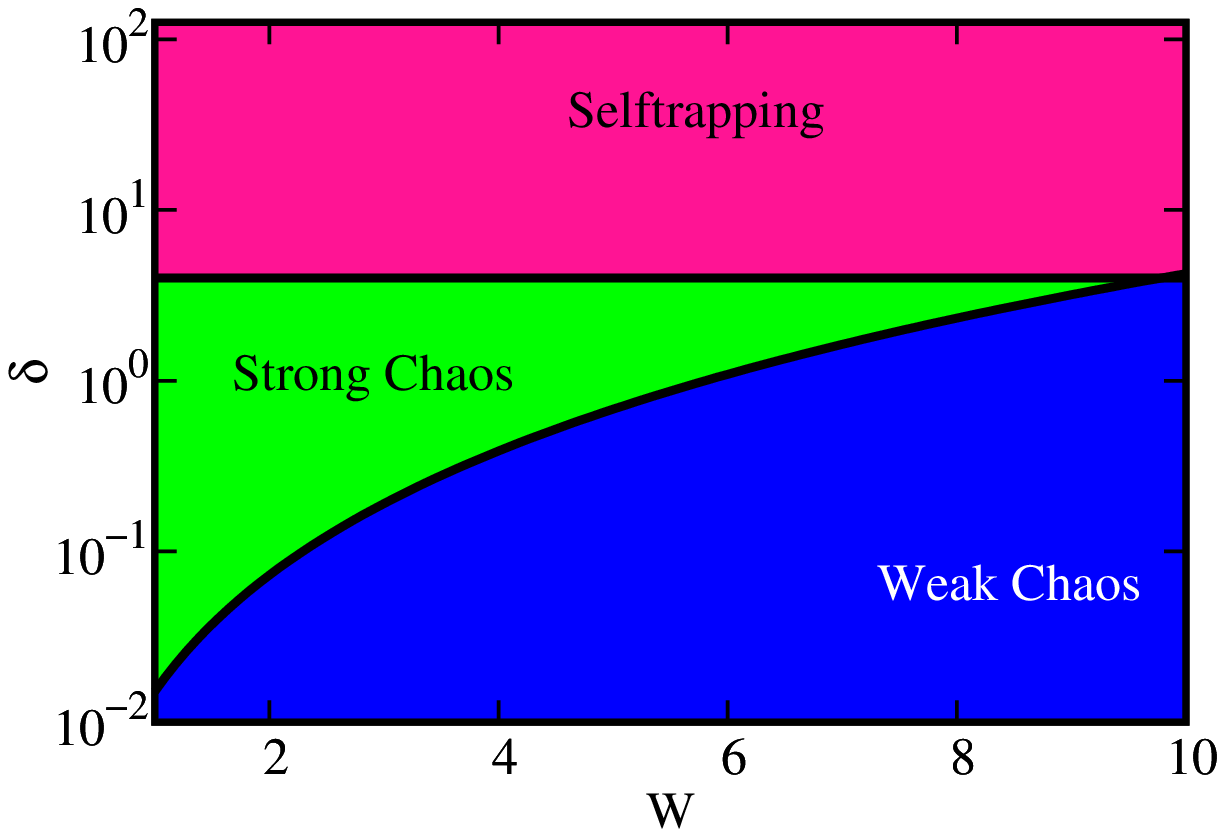}
    \put(-215, 100){\makebox(0, 0){\rotatebox[origin=c]{90}{$\overline{\zeta},~\overline{V},~\overline{P}$}}}
    \put(-100, -5){\makebox(0, 0){$W$}}
    \caption{Numerical estimation of the mean localization length $\overline{\zeta}$ through the transfer matrix approach [(b)lue], the mean localization volume $\overline{V}$ (see Eq.~\eqref{eq:loc_vol_m2}) [(r)red] and the mean participation number $\overline{P}$ (see Eq.~\eqref{eq:loc_vol_p}) [(g)reen] of the NMs with eigenfrequencies $E_\nu$ near the center of the frequency band of system~\eqref{eq:hamilton_eq_anderson_model} as a function of the disorder strength $W$.
    For moderate $W$, the values of $\overline{\zeta}$ and $\overline{V}$ are fitted by an expression of the form $a_\zeta/W^2$, depicted by dashed lines.
    In particular, the lower dashed line close to the blue curve corresponds to the function $100/W^2$, while the upper dashed line close to the red curve to $330/W^2$ (from~\citep{krimer2010statistics}).
    }
    \label{fig:V_vs_W_KF10}
\end{figure}

In Fig.~\ref{fig:V_vs_W_KF10}, we present results obtained in~\citep{krimer2010statistics} of the comparison between calculations of the average localization length $\overline{\zeta}$ using the transfer matrix approach (see (b)lue curve in Fig.~\ref{fig:V_vs_W_KF10}), the averaged localization volume $\overline{V}$ based on Eq.~\eqref{eq:loc_vol_m2} (see (r)ed curve in Fig.~\ref{fig:V_vs_W_KF10}) and of the average participation number $\overline{P}$ based on Eq.~\eqref{eq:loc_vol_p} (see (g)reen curve in Fig.~\ref{fig:V_vs_W_KF10}).
All these results are presented as function of the disorder strength $W$.
For small $W$ values, the proportionality between these quantities is evident as all of them exhibit the same trend, being $\propto W^{-2}$ in agreement with Eq.~\eqref{eq:loc_vol_1d_anderson_small_w}.
Although Eq.~\eqref{eq:loc_vol_1d_anderson_small_w} was derived for small $W$, the numerical calculations suggest that this relation is valid for $W \lesssim  4$~\citep{krimer2010statistics}.
In the limit of strong disorder $W \rightarrow \infty$, Eq.~\eqref{eq:loc_vol_1d_anderson_big_w} suggests that the localization length $\zeta_\nu \rightarrow 0$, while the definition~\eqref{eq:loc_vol_m2} implies the localization volume $V_\nu \rightarrow 1$ due to the discreteness of the model.
These tendencies are well captured by the results of Fig.~\ref{fig:V_vs_W_KF10}.

According to~\citep{kramer1993localization,krimer2010statistics,laptyeva2012subdiffusion,laptyeva2013nonlinear}, the width of the eigenvalue frequency spectrum $E_{\nu}$ of~\eqref{eq:eigenvalue-vectors_eq} is 
\begin{equation}
    \Delta = W + 4, \quad \text{with} \quad E_{\nu} \in \left[-2 - \frac{W}{2}, 2 + \frac{W}{2} \right].
    \label{eq:characteristics_eigfreq_anderson_model}
\end{equation}
Interestingly enough, this can be analytically deduced using simple algebraic manipulations (see Appendix~\ref{app:conv_linear_ddnls_dkg}) based on~\citep{kopidakis2008absence}.
Because we will use this quantity quite often through the rest of this thesis, it is worth defining what is called the {\it averaged spacing between neighboring NMs} $d$~\citep{shepelyansky1993delocalization,flach2009universal,flach2010spreading} as
\begin{equation}
    d \propto \frac{\Delta}{\Delta l},
    \label{eq:average_spacing_between_freq_anderson_model}
\end{equation}
where $\Delta l$ is the wave packet's width i.e.~the number of sites which have a non-negligible amount of norm/energy.

At this stage, we have characterized the dynamical features of the DDLS model~\eqref{eq:hamilton_anderson_model} based on three quantities: (i) the mean localization volume $\overline{V}$~\eqref{eq:loc_vol_m2}, (ii) the eigenfrequency bandwidth $\Delta$~\eqref{eq:characteristics_eigfreq_anderson_model} and (iii) the average spacing between NMs $d$~\eqref{eq:average_spacing_between_freq_anderson_model}.
It is worth mentioning that similar studies of the linear NMs have been performed for the disordered Klein-Gordon (DKG) chain of harmonic oscillators~\citep{senyange2020properties} whose nonlinear version is also mentioned throughout this thesis.

\section{\label{sec:destruction_and_loc}Adding nonlinearity}

We now introduce nonlinear interactions into the DDLS model~\eqref{eq:hamilton_anderson_model} and study the fate of the localized wave packet in this context.
As we are going to witness, the nonlinear interactions couple the system's NMs which in principle could exchange norm/energy.
This immediately blurs the fate of the wave packet.
Will the wave packet spread? 
If yes, is the spreading limited in time or will it last for ever?
These are example of questions that arise from this problem and that we intend to tackle in this section.
We describe an appropriate parameter space on which points represent wave packets with specific geometry and density characteristics and use these spaces to develop the theoretical framework governing the evolution of any initial wave packet.
In this way, we are able to establish boundaries which delimit sets of parameters leading to similar wave packet dynamics, as well as provide the expected wave packet statistical characteristics within each domain. 

\subsection{\label{subsubsec:ddnls_lattices}The disordered discrete nonlinear Schr\"{o}dinger equation}
We start by presenting our principal model of investigation, the 1D DDNLS model, which can be viewed as the DDLS lattice model~\eqref{eq:hamilton_anderson_model} into which we introduce an onsite nonlinear term, leading to the Hamiltonian function~\citep{molina1998transport,pikovsky2008destruction,kopidakis2008absence,flach2009universal,veksler2009spreading,iomin2010subdiffusion,mulansky2010spreading,iomin2017subdiffusion}
\begin{equation}
    \mathcal{H}_{1D} = \sum _{l = 1}^{N} \epsilon _l \lvert \psi _l \rvert ^2 + \frac{\beta }{2}\lvert \psi _l \rvert ^{4} - \left(\psi_{l+1} \psi _l ^\star + \psi_{l+1}^\star \psi _l\right),
    \label{eq:Hamilton_complex_dnls_1d}
\end{equation}
where again $\psi _l$ is a complex valued function representing the state of the wave packet at site $l$ and $\beta$ defines the nonlinearity strength.
As in Eq.~\eqref{eq:hamilton_anderson_model}, the onsite parameters $\epsilon _l$, $1\leq l \leq N$ are drawn from a uniform distribution $\mathcal{P}_W = 1/W$ in the interval $\left[-W/2, W/2\right]$.
Here $W$ is the disorder strength and $N$ is the total size of the lattice.
The equations of motion generated from Eq.~\eqref{eq:hamilton_eq_anderson_model} are
\begin{equation}
    i\dot{\psi} _l = \epsilon _l \psi _l + \beta \lvert\psi _{l}\rvert^2 \psi _l - \left(\psi_{l - 1} + \psi _{l+1} \right),
    \label{eq:eq_mot_1d_ddnls}
\end{equation}
and conserve both the total energy $\mathcal{H}_{1D}$~\eqref{eq:Hamilton_complex_dnls_1d}, as well as the total norm of the system
\begin{equation}
    \mathcal{S}_{1D} = \sum _{l=1}^{N} \lvert \psi _l \lvert^2.
    \label{eq:norm_complex_dnls_1d}
\end{equation}

It is straightforward to extend Eq.~\eqref{eq:Hamilton_complex_dnls_1d} to a 2D lattice.
In that case the Hamiltonian~\citep{garcia2009delocalization,laptyeva2012subdiffusion}
\begin{equation}
    \mathcal{H}_{2D} = \sum _{l=1, m=1}^{N, M} \epsilon _{l, m} \lvert \psi _{l, m} \rvert ^2 + \frac{\beta}{2}\lvert \psi _{l, m} \rvert ^{4} - \left(\psi_{l+1, m} \psi _{l, m} ^\star + \psi _{l,m+1}\psi _{l, m}^\star + \psi_{l+1, m}^\star \psi _{l, m} + \psi _{l,m+1}^\star\psi _{l, m} \right),
    \label{eq:hamilton_complex_ddnls_2d}
\end{equation}
represents the total energy of the system, where $\psi _{l, m} \in \mathbb{C}$ is the wave function at site $(l, m)$, $\beta$ is the nonlinear coefficient and $N\times M$ being the lattice size.
Again, the $\epsilon_{l, m}$, $1\leq l \leq N$, $1\leq m\leq M$, are onsite parameters randomly drawn from a uniform distribution  $\mathcal{P}_W = 1/W$ in the interval $\left[-W/2, W/2\right]$, with $W$ being the disorder strength.
The equations of motion obtained via 
\begin{equation}
    \dot{\psi} _{l, m} = \frac{\partial \mathcal{H}_{2D}}{\partial (i\psi _{l, m} ^\star)},
    \label{eq:ham_eq_motion_2d}    
\end{equation}
are 
\begin{equation}
    i\dot{\psi} _{l, m} = \epsilon _{l, m} \psi _{l, m} + \beta\lvert\psi _{l, m}\rvert^{2} \psi _{l, m} - \left(\psi_{l - 1, m} + \psi _{l+1, m} + \psi_{l, m - 1} + \psi_{l, m+1} \right),
    \label{eq:eq_mot_2d_ddnls}
\end{equation}
and conserve in addition to the total energy of the system~\eqref{eq:hamilton_complex_ddnls_2d} its total norm 
\begin{equation}
    \mathcal{S}_{2D} = \sum _{l=1, m=1}^{N, M} \lvert \psi _{l, m}\rvert ^2.
    \label{eq:norm_complex_dnls_2d}
\end{equation}

These models (Eqs.~\eqref{eq:Hamilton_complex_dnls_1d} and~\eqref{eq:hamilton_complex_ddnls_2d}) are for instance used to describe the conductivity of materials (see e.g.~\citep{sales2018sub}), the behavior of DNA molecules (see e.g.~\citep{peyrard2004nonlinear,dang2012nonlinear}), two-body interactions in ultracold atomic gases and the propagation of light through networks of coupled optical waveguides (see e.g.~\citep{rasmussen2000statistical,christodoulides2003discretizing,kivshar2003optical,johansson2004statistical,morsch2006dynamics,mithun2018weakly}) to name a few.

\subsection{\label{subsubsec:kg_lattices}The disordered Klein-Gordon chain of anharmonic oscillators}
We will also sporadically discuss the 1D and 2D DKG models.
The Hamiltonian function of the 1D version of this model~\citep{flach2009universal,laptyeva2010crossover,skokos2010spreading} reads
\begin{equation}
\mathcal{H}_{1K} = \sum _{l=1}^{N} \frac{p_l^2}{2} + \frac{\tilde{\epsilon} _l }{2} q_l ^2 + \frac{1}{4} q_l ^{4} + \frac{1}{2W} \left(q_{l+1} - q_{l} \right)^2,
\label{eq:hamilton_1d_dkg}
\end{equation}
where $\tilde{\epsilon}_l$ are random uncorrelated parameters drawn from a uniform distribution $\mathcal{P} = 1$ in the interval $\left[1/2, 3/2\right]$.
In addition, $q_{l}$ and $p_{l}$, $1\leq l\leq N$ are respectively the generalized displacements and momenta and $N$ is the total number of lattice sites.
The choice of the hopping coefficient $1/2W$ into Hamiltonian~\eqref{eq:hamilton_1d_dkg} is made such that the NMs of the 1D DDLS system~\eqref{eq:hamilton_anderson_model} and the 1D DKG model of harmonic oscillators ($q_l^4\rightarrow 0$ in $\mathcal{H}_{1K}$~\eqref{eq:hamilton_1d_dkg}) are the same (see Sec.~\ref{sec:linear_kg_model_app} in Appendix~\ref{app:conv_linear_ddnls_dkg} for further details).
Thus $W$ is considered as the strength of the disorder as it tunes the NMs localization volume.
The equations of motion are derived from
\begin{equation}
    \ddot{q}_l = - \frac{\partial \mathcal{H}_{1K}}{\partial q_l},    
\end{equation}
leading to  
\begin{equation}
    \ddot{q}_l = - \tilde{\epsilon }_l q_l + q_l ^{3} + \frac{1}{W} \left( q_{l-1} - 2q_l + q_{l+1}\right).
    \label{eq:eq_mot_1d_dkg}
\end{equation}
The set of equations \eqref{eq:eq_mot_1d_dkg} conserve the value of the Hamiltonian function~\eqref{eq:hamilton_1d_dkg}. 
Note that changing the energy value $\mathcal{H}_{1K}$ is equivalent to tuning the nonlinear term $q_l^{4}/4$, therefore the value of $\mathcal{H}_{1K}$ is used as a control parameter of the system.

The Hamiltonian function of the 1D DKG system can be extended to 2D lattices in the same fashion as done for the DDNLS model~\citep{laptyeva2012subdiffusion,senyange2018computational,manda2020chaotic}, leading to the Hamiltonian
\begin{equation}
    \mathcal{H}_{2K} = \sum _{l=1, m= 1}^{N, M} \frac{p_{l, m}^2}{2} + \frac{\tilde{\epsilon }_{l,m} }{2} q_{l,m} ^2 + \frac{1}{4} q_{l,m} ^{4} + \frac{1}{2W} \left[ \left(q_{l+1, m} - q_{l, m}\right)^2 + \left(q_{l, m+1} - q_{l, m}\right)^2 \right].
    \label{eq:hamilton_2d_dkg}
\end{equation}
In Eq.~\eqref{eq:hamilton_2d_dkg}, $\tilde{\epsilon}_{l, m}$ are random uncorrelated parameters drawn from a uniform distribution $\mathcal{P} = 1$ in the interval $\left[1/2, 3/2\right]$.
Again, $q_{l, m}$ and $p_{l,m}$ are the generalized displacement and momentum at site $(l, m)$. 
As previously, $W$ denotes the disorder strength and $N\times M$ is the total lattice size.
The equations of motion are obtained from
\begin{equation}
    \ddot{q}_{l, m} = - \frac{\partial \mathcal{H}_{2K}}{\partial q_{l, m}},
\end{equation}
and have the form 
\begin{equation}
    \ddot{q}_{l, m} = - \tilde{\epsilon }_{l, m} q_{l, m} + q_{l, m} ^3 + \frac{1}{W} \left( q_{l-1, m} + q_{l, m-1} - 4q_{l, m} + q_{l, m+1} + q_{l+1, m}\right).
    \label{eq:eq_mot_2d_dkg}
\end{equation}
As before, these equations conserve the value of the Hamiltonian $\mathcal{H}_{2K}$~\eqref{eq:hamilton_2d_dkg}, which can be used as tuning parameter of the nonlinearity in the system.

These models (see Eqs.~\eqref{eq:hamilton_1d_dkg} and~\eqref{eq:hamilton_2d_dkg}) are relevant for several physical processes like anharmonic vibrations in crystal lattices~\citep{ovchinnikov2012vibrational}.

\section{\label{sec:dynamical_regimes_chirikov}Expected dynamical regimes}
To understand the impact of the nonlinearity on the time evolution of the wave packet in the 1D DDNLS lattice model, we express the nonlinear wave function $\psi _l$ in the basis of the NMs $\bm{\Psi}_\nu = (A_{\nu, 1}, A_{\nu, 2}, \ldots, A_{\nu, N})$ with $A_{\nu, l} \propto \zeta^{-1/2} \exp \left[-\left(l - \overline{l}_\nu\right)/\zeta\right]$ of the 1D DDLS model~\eqref{eq:hamilton_anderson_model}, i.e.
\begin{equation}
\psi _l =  \sum _{\nu} \phi _\nu A_{\nu, l}, 
\end{equation}
where $\phi _\nu$ are time dependent complex amplitudes which define the contribution of mode $\nu$ to the state of site $l$.
The equations of motion~\eqref{eq:eq_mot_1d_ddnls} can be written in the NMs space~\citep{pikovsky2008destruction,kopidakis2008absence,milovanov2012localization} (see also Appendix~\ref{app:eq_mot_normal_mode_space}) as
\begin{equation}
i \frac{\partial \phi _\nu}{\partial t} = \lambda _\nu \phi _\nu + \beta \sum _{\nu _1, \nu_2, \nu_3} I_{\nu, \nu _1, \nu_2, \nu_3} \phi _{\nu _1}^\star \phi _{\nu _2} \phi _{\nu _3},
\label{eq:1d_DDNLS_eq_motion_normal_modes_space}
\end{equation}
with
\begin{equation}
I_{\nu, \nu _1, \nu_2, \nu_3} = \sum _{l} A_{\nu, l}^{\star}A_{\nu _1, l}^\star A_{\nu _2, l}A_{\nu _3, l},
\label{eq:overlap_integral_ddnls}
\end{equation}
being the so-called the {\it overlap-integral}.
From Eq.~\eqref{eq:1d_DDNLS_eq_motion_normal_modes_space}, we see that the presence of the nonlinearity when $\beta >0$ introduces complex interactions between NMs in the equations of motion through the term $\beta\sum _{\nu _1, \nu_2, \nu_3}I_{\nu, \nu_1, \nu_2, \nu_3} \phi_{\nu_1}^{\star}\phi_{\nu_2}\phi_{\nu_3}$.
The strength of the latter term depends on three parameters: (i) the norm of mode $\nu$, $s_\nu = \lvert \phi _\nu \rvert^2$, (ii) the nonlinear coefficient $\beta$ and (iii) the value of the overlap integral $I_{\nu, \nu_1, \nu_2, \nu_3}$. 
While the value of $s_\nu$ (and $\beta$) can be easily evaluated, the determination of the magnitude of $I_{\nu, \nu_1, \nu_2, \nu_3}$ is more complicated as we do not really have exact analytical forms for the NM expressions~\eqref{eq:asymtotic_behavior_normal_mode}.
Indeed, even if Eq.~\eqref{eq:asymtotic_behavior_normal_mode} predicts in a statistical manner the tail shape of most of the NMs, multi-peaked NMs having significant contributions in larger parts of the lattice do exist~\citep{krimer2010statistics}.
Nevertheless, we can develop some intuition about the behavior of $I_{\nu, \nu _1, \nu_2, \nu_3}$ following the approach of~\citep{shepelyansky1993delocalization,garcia2009delocalization}.
For instance, the fact that a NM $\bm{\Psi}_{\nu}$~\eqref{eq:asymtotic_behavior_normal_mode} spans a finite amount of lattice sites, suggests that the integral $I_{\nu, \nu_1, \nu_2, \nu_3}$ is bounded by a finite value and so we expect each mode to overlap/interact with a finite number of neighboring modes (such that a lattice site contributes to the energy of a finite number of NMs).
Consequently, the introduction of the nonlinearity induces a nonlinear frequency shift 
\begin{equation}
    \delta _\nu \propto \beta \lvert \phi_\nu \rvert^2 = \beta s_\nu,
    \label{eq:nonlinear_freq_shift}
\end{equation}
on the frequency of each mode $\nu$.

Depending on the strength of the frequency shift $\delta _\nu$, three different behaviors are expected: (I) if the nonlinear frequency shift $\delta _\nu$ is small enough that most of the modes of the system do not change much and still do not behave chaotically, then AL is preserved at least for a very long time interval.
This case corresponds to the so-called Kolmogorov-Arnol'd-Moser (KAM) regime~\citep{johansson2010kam,aubry2011kam,lichtenberg2013regular} where the system's orbits in the phase space remain on KAM tori and exhibit quasi-periodic behavior.
(II) On the other hand, if $\delta _\nu$ is large enough, the modes behave chaotically allowing energy/norm exchanges between themselves.
In this case, a clear expansion of the wave packet is noticed, which in turn will lead to a decreasing nonlinear frequency shift in time since the amplitude at each active mode is diminishing~\citep{shepelyansky1993delocalization,pikovsky2008destruction, garcia2009delocalization, flach2010spreading}. 
In this case, two potential scenarios are possible: (IIA) the wave packet spreads and after some time the modes' interactions become weak enough to be neglected and consequently, the system's trajectories converge to KAM-like tori and AL is restored or (IIB) the system remains chaotic for all times with a decreasing chaos strength (which is proportional to $\delta_\nu$), which nevertheless is enough to maintain the spreading of the wave packet for all times.
This means that unlike case (I) and scenario (IIA), in scenario (IIB) the system's trajectories do not converge to quasi-periodic orbits~\citep{skokos2013nonequilibrium}.

We note that this is the first time that the notion of chaos is mentioned in our work.
Our approach is not to give a strict definition of it (the interested reader could for example see~\citep{goldstein2002classical,skokos2010lyapunov,pikovsky2016lyapunov} for more information on the subject) but to loosely relate chaos with the disappearance of quasi-periodic motion (which is often referred to as regular motion).
A very useful criterion has been developed to determine the condition under which a system becomes chaotic: the so-called {\it Chirikov criterion}~\citep{chirikov1971research,chirikov1979universal} which has been already successfully implemented in many physical phenomena.
For example, determining the condition of appearance of the Fermi-Pasta-Ulam-Tsingou (FPUT)--recurrence (see~\citep{berman2005fermi,campbell2005introduction} and references therein), the dynamics of breathers~\citep{flach2008fermi,flach2008periodic} and the onset of chaotic behavior in quantum systems~\citep{shepelyansky1993delocalization,shepelyansky1995quantum}.
It is worth mentioning that a review of the various contributions of Chirikov in the theory of chaos can be found in~\citep{bellissard1998boris}. 
The theoretical attempts to describe the dynamical behavior of nonlinear disordered systems~\citep{shepelyansky1993delocalization,pikovsky2008destruction,garcia2009delocalization} were based on the application of the Chirikov criterion.
We outline them here: a disordered nonlinear system shows a {\it deterministic} chaotic behavior when its perturbation parameter $K$ satisfies  
\begin{equation}
K  = \mathcal{K}^2 \propto \left(\frac{\delta _{\nu}}{d_\nu}\right)^2 > 1.
\label{eq:chirikov_criterion}
\end{equation}
Note that $\mathcal{K}$ is often called the {\it resonance-overlap parameter}.
In Eq.~\eqref{eq:chirikov_criterion}, $\delta _\nu$ is the nonlinear frequency shift~\eqref{eq:nonlinear_freq_shift} and 
\begin{equation}
    d_\nu = \Delta/\Delta \nu,
\end{equation}
is the average frequency spacing between NMs, with $\Delta$ being the size of the linear frequency spectrum and $\Delta \nu$ denoting the wave packet extent in terms of the total number of activated NMs.
This approach allows the determination of the regions of the system's parameter space where similar dynamical behaviors are seen. 
Based on the value of $\mathcal{K}$~\eqref{eq:chirikov_criterion} we can see the following cases:
(a) $\mathcal{K} < 1$.
In this case, $\delta  _\nu < d_\nu$ and the nonlinear frequency shift is smaller than the average spacing between the frequencies of the NMs $d_\nu$, i.e. no initial resonance overlap between the excited modes happens.
Consequently, we expect the system to behave linearly, at least for a very long time after which eventually wave packet spreading will appear as energy exchanges between modes slowly start in Eq.~\eqref{eq:1d_DDNLS_eq_motion_normal_modes_space}
(b) $\mathcal{K} > 1$ giving $\delta _\nu > d_\nu$.
In this case initial resonance overlaps between modes and wave packet spreading is expected to occur immediately.
(c) $\mathcal{K}\gg 1$ such that $\delta _\nu \gg d_\nu$.
In this case the nonlinear frequency shift $\delta _\nu$ is very large compared to $d_\nu$.
This result to some of the nonlinear NMs to be tuned out of resonances with their frequencies shifted outside the linear frequency band.
This case will be discussed in more detail in Sec.~\ref{subsec:selftrapping_theorem_I}.

A difficulty in the determination of the above mentioned dynamical regimes [(a), (b) and (c)] is the estimation of quantities $\delta _\nu$ and $d_\nu$.
Thus, it is important to be able to define these regimes in terms of more accessible parameters, which will make easier the setting up of experiments or numerical simulations.
We will undertake this task in the following section.
It is also worth  mentioning that in the work of~\citep{shepelyansky1993delocalization} only two regimes [cases (a) and (b)] are considered with respect to Eq.~\eqref{eq:chirikov_criterion}.
For small resonance-overlaps (case (a) with $\delta _\nu < d_\nu$) no spreading and therefore localization for the initial wave packet was predicted, while for large resonance-overlaps (case (b) where $\delta _\nu > d_\nu$) extensive spreading is observed for the system with 
\begin{equation}
    \mathcal{K} \propto \frac{\beta \lvert \phi _\nu \rvert ^2}{d_\nu} \sim \frac{\beta/\Delta \nu}{1 / \Delta \nu} = \beta.
    \label{eq:resonant_overlap_calculation_1} 
\end{equation}
This prediction of constant resonance-overlap parameter of the mode interactions (as $\beta$ remains fixed) suggests that the spreading of the wave packets persist for all times.
We note that in the approximation above, we used the fact that $\lvert \phi _\nu\rvert ^2 \sim 1/\Delta \nu$ and $d_\nu \sim 1/\Delta \nu$ (since $\Delta$ always remains finite) in the limit of extended wave packets i.e. $\Delta \nu \rightarrow \infty$.

\subsection{\label{subsec:selftrapping_theorem_I}Selftrapping regime}
The analysis presented in the previous section does not provide enough information on the fate of the wave packet in the case $\delta_\nu \gg d_\nu$.
We will study this case in more detail here for the 1D DDNLS system~\eqref{eq:Hamilton_complex_dnls_1d}, following the work of~\citep{skokos2009delocalization} which was based on the results presented in~\citep{rasmussen2000statistical,johansson2004statistical,kopidakis2008absence} and recently discussed in~\citep{flach2016spreading,kati2020density}.
In particular, we will show that it is possible that the wave packet's evolution can lead to an asymptotically inhomogeneous behavior, where part of the wave packet remains trapped so that its amplitudes do not attain arbitrarily small values. 
Initial excitation leading to such behaviors are said to belong to the sefltrapping regime.
We note that this behavior contradicts the asymptotic homogeneous nonlinear wave packet whose existence was proposed in~\citep{shepelyansky1993delocalization,pikovsky2008destruction}. 

The participation number
\begin{equation}
    P = \frac{\mathcal{S}_{1D}^2}{\sum _{l} \lvert \psi _l \rvert ^4},
    \label{eq:pn_selftrapping_theorem}
\end{equation}
is a widely used quantity to measure the number of highly excited sites and consequently to measure the wave packet's inhomogeneity.
This quantity is bounded from below by $P = 1$ corresponding to the case where the whole norm/energy is concentrated at a single lattice site and has an upper bound $P = N$ when equipartition is reached. 

The Hamiltonian function~\eqref{eq:Hamilton_complex_dnls_1d} can be split into two terms
\begin{equation}
    \mathcal{H}_{1D} = \mathcal{H}_{L} + \mathcal{H}_{NL},
    \label{eq:hamilton_ddnls_splitting_l_and_nl}
\end{equation}
where 
\begin{equation}
    \mathcal{H}_{L} = \sum _{l = 1}^{N} \epsilon _l \lvert \psi _l\rvert ^2 - \left(\psi _l \psi _{l + 1}^\star + \psi _l ^\star\psi _{l + 1}\right),
    \label{eq:hamilton_ddnls_splitting_l}    
\end{equation}
represents the linear part of $\mathcal{H}_{1D}$~\eqref{eq:Hamilton_complex_dnls_1d} (see also Eq.~\eqref{eq:hamilton_anderson_model}) and 
\begin{equation}
    \mathcal{H}_{NL}= \sum _{l = 1}^{N} \frac{\beta}{2} \lvert \psi _l \rvert ^4,
    \label{eq:hamilton_ddnls_splitting_nl}
\end{equation}
includes the nonlinear terms of $\mathcal{H}_{1D}$~\eqref{eq:Hamilton_complex_dnls_1d}. 
The $\mathcal{H}_L$~\eqref{eq:hamilton_ddnls_splitting_l} is bounded such that~\citep{kopidakis2008absence,skokos2009delocalization} (see also Eq.~\eqref{eq:bounded_hamilton_anderson_full_app} in Appendix~\ref{app:conv_linear_ddnls_dkg}) 
\begin{equation}
    (-2 - \frac{W}{2})\mathcal{S}_{1D} \leq \mathcal{H}_{L} \leq (2 + \frac{W}{2})\mathcal{S}_{1D},
    \label{eq:cond_primary_lin_ham}
\end{equation}
where $\mathcal{S}_{1D} = \sum _l \lvert \psi _l \rvert ^2$ is the system's total norm~\eqref{eq:norm_complex_dnls_1d}.
Trying to identify under which condition the selftrapping can be avoided, let us assume that the wave packet asymptotically spreads toward zero amplitudes.
In that case, we have
\begin{equation}
    0 \le \lim _{t\rightarrow \infty} \left( \sum _{l} \lvert \psi _{l} \rvert ^4 \right) < \lim _{t\rightarrow \infty} (\sup _{l}\lvert \psi _l  \rvert^2 ) \left( \sum _{l} \lvert \psi _{l} \rvert ^2 \right) = 0.
    \label{eq:lim_cond_selftrapping}
\end{equation} 
Since $\mathcal{S}_{1D} = \sum _{l} \lvert \psi _{l} \rvert ^2 > 0$ is a constant of motion, the only possibility to achieve Eq.~\eqref{eq:lim_cond_selftrapping} is to set 
\begin{equation}
    \lim _{t\rightarrow \infty} \sup _{l}\lvert \psi _l \rvert^2 = 0.
\end{equation}
Whence $\mathcal{H}_{NL} = 0$ in the limit of $t \rightarrow \infty$.
This result and the inequality~\eqref{eq:cond_primary_lin_ham} imply 
\begin{equation}
    \mathcal{H}_{1D} \leq (2 + \frac{W}{2}) \mathcal{S}_{1D}.
    \label{ineq:h_s_selftrapping}
\end{equation}
As $\mathcal{H}_{1D}$ and $\mathcal{S}_{1D}$ are both constants of motion, the inequality~\eqref{ineq:h_s_selftrapping} has to be true for all times.
As we will see below, this is not always the case, indicating that selftrapping can happen.
Indeed, we will show that selftrapping appears when the norm $\mathcal{S}_{1D}$ exceeds a specific critical value. 
To demonstrate this behavior, we consider the simple case of an initial wave packet of size $L=1$ (single site excitation) of the form $\psi_{0} (0) = \sqrt{s}e^{i\theta_{0}}$.
Thus, the constants of motion $\mathcal{H}_{1D}$~\eqref{eq:Hamilton_complex_dnls_1d} and $\mathcal{S}_{1D}$~\eqref{eq:norm_complex_dnls_1d} are   
\begin{equation}
    \mathcal{H}_{1D} = \epsilon _0 s + \frac{\beta}{2}s^2, \quad \text{and} \quad \mathcal{S}_{1D} = s,
    \label{eq:ham_selftrapping_l=1}
\end{equation}
with $\epsilon _0$ being the value of the onsite parameter at the initially excited site.
In addition, this initial excitation at $t=0$ yields
\begin{equation}
    \mathcal{H}_L (t = 0) = \epsilon _0 s \qquad \mbox{ and } \qquad \mathcal{H}_{NL} (t = 0) = \frac{\beta}{2} s^2 = \frac{\beta}{2} s^2P^{-1} (t = 0).
    \label{eq:hamilton_ddnls_splitting_l_and_nl_2}
\end{equation} 

Let us now evaluate bounds for the participation number $P(t)$, $t>0$.
Inserting Eq.~\eqref{eq:ham_selftrapping_l=1} into the right part of Eq.~\eqref{eq:cond_primary_lin_ham}, using Eq.~\eqref{eq:hamilton_ddnls_splitting_l_and_nl} to evaluate $\mathcal{H}_L$, as well as the fact that Eqs.~\eqref{eq:hamilton_ddnls_splitting_nl} and~\eqref{eq:pn_selftrapping_theorem} give $\mathcal{H}_{NL}(t) = \frac{\beta}{2} s^2 P^{-1}(t)\ge 0$, we obtain
\begin{equation}
    \left(-2 - \frac{W}{2}\right)s + \frac{\beta}{2}s^2 + \epsilon _ 0 s \leq \frac{\beta}{2}s^2P^{-1} (t),
    \label{eq:bound_participation_number_selftrapping_1}
\end{equation} 
which allows us to find bounds for the participation number in the form of
\begin{equation}
    1 \leq P(t) \leq \frac{\beta s^2}{\beta s^2 - (\Delta - 2\epsilon_0)s},
    \label{ineq:part_num_selftrapping}
\end{equation}
where $\Delta = W + 4$~\eqref{eq:characteristics_eigfreq_anderson_model}.
We note that the upper bound of the inequality is valid only if 
\begin{equation}
   \beta s > \Delta - 2\epsilon _0.
    \label{eq:selftrapping_condition_2}
\end{equation}
For the 1D DDNLS system, the initial nonlinear frequency shift $\delta$ of the excited single site (see for example~\citep[Page 42]{laptyeva2013nonlinear}) is given by
\begin{equation}
    \delta = \beta s.
\end{equation}
Then from Eq.~\eqref{eq:selftrapping_condition_2}, we conclude that the $\delta$ value above which selftrapping occurs is 
\begin{equation}
    \delta = \Delta - 2\epsilon_0.
    \label{eq:selftrapping_condition_theorem}
\end{equation}

Equation~\eqref{ineq:part_num_selftrapping} tells us that in the regime of selftrapping, the value of the participation number is bounded from above, which taking into account that $s^2P^{-1}(t) = \sum _l \lvert \psi _l \rvert ^4 < \sup _l \lvert \psi _l \rvert ^2 s$~\citep{kopidakis2008absence,skokos2009delocalization} leads to
\begin{equation}
    \sup _l \lvert \psi _l \rvert ^2 > \frac{\beta s^2 - \left(\Delta - 2\epsilon_0\right)s}{\beta s}. 
    \label{eq:condition_sup_l_psi_selftrapping}
\end{equation}
The fulfillment of these conditions mean that the wave packet will never reach equipartition and that part of it remains trapped.
Nevertheless, a limitation of the above analysis is the lack of information about the spatiotemporal behavior of the trapped part of the wave packet, although we secure that a part of it will not spread, we do not know if the resulting inhomogeneity will remain constant in time and space.
We note that Eqs.~\eqref{ineq:part_num_selftrapping} and~\eqref{eq:condition_sup_l_psi_selftrapping} respectively correspond to Eqs.~(14) and~(15) of~\citep{skokos2009delocalization} for the case $s = 1$ and $\epsilon_0 = 0$ considered in that paper. 

\subsection{\label{subsec:spreading_regimes}Spreading regimes}
The Chirikov criterion and the analysis of the selftrapping regime of Sec.~\ref{subsec:selftrapping_theorem_I} give a very good basis for the classification of the dynamical regimes occurring in disordered systems.
Although the Chirikov criterion uses the NMs' nonlinear frequency shift and in particular when the resonance-overlap parameter $\mathcal{K}\gg 1$, leading to the inequality $\delta _\nu \gg d_\nu$ of case (c) in Sec.~\ref{sec:dynamical_regimes_chirikov} to roughly predict the appearance of the seftrapping regime, we see that its prediction is in good agreement with the results obtained using the inequality $\delta _l > \Delta$~\eqref{eq:selftrapping_condition_theorem} [with $\Delta \gg d_\nu$] which is based on the nonlinear frequency shift of the sites' oscillations.
Thus we foresee a relation of proportionality between $\delta _\nu$ and $\delta _l$.  
Indeed, if we excite a single mode, say mode $\nu$, all the lattice sites within its localization volume $P_\nu$ become thermalized.
Therefore the sum of norms/energies of each lattice sites within $P_\nu$ gives the total norm/energy of mode $\nu$ since the energy is an extensive parameter.
As the energy is equal to the frequency here, it means that the nonlinear frequency shift of the mode $\nu$ is the cumulative nonlinear frequency shifts of all lattice sites within $P_\nu$.
Thus 
\begin{equation}
\delta _\nu  \propto P_\nu \delta _l,
\end{equation}
where $\delta _l \approx \beta \sup _l \lvert \psi _l \rvert ^2$ approximates the nonlinar frequency shift at every lattice site within $P_\nu$.
In addition, the relation between the width of the wave packet in both the NMs $\Delta \nu$ and real $\Delta l$ spaces is of the form 
\begin{equation}
    \Delta \nu \propto \frac{\Delta l}{P_\nu}.
\end{equation}
Consequently the resonance-overlap parameter~\eqref{eq:chirikov_criterion} becomes 
\begin{equation}
    \mathcal{K} \propto \delta _\nu \cdot \frac{1}{d_\nu} \propto P_\nu \delta _l \cdot \frac{\Delta l}{P_\nu \Delta} = \frac{\delta _l}{d},
    \label{eq:chirikov_criterion_sites}
\end{equation}
where $d = \Delta / \Delta l$ was mentioned in Eq.~\eqref{eq:average_spacing_between_freq_anderson_model}, $\Delta l$ being the number of sites with non-negligible amount of norm.
\subsubsection{\label{subsubsec:1d_ddnls_spreading}The 1D DDNLS system}
We now apply the machinery developed above to determine the boundaries of the different dynamical regimes in the 1D DDNLS model.
In doing so, we consider a compact, initial wave packet over $L$ consecutive sites, with each site having the same norm $s$ so that the total norm~\eqref{eq:norm_complex_dnls_1d} is $\mathcal{S}_{1D} = L\cdot s$.
Thus, $\delta _l = \delta = \beta s$ for each initially excited site. 
If $\beta s > \Delta$ a part of the wave packet remains trapped.
The nonlinear frequency shift of every initial excited site is tuned out of resonance with the non excited surrounding and the selftrapping appears.    
If on the other hand, selftrapping is avoided (i.e. $\beta s < \Delta$), the wave packet starts spreading.
We can distinguish between various cases.
First if the initial wave packet size is smaller than the localization volume i.e. $L<\overline{V}$, it starts spreading over the sites within $\overline{V}$ during a time $\tau _{in}$.
After that time the situation can be considered as having a new initial wave packet of size $\overline{V}$ and {\it almost} uniform norm at each lattice excited sites $s(\tau_{in}) = sL/\overline{V}$.
For $L\approx \overline{V}$ the norm at each site within the wave packet will not change much and $s(\tau _{in}) \approx s$.
The obtained nonlinear frequency shift $\delta = \beta s(\tau_{in})$ at time $\tau_{in}$ can then be compared to the average spacing between frequencies $d$.
Thus, if $\beta s (\tau _{in}) > d$, all the NMs in the wave packet are {\it strongly} resonantly interacting (this behavior corresponds to case (b) in Sec.~\ref{sec:dynamical_regimes_chirikov}).
This regime is referred to as the {\it strong chaos regime}~\citep{flach2010spreading,laptyeva2010crossover}.
On the other hand, if $\beta s (\tau _{in}) < d$ (this behavior is equivalent to case (a) above), the modes are weakly interacting and this regime is called the {\it weak chaos regime}~\citep{flach2010spreading,laptyeva2010crossover}.
In summary based on the various initial wave packet excitations, the different possible dynamical regimes are
\begin{align}
    \nonumber
    \mbox{weak chaos} \quad &: \qquad \beta s \frac{L}{\overline{V}} < d, \\
    \mbox{strong chaos} \quad &: \qquad \beta s \frac{L}{\overline{V}} > d,  \label{eq:spreading_regime_1d_ddnls_1}\\
    \nonumber
    \mbox{selftrapping} \quad &: \qquad \beta s > \Delta.
\end{align}
It is easy to see that in the case of single site initial excitation where $L=1$, we cannot observe the strong chaos regime as $d = \Delta$ and the relations of Eq.~\eqref{eq:spreading_regime_1d_ddnls_1} become
\begin{align}
    \nonumber
    \mbox{weak chaos} \quad &: \qquad \beta s < \Delta, \\
    \mbox{selftrapping} \quad &: \qquad \beta s > \Delta. \label{eq:spreading_regime_1d_ddnls_3}
\end{align}

\begin{figure}[!htb]
\centering 
\includegraphics[width=0.5\textwidth, height=0.5\linewidth]{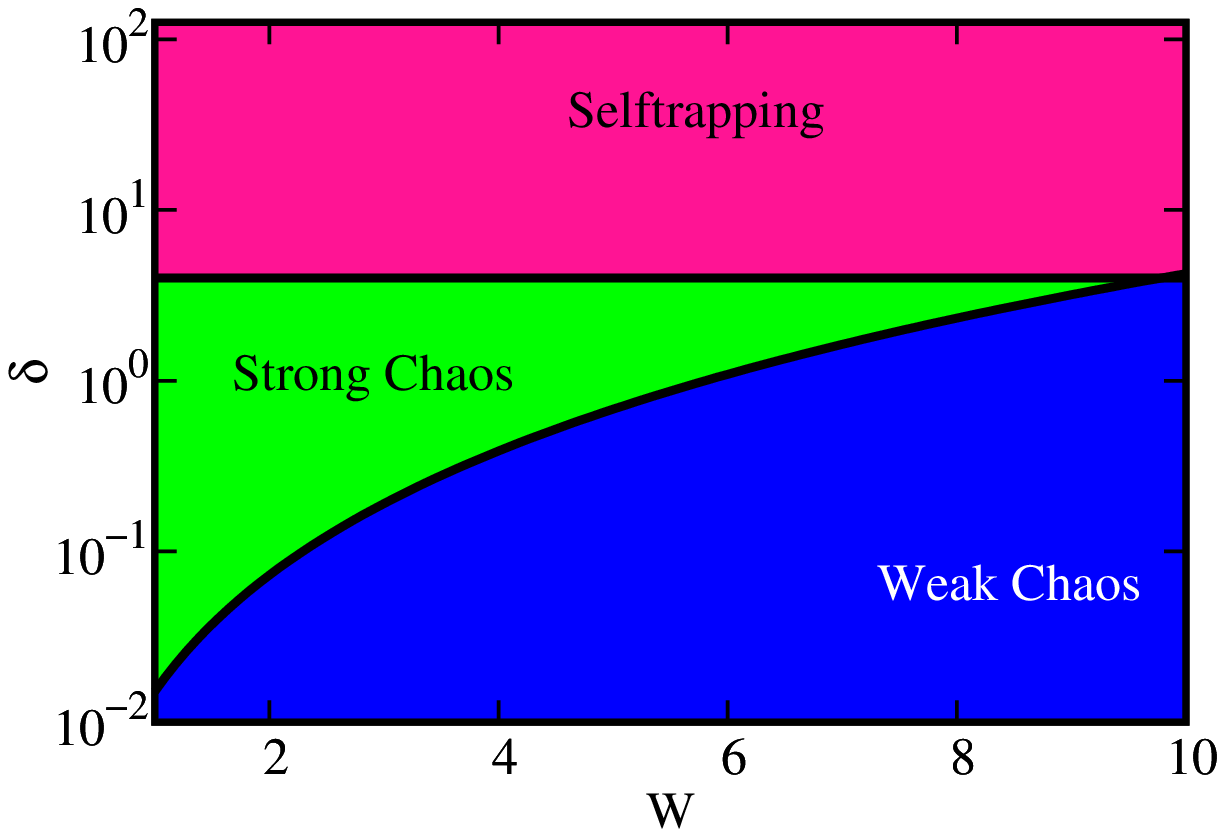}
\caption{Location of the different dynamical regimes in the parameter space $(W, \delta)$ of the 1D DDNLS Hamiltonian~\eqref{eq:Hamilton_complex_dnls_1d} for the case of an initial wave packet of size $L = \overline{V}$.
        The upper curve corresponds to $\delta = 4$~\eqref{eq:condition_for_selftrapping_multi_sites_wave_packet_1d_ddnls} and the lower to $\delta = \left(W + 4\right)W^2/330$~\eqref{eq:boundary_weak_strong_1d_spreading}.
        }
\label{fig:parameter_space_dnls_1d_delta_W}
\end{figure}

Figure~\ref{fig:parameter_space_dnls_1d_delta_W} depicts the location of each dynamical regime in the $(W, \delta)$ parameter space, when multi-site initial wave packets with width $L = \overline{V}$ are considered.
The curve
\begin{equation}
    \delta = \frac{\Delta}{\overline{V}} \approx \frac{W^2\left(W + 4\right)}{330},
    \label{eq:boundary_weak_strong_1d_spreading}
\end{equation}
denotes the boundary between the weak and strong chaos spreading regions and is obtained by estimating the localization volume as $\overline{V} \approx 330/ W^{2}$ according to the fit presented in Fig.~\ref{fig:V_vs_W_KF10}.

It is worth noting that in the case of a multi-site excited initial wave packet, the selftrapping in Eq.~\eqref{eq:spreading_regime_1d_ddnls_1} can occur even faster as all we need is one activated site to be trapped.
Indeed, at each site within the wave packet, the general condition $\delta = \Delta - 2\epsilon_l$~\eqref{eq:selftrapping_condition_theorem} is tested.
In particular, for lattice's sites with $\epsilon _l \approx W/2$, the condition for selftrapping becomes  
\begin{equation}
    \delta \approx W + 4 - 2 \frac{W}{2} = 4,
    \label{eq:condition_for_selftrapping_multi_sites_wave_packet_1d_ddnls}
\end{equation} 
similar to the one found in~\citep{laptyeva2010crossover}.

Let us discuss the behavior of the system's orbits in the parameter space of Fig.~\ref{fig:parameter_space_dnls_1d_delta_W}.
During the evolution of an initial wave packet, the value of $W$ remains always fixed.
As the total norm (which also remains constant as it is an integral of motion) is shared among more and more sites as the wave packet's $\Delta l$ increases, the norm's amplitude at each site $s_l = \lvert \psi _l \rvert ^2$  decreases in time and subsequently the related nonlinear frequency shift $\delta_l = \beta s_l$ decreases.
Therefore this dynamical evolution is represented by points which move downward in the parameter space ($W$, $\delta$) along lines perpendicular to the $W$ axis.
Thus a wave packet launched in the regime of weak chaos will remain there for all time.
On the other hand, a wave packet initiated in the strong chaos regime (i.e.~$\delta > d$) will remain in that region for some finite time $\tau_s$, and eventually will cross over to the weak chaos regime (i.e. $\delta < d$) as $\delta$ is decreasing~\citep{flach2010spreading}.
Finally, an initial wave packet launched in the selftrapping regime (i.e.~$\delta > 4$) will always have part of its norm remaining trapped at some sites, a behavior corresponding to regions of quasi-periodic motion in the system's phase space.

\subsubsection{\label{subsubsec:1d_dkg_spreading}The 1D DKG system}

For the sake of completeness we will briefly discuss the manifestation of the various dynamical regimes in the 1D DKG system~\eqref{eq:hamilton_1d_dkg}.
More details on the subject can be found in~\citep{senyange2020chaotic}. 
A correspondence between the 1D DDNLS system and the 1D DKG model described by $\sum _l \beta s_l \approx \sum _l 3W \tilde{h}_l$ is valid for small amplitude excitation of the 1D DDNLS model~\citep{kivshar1992modulational,mackay1994proof,johansson2004statistical, johansson2006discrete,laptyeva2010crossover,laptyeva2012subdiffusion} where $s_l = \lvert \psi _l \rvert ^2$ and $\tilde{h}_{l} = p_l^2/2 + \tilde{\epsilon}_l q^4/4 + [(q_{l} - q_{l - 1})^2 + (q_{l + 1} - q_l)^2]/4W$ are respectively the norm for the 1D DDNLS system and the energy for the 1D DKG model at site $l$.
Considering uniform norm (energy) excitations of adjacent sites for the 1D DDNLS (1D DKG) model so that each excited site has initially norm (energy) $s$ ($\tilde{h}$), the various dynamical regimes of ~\eqref{eq:spreading_regime_1d_ddnls_1} correspond to the following cases for the DKG system~\eqref{eq:hamilton_1d_dkg} 
\begin{align}
    \nonumber
    \mbox{weak chaos} \quad &: \qquad \tilde{h} \frac{L}{\overline{V}} < d/3W, \\
    \mbox{strong chaos} \quad &: \qquad \tilde{h} \frac{L}{\overline{V}} > d/3W, \label{eq:spreading_regime_1d_dkg_1} \\
    \nonumber
    \mbox{selftrapping} \quad &: \qquad \tilde{h} > \Delta_K/3,
\end{align}
where $\Delta _K$ is the width of the squared frequency spectrum of the linear DKG model.
Note that in Eq.~\eqref{eq:spreading_regime_1d_dkg_1}, we use the fact that the frequency bandwidth of the 1D DDLS model~\eqref{eq:hamilton_anderson_model} corresponds to $\Delta _KW$ (see Appendix~\ref{app:conv_linear_ddnls_dkg} for the derivation of this relation) and that the NMs average spacing $d$ is the same for both 1D models.
Again, we see that in the case of a single site excitation ($L = 1$), the strong chaos regime does not appear as $d/3W = \Delta _K/3$.

\subsubsection{\label{susubsec:2d_ddnls_dkg_spreading}The 2D DDNLS system}
\begin{figure}[!hhtb]
    \centering 
    \includegraphics[width=0.5\textwidth, height=0.5\linewidth]{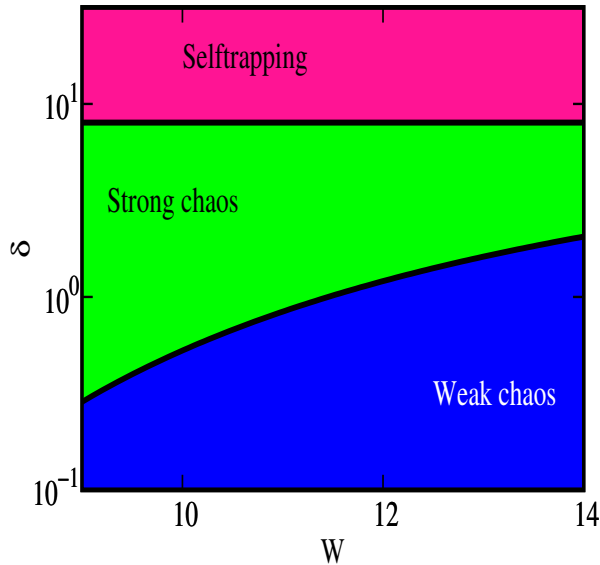}
    \caption{Location of the different dynamical regimes in the parameter space $(W, \delta)$ of the 2D DDNLS Hamiltonian~\eqref{eq:hamilton_complex_ddnls_2d} for the case of an initial wave packet of size $L^2 = \overline{V}$.
    The upper curve corresponds to $\delta = 8$ using a similar method like in Eq.~\eqref{eq:condition_for_selftrapping_multi_sites_wave_packet_1d_ddnls} and the lower to $\delta \approx \left(W + 8\right)/\left(3.2e^{237/W^2}\right)$.}
\label{fig:parameter_space_dnls_2d_delta_W}
\end{figure}

The dynamical regimes underlined in Eq.~\eqref{eq:spreading_regime_1d_ddnls_1} are valid for the 2D DDNLS where $L$ has to be replaced by $L^2$ (square excitation).
Defining the different dynamical regimes in Eqs.~\eqref{eq:spreading_regime_1d_ddnls_1} and~\eqref{eq:spreading_regime_1d_ddnls_3}, we had to find $d$ and $\overline{V}$ and the relation between them, something which can be done also for the 2D DDNLS system.
Unfortunately in this case, the relation between the localization volume $\overline{V}$ and the disorder strength $W$ is not well established.
This is mainly because investigating the statistical properties of the system's NMs through the diagonalization (or the transfer matrix approach) of very large matrices is a hard computational task.
Nevertheless, a rough estimation of $\overline{V}$ can be obtained by evolving single site excitations in the linear version of the 2D lattice for very long times until the halting of the wave packet spreading is observed as a result of reaching the extent of the localization volume $\overline{V}$.
Then according to~\citep{laptyeva2012subdiffusion,manda2020chaotic}, the wave packet participation number averaged over several disorder realizations can be considered as a good estimator of $\overline{V}$.
Assuming the relation,
\begin{equation}
    \overline{V} \approx a_\zeta \overline{\zeta}^2, \quad \overline{\zeta} \approx \exp \left(\frac{b_\zeta}{W^2}\right)
    \label{eq:loc_vol_loc_length_2d_ddnls_estimation}
\end{equation}
between the averaged localization volume $\overline{V}$ and localization length $\overline{\zeta}$, we obtain fitting the data of the average participation number over $10$ disorder realizations in~\citep{manda2020chaotic} (see also Table~\ref{tab:} of Chap.~\ref{chap:chaotic_two_dimensional}) $a_\zeta \approx 3.2$ and $b_\zeta \approx 118.3$ for $W$ values around $W = 10$ which is in the range of our interest.
In Eq.~\eqref{eq:loc_vol_loc_length_2d_ddnls_estimation} the relation $\ln \zeta \sim 1/W^2$ was used to express the localization length in the 2D DDNLS model~\citep{kramer1993localization,garcia2009delocalization}.  
Since the width of the linear frequency spectrum is $\Delta = W + 8$~\citep{laptyeva2012subdiffusion}, we can easily determine the boundaries of the various spreading regimes in the $(W, \delta)$ parameter space (see Fig.~\ref{fig:parameter_space_dnls_2d_delta_W}).
In particular, the line $\delta = 8$ separates the selftrapping regime from the spreading ones and $\delta = \left(W + 8\right)/\left(3.2e^{237/W^2}\right)$
defines the boundary between the weak and strong chaos spreading regimes as shown in Fig.~\ref{fig:parameter_space_dnls_2d_delta_W}.

\section{\label{sec:statistical_physics_spreading}A statistical physics approach in determining the various dynamical regimes}
So far, following~\citep{shepelyansky1993delocalization,pikovsky2008destruction,flach2009universal,skokos2009delocalization,flach2010spreading,laptyeva2010crossover,laptyeva2012subdiffusion,bodyfelt2011nonlinear,laptyeva2013nonlinear,laptyeva2014nonlinear} we characterized the different dynamical regimes of the DDNLS system in one and two spatial dimensions in the parameter spaces $(W, \delta)$ based on only one integral of motion of the system, namely the total norm.
In what follows, we will use the existence of the system's two integrals of motion, i.e. the total energy and the total norm, to investigate the existence of the various spreading regimes.
We will do that by implementing the theory of the {\it grand canonical ensemble} of statistical physics~\citep{landau1958lifshitz,tong2012statistical} which describes the thermal states of a system conserving both the total energy and the total number of particles.
The latter, in the case of the DDNLS models, corresponds to the total norm.
Our analysis is based on the results presented in~\citep{rasmussen2000statistical,johansson2004statistical,flach2016spreading,mithun2018weakly,kati2020density}.

\subsection{\label{subsec:thermal_non_thermal_regions_stat}Thermal and non-thermal regions}
We will first describe the energy-norm parameter space for the DDNLS models and identify there the regions corresponding to the different dynamical spreading regimes we discussed so far.
We relate to the thermal or Gibbs phase the region of the energy-norm parameter space whose system's states have finite temperatures, denoting as non-thermal or non-Gibbs phase the remaining regions within the system's allowed energy values.
In this section, we consider the 1D DDNLS model whose energy and norm are given by Eq.~\eqref{eq:Hamilton_complex_dnls_1d} and Eq.~\eqref{eq:norm_complex_dnls_1d} respectively.

Further, for the rest of this section, we consider the thermodynamic limit i.e. $N \rightarrow \infty$.
Since the disorder parameter are randomly distributed with a uniform probability $\mathcal{P}_W = 1/W$ in the interval $[-W/2, W/2]$,
we have~\citep{flach2016spreading}\footnote{The distribution of the disorder parameter is similar to the ones of the thermodynamics states in the microcanonical ensemble~\citep{landau1958lifshitz,tong2012statistical}.}
\begin{equation}
    \epsilon = \lim _{N \rightarrow \infty} \frac{1}{\sum_{l=1}^N \mathcal{P}_W} \sum_{l=1}^N \mathcal{P}_W \epsilon_l =  \lim _{N \rightarrow \infty} \frac{1}{N}\sum _{l = 1}^{N} \epsilon _l  = 0,
    \label{eq:av_random_on_site_disorder_1}
\end{equation} 
with the finite standard deviation 
\begin{equation}
    \sigma _\epsilon = \lim _{N \rightarrow \infty} \frac{1}{\sum_{l=1}^N \mathcal{P}_W} \sum_{l=1}^N \mathcal{P}_W \epsilon_l^2  = \lim _{N \rightarrow \infty} \frac{1}{N} \sum _{l = 1}^{N} \epsilon _l^2.
    \label{eq:var_random_on_site_disorder_1}
\end{equation}
We write the wave function as
\begin{equation}
    \psi _l = \sqrt{s_l} e^{i\theta _l},
    \label{eq:wave_function_site_l_stat}
\end{equation} 
where $s_l = \lvert \psi _l \rvert^2$ is the norm and $\theta _l$ being the local phase which contains information about the local energy $h_l$ at site $l$.
Substituting Eq.~\eqref{eq:wave_function_site_l_stat} into the norm reduces to
\begin{equation}
    \mathcal{S}_{1D} = \sum _{l = 1}^{N}s_l,
    \label{eq:norm_gibss_stat_ddnls}
\end{equation}
and the energy~\eqref{eq:Hamilton_complex_dnls_1d} reads
\begin{equation}
    \mathcal{H}_{1D} = \sum _{l = 1}^{N} \epsilon _l s_l + \frac{\beta}{2}s_l ^2 - 2\sqrt{s_l s_{l + 1}} \cos \left(\theta _l - \theta _{l + 1}\right).
    \label{eq:hamilton_gibbs_stat_ddnls}
\end{equation}
In addition, we respectively denote the ensemble average energy and norm densities  
\begin{equation}
    h = \frac{\mathcal{H}_{1D}}{N}, \quad s =  \frac{\mathcal{S} _{1D}}{N}.
    \label{eq:relation_of_densities_stat_1}
\end{equation}
Furthermore, following~\citep{flach2016spreading,mithun2019dynamical} we define for convenience the rescaled averaged norm and energy densities 
\begin{equation}
    x = \beta s, \quad y = \beta h,
    \label{eq:rescaled_densities_stat}
\end{equation}
respectively and use them to span the energy-norm parameter space.
We first start searching the location in the $(x, y)$ parameter space of thermal states which are bounded from below by zero ($T = 0$) and above by infinite ($T = \infty$) temperature states.
If we set $\tilde{\nu} = 1/T$, then we denote the boundary of zero temperature states by the curve $y_{\tilde{\nu} \rightarrow \infty}(x)$ and the one of infinite temperature states with $y_{\tilde{\nu} = 0}(x)$, where the subscript indicates the temperature of the system along each curve.
Within the thermal region, the probability $\rho_G$ to find a system with temperature $T$, energy $\mathcal{H}_{1D}$ and norm $\mathcal{S}_{1D}$  is given by~\citep{rasmussen2000statistical,johansson2004statistical,flach2016spreading,mithun2018weakly} 
\begin{equation}
    \rho_G = \frac{1}{\mathcal{Z}} \exp \left\{-\tilde{\nu} \left(\mathcal{H}_{1D} + \mu \mathcal{S}_{1D}\right)\right\},
    \label{eq:Gibb_prob_distr_function}
\end{equation}
where $\mathcal{Z}$ is the so-called {\it the partition function}, $\tilde{\nu}$ being the inverse temperature.
The variable $\mu$ is equivalent to a chemical potential in the normal grand canonical ensemble~\citep{landau1958lifshitz,tong2012statistical}.


The thermal states with zero temperature possess the minimum of energy.
In expression~\eqref{eq:hamilton_gibbs_stat_ddnls}, such a value of the energy is reached for wave functions with $\theta _{l + 1} - \theta_l = 0$ and $s_l = s$ 
\begin{equation}
    E^{(min)} = \sum _{l = 1}^{N} \left(\epsilon _l - 2\right) s + \frac{\beta}{2} s^2,
    \label{eq:energy_min_stat}
\end{equation} 
and depends on the disorder realizations through the values of $\epsilon_l$~\citep{flach2016spreading,kati2020density}.
However in the thermodynamic limit (averaging over lattice sites)
\begin{equation}
    h = \frac{E^{(min)}}{N}  = - 2 s + \frac{\beta}{2} s^2,
    \label{eq:curve_zero_temperature_a_h}
\end{equation} 
which is a quantity independent of the considered disorder realization.
Consequently the zero temperature states belong to the curve 
\begin{equation}
    y _{\tilde{\nu} \rightarrow \infty} = -2x + \frac{x^2}{2}.
    \label{eq:curve_zero_temperature_x_y}
\end{equation}
In the case of a finite number of sites $N$ and non-negligible value of $W$, the curve of zero temperature~\eqref{eq:curve_zero_temperature_x_y} is bounded as~\citep{flach2016spreading}
\begin{equation}
    - (2 + \frac{W}{2})x + \frac{x^2}{2} \le y_{\tilde{\nu} \rightarrow \infty}(x, \epsilon_l) \le (-2 + \frac{W}{2})x + \frac{x^2}{2}, 
    \label{eq:max_bound_zero_temperature_stat}
\end{equation}
and can be located within this interval using numerical computations~\citep{kati2020density}.

We now look at the boundary curve which contains infinite temperature states $y_{\tilde{\nu} = 0}$.
In the same fashion as above, such states correspond to the case with maximal energy.
Unfortunately, the Hamiltonian~\eqref{eq:hamilton_gibbs_stat_ddnls} is not bounded from above as it diverges for $s_l \rightarrow \infty$.
Instead, we use the fact that if one knows the analytical form of $\mathcal{Z}$ then 
\begin{equation}
    \mathcal{S}_{1D} = - \frac{1}{\tilde{\nu}}\frac{\partial}{\partial \mu} \ln \mathcal{Z}, \quad \mathcal{H}_{1D} + \mu \mathcal{S}_{1D} = - \frac{\partial}{\partial \tilde{\nu}} \ln \mathcal{Z}.
    \label{eq:energy_norm_general_stat_1}
\end{equation}

The partition function~\citep{landau1958lifshitz,tong2012statistical,rasmussen2000statistical,johansson2004statistical,mithun2018weakly}
\begin{equation}
    \mathcal{Z} = \int _{0}^{\infty} \int _{0}^{2\pi} \prod _{l = 1}^{N} d\theta _l ds_l \exp \left\{-\tilde{\nu} \left(\mathcal{H}_{1D} + \mu \mathcal{S}_{1D}\right)\right\},
    \label{eq:partition_function_stat_1}
\end{equation}
contains all the thermodynamical information of the states of the system. 
Substituting $\mathcal{H}_{1D}$~\eqref{eq:hamilton_gibbs_stat_ddnls}  in~\eqref{eq:partition_function_stat_1} and integrating with respect to the phases $\theta _l$, we get 
\begin{equation}
    \mathcal{Z} = \left(2\pi\right)^{N} \int _{0}^{\infty} \prod _{l = 1}^{N} ds_l I_0 \left(2\tilde{\nu} \sqrt{s_l s_{l + 1}}\right) \exp\left\{-\tilde{\nu} \sum_{l=1}^{N}\left( \mu _l s_l + \frac{\beta}{2} s_l ^2\right)\right\},
    \label{eq:part_function_z_2}
\end{equation}
where $\mu_l = \epsilon_l + \mu$ is a disorder realization dependent constant and
\begin{equation}
    I_0 (x) = \frac{1}{\pi} \int _{0}^{\pi} e^{x \cos \theta} d\theta,
\end{equation}
is the modified Bessel function of order zero.
As we are solely interested in the limit for large temperatures $T\rightarrow \infty$, we can find an analytical approximation of the partition function~\eqref{eq:part_function_z_2} for $\tilde{\nu} \rightarrow 0$.
The first step to achieve this goal is to approximate the modified Bessel function~\citep{rasmussen2000statistical,johansson2004statistical,flach2016spreading,mithun2018weakly}
\begin{equation}
    I_0 \left(2\tilde{\nu} \sqrt{s_l s_{l + 1}}\right) \approx I_0 (0) \approx 1,
\end{equation}   
as its value is slowly changing around the origin.
Thus
\begin{equation}
    \mathcal{Z} \propto  \left(2\pi \right)^N \prod _{l =1}^N \tilde{y}_{\beta} \left(\tilde{\nu}, \mu _l\right),
    \label{eq:partition_function_z_3}
\end{equation}
and the problem is reduced to finding an approximate closed expression for an integral of the form  
\begin{equation}
    \tilde{y}_{\beta} \left(\tilde{\nu}, \mu _l\right) = \int _{0}^{\infty} dx e^{-\tilde{\nu} \mu _l x} e^{-\tilde{\nu} \frac{\beta}{2} x^2}.
    \label{eq:integral_over_a_norm_coord}
\end{equation}
The details of these calculations can be found in Appendix~\ref{sec:app_yb}.
Here, we give only the final result  
\begin{equation}
    \tilde{y}_{\beta} \left(\tilde{\nu}, \mu _l\right) \approx \frac{1}{\tilde{\nu} \mu _l} \left(1  - \frac{\tilde{\nu} \beta }{\left(\tilde{\nu} \mu _l\right)^2}\right).
    \label{eq:calculated_y_stats}
\end{equation}
Inserting Eq.~\eqref{eq:calculated_y_stats} into Eq.~\eqref{eq:partition_function_z_3}, we get 
\begin{equation}
    \ln \mathcal{Z} \approx N \ln 2\pi - \sum _{l =1}^{N} \left[\ln\left(\tilde{\nu} \mu _l\right) + \frac{\tilde{\nu} \beta}{\left(\tilde{\nu} \mu _l\right)^2} \right],
    \label{eq:partition_function_z_ln}
\end{equation}
by assuming that $\beta \tilde{\nu}/(\tilde{\nu} \mu _l)^2 \ll 1$ and applying the approximation $\ln (1 - x) \approx -x$.
From the value of the partition function~\eqref{eq:partition_function_z_ln}, we estimate the ensemble average energy and norm~\citep{rasmussen2000statistical,johansson2004statistical,tong2012statistical,flach2016spreading} 
\begin{equation}
    \mathcal{S}_{1D} = -\frac{1}{\tilde{\nu}} \frac{\partial}{\partial \mu} \ln \mathcal{Z}, \qquad  \mathcal{H}_{1D} = \left(\frac{\mu}{\tilde{\nu}}\frac{\partial }{\partial \mu} - \frac{\partial}{\partial \tilde{\nu}}\right) \ln \mathcal{Z},
\end{equation}
for system with infinite temperature.
We obtain 
\begin{equation}
    s = \frac{\mathcal{S}_{1D}}{N} = \frac{1}{\tilde{\nu} \mu}, \qquad  h = \frac{\mathcal{H}_{1D}}{N} = \frac{\beta}{\tilde{\nu} ^2 \mu ^2}. 
    \label{eq:ham_norm_inf_temp}
\end{equation}
Again, we refer the reader to Appendix~\ref{secapp:ensemble_average} for the detailed calculations.
It follows from Eq.~\eqref{eq:ham_norm_inf_temp} that the infinite temperature states lie on the curve 
\begin{equation}
    y_{\tilde{\nu} = 0} = x^2,
\end{equation}
and correspond to chains of uncoupled harmonic oscillators~\citep{rasmussen2000statistical,johansson2004statistical,flach2016spreading,mithun2018weakly}.
As the energy~\eqref{eq:hamilton_gibbs_stat_ddnls} is not bounded from above, it means that every state in the $(x, y)$ plane with $y > y_{\tilde{\nu} =0}$ cannot be described by the Gibbs probability distribution function $\rho_G$~\eqref{eq:Gibb_prob_distr_function} and consequently it corresponds to the non-thermal region of the parameter space.
In Fig.~\ref{fig:x_y_thermal_selftrapping_phases}(a) we recapitulate the location of each statistical domain in the thermodynamic limit.
Below the thermal phase is an unaccessible region for the system's energy and norm parameters.

\subsection{\label{subsec:weak_strong_spreading}Weak, strong chaos and selftrapping regimes}
In the context developed in the previous section let us now investigate the spreading of an initial block excitation of $L$ central sites, so that $\psi _l = \sqrt{s}e^{i\theta _l}$ inside the excited part of the initial wave packet and $0$ otherwise, where $s$ is the amplitude and $0 \leq \theta_l \leq 2\pi$ the phase of the wave function at site $l$.
As discussed in the previous section, in-phase wave packets i.e. $\theta _{l + 1} = \theta _l$ attain the minimum possible energy $E^{(min)}$ [see Eq.~\eqref{eq:energy_min_stat}].
Therefore, such initial wave packets will always belong to the Gibbs region.
More specifically, such wave packets were encompassed within a regime named Lifshits region in~\citep{kati2020density} whose maximal size is defined by inequality~\eqref{eq:max_bound_zero_temperature_stat}. 

In case of an out-of-phase initial excitation such that  $\theta _{l + 1} = \theta _l + \pi\mod{2\pi}$, the energy of the system is given by 
\begin{equation}
    \mathcal{H}_{1D} = \sum_{l = 1}^{N}\left(2 + \epsilon _l \right)s + \frac{\beta}{2}s^2,
\end{equation}
which also depends on the disorder realization.
These system states in the $(x, y)$ parameter space lie on the curve $y(x, \epsilon_l)$ which is bounded as  
\begin{equation}
    \left(2 - \frac{W}{2}\right)x + \frac{x^2}{2} \le y(x, \epsilon_l) \le  \left(2 + \frac{W}{2}\right)x + \frac{x^2}{2}.
\end{equation}
Consequently for $x > \Delta$ all these initial wave packets belong to the Gibbs regime, while for $x < \Delta$ some of these initial excitations may lie within the non-Gibbs regime. 

An important factor for our analysis is that in the Gibbs region, thermalization is well defined and therefore we expect the regimes of weak and strong chaos spreading to be located there.
Let us also note that for the nonlinear frequency shift of a single oscillator we get 
\begin{equation}
    \delta _l = \beta s \approx x.
    \label{eq:relation_freq_spacing_vs_x_stat}
\end{equation}
Therefore, we compare the value of $x$ to the average spacing between linear frequencies $d$~\eqref{eq:average_spacing_between_freq_anderson_model} of the initial wave packet excitation in order to delimit the various dynamical regimes.
Thus for our homogeneous finite size initial wave packet of size $L = \overline{V}$, the following outcomes are expected: if $x \ll d$, the localization of the wave packet is expected to persist at least for a long time.
In addition if  $x \lesssim d$ a subdiffusive spreading of the wave packet in the weak chaos regime will take place, while for $x \gtrsim d$ that spreading happens in the strong chaos regime.

\begin{figure}[!htbp]
    \centering 
    \includegraphics[width=0.49\textwidth, height=0.5\linewidth]{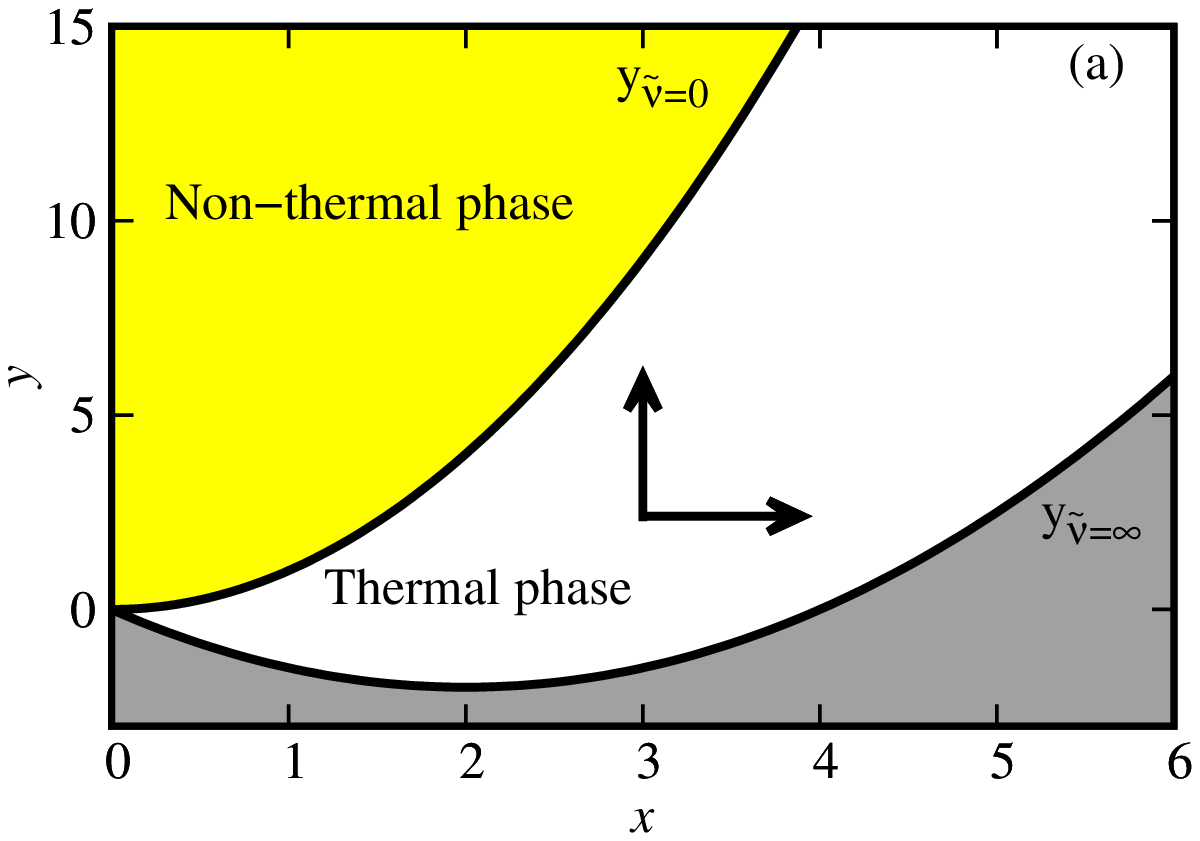}
    \includegraphics[width=0.49\textwidth, height=0.5\linewidth]{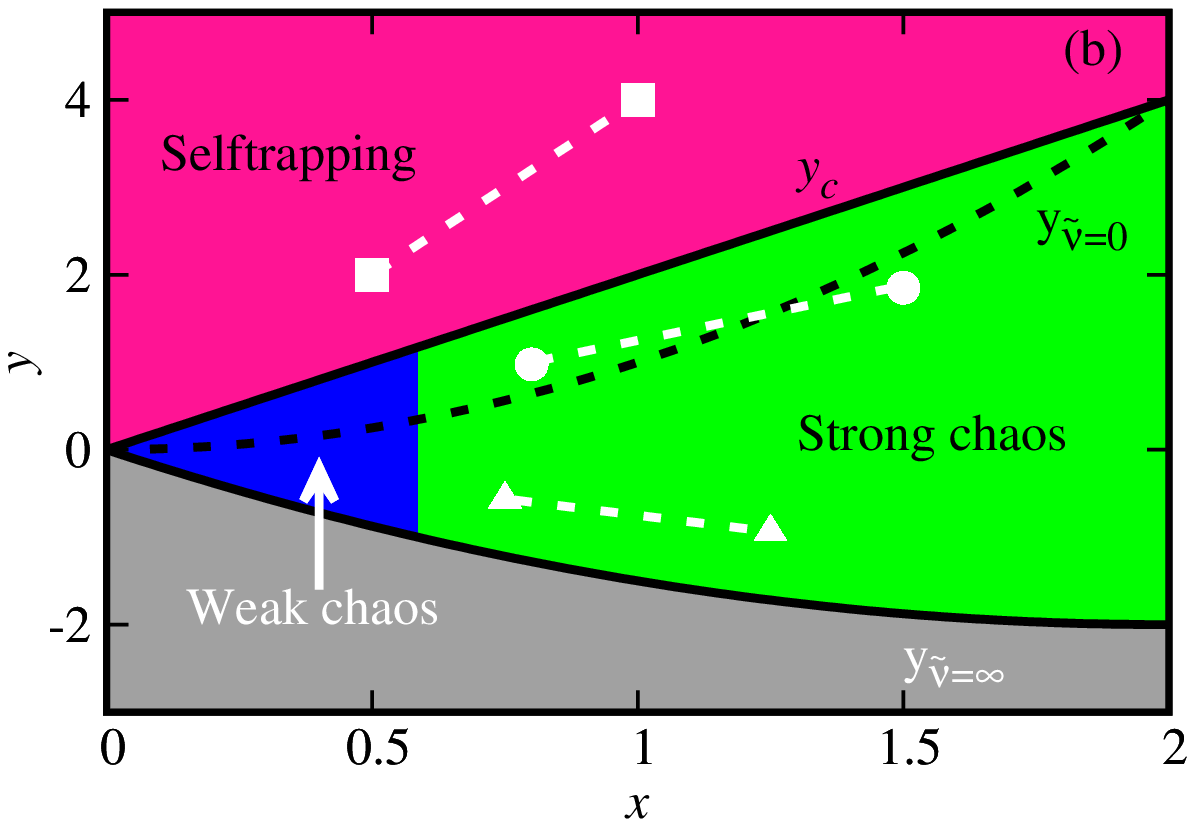}
    \caption{Parametric space of average norm $x$, {\it vs.} average energy $y$ per site of the 1D DDNLS model~\eqref{eq:Hamilton_complex_dnls_1d} in the thermodynamics limit. 
        (a) The bold lines $y_{\tilde{\nu} = \infty} = -2x + x^2/2$ and $y_{\tilde{\nu} = 0} = x^2$ delimit the thermal phase in the Gibbsean sense.
        The arrows inside the thermal regimes report the directions toward the seftrapping regime.
        (b) The three colored areas represent the weak chaos [blue], strong chaos [green] and sefftrapping [pink] regimes.
        The black bold curves $y_{\tilde{\nu} = \infty} = -2x + x^2/2$  and $y_c = 2x$ denotes the limit of the spreading regimes.
        The black dashed curve $y_{\tilde{\nu} = \infty} = x^2$ is the upper limit of thermal states in the sense of the grand-canonical distribution.
        The white dashed lines represent examples of trajectories followed by a system starting from the distant points when its state belong to the thermal phase with negative [triangles] and positive [circles] energies, and in the non-thermal phase [squares].
        }
    \label{fig:x_y_thermal_selftrapping_phases}
\end{figure}

We now study the various locations of the selftrapping regime in the $(x, y)$ parameter space of the 1D DDNLS model.
From Eq.~\eqref{eq:spreading_regime_1d_ddnls_1}, depending on whether or not
\begin{equation}
    x > \Delta,  
    \label{eq:selftrapping_x_y_stat}
\end{equation}
the initial wave packet will be almost fully trapped.
Since Eq.~\eqref{eq:selftrapping_x_y_stat} does not specify conditions on $y$ values, we loosely conclude that the selftrapping regime incorporate both Gibbs and non-Gibbs phases for large values of $x$ (the norm) in the energy-norm parameter space.
The question is to know if selftrapped wave packets also occur in the region $x<\Delta$.
The first clue comes from the fact that, in the non-Gibbs region i.e. $y > x^2$, long time localized nonlinear solutions (breathers) appear in the DNLS model which may support the onset of selftrapping of wave packets in the DDNLS systems~\citep{tsironis1996slow,mackay1994proof,flach1998discrete,rasmussen2000statistical,campbell2004localizing,johansson2004statistical,kopidakis2008absence,flach2008discrete,flach2016spreading,mithun2018weakly}. 
To be more specific, based on the work of~\citep{kopidakis2008absence,skokos2009delocalization} (see Sec.~\ref{subsec:selftrapping_theorem_I}), Eq.~\eqref{ineq:h_s_selftrapping} gives a rough boundary above which such selftrapping regime can be seen  
\begin{equation}
    y > 2x,
    \label{eq:selftrapping_x_y_3_stat}
\end{equation}
in the thermodynamic limit. 
Consequently, a part of the weak and strong chaos regimes when $x \approx d$ belongs to the non-Gibbs phase as depicted in Fig.~\ref{fig:x_y_thermal_selftrapping_phases}(b).
The best explanation for this, so far comes from~\citep{mithun2018weakly} where it was suggested that once crossing the curve $y_{\tilde{\nu} = 0}$ the dynamics of the system do not directly becomes fully non-ergodic.
Whence, spreading is still possible in that region.
Here, it is easy for us to relate the ergodicity of a dynamical system's orbits in its phase space to its chaoticity.
Roughly speaking, a regular orbit of the system is non-ergodic and a chaotic one is ergodic.   

\subsection{\label{subsec:wave_evolution_stat}Wave packet evolution in the energy-norm parameter space}
Let us look now at how trajectories of systems are represented in the $(x, y)$ parameter space (see Figs.~\ref{fig:x_y_thermal_selftrapping_phases}(a) and~(b)).
In the course of the wave packet spreading both the constant total energy and norm are shared to more sites.
Therefore, we expect the coordinates $x (t)$ and $y (t)$ of the system to decrease in time $t>0$ as the system is getting closer to the integrable limit.
A way to find the rate at which this happens is via a simple algebraic manipulation based on the conservation of the integrals of motions.
Indeed,
\begin{equation}
    \mathcal{H}_{1D}  = h(0) L = h(t) \Delta l(t) \quad \text{and} \quad \mathcal{S}_{1D} = s (0) L = s (t) \Delta l(t),
    \label{eq:rel_en_norm_xy_space}
\end{equation}  
Dividing the two equations in~\eqref{eq:rel_en_norm_xy_space} leads to~\citep{flach2016spreading}
\begin{equation}
    y(t) = \frac{y_0}{x_0}x(t),
    \label{eq:rel_x_y_xy_space}
\end{equation}
where in Eq.~\eqref{eq:rel_x_y_xy_space} the subscript `$0$' means that the observable is taken at time $t = 0$. 
From Eq.~\eqref{eq:rel_x_y_xy_space} we see that the system's evolution follows a line starting from $(x_0, y_0)$ toward the origin $(0, 0)$ at a constant rate $y_0/x_0$.
Three arguments arise.
Indeed, if the slope $y_0/x_0 \ge 2$ (see squared points in Fig.~\ref{fig:x_y_thermal_selftrapping_phases}(b))
the wave packet belongs to the selftrapping regime.
Further if $y_0/x_0 < 2$, two scenarios appear: for states with positive average energy per site (see circle points in Fig.~\ref{fig:x_y_thermal_selftrapping_phases}(b)) launched in the Gibbs region enters the non-Gibbs region, while those with negative average energy per site (see triangle points in Fig.~\ref{fig:x_y_thermal_selftrapping_phases}(b)) launched within the Gibbs region does not crossover to the non-Gibbs region as they never intersect the curve $y_{\tilde{\nu} = 0} = x^2$. 

\section{\label{sec:spreading_mechanism}Characteristic measures of wave packet spreading}
We now discuss some characteristics of wave packet spreading in the weak and strong chaos dynamical regimes.
In what follows we present results for the DDNLS models in one and two spatial dimensions which can be easily transferred to the DKG models.

\subsection{\label{subsec:spreading_mechamism_1d_sigma_2}The 1D DDNLS model}
As we have seen in Sec.~\ref{sec:dynamical_regimes_chirikov}, the interactions of the NMs in the 1D DDNLS system is governed by~\eqref{eq:1d_DDNLS_eq_motion_normal_modes_space}
\begin{equation}
    i \frac{\partial \phi _\nu}{\partial t} = E _\nu \phi _\nu + \beta \sum _{\nu _1, \nu_2, \nu_3} I_{\nu, \nu _1, \nu_2, \nu_3} \phi _{\nu _1}^\star \phi _{\nu _2} \phi _{\nu _3},
    \label{eq:1d_DDNLS_eq_motion_normal_modes_space_2}
\end{equation}
where we expanded the wave packet $\psi _l = \sum _\nu \phi _\nu A_{\nu, l}$ in the basis of the normal modes $A_{\nu, l} \propto \overline{\zeta} ^{-1/2}\exp (-(l -l_\nu)/\overline{\zeta})$ and  the prefactor $\overline{\zeta} ^{-1/2}$ ensures that the NM is normalized i.e $\sum _{l} A_{\nu, l}^2 = 1$.
Let us simplify Eq.~\eqref{eq:1d_DDNLS_eq_motion_normal_modes_space_2} in order to capture its main features.
The complex overlap between modes $I_{\nu, \nu _1, \nu_2, \nu_3} = \sum _{l} A_{\nu, l}^{\star}A_{\nu _1, l}^\star A_{\nu _2, l}A_{\nu _3, l } \propto \overline{\zeta}^{-3/2}\sum _{l} A_{\nu, l}^{\star} \propto \zeta^{-3/2}$~\citep{shepelyansky1993delocalization}.
Due to the localization of NMs we expect the sum in the right hand side of Eq.~\eqref{eq:1d_DDNLS_eq_motion_normal_modes_space_2} to possess approximately $\overline{\zeta}^3$ terms as we are summing over three modes whose only sites' contribution within the localization volumes are non-negligible.
Consequently the evolution of the amplitude of an exterior mode $\nu$ to the wave packet is asymptotically governed by~\citep{shepelyansky1993delocalization}
\begin{equation}
    i\frac{\partial \phi_\nu }{\partial t} \sim E_\nu \phi_\nu + \beta \phi_{\nu_1}^3.
    \label{eq:evol_transmission}
\end{equation}
From Eq.~\eqref{eq:evol_transmission}, we get 
\begin{equation}
    \left\lvert \dot{\phi}_\nu \right\rvert ^2 \sim \beta ^2 \lvert \phi_{\nu_1} \rvert ^6.
    \label{eq:evol_norm_01_shepelyansky}
\end{equation}
which gives us an estimation of the diffusion coefficient (also called momentary diffusion rate) 
\begin{equation}
    D \sim \beta^2 \lvert \phi _{\nu _1} \rvert^6,    
    \label{eq:diffusion_rate_shepelyansky}
\end{equation}
of influx of the norm from thermalized modes $\nu_1$ to non-thermalized ones $\nu$. 
It is worth pointing out that the diffusion coefficient $D$~\eqref{eq:diffusion_rate_shepelyansky} is time dependent (i.e. $D = D(t)$) through the mode's norm $\lvert \phi _{\nu_1}\rvert^2$ (see Eq.~\eqref{eq:evol_norm_01_shepelyansky}).
For the rest of our work, we will always consider that the diffusion process of the wave packet is well defined by the relation
\begin{equation}
    m_2 = Dt,
\end{equation}
with $m_2$ being the second moment of the wave wave packet and $D$ its momentary diffusion rate~\citep{shepelyansky1993delocalization,pikovsky2008destruction,flach2009universal,skokos2009delocalization,skokos2013nonequilibrium,senyange2018characteristics,manda2020chaotic}.
If we further assume that during the time for which the approximation~\eqref{eq:evol_transmission} holds, the wave packet width is $\Delta l$ such that the norm $\lvert \phi _{\nu _1}\rvert ^2$ is more or less equally distributed among all excited modes, the normalization condition 
\begin{equation}
    \Delta l \lvert \phi _{\nu_1}\rvert^2 \sim 1,
    \label{eq:normalization_condition_chap_02}
\end{equation}
leads to the relation $\lvert \phi _{\nu_1}\rvert^2 \sim 1/\Delta l$.
In addition, the wave packet second moment $m_2$ and participation number $P$ are related to $\Delta l$ as
\begin{equation}
    m_2 \propto (\Delta l)^2 \sim \frac{1}{\lvert \phi_{\nu_1}\rvert^4}, \quad P \propto \Delta l \sim \frac{1}{\lvert \phi_{\nu_1}\rvert^2},
    \label{eq:m2_p_characteristics_spreading}
\end{equation}
respectively.
Consequently, the wave packet second moment is 
\begin{equation}
    m_2 = Dt \sim \frac{1}{\lvert \phi _{\nu_1 }\rvert ^4},
    \label{eq:second_moment_general_1}
\end{equation}
so that 
\begin{equation}
    \frac{1}{\lvert \phi _{\nu_1} \rvert^4} \sim Dt \sim \beta^2 \lvert \phi _{\nu_1} \rvert^6 t \quad \Rightarrow \quad \frac{1}{\lvert \phi _{\nu_1} \rvert^{10}} \sim \beta^2 t \quad \Rightarrow \quad m_2 \sim \frac{1}{\lvert \phi _{\nu_1} \rvert^4} \sim \beta ^{4/5}t^{2/5}.
    \label{eq:diff_eq_schepelyansky_1}
\end{equation}
Thus we obtain
\begin{equation}
    m_2 \sim \beta ^{4/5} t^{2/5}, \quad  P \sim \beta ^{2/5} t^{1/5},
    \label{eq:shepelyansky_spreading_characteristic}
\end{equation}
which is exactly the estimation in~\citep{shepelyansky1993delocalization}.
It is important to note that the result $m_2\sim t^{2/5}$ is not in agreement with our numerical results (and many others~\citep[see~e.g.][]{flach2009universal,skokos2009delocalization,flach2010spreading,laptyeva2010crossover,skokos2010spreading,ivanchenko2014quantum,iomin2017subdiffusion,vakulchyk2019wave,kati2020density}).

The analysis presented by~\citep{shepelyansky1993delocalization,pikovsky2008destruction} was improved by~\citep{flach2009universal,flach2010spreading,krimer2010statistics}.
Indeed, based on the Chirikov criterion, resonant modes within the wave packet behave chaotically.
Thus, we can write
\begin{equation}
    \phi_{\nu_1}  = \phi _{\nu_1}^{r} + \phi _{\nu_1}^{c},
    \label{eq:superposition_mode_chaotic_regular_1}
\end{equation}
with $\phi _{\nu_1}^{r}$ and $\phi_{\nu_1} ^{c}$ being the regular and chaotic contributions to the wave function dynamics of mode $\nu_1$ respectively.
According to~\citep{flach2010spreading,krimer2010statistics}, only the modes which possess a non-negligible chaotic contribution are resonantly interacting~\citep{flach2009universal,flach2010spreading}.
We note that as the wave packet spreads, $\lvert \phi_{\nu_1} \rvert  \rightarrow 0$ taking the system toward the integrable limit as the nonlinear interaction diminishes, we expect the regular part to dominate and the chaotic contribution of resonant modes to decrease.
Furthermore, the probability $\mathcal{P}$ of a mode to be resonant depends on the nonlinear frequency shift i.e.~$\mathcal{P} (\beta \lvert \phi_{\nu_1}\rvert ^2)$.
Since the $\mathcal{P}$ value is between $0$ and $1$, this probability also represents the average fraction of chaotic modes within the wave packet
\begin{equation}
    \mathcal{P} (\beta  \lvert \phi_{\nu_1} \rvert ^2) \propto \frac{ \lvert \phi_{\nu_1} ^{c} \rvert }{ \lvert \phi_{\nu_1} \rvert }.
    \label{eq:fraction_chaotic_modes}
\end{equation}
Consequently, we rewrite Eq.~\eqref{eq:evol_transmission}
\begin{equation}
    i\frac{\partial \phi_\nu }{\partial t} \sim E_\nu \phi_\nu + \beta \mathcal{P} \left(\beta  \lvert \phi_{\nu_1} \rvert ^2\right) \phi_{\nu_1}^3f(t),
    \label{eq:evol_transmission_flach}
\end{equation}
where $f(t)$ accounts for the continuous spectrum of the chaotic fraction of thermalized modes $\mathcal{P} (\beta  \lvert \phi_{\nu_1} \rvert ^2)$ and corresponds to a Gaussian white noise so that $\langle f(t), f(t^\prime) \rangle = \delta (t -  t^\prime)$ (in opposition to the discrete spectrum of the linear background $E_\nu$).
The above equation~\eqref{eq:evol_transmission_flach} leads to the time-dependent norm influx from the thermalized mode $\nu_1$ to the cold one $\nu$
\begin{equation}
    \left\lvert \dot{\phi}_\nu \right\rvert ^2 \sim \beta ^2 \lvert \phi_{\nu _1} \rvert ^6 \left[\mathcal{P} \left(\beta \lvert \phi _{\nu_1} \rvert ^2\right)\right]^2 \quad \Rightarrow \quad \lvert \phi _{\nu}\rvert ^2 \sim \beta ^2 \lvert \phi_{\nu _1} \rvert ^6 \left[\mathcal{P} \left(\beta \lvert \phi _{\nu_1} \rvert ^2\right)\right]^2 t.
\end{equation}
Thus the mode $\nu$ reaches the same norm as mode $\nu_1$ i.e. $\lvert \phi _{\nu_1} \rvert^2$ after 
\begin{equation}
    T_M \sim \left(\beta ^2 \lvert \phi _{\nu_1} \rvert ^4 \left[\mathcal{P} \left(\beta \lvert \phi _{\nu_1} \rvert  ^2\right)\right]^2\right)^{-1},
\end{equation}
which is a characteristic timescale for the wave packet diffusion, such that its inverse defines a diffusion rate (see Secs.~\ref{subsec:mle_num},~\ref{sec:discussion_1d_spreading} and~\ref{sec:summary_2d_ddnls} for further details)
\begin{equation}
    T_M \sim \frac{1}{D} \quad \Rightarrow \quad  D \sim \beta ^2 \lvert \phi _{\nu_1} \rvert ^4 \left[\mathcal{P} \left(\beta \lvert \phi _{\nu_1} \rvert  ^2\right)\right]^2, 
    \label{eq:difussion_rate_flach}
\end{equation}
which depends on the portion of chaotically resonant modes $\mathcal{P}$ inside the wave packet.

Let us see now the result of the analysis for the strong and weak chaos spreading regimes. 
\begin{enumerate}
    \item[a)] {\bf Strong chaos}.
        As discussed in Sec.~\ref{sec:dynamical_regimes_chirikov}, in this case all the modes in the wave packet are chaotic, which means that $\mathcal{P} \approx 1$, we get from Eq.~\eqref{eq:difussion_rate_flach}
        \begin{equation}
            D \sim \beta ^2 \lvert \phi _{\nu _1} \rvert ^4.
            \label{eq:diffusion_ceofficient_strong_chaos}
        \end{equation}
        Then from Eqs.~\eqref{eq:second_moment_general_1} and~\eqref{eq:diffusion_ceofficient_strong_chaos} we obtain the estimation 
        \begin{equation}
            \frac{1}{\lvert \phi _{\nu_1} \rvert^4} \sim Dt \sim \beta ^2 \lvert \phi _{\nu_1} \rvert^4t \quad \Rightarrow \quad \frac{1}{\lvert \phi _{\nu_1} \rvert^8} \sim \beta^2 t \quad \Rightarrow \quad m_2 \sim \frac{1}{\lvert \phi _{\nu_1} \rvert^4} \sim \beta t^{1/2}.
        \end{equation} 
        Thus we get $m_2 \sim \beta t^{1/2}$ and $P \sim \beta^{1/2} t^{1/4}$ found in~\citep{flach2009universal,flach2010spreading,laptyeva2010crossover,iomin2017subdiffusion}.
    \item[b)] {\bf Weak chaos}.
        In this case, not all the modes within the wave packet interact chaotically.
        In particular $\mathcal{P} \approx C \beta \lvert \phi _{\nu_1} \rvert^2$, where $C$ is a prefactor~\citep{flach2010spreading,krimer2010statistics,laptyeva2010crossover}. 
        Following a similar approach as in the above case a), we obtain the diffusion coefficient 
        \begin{equation}
            D \sim C^2 \beta ^4 \lvert \phi_{\nu _1} \rvert^8.
        \end{equation}
        Consequently 
        \begin{equation}
            \frac{1}{\lvert \phi_{\nu _1} \rvert^4} \sim Dt \sim C^2 \beta^4 \lvert \phi_{\nu _1} \rvert^8 t \quad \Rightarrow \quad \frac{1}{\lvert \phi_{\nu _1} \rvert^{12}} \sim C^2 \beta^4 t \quad \Rightarrow \quad m_2 \sim \frac{1}{\lvert \phi_{\nu _1} \rvert^4} \sim C^{2/3}\beta^{4/3}t^{1/3}.
        \end{equation}
        Therefore $m_2 \sim t^{1/3}$ and $P\sim t^{1/6}$ exactly the estimation found in~\citep{flach2010spreading,laptyeva2010crossover}.
        \item[c)] {\bf Selftrapping}.
            In the case of selftrapping regime, a substantial part of the wave packet remain localized while the other portion spreads infinitly.
            Thus, we should expect the wave packet second moment to always increase as $m_2\sim t^{a_m}$ ($0< a_m< 1$) following the processes described above.
            However, when the norm of the wave packet background is small enough, the value of the participation number which indicates the number of highly excited sites saturates at a value which gives the number of sites effectively trapped $P_{tr}$.
            Therefore, we expect $P \sim P_{tr}t^{0}$ in the limit of very large wave packets. 

\end{enumerate}

\subsection{\label{subsec:spreading_mechamism_any_dimension_and_sigma}The 2D DDNLS model}
The analysis of Sec.~\ref{subsec:spreading_mechamism_1d_sigma_2} can be extended to any number of spatial dimensions, but we will consider only the 2D case~\citep{garcia2009delocalization,flach2010spreading,laptyeva2012subdiffusion}.
Indeed, the system's wave function can be written as $\psi _{l, m} =\sum_{\nu} \phi _{\nu} A_{\nu, l, m}$ where $A_{\nu, l, m} \propto \overline{\zeta} ^{-1} \exp\left(-\left[(l - \overline{l}_\nu )^2 - (m -  \overline{m}_\nu )^2\right]^{1/2}/\overline{\zeta}\right)$ with $(\overline{l}_\nu , \overline{m}_\nu)$ the spatial center of mode $\bm{\Psi}_\nu = \left(A_{\nu, 1, 1}, A_{\nu, 1, 2}, \ldots, A_{\nu, N, M}\right)$ and $\overline{\zeta}$ the average localization length of NMs over modes and disorder realizations (note that in 2D the $\overline{\zeta}$ can be considered as the radius or the side of a NM's volume).
Thus the evolution of the 2D DDNLS model in the NM space $A_{\nu, l, m}$ is still described by Eq.~\eqref{eq:1d_DDNLS_eq_motion_normal_modes_space_2} but now the overlap-integral is $I_{\nu, \nu _1, \nu_2, \nu_3} = \sum _{l, m} A_{\nu, l, m}^{\star}A_{\nu _1, l, m}^\star A_{\nu _2, l, m}A_{\nu _3, l, m} \propto \overline{\zeta}^{-3}\sum _{l, m} A_{\nu, l, m}^{\star} \propto \overline{\zeta}^{-3}$ and the number of terms in the sum in the right hand side of Eq.~\eqref{eq:1d_DDNLS_eq_motion_normal_modes_space} being $\overline{\zeta} ^{3\times 2}$ due to the localization of the modes $\nu_1$, $\nu_2$ and $\nu_3$~\citep{garcia2009delocalization}. 
This leads again in the approximations~\eqref{eq:evol_transmission_flach} to~\eqref{eq:difussion_rate_flach}.

The difference with the 1D case comes from the expression of the geometrical characteristics of the wave packet.
The normalization of the system's wave packet implies $\Delta l \lvert \phi_{\nu_1} \rvert^2 \sim 1$ where $\Delta l$ is the volume of the wave packet in 2 dimensions.
We express $\Delta l$ as
\begin{equation}
    \Delta l \propto \Delta R^2,
\end{equation}
where $\Delta R$ can be viewed as the side or the radius of the wave packet.
The value of $\Delta R$ is related to the second moment and participation number of the wave packet by~\citep{garcia2009delocalization,flach2010spreading}
\begin{equation}
    m_2 \propto \left(\Delta R\right)^2 \sim \frac{1}{\lvert \phi_{\nu_1}\rvert ^2}, \quad P \propto \Delta l = \left(\Delta R\right)^2\sim \frac{1}{\lvert \phi_{\nu_1}\rvert ^2}.
    \label{eq:m_2_P_2D_flach}
\end{equation}
From Eq.~\eqref{eq:m_2_P_2D_flach}, we see that the time dependence of the $m_2$ and $P$ are the same in 2D lattice. 
So in this case, we have the following behaviors.
\begin{enumerate}
    \item[a)] {\bf Strong chaos}. Repeating the same process as in case a) of Sec.~\ref{subsec:spreading_mechamism_1d_sigma_2} with $D \sim \beta \lvert \phi_{\nu_1} \rvert^4$, we get
        \begin{equation}
            \frac{1}{\lvert \phi_{\nu_1} \rvert^2} \sim Dt \sim \beta^2 \lvert \phi_{\nu_1} \rvert^4 t \quad \Rightarrow \quad \frac{1}{\lvert \phi_{\nu_1} \rvert^6} \sim \beta^2 t \quad \Rightarrow \quad m_2 \sim \frac{1}{\lvert \phi_{\nu_1} \rvert^2} \sim \beta^{2/3}t^{1/3}.
        \end{equation} 
        Consequently, we find the estimations $m_2 \sim t^{1/3}$ and $P\sim t^{1/3}$ of~\citep{flach2010spreading,laptyeva2012subdiffusion}.        
        \item[b)] {\bf Weak chaos}. In analogy to case b) in Sec.~\ref{subsec:spreading_mechamism_1d_sigma_2} where $D \sim C^2 \beta^4 \lvert \phi_{\nu_1} \rvert^8$, we obtain 
        \begin{equation}
            \frac{1}{\lvert \phi_{\nu_1} \rvert^2} \sim Dt \sim C^2 \beta^4 \lvert \phi_{\nu_1} \rvert^8 t \quad \Rightarrow \quad \frac{1}{\lvert \phi_{\nu_1} \rvert^{10}} \sim C^2\beta^4 t \quad \Rightarrow \quad m_2 \sim \frac{1}{\lvert \phi_{\nu_1} \rvert ^2} \sim C^{2/5}\beta^{4/5}t^{1/5}.
        \end{equation} 
        Thus we get $m_2 \sim t^{1/5}$ and $P\sim t^{1/5}$ which are the estimations obtained in~\citep{flach2010spreading,laptyeva2012subdiffusion}.
        \item[c)] {\bf Selftrapping}. Similar arguments as in case c) of Sec.~\ref{subsec:spreading_mechamism_1d_sigma_2} hold. 
            In addition, as the 2D NMs are more extended compared to the 1D counterpart, the spreading is faster. 
            Whence, we expect the saturation $P \sim P_{tr}t^{0}$ to be reached faster in the former system. 
\end{enumerate}

\section{\label{sec:incommensureate_wave_packet}Dynamics of extended initial wave packet}
At last we would like to build a formalism for a special case of initial wave packets which proved to be very useful for numerical experiments, as it constitutes an effective way for obtaining long lasting strong chaos behavior.
This is especially useful in 2D lattices where the final integration time is limited by computer capabilities and the boundaries of the strong chaos dynamical spreading regime are not very well defined~\citep{laptyeva2012subdiffusion,manda2020chaotic}.
Indeed all the results above (see Figs.~\ref{fig:x_y_thermal_selftrapping_phases},~\ref{fig:parameter_space_dnls_1d_delta_W} and~\ref{fig:parameter_space_dnls_2d_delta_W} and Sec.~\ref{subsec:spreading_regimes}) are presented for either single-site i.e.~$L = 1$ or block-site excitations of about the localization volume i.e.~$L \approx \overline{V}$ wave packets.
Now let us ask ourselves what happens for much larger initial wave packet excitations i.e.~$L \gg \overline{V}$ when selftrapping is avoided?
First, if such an extended wave packet is given a large average norm $s$ per site, we expect it to exhibit the characteristics of the strong chaos regime as the majority of the thermalized modes will chaotically interact (the effective nonlinearity per site is high).
On the other hand, if $s$ is small, both the strong and weak chaos spreading regimes are possible. 
More specifically, for the strong chaos spreading to be visible, the sum of the fractions of chaos within each wave packet's NMs should sum up to unity.
For the 1D DDNLS model~\eqref{eq:Hamilton_complex_dnls_1d}, we get 
\begin{equation}
    \sum_l C\beta s \sim 1 \Rightarrow x L \sim\frac{1}{C},
    \label{eq:counting_nm_incommensurate}
\end{equation}
where $x = \beta s$, with $s$ being the constant norm at each excited site and the prefactor $C$ introduces a disorder dependence on the fraction of chaotic NMs~\citep{skokos2009delocalization,flach2010spreading}. 
Note that here, we assume that the number of NMs within the wave packet is approximately equal to the total number of excited sites.
This leads to the estimation 
\begin{equation}
    L \sim \frac{W}{x},
    \label{eq:size_incommensurate_wave}
\end{equation}
with $C \sim 1/W$ for the 1D DDNLS model~\citep{krimer2010statistics,flach2010spreading}.
In the 2D DDNLS model~\eqref{eq:hamilton_complex_ddnls_2d}, we obtain 
\begin{equation}
    \sum_{l,m} C\beta s \sim 1 \Rightarrow L^2 \sim \frac{1}{Cx}.
    \label{eq:size_incommensurate_wave_2d}
\end{equation}
In Eq.~\eqref{eq:size_incommensurate_wave_2d} the dependence of $C$ on the disorder parameter $W$ is not known.
Nevertheless, we can expect $C = C(1/W)$ similar to the 1D DDNLS model~\eqref{eq:size_incommensurate_wave} [$C$ decreases with increasing disorder strength].
In this context, for large enough disorder parameter $W$, we can loosely assume that $C \sim 1/W$.

\section{\label{sec:conclusion_anderson_localization_spreading}Summary}
In this chapter, we reviewed the notion of AL and analyzed both quantitatively and qualitatively its fate when nonlinearity is introduced.
We focussed on the 1D and 2D DDNLS lattice models, but connections to the DKG systems were also briefly presented.
When AL is destroyed, three dynamical behaviors are observed: the weak and strong chaos spreading regimes, as well as selftrapping case where a substantial part of the wave packet remains localized.
The characteristics of these regimes were analyzed based on the Chirikov criterion about the onset of chaos in nonlinear systems, as well as the study of selftrapping presented in~\citep{kopidakis2008absence} and the location of these regimes in an appropriate parameter space (namely disorder strength $W$ and average nonlinear frequency shift $\delta$) was shown.

A further analysis based on statistical physics ideas was conducted in the parameter space of the rescaled average norm $x = \beta s$ and energy $y = \beta h$ densities where we identified the existence of Gibbs and non-Gibbs regions and their relation with the various dynamical regimes.
Then, we presented the analytic derivation of the evolution of the wave packet's second moment $m_2$ and participation number $P$.
In particular we have shown that 
\begin{equation}
    m_2 \sim 
        \left\{
            \begin{array}{l}
                C^{2/3}\beta^{4/3} t^{1/3}  \\
                \beta t^{1/2} 
            \end{array}
        \right.
        , \quad 
        P \sim 
        \left\{
            \begin{array}{ll}
                C^{1/3}\beta^{2/3} t^{1/6} & \mbox{ weak chaos,} \\
                \beta^{1/2} t^{1/4} & \mbox{ strong chaos,}
            \end{array}
        \right.
\end{equation}
in the 1D case, while for its 2D counterpart, we obtained 
\begin{equation}
        m_2 = P \sim 
        \left\{
            \begin{array}{ll}
                C^{2/5}\beta^{4/5} t^{1/5} & \mbox{ weak chaos,} \\
                \beta^{2/3} t^{1/3} & \mbox{ strong chaos.}
            \end{array}
        \right.
\end{equation}
Furthermore, in the selftrapping regime of both the 1D and 2D systems, we asymptotically expect 
\begin{align}
    m_2 \sim t^{a_m}, \quad P \sim P_{tr}t^{0},  \\ 
\end{align}
where $a_m$ depends on the characteristics of chaos of the spreading part of the wave packet, while $P_{tr}$ is the number of sites which remain highly thermalized (the number of sites the system's energy/norm is effectively trapped) in comparison to the wave packet's background.

Finally we also discussed the case of an initial wave packet which extends much further than the system's localization volume $\overline{V}$ and has a small average norm $s$ per site, showing that such an excitation can still evolve in the strong chaos regime.
This result will be of practical use to us, especially when we will discuss the dynamics of the 2D DDNLS system in Chap.~\ref{chap:chaotic_two_dimensional}.

\chapter{\label{chapter:num_integration}Computational techniques}
\pagestyle{fancy}
\fancyhf{}
\fancyhead[OC]{\leftmark}
\fancyhead[EC]{\rightmark}
\cfoot{\thepage}

\section{\label{sec:intro_num_integration}Introduction}
In this chapter, we present the main numerical techniques we use in this thesis.
More specifically, we discuss the wave packet's initialization and present the quantities we measure to characterize its evolution along with the system's chaoticity.
A large part of this chapter is devoted to designing numerical integration schemes which efficiently follow the time evolution of wave packets.
A particular emphasis is placed on a special type of numerical method called symplectic integrators (SIs) which preserve the phase space volume of conservative Hamiltonian systems, a property which is translated to the accurate conservation of the system's energy (i.e.~the value of the Hamiltonian function itself) as its numerical error remains bounded for all times.
This behavior makes SIs ideal numerical schemes for long time evolution of autonomous (conservative) Hamiltonian systems.
In addition, the composition of simple low order SIs leads to new SIs with improved accuracy.
We investigate the efficiency of these methods for the evolution of the equations of motion and the variational equations of the 1D and 2D DDNLS models.
A large part of the material presented in this chapter is based on results published in~\citep{danieli2019computational}.

The chapter is organized as follows: In Sec.~\ref{sec:wavepacket_characterisation} we define the quantities which characterize the wave packet's spatial extent, while in Sec.~\ref{sec:initialization_wave} we discuss how we set up the initial conditions for our numerical simulations.
The framework of our numerical integration techniques along with the notions of the MLE and DVD are presented in Sec.~\ref{sec:tangent_map_method}.
In Sec.~\ref{sec:slope_determination_num}, we explain the method used to obtain the power law exponents from scattered data sets.
Section ~\ref{sec:nonSI_num} is devoted to the presentation of standard integration techniques which can be used to integrate a general system of differential equations, while in Sec.~\ref{sec:sis_num} we develop various SI schemes which we implement for integrating simultaneously the equations of motion and the variational equations of both the 1D and 2D DDNLS systems.
Then in Sec.~\ref{sec:num_comparison_num} all integrators are compared and ranked according to their accuracy, efficiency and speed.
Finally in Sec.~\ref{sec:conclusion_num} we summarize the content of this chapter.

\section{\label{sec:wavepacket_characterisation}Wave packet characterization}
For the DDNLS model in one~\eqref{eq:Hamilton_complex_dnls_1d} and two~\eqref{eq:hamilton_complex_ddnls_2d} spatial dimensions, the wave packet extent is represented by the normalized norm density distribution~\citep{flach2009universal,senyange2018characteristics,manda2020chaotic}
\begin{equation}
    \xi _{\bm{r}} = \frac{s_{\bm{r}}}{\sum _{\bm{r}} s_{\bm{r}}},
    \label{eq:norm_density_distribution_num}
\end{equation}
where $s_{\bm{r}} = \lvert \psi_{\bm{r}}\rvert^2$ is the local norm at site $\bm{r} = l$ in the 1D case, and $\bm{r} = (l, m)$ in the 2D lattice.
If the norm is normalized so that $\sum _{\bm{r}} s_{\bm{r}} = 1$, the $\xi _{\bm{r}}$ and $s_{\bm{r}}$ are equal, meaning that the two quantities are equivalent.
So in many place in this thesis we refer to $s _{\bm{r}}$ as simply the norm density.
For the sake of completeness, let us note that for the DKG system, the normalized energy density~\citep{skokos2009delocalization,laptyeva2010crossover,senyange2018characteristics,manda2020chaotic} 
\begin{equation}
    \xi _{\bm{r}} = \frac{h_{\bm{r}}}{\sum _{\bm{r}} h_{\bm{r}}},
    \label{eq:norm_density_distribution_num_02}
\end{equation}
is considered, where 
\begin{equation}
    h _{\bm{r}} = \frac{p_{\bm{r}}^2}{2} + \frac{\tilde{\epsilon} q_{\bm{r}}^2}{2} + \frac{q_{\bm{r}}^4}{4} + \frac{1}{4W} \sum _{\bm{n}} \left(q_{\bm{n}} - q_{\bm{r}}\right)^2,
\end{equation}
is the local energy at site $\bm{r}= l$ [$\bm{r} = (l, m)$] in the 1D [2D] model and $\sum _{\bm{n}}$  represents a summation along all nearest neighbors of site $\bm{r}$. 

We find the position of the center of the wave packet i.e. the first moment of the norm distribution as 
\begin{equation}
    \overline{\bm{r}} = \sum _{\bm{r}} \bm{r} \xi _{\bm{r}},
    \label{eq:first_moment_Chap_03}    
\end{equation}
while the squared distance from the center~\eqref{eq:first_moment_Chap_03} i.e.~the second moment
\begin{equation}
    m_2  = \sum _{\bm{r}} \left\lVert \bm{r} - \overline{\bm{r}}\right\rVert ^2 \xi _{\bm{r}},
    \label{eq:second_moment_distr_num}
\end{equation}
of the norm distribution is used to quantify the extent of the wave packet.
As an alternative way to quantify the wave packet size, we also compute the so-called participation number (see also Eqs.~\eqref{eq:loc_vol_p} and~\eqref{eq:pn_selftrapping_theorem} in Chap.~\ref{chap:spreading})
\begin{equation}
    P  = \frac{1}{\sum _{\bm{r}} \xi _{\bm{r}} ^2},
    \label{eq:participation_ratio_num}
\end{equation}
which measures the number of highly excited sites.

\section{\label{sec:initialization_wave}Wave packet initialization}
The fact that the DDNLS model in one and two spatial dimensions admits two integrals of motion, namely the total energy (Eqs.~\eqref{eq:Hamilton_complex_dnls_1d} and~\eqref{eq:hamilton_complex_ddnls_2d} for 1D and 2D systems respectively) and norm (Eqs.~\eqref{eq:norm_complex_dnls_1d} and~\eqref{eq:norm_complex_dnls_2d} for 1D and 2D systems respectively) requires some special care in setting up the the packet's initialization in order to create our initial condition at a particular point in the $(x, y)$ plane of Figs.~\ref{sec:statistical_physics_spreading} and~\ref{fig:x_y_thermal_selftrapping_phases} (1D DDNLS model) and Fig.~\ref{fig:parameter_space_dnls_2d_delta_W} (2D DDNLS system), where $x = \beta s$~\eqref{eq:rescaled_densities_stat} is the rescaled average norm density; which is equivalent to the average nonlinear frequency shift at each oscillator and $y = \beta h$~\eqref{eq:rescaled_densities_stat} the rescaled energy density. 
Let us focus on the 1D DDNLS system.
In our numerical computation (and as suggested in Secs.~\ref{sec:dynamical_regimes_chirikov} and~\ref{sec:statistical_physics_spreading}), we want to initialize a wave packet of $L$ consecutive sites with $\psi _l = \sqrt{s}\exp i\theta_l$ at the center of our lattice such that the norm~\eqref{eq:norm_complex_dnls_1d} and the Hamiltonian~\eqref{eq:Hamilton_complex_dnls_1d} give
\begin{equation}
    \mathcal{S}_{1D} = Ls, \quad 
    \mathcal{H}_{1D} = \sum _{l} \epsilon _l s + \frac{\beta}{2} s^2 - 2s \cos \Delta \theta _l,
    \label{eq:hamilton_1d_ddnls_stat_mech_num}
\end{equation}
with $\Delta \theta _l = \theta _{l + 1} - \theta _l$ being the dephasing between adjacent sites and the summation in $\mathcal{H}_{1D}$ counts only the $L$ initially excited sites.
In what follow, we describe two approaches in obtaining such initial states for specific $x$ and $y$ values. 

\subsection{\label{subsec:method_1_initialization_1d_ddnls_num}Approach 1: Thermodynamical averaging}
In this approach, we approximate the Hamiltonian in Eq.~\eqref{eq:hamilton_1d_ddnls_stat_mech_num} in the thermodynamic limit~\citep{flach2016spreading,kati2020density}, $\epsilon = \lim _{N\rightarrow \infty} N^{-1}\epsilon_l=0$ and $\Delta \theta= \lim_{N\rightarrow \infty} \left(\sum_{l}\mathcal{P}_\theta \right)^{-1} \mathcal{P}_\theta \Delta \theta _l$ with $\mathcal{P}_\theta$ the probability distribution of the phases.
Consequently from Eq.~\eqref{eq:hamilton_1d_ddnls_stat_mech_num}, we obtain 
\begin{equation}
    h = \frac{\beta}{2}s^2 - 2s \cos \Delta \theta \quad \Rightarrow \quad y =  \frac{1}{2}x^2 - 2x \cos \Delta \theta.
    \label{eq:a_h_equation_1d_ddnls_method_1_num}
\end{equation}
In order to obtain a specific set of ($x$, $y$) values, we have to set the dephasing angle $\Delta \theta$ to the constant value
\begin{equation}
    \Delta \theta = \arccos \left[\frac{x}{4} - \frac{y}{2x}\right].
    \label{eq:dephasing_method_1_constant_numerical_method}
\end{equation}
This arrangement leads to the following relation for the phases  $\theta _l$ of initially excited sites
\begin{equation}
    \theta _l \approx l \Delta \theta + \theta _0,
\end{equation} 
with $\theta_0$ being the phase at the first site.

This method for setting our initial condition is quite effective for extended initial wave packet which can be considered to be close to the thermodynamic limit.
For not-too-extended wave packets, we cannot neglect the contribution of the disorder parameter values at each site and the approximation of Eq.~\eqref{eq:a_h_equation_1d_ddnls_method_1_num} is not very accurate.
In this case, we find an initial condition leading to a point $(x, y^\prime)$ which is close to the initially desired $(x, y)$.
We consider different disorder realizations, i.e.~different sets of $\epsilon_l$ values which generate a plethora of points $(x, y^\prime)$ and keep in our analysis the ones which are the closest to the desired point $(x, y)$.

\subsection{\label{subsec:method_2_initialization_1d_ddnls_num}Approach 2: Setting specific energy per site}
In order to overcome the problem of not accurately obtaining a specific $(x, y)$ point, which as we already explained becomes more important for not very extended initial wave packets, we will take into account here the exact values of the $\epsilon_l$ and $\theta_l$ and accordingly tune the energy $h_l$ of each site individually.
For the initial block-site excitation $\psi_l = \sqrt{s}\exp i\theta_l$ of size $L$, described in the first lines of Sec.~\ref{subsec:method_1_initialization_1d_ddnls_num}, we require the $h_l$ values to satisfy Eq.~\eqref{eq:hamilton_1d_ddnls_stat_mech_num} i.e~for a given norm $s$ per site
\begin{equation}
    \sum _l \epsilon_l s + \frac{\beta}{2}s^2 - 2s\cos \Delta \theta _l = \sum _l h_l,  
    \label{eq:hamilton_identification_method_2_1d_ddnls_num}
\end{equation}
where again, the index $l$ corresponds to only initially excited sites.

We can have infinitely many choices of $h_l$ so that Eq.~\eqref{eq:hamilton_identification_method_2_1d_ddnls_num} is satisfied.
The simplest one is to set all these values equal to a constant term i.e~$h_l = h$.
Inserting this choice in Eq.~\eqref{eq:hamilton_identification_method_2_1d_ddnls_num} and multiplying by $\beta$, we get the following set of nonlinear equations 
\begin{equation}
    \sum _l \epsilon_l x + \frac{x^2}{2} - 2x\cos \left(\theta _{l+1} - \theta _l\right)  = \sum _l y. 
    \label{eq:hamilton_identification_method_2_1d_ddnls_num_2}
\end{equation}
Since the desired $x$ and $y$ values are given along with the parameters $\epsilon_l$ defining the specific disordered realization, system~\eqref{eq:hamilton_identification_method_2_1d_ddnls_num_2} corresponds to a set of $L$ equations with $L$ unknowns, the angles $\theta_l$.
Consequently, a root finding algorithm (e.g. Newton-Raphson) could be used to solve~\eqref{eq:hamilton_identification_method_2_1d_ddnls_num_2} in order to get the initial wave packet of the system.
Since these angles appear in~\eqref{eq:hamilton_identification_method_2_1d_ddnls_num_2} only through their difference $\theta_{l + 1} - \theta_l$, we have some flexibility in their determination.
For example, setting $\theta _1$ to be equal to a specific value, let's say $\varTheta _1$, the solution of set~\eqref{eq:hamilton_identification_method_2_1d_ddnls_num_2} is straightforward
\begin{equation}
    \begin{split}
        \theta_1 &= \varTheta _1,\\
        \theta _ 2 &= \arccos \left[\frac{x}{4} - \frac{y}{2x} + \frac{\epsilon _2}{2}\right] + \varTheta_1, \\
        \theta _ 3 &= \arccos \left[\frac{x}{4} - \frac{y}{2x} + \frac{\epsilon _3}{2}\right] + \varTheta _2, \\
        \vdots &= \vdots \\
        \theta _ L &= \arccos \left[\frac{x}{4} - \frac{y}{2x} + \frac{\epsilon _L}{2}\right] + \varTheta _{L - 1}.
    \end{split}
    \label{eq:solution_of_eq_method_2_1d_ddnls_num}
\end{equation}
In our computer code, we are using Eq.~\eqref{eq:solution_of_eq_method_2_1d_ddnls_num}.
In practice, we still have to iterate through several sets of disorder realizations, i.e.~$\epsilon_l$ sequences, in order to find the ones which match the best with the solution~\eqref{eq:solution_of_eq_method_2_1d_ddnls_num}.
This is because the way we are treating endpoints in Eq.~\eqref{eq:solution_of_eq_method_2_1d_ddnls_num} matters.
The difference with the previous method is the number of iterations, which appears to be drastically smaller for some types of initial conditions.
This method needs still to be refined, something that we intend to do in the future.

The approach 1 in Sec.~\ref{subsec:method_1_initialization_1d_ddnls_num} can be easily translated to the 2D DDNLS system, while in the approach 2 of Sec.~\ref{subsec:method_2_initialization_1d_ddnls_num} the size of the system~\eqref{eq:hamilton_identification_method_2_1d_ddnls_num_2} increases exponentially with the initial wave packet volume and the form of the solution~\eqref{eq:solution_of_eq_method_2_1d_ddnls_num} is more subtle.
Consequently, in most of the work performed within our simulations for the lattice initialization, we use the approach 1 of Sec.~\ref{subsec:method_1_initialization_1d_ddnls_num} for the 2D DDNLS system and the approach 2 presented in Sec.~\ref{subsec:method_2_initialization_1d_ddnls_num} in the  case of the 1D DDNLS model.

\section{\label{sec:tangent_map_method}Variational equations}
Let us briefly discuss the time evolution of small perturbations to an orbit of a Hamiltonian system needed for the computation of various chaos indicators like the MLE that we will implement in our study.
The Hamilton equations of motion, for an autonomous Hamiltonian system of $N$ degrees of freedom have the form
\begin{equation}
    \frac{d\bm{q}}{dt} = \frac{\partial \mathcal{H}}{\partial \bm{p}}, \qquad \frac{d\bm{p}}{dt} = -\frac{\partial \mathcal{H}}{\partial \bm{q}} ,
    \label{eq:eq_mot_general_var_eq_1}
\end{equation}
where $\mathcal{H} = \mathcal{H}\left( \bm{q}, \bm{p}\right)$ is the system's Hamiltonian function, with $\bm{q} = \left(q_1~q_2 \ldots q_N\right)^T$ and $\bm{p} = \left(p_1~p_2 \ldots p_N\right)^T$  being respectively the generalized positions and momenta. 
Here the superscript $({}^{T})$ denotes the matrix transpose. 
The equations of motion~\eqref{eq:eq_mot_general_var_eq_1} can be expressed in the general setting of ordinary differential equations (ODEs) as 
\begin{equation}
    \frac{d\bm{z}}{dt} =\dot{\bm{z}} = \bm{J}_{2N} \cdot \bm{D}_{\mathcal{H}} \left(\bm{z}(t) \right),
    \label{eq:generalz}
\end{equation}
where $\bm{z}=\left(\bm{q}~\bm{p}\right)^T = \left(z_1~z_2~\ldots~z_N~z_{N + 1}~\ldots~z_{2N}\right)^T = \left(q_1~q_2~\ldots~q_N~p_1~\ldots~p_N\right)^T$ is a vector representing the position of the system in its phase space and $(~\dot{ }~)$ denotes differentiation with respect to the time $t$. 
In Eq.~\eqref{eq:generalz}
\begin{equation}
\bm{J}_{2N} = 
\begin{pmatrix}
\bm{O}_N & \bm{I}_N \\
- \bm{I}_N & \bm{O}_N
\end{pmatrix},
\end{equation}
is the symplectic matrix with $\bm{I}_N$ and $\bm{O}_N$ being the $N\times N$ identity and null matrices respectively, and 
\begin{equation}
\bm{D}_\mathcal{H} = \begin{pmatrix}
\frac{\partial \mathcal{H}}{\partial q_1}~\frac{\partial \mathcal{H}}{\partial q_2}~\ldots~\frac{\partial \mathcal{H}}{\partial q_N}~\frac{\partial \mathcal{H}}{\partial p_1}~\ldots~\frac{\partial \mathcal{H}}{\partial p_N}
\end{pmatrix}^T.
\end{equation}

The evolution of a small perturbation $\bm{w}(0) = \delta \bm{z} (0) =  (\delta z_1(0)~\delta z_2(0)~\ldots~\delta z_N(0)~\delta z_{N + 1}(0) \allowbreak \ldots~\delta z_{2N(0)}) = (\delta q_1(0)~\delta q_2(0)~\ldots~\delta q_N(0)~\delta  p_1(0)~\ldots~\delta p_N(0))$ which is also refereed to as the deviation vector (DV) to the system's orbit $\bm{z}(0)$ evolves in the system's {\it tangent space}.
The system's {\it tangent dynamics} describing the evolution of $\bm{w}(t)$ is governed by the so called variational equations (see e.g.~\citep{skokos2010lyapunov, skokos2010numerical})
\begin{equation}
\frac{d \bm{w}}{dt}(t) = \dot{\bm{w}}(t) = \left[ \bm{J}_{2N}\cdot \bm{D}^2_{\mathcal{H}} \left( \bm{z} (t)\right) \right] \bm{w}(t)
\label{eq:generalw},
\end{equation}
where
\begin{equation}
 \bm{D}^2_{\mathcal{H}} \left( \bm{z} (t)\right) _{i, j} = \left.\frac{\partial ^2 \mathcal{H}}{\partial z_i \partial z_j} \right\rvert _{\bm{z} (t)},
\end{equation}
is the $2N\times 2N$ Hessian matrix of the system evaluated at the phase space trajectory $\bm{z} (t)$.
We emphasize that Eq.~\eqref{eq:generalw} is linear in $\bm{w}(t)$ with coefficients depending on the system's trajectory $\bm{z}(t)$.
Therefore, one has to integrate the variational equations~\eqref{eq:generalw} along with the equations of motion~\eqref{eq:generalz} in order to follow the evolution of the DV. 
In practice, we create an extended vector $\bm{Z} (t) = \left( \bm{z}(t)~\delta \bm{z}(t)\right)^T$ and numerically solve the unified system of ODEs~\citep{skokos2010numerical,gerlach2011dynamical,gerlach2012efficient}
\begin{equation}
    \frac{d\boldsymbol{Z}}{dt} = 
    \bm{f}\left(\bm{z}, \delta \bm{z}\right)
    = 
    \begin{pmatrix}
        \dot{\bm{z}} \\ 
        \dot{\delta \bm{z}} 
    \end{pmatrix}
    =
    \begin{pmatrix}
        \bm{J}_{2N} \cdot D_\mathcal{H} \left( \bm{z}(t) \right) \\
        \left(\bm{J}_{2N} \cdot \bm{D}^2_{\mathcal{H}} \right) \cdot \delta \bm{z}(t)
    \end{pmatrix}.
    \label{eq:generalZ}
\end{equation}
The vector function $\bm{f}$ is often called the vector field of~\eqref{eq:generalZ}, and is tangent at every point to the system's orbit $\bm{Z}(t)$~\citep{schroers2011ordinary}.

\subsection{\label{subsec:mle_num}Maximum Lyapunov exponent}
The variational equations constitute a very important tool in our effort to characterize the chaoticity of Hamiltonian systems. 
A system's chaotic behavior is associated with an exponential rate of growth of the DV. 
This is quantified by computing the so-called finite time maximum Lyapunov exponent (ftMLE)~\citep{benettin1980lyapunov,skokos2010lyapunov,pikovsky2016lyapunov} 
\begin{align}
    \Lambda (t) = \frac{1}{t} \ln\left( \frac{\lVert \bm{w}(t) \rVert}{\lVert \bm{w}(0) \rVert} \right),
    \label{eq:finite_mle}
\end{align}
at each time $t$ of the evolution, with $\bm{w}(t)$ and $\bm{w}(0)$ being respectively the DV at time $t > 0$ and $t = 0$. 
Therefore, the MLE $\Lambda_1$ is 
\begin{equation}
    \Lambda _1 = \lim _{t \rightarrow \infty} \Lambda  (t).
    \label{eq:mle}
\end{equation}
The time evolution of $\Lambda (t)$ can discriminate between regular and chaotic orbits in a rather straightforward way. 
Indeed, for regular orbits we expect the evolution of $\bm{w}(t)$ to be sub-exponential so that $\Lambda (t)$ tends to zero following a power law~\citep{benettin1980lyapunov, skokos2010lyapunov}
\begin{equation}
\Lambda (t) \propto t ^{-1},
\label{eq:fin_lyap_regular}
\end{equation}
otherwise the orbit is said to be chaotic.
In the latter case, we can define approximately how much time the system needs to become chaotic~\citep{skokos2010lyapunov,pikovsky2016lyapunov},
\begin{equation}
    T_L = \frac{1}{\Lambda _1 },
    \label{eq:lyapunov_time_chap_03}
\end{equation}
refered to as Lyapunov time.

\subsection{\label{subsec:dvd_num}Deviation vector distribution}
As already mentioned, the ftMLE measures the average exponential rate of growth (or shrinking) of perturbations of an orbit and it is influenced by the global dynamics of the system.
Therefore is a macroscopic observable, influenced by the overall dynamics.
Trying to define a quantity which measures the local contribution of perturbations to the chaotic dynamics of the system, we compute what is called DVD~\citep{skokos2013nonequilibrium,senyange2018characteristics,hillebrand2019heterogeneity,miranda2019contribution,ngapasare2019chaos,manda2020chaotic} defined as the distribution of quantities
\begin{equation}
    \xi _l ^D = \frac{\delta q _l^2 + \delta p _l^2  }{\sum _{l} \delta q _l^2 + \delta p _l^2 },
    \label{eq:DVD_definition_num}
\end{equation}
where $\delta q_l$ and $\delta p_l$ are the components of the deviation vector $\bm{w}$ and $l$ is the lattice's site.
This distribution attains its highest values at sites which are more sensitive to small changes of initial conditions i.e.~more chaotic.
Consequently, it is a very useful tool to visualize the location of chaotic contributions (see the papers cited above).

Similarly to what we did for the normalized norm/energy density $\xi$, during our simulations we also calculate the statistical properties of the DVD $\xi ^{D}$.
We compute the first $\overline{l}^{D}$ and second $m_2^D$ moments which measure respectively the center's coordinates and the extent of the distribution.
In addition, we also evaluate the participation number $P^D$ of the DVD which gives the number of highly chaotic sites.
It is worth noting that $P^D$ has already been used to study the localization of chaos in the FPUT lattice~\citep{dauxois1997modulational} and DNA models~\citep{barre2001lyapunov,hillebrand2019heterogeneity}.

\section{\label{sec:slope_determination_num}Numerical determination of power law exponents}
As we will see in the two next chapters in several places we will need to accurately determine the power law exponent of the time evolution of dynamical quantities.
Since the numerical data will not be smoothly located on a curve, we need a reliable numerical approach to determine such exponents in expected behavior of the form $Q(t) = Q_0t^{\alpha _Q}$.
The methodology we briefly describe here has been implemented successfully in many studies~\citep{laptyeva2010crossover,bodyfelt2011nonlinear,bodyfelt2011wave,laptyeva2012subdiffusion,laptyeva2013nonlinear,skokos2013nonequilibrium,laptyeva2014nonlinear,yusipov2017quantum,senyange2018characteristics,vakulchyk2019wave,manda2020chaotic}.

Let us present the whole process by considering that our observable $Q$ grows as $Q (t) = t^{2/5}$.
As mentioned, the raw data obtained from the numerical simulations are not smooth.
In order to mimic this behavior in this simple example, we create a time series of data by introducing a random noise to the exact values obtained by the law $t^{2/5}$ whose magnitude depends on the expected $Q$ values.
\begin{figure}[!htb]
    \centering 
    \includegraphics[width=0.49\textwidth, height=0.5\linewidth]{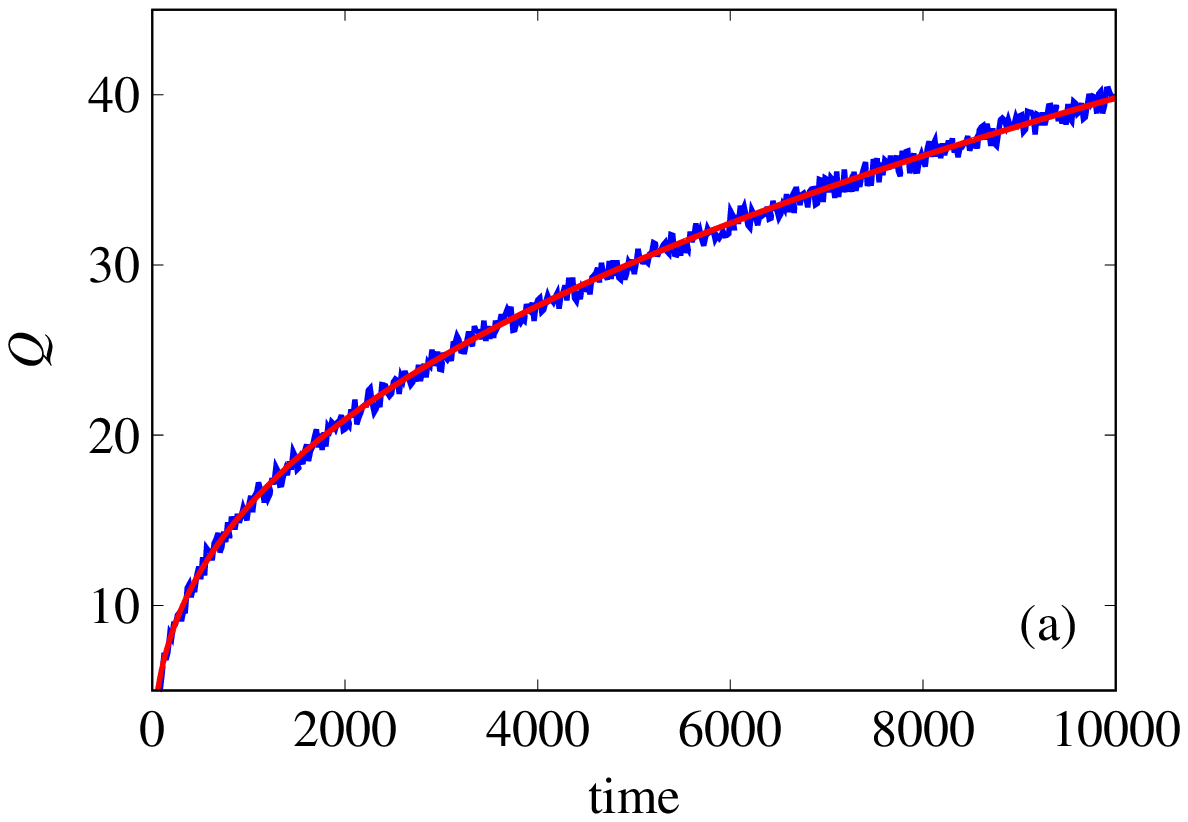}
    \includegraphics[width=0.49\textwidth, height=0.5\linewidth]{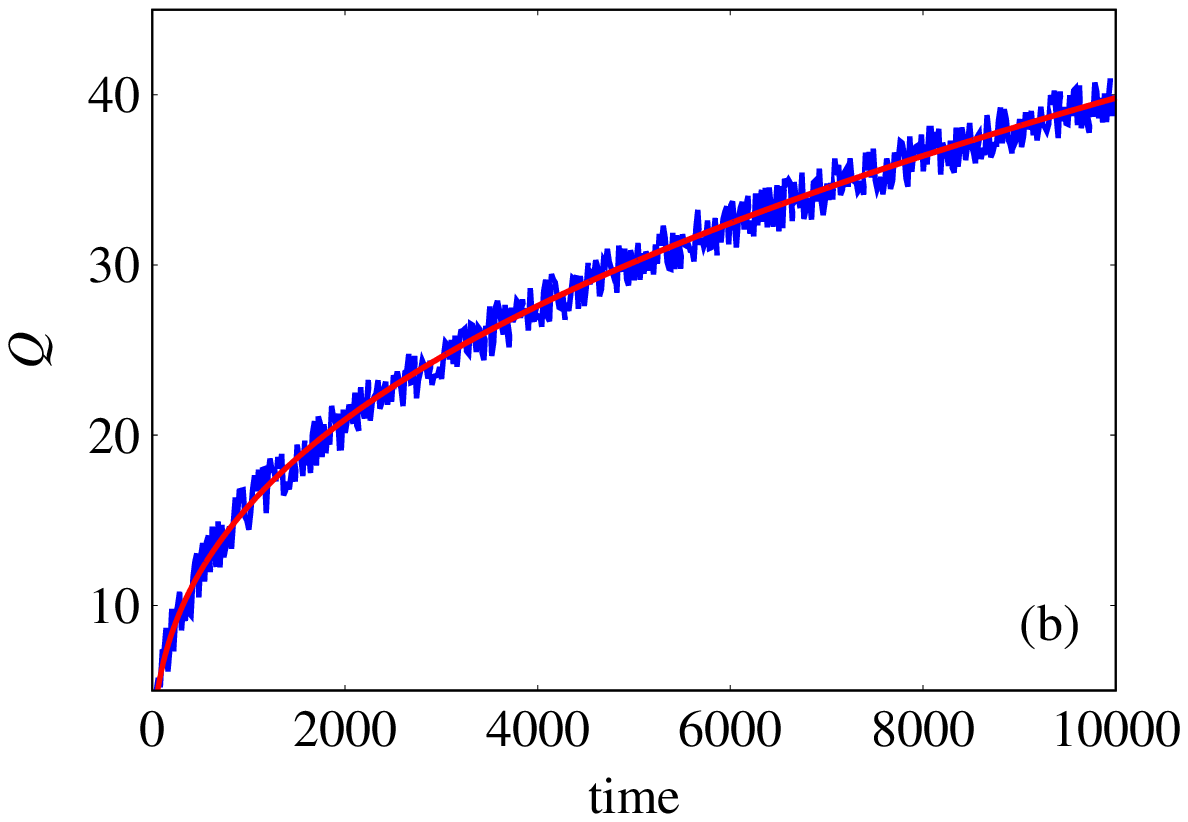}
    \caption{Randomly created values [(b)lue line connected points] around an artificial  function $Q(t) = t^{2/5}$ [(r)ed curves]. 
        The amplitude of the randomness is $\frac{1}{50}Q(t)$ (a) and $\frac{1}{25}Q(t)$ (b). 
        }
    \label{fig:raw_data_smth_num}
\end{figure}
Fig.~\ref{fig:raw_data_smth_num} presents the evolution of such noisy outcome of $Q(t)$ for two different amplitudes of the noise.
For example, in Fig.~\ref{fig:raw_data_smth_num} we see such a data set created by random additions (or subtractions) to the $Q(t) = t^{2/5}$ values when the magnitude of this randomness is $\frac{1}{50}Q(t)$ [panel (a)] and $\frac{1}{25}Q(t)$ [panel (b)]. 
Now, the question is how can we obtain the correct exponent ($2/5$ in our simple example) from the randomly scattered points. 

More generally, if we want to estimate the exponent $\alpha_Q$ of the expression $Q(t)=Q_0t^{\alpha_Q}$, it is better to see this law in the $\log-\log$ scale as 
\begin{equation}
    \log _{10}Q = \alpha_Q \log_{10}t + \log_{10}Q_0,
    \label{eq:q_general_log_log_scale}
\end{equation}
and determine $\alpha_Q$ as
\begin{equation}
    \alpha _{Q} = \frac{d\log_{10} Q}{d\log_{10} t}.
    \label{eq:slope_theoretical_num}
\end{equation}
Since by line connecting the noisy data of Fig.~\ref{fig:raw_data_smth_num} we do not get a differentiable curve, we first need to smooth the raw data.
We do so by applying a locally weighted regression method developed in~\citep{cleveland1988locally}, then use finite difference methods to evaluate the slope~\eqref{eq:slope_theoretical_num}~\citep{fornberg1988generation,fornberg1998classroom}.
More specifically in this work, we adapt the computer codes developed by Hilboll~\citep{hilboll2007lowess} for the regression process and by Xavier~\citep{timmes2017fdcoef} for the finite difference techniques.

\begin{figure}[!htb]
    \centering 
    \includegraphics[width=0.49\textwidth, height=0.5\linewidth]{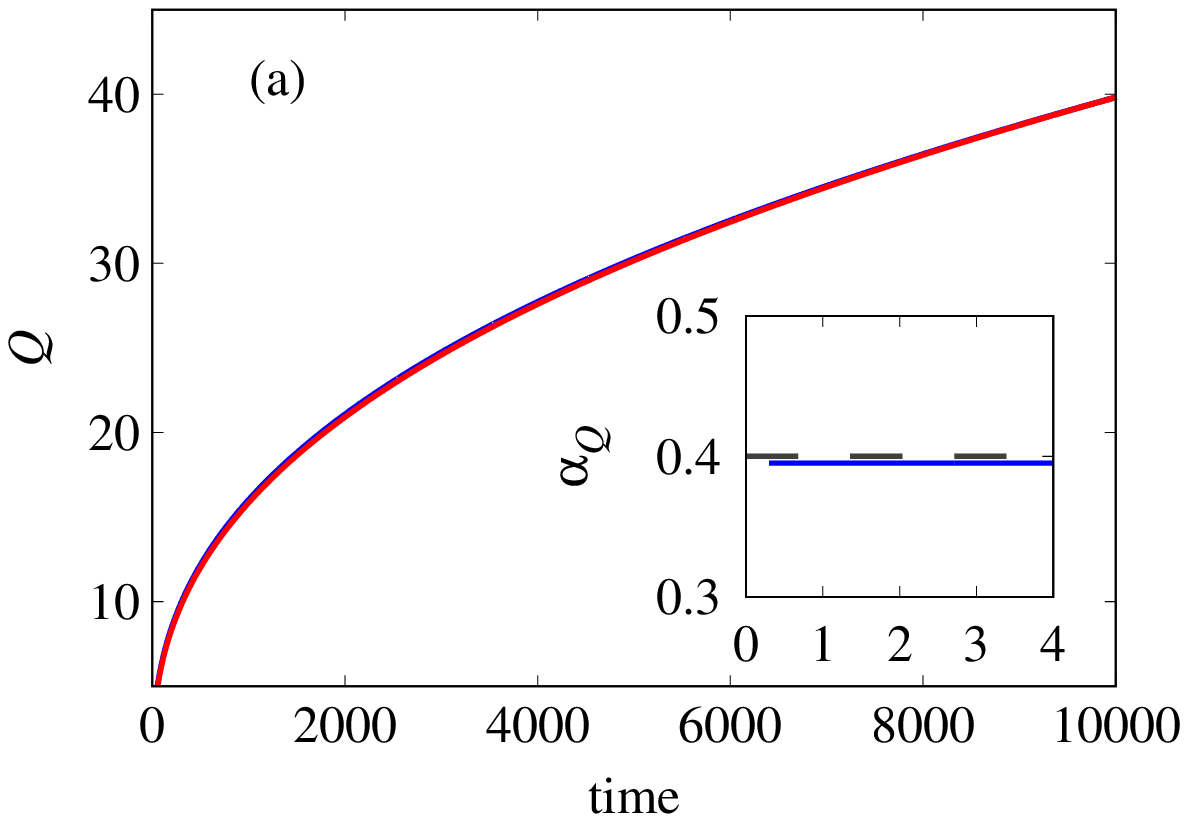}
    \includegraphics[width=0.49\textwidth, height=0.5\linewidth]{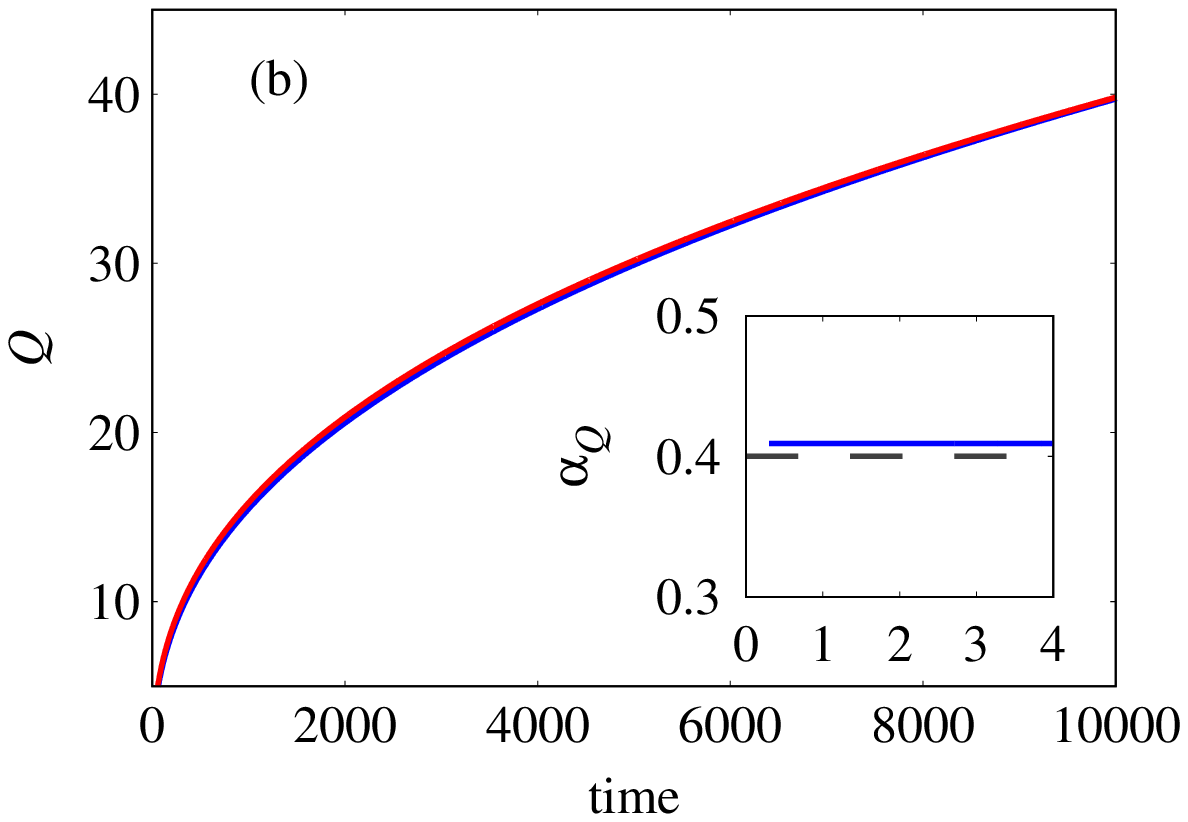}
    \caption{The smoothed data [(b)lue line connected points] of Fig.~\ref{fig:raw_data_smth_num} along with the function $Q = t^{2/5}$ [(g)reen curves].
            The insets show the time dependence in semi-log scale of the value of the numerically obtained exponent $\alpha_Q$~\eqref{eq:slope_theoretical_num} for the smoothed data [(b)lue curves].
            The thick dashed lines denote the value $0.4$.}
    \label{fig:smth_slopes_data_smth_num}
\end{figure}

We present in Fig.~\ref{fig:smth_slopes_data_smth_num} the results of the smoothing procedure for the data of Fig.~\ref{fig:raw_data_smth_num} [(b)lue line connected points] along with original function $Q = t^{2/5}$ [(r)ed curves].
There is a good agreement between the two, further supported by the value of the numerically computed exponent $\alpha_Q$ of the smoothed data presented in the insets of Fig.~\ref{fig:smth_slopes_data_smth_num} whose values are very close to $2/5$.
Obviously, the deviation from the `actual' value $\alpha_Q = 2/5 = 0.4$ is larger in the case of the more noisy points (Fig.~\ref{fig:smth_slopes_data_smth_num}(b)).

\section{\label{sec:nonSI_num}General purpose integration schemes}
Let us now briefly present some possible ways to numerically integrate the system of ODEs~\eqref{eq:generalZ}.
We will first discuss some general purpose integrators.

\subsection{\label{subsec:explicit_euler_demo_num}The explicit Euler method}
Let us first describe the so-called explicit Euler method which is a basic numerical technique for solving system~\eqref{eq:generalZ}.
So, starting with the initial condition, $\bm{Z} (t)$ at $t \ge 0$, we propagate the Hamiltonian flow and the variational equations to $t + \tau$ using a linear interpolation
\begin{equation}
    \bm{Z}(t + \tau) = \bm{Z}(t) + \tau \bm{f}(\bm{Z}(t), t),
    \label{eq:explicit_euler_scheme}
\end{equation}
where $\tau$ is the integration time increment.
In what follows we will occasionally use $\bm{Z}^\prime$ to denote $\bm{Z}(t + \tau)$ and $\bm{Z}$ for $\bm{Z}(t)$.
Following the process of~\citep{yoshida1993recent}, in order to illustrate the applicability and the limitations of this method, let us use as a `toy model' the Hamiltonian of a single harmonic oscillator
\begin{equation}
    \mathcal{H} = \frac{\varepsilon}{2}\left( q^2 + p^2\right),
    \label{eq:ham_harm_osc}
\end{equation}
where $q$ and $p$ are the canonical generalized position and momentum respectively and $\varepsilon$ is the proper frequency of the oscillator.
Its tangent dynamics is described by the Hamiltonian function of the form~\citep{skokos2010numerical} 
\begin{equation}
    \mathcal{H}_{V} = \varepsilon \left( q \delta q + p \delta p \right),
    \label{eq:tangent_dynamic_ham_harm_osc}
\end{equation}
with canonical coordinates $\delta p$ and $\delta q$.
$\mathcal{H}_V$ depends explicitly on time $t$ through the presence of $q(t)$ and $p(t)$.
In this case Eq.~\eqref{eq:generalZ} take the form
\begin{equation}
    \begin{pmatrix}
        \dot{q} \\
        \dot{p} \\
        \dot{\delta q} \\
        \dot{\delta p} 
    \end{pmatrix}
    = \bm{f}(q, p, \delta q,\delta p) = 
    \begin{pmatrix}
        \varepsilon   p \\
        - \varepsilon q \\
        \varepsilon  \delta p \\
        -\varepsilon  \delta q
    \end{pmatrix}.
    \label{eq:ham_eqmot_harm_osc}
\end{equation}
Consequently, the explicit Euler scheme~\eqref{eq:explicit_euler_scheme} evolves Hamiltonians~\eqref{eq:ham_harm_osc} and~\eqref{eq:tangent_dynamic_ham_harm_osc} through 
\begin{equation}
    \begin{pmatrix}
        q^\prime \\
        p^\prime\\
        \delta q ^\prime\\
        \delta p ^\prime
    \end{pmatrix}
    = 
    \begin{pmatrix}
        1	&	\varepsilon \tau & 0 & 0\\
        - \varepsilon \tau & 1 & 0 & 0 \\
        0   &    0  & 1 & \varepsilon  \tau \\
        0 &		0 	& - \varepsilon  \tau & 1 \\
    \end{pmatrix}
    \begin{pmatrix}
        q \\
        p \\
        \delta q \\
        \delta p 
    \end{pmatrix}.
    \label{eq:num_harm_osc_sol_euler_scheme}
\end{equation}
Let us now, compute these Hamiltonian functions after one integration step $\tau$, getting
\begin{equation}
    \mathcal{H}^\prime = \frac{\varepsilon}{2}\left(q ^{\prime~2} + p^{\prime~2} \right) = \frac{\varepsilon}{2}\left(1 + \varepsilon  ^2\tau ^2 \right)\left(q^2 + p^2 \right), 
\end{equation}
and 
\begin{equation}
    \mathcal{H}_{V}^\prime = \varepsilon \left(q ^\prime \delta q \prime + p ^\prime \delta p^\prime \right) = \varepsilon\left( 1 + \varepsilon  ^2 \tau ^2 \right) \left( q  \delta q  + p  \delta p  \right).
\end{equation}
We see that the new Hamiltonians for both the phase space orbit and the related deviation vector have some additional terms as they now have the form $\mathcal{H}^\prime = \mathcal{H} + \varepsilon ^2\tau ^2 \mathcal{H}$ and  $\mathcal{H}_{V}^\prime = \mathcal{H}_{V} + \varepsilon ^2 \tau ^2 \mathcal{H}_{V}$.
This means that, an artificial drifting force proportional to $\tau ^2$ has been introduced introduced to the system whose origin is purely numerical.
Furthermore at each iteration, the cumulative drift from the initial Hamiltonian will become larger up to the point where it could not be anymore neglected in comparison to the original system. 
This deviation could lead to the appearance of an artificial chaotic behavior.
A common practice to try to address this problem is the use of small integration time steps $\tau$  along with the development of time adaptative step size integrators, which is not very efficient for very long time integrations as the ones needed for our study.

\subsection{\label{subsec:taylor_series}The Taylor/Lie series method}
In general, the solution of Eq.~\eqref{eq:generalZ} $\bm{Z} (t + \tau)$ at time $t + \tau$ can be expressed as a  Taylor series of $\bm{Z} (t)$ 
\begin{equation}
\bm{Z} (t + \tau) = \bm{Z} (t) + \tau \frac{d\bm{Z} (t)}{dt} + \frac{\tau ^2}{2} \frac{d^2\bm{Z}(t)}{dt^2} + \ldots +\frac{\tau ^n}{n!} \frac{d^n \bm{Z} (t)}{dt^n} + \mathcal{O}\left(\frac{\tau ^{n + 1}}{(n + 1)!}\frac{d^{n + 1} \bm{Z} (t)}{dt^{n + 1}} \right).
\label{eq:taylor_series_form_general}
\end{equation}
If we consider the first two terms, we obtain the explicit Euler method presented in Sec.~\ref{subsec:explicit_euler_demo_num}.
Thus an artificial drift from the primary system's dynamics is also present. 
Obviously, increasing the number of terms considered in the expansion~\eqref{eq:taylor_series_form_general} this artificial drift reduces. 
The scheme presented in Eq.~\eqref{eq:taylor_series_form_general} where the first $n$ terms of the expansion are used is said to be of order $n$. 
Further information about this method is presented in~\citep[Sec. I.8]{hairer1993solving} and references therein, while a detailed discussion of the performances of this technique in the present context can be found for example in~\citep{gerlach2012efficient,danieli2019computational}.

Following~\citep{grobner1967lie,hanslmeier1984numerical, eggl2010introduction, blanes2016concise}, we will discuss an elegant way to present the Taylor series using Lie algebra. 
Indeed, Eq.~\eqref{eq:generalZ} can be written in terms of Lie operators as
\begin{equation}
\frac{d\bm{Z} (t)}{dt} = \bm{L}_{\mathcal{H}_V} \bm{Z} (t),
\label{eq:eq_lie_formalism}
\end{equation}
where the Lie operator $\bm{L}_{\mathcal{H}_V}$ is defined as 
\begin{equation}
 \bm{L}_{\mathcal{H}_V} = \sum _{i=1}^{2N} f_i \frac{\partial }{\partial z_i} + f_{i + 2N} \frac{\partial}{\partial \delta z_i}.
\end{equation}
Then the solution of Eq.~\eqref{eq:eq_lie_formalism} can be formally written as
\begin{equation}
\bm{Z} (t + \tau ) = e^{\tau \bm{L}_{\mathcal{H}_V}}\bm{Z} (t),
\label{eq:sol_gen_lie_formalism}
\end{equation}
where the exponential differential operator $e^{\tau \bm{L}_{\mathcal{H}_V}}$ plays the role of the propagator of the solution as its application on the initial state $\bm{Z}(t)$ advances the solution by time step $\tau$ i.e.~to $\bm{Z} (t+\tau)$. 
In the case of integrable systems, such operators represent the exact solution of the equations of motion and the variational equations.
However in the case of non-integrable systems, the Lie series integrator approximates the solution through a series expansion of the operator $\bm{L}_{\mathcal{H}_V}$
\begin{equation}
\bm{Z} (t + \tau) =  \bm{L}_{\mathcal{H}_V}^0\bm{Z} (t) + \tau  \bm{L}_{\mathcal{H}_V}\bm{Z} (t)  + \frac{\tau ^2}{2}  \bm{L}_{\mathcal{H}_V}^2\bm{Z} (t) + \ldots + \frac{\tau ^n}{n!}\bm{L}_{\mathcal{H}_V}^n\bm{Z} (t) + \mathcal{O}\left(\frac{\tau ^{n + 1}}{(n + 1)!} \bm{L}_{\mathcal{H}_V}^{n + 1} \bm{Z} (t)\right).
\label{eq:lie_series_method_gen}
\end{equation}
The next step in order to practically implement Eq.~\eqref{eq:lie_series_method_gen} is to find an analytical expression for the action of $\bm{L}_{\mathcal{H}_V}$ and its powers $\bm{L}_{\mathcal{H}_V}^n$ on the system state $\bm{Z}$ at time $t$.
Equation~\eqref{eq:lie_series_method_gen} corresponds to the Lie series method of order $n$ and is equivalent to expression~\eqref{eq:taylor_series_form_general} by considering $\bm{L}_{\mathcal{H}_V}^0 \bm{Z} = \bm{Z}$, $\bm{L}_{\mathcal{H}_V} \bm{Z} = \frac{d\bm{Z}}{dt}$, $\bm{L}_{\mathcal{H}_V}^2 \bm{Z} = \frac{d^2\bm{Z}}{dt^2}$ etc. 

Actually, the determination of an analytical closed form of $\bm{L}_{\mathcal{H}_V}^n \bm{Z}$ is a very challenging task even for systems with few degrees of freedom~\citep{eggl2010introduction}, and becoming almost impossible for multidimensional systems.
Nevertheless the application of modern computer algebra softwares, can automate successive differentiations of the force field vector $\bm{f}$ (see e.g.~\citep{barrio2005performance}).
This approach has been followed for instance using the software {\it Mathematica}\footnote{Wolfram Mathematica: Modern technical computing.~\url{https://www.wolfram.com/mathematica/}.} in the freely available package called $TIDES$~\citep{abad2014tides}. 
The $TIDES$ package has already been used in several investigations of dynamical systems (see e.g.~\citep{barrio2011parameter,gerlach2012efficient,barrio2015database,wilczak2016coexistence,danieli2019computational}). 
It requires as input the set of differential equations to be integrated along with a set of initial conditions, as well as the final time of integration.
It creates internally a Fortran77 or C code which is immediately compiled producing an executable program which generates the numerical solution. 
The accuracy of the performed integration does not only depend on the integration time step $\tau$ but also on a parameter (denoted as $\delta_a$) defining the integrator's single step precision. 
As a consequence, the $TIDES$ package creates an adaptive step size integrator.  

\subsection{\label{subsec:rk_family_num}The General purpose DOP853 Runge-Kutta integrator}

The Runge-Kutta (RK) numerical integration methods are some of the most used numerical schemes.
Their simplicity and robustness make them widely used in several scientific fields such as physics, astrophysics, population dynamics, etc.
Details on this family of integrators can be found in~\citep{hairer1993solving, hairer2006geometric,eggl2010introduction}. 
A~\textit{p-stage} RK method for Eq.~\eqref{eq:generalZ}, is described by the scheme
\begin{align}
\bm{Z}(t + \tau) &= \bm{Z}(t) + \tau \sum_{i=1}^{p} b_i\bm{k}_i, \quad \text{with} \\
\bm{k}_i &= \bm{f}\left( t + c_i\tau, \bm{Z} (t) + \tau \sum_{j = 1}^{p - 1} a_{ij}\bm{k}_j \right),
\end{align}
where the values of the coefficients $a_{i, j}$, $b_i$ and $c_i$ depend on the desired accuracy. 
In order to obtain numerical values for these coefficients, one  has to perform Taylor series expansions and deduce order conditions which are algebraic equations on $a_{ij}$, $b_i$ and $c_i$. 
As seen previously, Taylor series expansions introduce an artificial drift which is also expected to appear for this method.  

In the rest of this thesis, unless otherwise state, we implement a $12$-stage explicit RK algorithm called DOP$853$ which is freely available online~\citep{hairer2009dop853} and is based on the Dormand and Prince method~\citep{hairer1993solving}. 
The DOP$853$ has already been extensively used in lattice dynamics (see e.g.~\citep{mulansky2009dynamical, sales2018sub}).

\section{\label{sec:sis_num}Symplectic integrators}

Based on the nature of the physical phenomena they describe, Hamiltonian models should conserve certain macroscopic observables such as mass, energy and number of particles.
These constants are called {\it integrals of motion} and, in principle, described by analytical functions whose values remain constant throughout the system's evolution.
Since the basic principle of finding  accurate numerical solutions is to get as close as possible to the properties of the real analytical solutions, finding numerical integration schemes which preserve the integrals of motion is highly desirable. 
%
%
SIs are numerical integration schemes specifically designed for autonomous Hamiltonian systems as they preserve by default the system's phase space volume $d\mathfrak{V} = d\bm{q} \wedge d\bm{p}$.
The numerical schemes presented in Secs.~\ref{subsec:taylor_series} and~\ref{subsec:rk_family_num} were losing the preservation of symplecticity once we considered that the action of the operator $e^{L_{\mathcal{H}_V}}$ was approximated by a truncated Taylor series expansion~\eqref{eq:lie_series_method_gen} (see for example Sec.~\ref{subsec:explicit_euler_demo_num}). 
Let us consider a particular class of Hamiltonian functions which is encountered in many applications, systems whose Hamiltonian can be expressed as a sum of a {\it kinetic} energy $A(\bm{p})$ depending only on the momentum variables $\bm{p}$ and a {\it potential} energy $B(\bm{q})$ term depending only on the position variables $\bm{q}$
\begin{equation}
    \mathcal{H} (\bm{q}, \bm{p}) = A(\bm{p}) + B(\bm{q}).
    \label{eq:hamiltonian_splitted_2ps_sis_sec}
\end{equation}
The formal solution of Eq~\eqref{eq:generalZ} (see ~\eqref{eq:sol_gen_lie_formalism}) becomes
\begin{equation}
\bm{Z} (t + \tau ) = e^{\tau ( \bm{L}_{A_V} + \bm{L}_{B_V})}\bm{Z} (t).
\label{eq:sol_gen_form_symplectic}
\end{equation}
Since each part $A(\bm{p})$ and $B(\bm{q})$ possess $N$ cyclic coordinates which act as integrals of motion, the analytical forms of the operators $e^{\tau \bm{L}_{A_V}}$ and $e^{\tau \bm{L}_{B_V}}$
are explicitly known.
These operators represent the exact solutions of the equations of motion and the related variational equations of the respectively simple Hamiltonian functions $A(\bm{p})$ and $B(\bm{q})$ 
\begin{figure}
    \centering
    \includegraphics[width=0.45\textwidth]{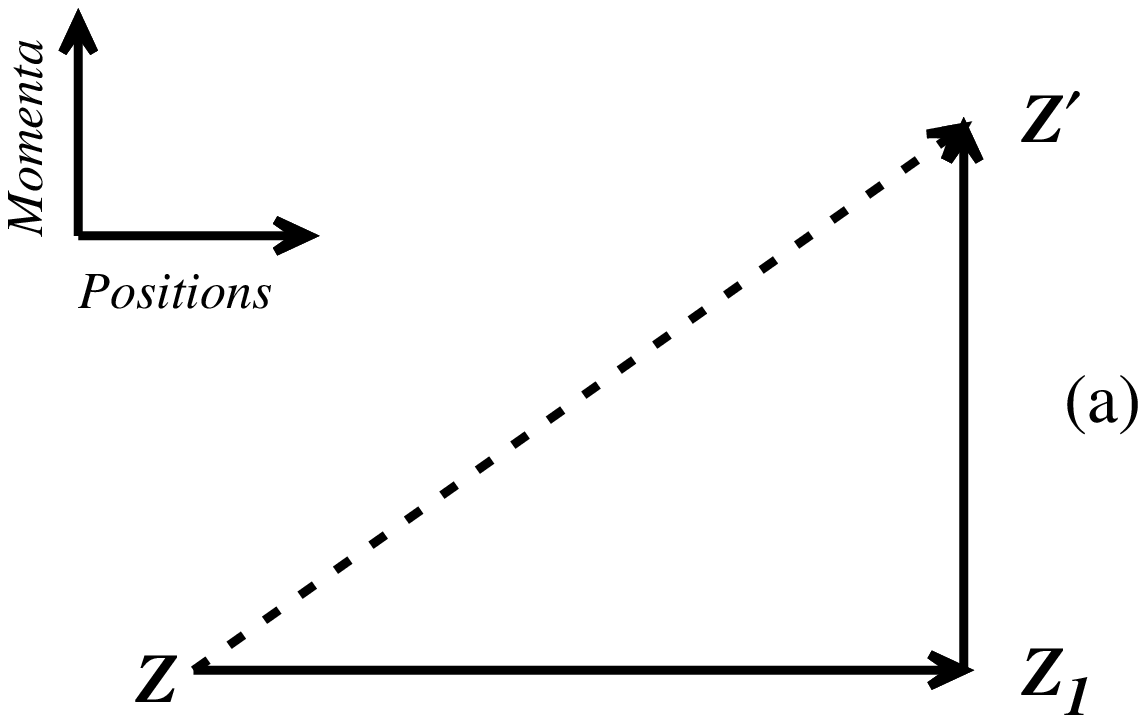}
    \includegraphics[width=0.45\textwidth]{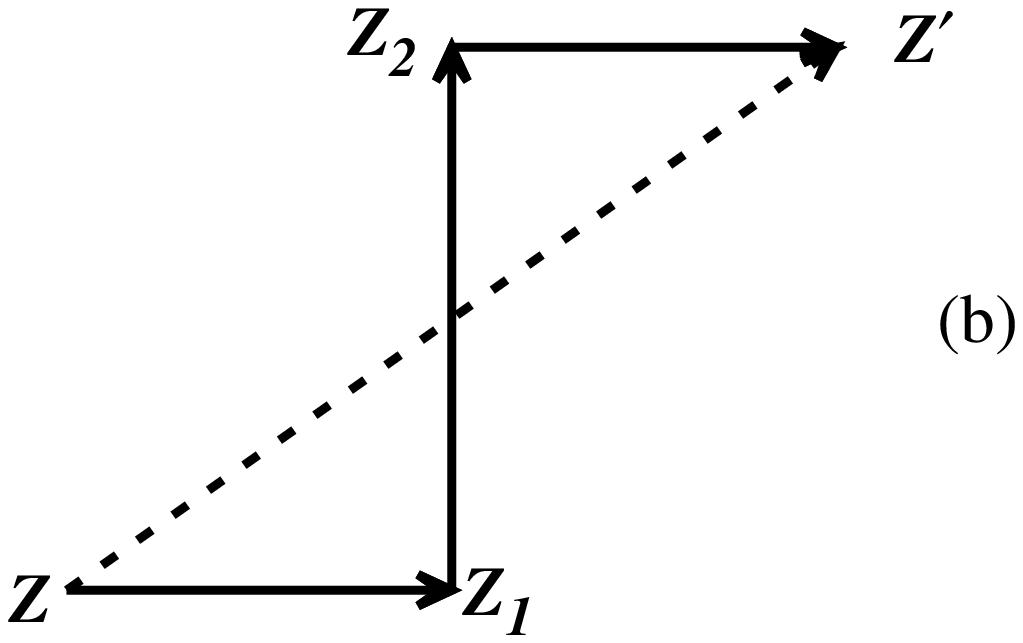}
    \caption{Geometric representations of the symplectic explicit Euler~\eqref{eq:explicit_euler_scheme} (a) and leapfrog (LF)~\eqref{eq:leapfrog_strormer-verlet} (b) methods. 
            The bold arrows represent the paths followed by the successive steps within the numerical scheme and the dashed arrows joins the initial and final integration points.
            In (a) $\bm{Z}_1 = e^{\tau \bm{L}_{A_V}}\bm{Z}$ and $\bm{Z}^\prime = e^{\tau \bm{L}_{B_V}}\bm{Z}_1$ respectively, giving a $2$ step SI. 
            In (b) $\bm{Z}_1 = e^{\tau \bm{L}_{A_V}/2}\bm{Z}$, $\bm{Z}_2 = e^{\tau \bm{L}_{B_V}}\bm{Z}_1$ and $\bm{Z}^\prime = e^{\tau \bm{L}_{A_V}/2}\bm{Z}_2$ such that the LF corresponds to a $3$ step SI.
        }
    \label{fig:euler_symplec_geometric}
\end{figure}
\begin{equation}
    e^{\tau \bm{L}_{A_V}}\colon
    \begin{dcases}
    \bm{q}^\prime & =  \bm{q} + \tau \left. \frac{\partial A}{\partial \bm{p}} \right\rvert _{\substack{\bm{p}=\bm{p} \\ \delta \bm{p}=\delta \bm{p}}} \\
    \delta \bm{q}^\prime & =  \delta \bm{q} + \tau  \left[ \delta \bm{p} \cdot \bm{D}^2 A \right] _{\substack{\bm{p}=\bm{p} \\ \delta \bm{p}=\delta \bm{p}}} 
    \end{dcases}
    ,
    \quad 
    e^{\tau \bm{L}_{B_V}}\colon 
    \begin{dcases}
    \bm{p}^\prime & =  \bm{p} - \tau \left. \frac{\partial B}{\partial \bm{q}} \right\rvert _{\substack{\bm{q}=\bm{q} \\ \delta \bm{q}=\delta \bm{q}}} \\
    \delta \bm{p}^\prime & =  \delta \bm{p} - \tau  \left[ \delta q \cdot \bm{D}^2 B \right] _{\substack{\bm{q}=\bm{q} \\ \delta \bm{q}=\delta \bm{q}}}
    \end{dcases}
    .
    \label{eq:explicit_euler_operators_1}
\end{equation}
Here $\bm{D}^2$ represents a multi-variable derivative operator of order two with respect to the momenta ($\bm{D}^2 A$ in $e^{\tau \bm{L}_{A_V}}$) or the positions ($\bm{D}^2 B$ in $e^{\tau \bm{L}_{B_V}}$).
Further assuming that $\bm{L}_{A_V}$ and $\bm{L}_{B_V}$ commute (something which is generally not true), we can write 
\begin{equation}
    e^{\tau \bm{L}_{\mathcal{H}_V}} =   e^{\tau \bm{L}_{B_V} } e^{\tau \bm{L}_{A_V}},
    \label{eq:comp_euler_2_num}
\end{equation}
such that 
\begin{equation}
    \bm{Z} ( t + \tau) = e^{\tau \bm{L}_{B_V} }  e^{\tau \bm{L}_{A_V}}\bm{Z}(t).
    \label{eq:comp_euler_1_num}
\end{equation}
The geometrical representation of Eq.~\eqref{eq:comp_euler_1_num} is presented in Fig.~\ref{fig:euler_symplec_geometric}(a).
We see that $\bm{Z}$ and $\bm{Z}^\prime$ are related through the application of a two step process involving two perpendicular mappings.
Thus using Eq.~\eqref{eq:comp_euler_2_num}, we approximate the exact solution described by $e^{\tau \bm{L}_{\mathcal{H}_V}}$, by the composition of two operators $e^{\tau \bm{L}_{A_V}}$ and $e^{\tau \bm{L}_{B_V}}$.
Since each operator is symplectic, so is the composition. 
Thus we have constructed a very simple SI
\begin{equation}
    S_1 (\tau) =  e^{\tau \bm{L}_{B_V}}e^{\tau \bm{L}_{A_V} } ,
    \label{eq:explicit_euler_scheme_SI}
\end{equation} 
which is called the {\it symplectic explicit Euler method}.

For the sake of completeness, let us apply the SI~\eqref{eq:explicit_euler_scheme_SI} to the single harmonic oscillator~\eqref{eq:ham_harm_osc} as done in~\citep{yoshida1993recent}.
Indeed, Hamiltonain~\eqref{eq:ham_harm_osc} can be splitted in two parts $A(p)=\varepsilon p^2/2$ and $B(q) = \varepsilon q^2/2$. 
Thus Eq.~\eqref{eq:explicit_euler_operators_1}, gives 
\begin{equation}
    e^{\bm{L}_{A_V}} \bm{Z}(t) \colon 
    \left\{ 
        \begin{array}{ll}
            q^\prime &= q + \tau \varepsilon p \\
            p^\prime &= p \\
            \delta q^\prime &= \delta q + \tau \varepsilon\delta  p \\
            \delta p^\prime &= \delta p \\
        \end{array}
    \right., 
    \quad \mbox{and} \quad
    e^{\bm{L}_{B_V}} \bm{Z}(t) \colon 
    \left\{ 
        \begin{array}{ll}
            q^\prime &= q \\
            p^\prime &= p - \tau \varepsilon p \\
            \delta q^\prime &= \delta q  \\
            \delta p^\prime &= \delta p - \tau \varepsilon \delta p \\
        \end{array}
    \right.,
    \label{eq:harmonic_oscillator_elav_elbv}
\end{equation}  
such that the numerical scheme $S_1 (\tau)$~\eqref{eq:explicit_euler_scheme_SI} gives 
\begin{equation}
    \begin{pmatrix}
        q^\prime \\
        p^\prime\\
        \delta q ^\prime\\
        \delta p ^\prime
    \end{pmatrix}
    = 
    \begin{pmatrix}
        1	&	\varepsilon \tau & 0 & 0\\
        - \varepsilon \tau & 1 - \varepsilon ^2 \tau ^2 & 0 & 0 \\
        0   &    0  & 1	&	\varepsilon \tau\\
        0 &		0 	& - \varepsilon \tau & 1 - \varepsilon ^2 \tau ^2 \\
    \end{pmatrix}
    \begin{pmatrix}
        q \\
        p \\
        \delta q \\
        \delta p 
    \end{pmatrix}.
    \label{eq:num_harm_osc_sol_symplectic_euler_scheme}
\end{equation}
The computation of the energy~\eqref{eq:ham_harm_osc} after one time step $\tau$
\begin{equation}
    \mathcal{H}^\prime = \mathcal{H} + \tau^2 \frac{\varepsilon^3}{2} \left(q^2 - p^2\right) + \tau ^3 \frac{\varepsilon^4}{2} qp + \tau ^4 \frac{\varepsilon^5}{2} p^2,
    \label{eq:energy_one_time_step_si_euler_scheme_harmonic_osc}
\end{equation}
depicts a non-conserved total energy of the system.
Nevertheless, the presence of a negative term in $\tau^2$, ensures that there is no monotonic increase of the energy error which is bounded by $\mathcal{O}\left(\tau\right)$. 
This observation is further explained by the fact the mapping~\eqref{eq:num_harm_osc_sol_symplectic_euler_scheme} has an integral of motion~\citep{yoshida1993recent} 
\begin{equation}
    \frac{\varepsilon}{2}\left(q^2 + p^2\right) + \frac{\tau}{2}\varepsilon qp = const.
    \label{eq:trajectory_si_euler_method}
\end{equation}
Indeed, if we start with a natural initial condition $(0, 1)$, all the points generated iterating the mapping~\eqref{eq:num_harm_osc_sol_symplectic_euler_scheme} lie on the conic section
$q^2 + p^2 + \tau qp = 1$, which deviates from the exact orbit $q^2 + p^2 = 1$ by a term of the order of $\tau$.
Consequently the error in energy in~\eqref{eq:energy_one_time_step_si_euler_scheme_harmonic_osc} cannot grow in time.
Further, for small enough values of $\tau$, the orbit of the system's perturbed trajectory also takes the form~\eqref{eq:trajectory_si_euler_method}, meaning that no exponenital growth of the DV is possible.
Thus, the system's MLE will always exhibit a regular behavior $\Lambda \propto t^{-1}$ as expected from the harmonic oscillator~\eqref{eq:ham_harm_osc}.
Obviously, the computer precision also has to be taken into account on the above argumentation.

In general, any composition of the operators $e^{\tau \bm{L}_{A_V}}$ and $e^{\tau \bm{L}_{B_V}}$ leads to a symplectic numerical scheme.
A SI of order $n$ is 
\begin{equation}
    S_{n} (\tau ) = \prod _{i=1}^{k} e^{\tau c_i \bm{L}_{A_V}}e^{\tau d_i \bm{L}_{B_V}} + \mathcal{O} (\tau ^{n + 1}).
    \label{eq:elh_elab_general}
\end{equation}
We note that the SI~\eqref{eq:explicit_euler_scheme_SI} is of order $1$ as it is indicated by the subscript `$1$' in $S_1$.
The problem of constructing the SIs in~\eqref{eq:elh_elab_general} is reduced in finding the coefficients $c_i$ and $d_i$ with $i = 1, 2, \ldots, k \leq n$ such that Eq.~\eqref{eq:elh_elab_general} approximates $e^{\tau (A_V + \bm{L}_{B_V})}$ up to a certain order of precision.
This problem, was initially set and tackled in several pioneering papers~\citep{neri1990lie,suzuki1990fractal,yoshida1990construction,suzuki1991general}.

From Eqs~\eqref{eq:explicit_euler_scheme_SI} and~\eqref{eq:elh_elab_general} we see that the symplectic explicit Euler method corresponds to $n = 1$ and $k = 1$ which leads to the trivial values of the coefficients $c_1 = 1$ and $d_1 =1$. 
For $n =2$ and $k = 2$ we get the second order SI~\citep{hairer1993solving,hairer2006geometric,blanes2016concise}
\begin{equation}
    S_2 (\tau) = e^{\tau c_1 \bm{L}_{A_V}}e^{\tau d_1 \bm{L}_{B_V}} e^{\tau c_2 \bm{L}_{A_V}}e^{\tau d_2 \bm{L}_{B_V}}.
    \label{eq:leapfrog_s2_1}
\end{equation}
For this scheme the coefficients $c_1$, $c_2$, $d_1$ and $d_2$ should satisfy the order conditions 
\begin{equation}
    \label{eq:sample_order_condition_s2}
    \left\{
        \begin{matrix}
            c_1 + c_2 = 1 \\
            d_1 + d_2 = 1 \\
        \end{matrix}
    \right.,
\end{equation}
whose solution is not uniquely determined.
For example the choice  
\[ c_1 = c_2 = \frac{1}{2}, \quad d_1=1 \quad \mbox{and} \quad d_2 = 0,\]
satisfies the order conditions~\eqref{eq:sample_order_condition_s2}, leading to the SI
\begin{equation}
    S_2^{[1]} (\tau) = e^{\tau \frac{1}{2} \bm{L}_{A_V}}e^{\tau \bm{L}_{B_V}} e^{\tau  \frac{1}{2} \bm{L}_{A_V}} + \mathcal{O} (\tau ^{3}).
    \label{eq:leapfrog_strormer-verlet}
\end{equation}
This integrator is the symplectic {\it leapfrog} or {\it St\"{o}rmer-Verlet} scheme~\citep{hairer1993solving}. 
The geometrical interpretation of the leapfrog scheme~\eqref{eq:leapfrog_strormer-verlet} is depicted in Fig.~\ref{fig:euler_symplec_geometric}~(b).
We see that $\bm{Z}^\prime$ and $\bm{Z}$ are related through $3$ different steps each conserving the local constants of motion.
It is worth mentioning that not all choices of the parameters $c_1$, $c_2$, $d_1$ and $d_2$ satisfying the order conditions~\eqref{eq:sample_order_condition_s2} lead to SIs. 
For example, the choice
\[
    c_1 = c_2 = d_1 = d_2 = \frac{1}{2},
\] 
leads to the numerical scheme
\begin{equation}
    S_2^{[2]} (\tau) = e^{\tau \frac{1}{2} \bm{L}_{A_V}}e^{\tau \frac{1}{2} \bm{L}_{B_V}} e^{\tau  \frac{1}{2} \bm{L}_{A_V}}e^{\tau \frac{1}{2} \bm{L}_{B_V}}  + \mathcal{O} (\tau ^{3}).
    \label{eq:other_si_order_2_not_rev}
\end{equation}
Although this integrator is also symplectic, it is not {\it time reversible} as is the leapfrog method $S_{2}^{[1]}$~\eqref{eq:leapfrog_strormer-verlet}.
This means that making a time step $\tau$ with this integrator followed by another step $-\tau$ i.e. going back and forth (and vise versa) in time for one time step will not lead us to our initial state:
\begin{equation}
    S_2^{[2]} (\tau) S_2^{[2]} (-\tau) \bm{Z} (t) \neq S_2^{[2]} (-\tau) S_2^{[2]} (\tau)\bm{Z} (t) \neq \bm{Z} (t).
    \label{eq:not_reversibility}
\end{equation}
Since the property of time reversibility is quite natural, we will restrict our study only to SIs which have this property.

\subsection{\label{subsec:bch}The Baker-Campbell-Hausdorff formula}
Let us now see how we can use the Baker-Campbell-Hausdorff (BCH) formula to obtain SIs.
The BCH formula gives the solution for operator $W_1$ in the equation
\begin{equation}
e^X e^Y = e^{W_1},
\label{eq:bch_general_exp_form}
\end{equation}
when $X$ and $Y$ are non-commutative elements. 
In particular, we get 
\begin{eqnarray}
W_1 &=& X + Y + \frac{1}{2}[X, Y] + \frac{1}{12} ([X, X, Y] + [Y, Y, X]) + \frac{1}{24}[X, Y, Y, X] + \ldots
\label{eq:bch_w1}
\end{eqnarray}
where $[\cdot, \cdot]$ and $[\cdot, \cdot, \cdot]$ are commutators such that $[X, Y] = XY - YX$, $[X, X, Y] = [X, [X, Y]]$ etc. 

Setting $X = \tau \bm{L}_{A_V}$ and $Y = \tau \bm{L}_{B_V}$ in~\eqref{eq:bch_general_exp_form} and~\eqref{eq:bch_w1}, we obtain 
\begin{align}
    \nonumber
    W_1 &= \tau \bm{L}_{A_V} + \tau \bm{L}_{B_V} + \frac{1}{2} \left[\tau \bm{L}_{A_V}, \tau \bm{L}_{B_V}\right] + \frac{1}{12}\left(\left[\tau \bm{L}_{A_V}, \tau \bm{L}_{A_V}, \tau \bm{L}_{B_V}\right] + \left[\tau \bm{L}_{B_V}, \tau \bm{L}_{B_V}, \tau \bm{L}_{A_V}\right] \right) + \ldots, \\
        &= \tau \bm{L}_{\mathcal{H}_V} + \frac{\tau^2}{2} \left[\bm{L}_{A_V}, \bm{L}_{B_V}\right] + \frac{\tau^3}{12} \left(\left[\bm{L}_{A_V}, \bm{L}_{A_V}, \bm{L}_{B_V}\right] + \left[\bm{L}_{B_V}, \bm{L}_{B_V}, \bm{L}_{A_V}\right]\right) + \ldots.
    \label{eq:bch_w1_elab_1}
\end{align}
So neglecting the terms of order $2$ and above in $\tau$, we obtain the symplectic explicit Euler integrator of Eq.~\eqref{eq:explicit_euler_scheme_SI}
\begin{equation}
    S_1 \left(\tau\right) = e^{\tau \bm{L}_{A_V}}e^{\tau \bm{L}_{B_V}} \approx e^{\tau \bm{L}_{\mathcal{H}_V}}.
    \label{eq:bch_euler_2}
\end{equation}

Let us now implement the BCH formula in order to obtain the leapfrog integrator of Eq.~\eqref{eq:leapfrog_strormer-verlet}, setting 
\begin{equation}
    e^X e^Y e^X = e^{W_1}e^{X} = e^{W_2},
    \label{eq:bch_leapfrog}
\end{equation}
and applying again~\eqref{eq:bch_w1}, we get
\begin{align}
    \nonumber
    W_2 &= 2 \frac{\tau}{2}\bm{L}_{A_V} + \tau \bm{L}_{B_V} + \frac{1}{6}\left(\left[\tau \bm{L}_{B_V}, \tau \bm{L}_{B_V}, \frac{\tau}{2}\bm{L}_{A_V}\right] - \left[\frac{\tau}{2}\bm{L}_{A_V}, \frac{\tau}{2}\bm{L}_{A_V}, \tau \bm{L}_{B_V}\right]\right) + \ldots, \\
        & = \tau \bm{L}_{\mathcal{H}_V} + \frac{\tau^3}{12} \left(\left[\bm{L}_{B_V}, \bm{L}_{B_V}, \bm{L}_{A_V}\right] - \frac{1}{2}\left[\bm{L}_{A_V}, \bm{L}_{A_V}, \bm{L}_{B_V}\right]\right) + \ldots.
    \label{eq:bch_w2_2}
\end{align}
So neglecting terms of order $3$ and above in $\tau$ we obtain the leapfrog integrator of Eq.~\eqref{eq:leapfrog_strormer-verlet} 
\begin{equation}
    S_2(\tau)  = e^{\frac{1}{2}\tau \bm{L}_{A_V}}e^{\tau \bm{L}_{B_V}}e^{\frac{1}{2}\tau \bm{L}_{A_V}} \approx e^{\tau \bm{L}_{\mathcal{H}_V}}.
    \label{eq:leapfrog_s-v_2}
\end{equation}

This process can be followed in order to build higher order SIs.
The error of such integrator is defined by the power of the integration time step $\tau$ encountered in the first nonlinear term.
For Eq.~\eqref{eq:bch_w1_elab_1}, we see that this term is $\tau^2$ for the symplectic explicit Euler method, which according to~\eqref{eq:elh_elab_general} means that this integrator is of order $1$, while Eq.~\eqref{eq:bch_w2_2} denotes that the leapfrog integrator is of order $2$. 
The approach describes here is based on Lie algebra concepts and is more complex as we go to higher order integrators. 
The interested reader can get a very concise and complete overview on these method in~\citep{hairer2006geometric,blanes2016concise}.

\subsection{\label{subsec:yosida_technique_num}The Yoshida composition technique}
Let us now see how we can use these ideas in order to build higher order methods in a repeatable way.
Let us rewrite Eq.~\eqref{eq:bch_w2_2} as 
\begin{equation}
    W_2 = \tau \alpha _1 + \tau^3 \alpha _3 + \tau^5 \alpha _5 + \ldots,
    \label{eq:log_s2} 
\end{equation}
with $\alpha _ 1 = \bm{L}_{A_V} + \bm{L}_{B_V}$, $\alpha _ 3 = \frac{1}{12}[\bm{L}_{B_V}, \bm{L}_{B_V}, \bm{L}_{A_V}] - \frac{1}{24} [\bm{L}_{A_V}, \bm{L}_{A_V}, \bm{L}_{B_V}]$, etc.
Expression~\eqref{eq:log_s2} stands at the basis of what is called {\it Yoshida composition} technique~\citep{yoshida1990construction}. 
According to that technique, we can build higher order {\it symmetric} SIs using ~\textit{palindromic} composition of lower order SIs.
For instance, let us try to construct a SI of order $4$ of the form
\begin{equation}
    S_4 (\tau) = S_2\left(x_1\tau\right)S_2\left(x_0\tau\right)S_2\left(x_1\tau\right),
    \label{eq:s4_e}
\end{equation}
where $x_0$, $x_1$ are real coefficients we have to define.
We get from~\eqref{eq:log_s2}
\begin{equation}
    W_2 =\tau \left(x_0 + 2x_1\right) \alpha _1 + \tau ^3 \left(x_0 ^3 + 2x_1^3\right)\alpha _ 3 + \tau ^5 \left(x_0 ^5 + 2x_1 ^5\right)\alpha _5 + \ldots. 
    \label{eq:yosh_w2}
\end{equation}
In order to achieve the desired order $4$, we want the first nonlinear term in $\tau$ we encounter in~\eqref{eq:yosh_w2} to be $\tau^5$.
This means that we want to kill the order $3$ term by setting $x_0^3 + 2x_1^3 = 0$ and at the same time keeping the $\alpha _1 = \bm{L}_{\mathcal{H}_V}$ term by setting $x_0 + 2x_1 = 1$.
Thus we obtain the conditions
\begin{equation}
    \left\{
        \begin{matrix}
            x_0 + 2 x_1 &= 1 \\
            x_0 ^3 + 2 x_1 ^3 & = 0
        \end{matrix}
    \right.,
    \label{eq:eq_of_coef_fr4}
\end{equation}
which can be solved analytically to give~\citep{yoshida1990construction}
\begin{equation}
x_0 = -\frac{2^{1/3}}{2 - 2 ^{1/3}}, \quad x_1 = \frac{1}{2 - 2^{1/3}}.
\end{equation}
In fact this process can be repeated over and over again in order to create higher order SIs.
Its generalization states that from any symmetric $n$th order SI we build a $(n + 2)$th scheme using the composition (see also~\citep{iserles2018geometric} and references therein) 
\begin{equation}
    S_{n + 2} (\tau) = S_n (x_1\tau)S_n (x_0\tau)  S_n (x_1\tau), 
    \label{eq:yoshida_composition_formula_general_num}
\end{equation}
with coefficients
\begin{equation}
    x_ 0 = - \frac{2^{1/(2n + 1)}}{2 - 2^{1/(2n + 1)}}, \quad x_1 = \frac{1}{2 - 2^{1/(2n + 1)}}.
    \label{eq:yoshida_composition_formula_general_num_coefficients}
\end{equation}
Equation~\eqref{eq:yoshida_composition_formula_general_num} is usually referred to as the~\textit{Yoshida composition}.
Let us make an additional remark here.
The palindromic composition as in~\eqref{eq:yoshida_composition_formula_general_num} ensures that the SIs built using the Yoshida composition technique are always time reversible. 
It is worth noting that there also exist other composition schemes which follow the same basic principle of Yoshida's method having some differences on the way the Lie algebra is applied (see  e.g.~\citep{suzuki1991general}), based sometimes on the particular type of physical problem they were initially created to solve (see e.g.~\citep{ruth1983canonical,laskar2001high,laskar2019dedicated}).

\subsection{\label{subsec:symp_2ps_num}Two-part split symplectic integrators}
In this section we present several two-part split SIs i.e.~schemes which can be used in case the Hamiltonian function can be split into two integrable parts like in Eq.~\eqref{eq:hamiltonian_splitted_2ps_sis_sec}, which have been recently used in studies of multidimensional Hamiltonian models.
These studies considered for example models such as the DKG chain of anharmonic oscillators~\citep{skokos2013nonequilibrium,danieli2017intermittent,senyange2018characteristics,manda2020chaotic}, the DDNLS system in one~\citep{gerlach2016symplectic,mithun2018weakly,mallick2020quench} and two~\citep{manda2020chaotic} spatial dimensions, the Josephson junction chain~\citep{mithun2019dynamical}, the FPUT system~\citep{danieli2017intermittent,danieli2019computational}, the Peyrard-Bishop-Dauxois model of the DNA~\citep{hillebrand2019heterogeneity}, as well as graphene models~\citep{bountis2020stability,hillebrand2020chaotic} to name a few.

\subsubsection{\label{subsubsec:si_2_num}Symplectic integrators of order two.}
\begin{enumerate}
    \item[] {\bf The leapfrog (LF) method.} 
        We explicitly presented this scheme in Sec.~\ref{subsec:bch}.
        More details about the LF scheme can be found in~\citep{ruth1983canonical} and~\citep[Sect. VI.3]{hairer2006geometric}.
        This integrator has $3$ individual steps i.e.~the number of applications of simple operators $e^{\tau \bm{L}_{A_V}}$, $e^{\tau \bm{L}_{B_V}}$ (see also Fig.~\ref{fig:euler_symplec_geometric}(b)) and is given by the relation
        \begin{equation}
            LF(\tau ) = e^{a_1 \tau \bm{L}_{A_V}} e^{b_1 \tau \bm{L}_{B_V}}e^{a_1 \tau \bm{L}_{A_V}},
            \label{eq:leapfrog}
        \end{equation}
        with $a_1 = \frac{1}{2}$ and $b_1 = 1$.

    \item[] {\bf The $SABA_2/SBAB_2$ integrators.}
        These SIs possess $5$ individual steps and have the following expressions
        \begin{equation}
            SABA_2(\tau )= e^{a_1 \tau \bm{L}_{A_V}}e^{b_1 \tau \bm{L}_{B_V}}e^{a_2 \tau \bm{L}_{A_V}}e^{b_1 \tau \bm{L}_{B_V}}e^{a_1 \tau \bm{L}_{A_V}},
            \label{eq:saba2}
        \end{equation}
        with $a_1 = \frac{1}{2} - \frac{1}{2\sqrt{3}}$, $a_2 =  \frac{1}{\sqrt{3}}$ and $b_1 = \frac{1}{2}$, and 
        \begin{equation}
            SBAB_2(\tau) = e^{b_1 \tau \bm{L}_{B_V}}e^{a_1 \tau \bm{L}_{A_V}}e^{b_2 \tau \bm{L}_{B_V}}e^{a_1 \tau \bm{L}_{A_V}}e^{b_1 \tau \bm{L}_{B_V}},
            \label{eq:sbab2}
        \end{equation}
        with $a_1 = \frac{1}{2}$, $b_1 = \frac{1}{6}$ and $b_2 = \frac{2}{3}$.
        This class of schemes was studied in~\citep{mclachlan1995composition}, where they were referred to as the $(4,2)$ methods.
        In addition, they have been employed in~\citep{laskar2001high} to study Hamiltonian functions of the form $A(\bm{p}) + \varrho B(\bm{q})$ or $A(\bm{q},\bm{p}) + \varrho B(\bm{q},\bm{p})$, where the contribution $\varrho B$ was considered small since $\varrho \ll 1$.
        Nevertheless these integrators have shown good performance also for cases $\varrho = 1$ (see e.g.~\citep{skokos2009delocalization,veksler2009spreading,danieli2017intermittent,mithun2018weakly,kati2020density,mallick2020quench}). 
    
        \item[] {\bf The $ABA82$ scheme.} 
            This integrator is given by~\citep{mclachlan1995composition,farres2013high}
            \begin{equation}
                ABA82(\tau ) = e^{a_1 \tau \bm{L}_{A_V}}e^{b_1 \tau \bm{L}_{B_V}}e^{a_2 \tau \bm{L}_{A_V}}e^{b_2 \tau \bm{L}_{B_V}}e^{a_3 \tau \bm{L}_{A_V}}e^{b_2 \tau \bm{L}_{B_V}}e^{a_2 \tau \bm{L}_{A_V}}e^{b_1 \tau \bm{L}_{B_V}}e^{a_1 \tau \bm{L}_{A_V}},
                \label{eq:aba82}
            \end{equation}
            and has $9$ individual steps.
            The values of the coefficients $a_i, b_i$, $i=1, 2, 3$ in Eq.~\eqref{eq:aba82} are given in Table 2 of~\citep{farres2013high}.
            It is worth mentioning that the $ABA82$ SI was also used in~\citep{laskar2001high} where it was named $SABA_4$.
\end{enumerate}

\subsubsection{\label{par:si_4_num}Symplectic integrators of order four.}

\begin{enumerate}
    \item[] {\bf The $ABA864/ABAH864$ integrators.}
        These SIs were introduced in~\citep{blanes2013new,farres2013high}.
        The first scheme has the form
        \begin{align}
            \nonumber
            ABA864(\tau ) = & e^{a_1 \tau \bm{L}_{A_V}}e^{b_1 \tau \bm{L}_{B_V}}e^{a_2 \tau \bm{L}_{A_V}}e^{b_2 \tau \bm{L}_{B_V}}  e^{a_3 \tau \bm{L}_{A_V}}e^{b_3 \tau \bm{L}_{B_V}}e^{a_4 \tau \bm{L}_{A_V}} \\
                & \times e^{b_4 \tau \bm{L}_{B_V}}e^{a_4 \tau \bm{L}_{A_V}}e^{b_3 \tau \bm{L}_{B_V}} e^{a_3 \tau \bm{L}_{A_V}}e^{b_2 \tau \bm{L}_{B_V}}e^{a_2 \tau \bm{L}_{A_V}}e^{b_1 \tau \bm{L}_{B_V}}e^{a_1 \tau \bm{L}_{A_V}},
            \label{eq:aba864}
        \end{align}
        and includes $15$ individual steps.
        The coefficients $a_i, b_i~i=1, 2, 3, 4$ of Eq.~\eqref{eq:aba864} are listed in Table 3 of~\citep{blanes2013new}.
        The form of the second SI is 
        \begin{align}
            \nonumber
            ABAH864(\tau ) = & e^{a_1 \tau \bm{L}_{A_V}}e^{b_1 \tau \bm{L}_{B_V}}e^{a_2 \tau \bm{L}_{A_V}}e^{b_2 \tau \bm{L}_{B_V}} e^{a_3 \tau \bm{L}_{A_V}}e^{b_3 \tau \bm{L}_{B_V}}e^{a_4 \tau \bm{L}_{A_V}}e^{b_4 \tau \bm{L}_{B_V}}e^{a_5 \tau \bm{L}_{A_V}} \\
                & \times e^{b_4 \tau \bm{L}_{B_V}}e^{a_4 \tau \bm{L}_{A_V}}e^{b_3 \tau \bm{L}_{B_V}} e^{a_3 \tau \bm{L}_{A_V}}e^{b_2 \tau \bm{L}_{B_V}}e^{a_2 \tau \bm{L}_{A_V}}e^{b_1 \tau \bm{L}_{B_V}}e^{a_1 \tau \bm{L}_{A_V}}.
            \label{eq:abah864}
        \end{align}
        This integrator has $17$ steps and its constants $a_i, b _i, ~i =1,2,3,4,5$ are presented in Table 4 of~\citep{blanes2013new}.
        The main difference between the two schemes~\eqref{eq:aba864} and~\eqref{eq:abah864} is the fact that the $ABAH864$ integrator was constructed for Hamiltonians for which one of their parts $A$ or $B$ is not directly integrable and its action can be approximated by an order two SI.
        Thus it can be considered as a two, or a three-part split SI.
        
\end{enumerate}

\subsubsection{\label{subsubsec:2ps_si_order_6}Symplectic integrators of order six.}
\begin{enumerate}
    \item[] {\bf The $SABA_2Y6$ integrator.}
        As mentioned earlier in Sec.~\ref{subsec:yosida_technique_num}, apart from the Yoshida composition technique~\eqref{eq:yoshida_composition_formula_general_num}, there exist other composition techniques leading to higher order SIs.
        A composition method for creating a sixth order SI, starting from a second order scheme $S_2$ was presented in~\citep{yoshida1990construction} having the form 
        \begin{equation}
            S_6(\tau) = S_2(w_3 \tau)S_2(w_2 \tau)S_2(w_1 \tau)S_2(w_0 \tau)S_2(w_1 \tau)S_2(w_2 \tau)S_2(w_3 \tau).
            \label{eq:yoshida_composition_formula_general_num_6_order}
        \end{equation}
        It is worth mentioning that the numerical scheme~\eqref{eq:yoshida_composition_formula_general_num_6_order} was also presented in~\citep{kahan1997composition} where it was called the $s6odr6$ method.
        The constants $w_i, ~i=0, 1, 2, 3$ of~\eqref{eq:yoshida_composition_formula_general_num_6_order} are listed in Table 1 of~\citep{yoshida1990construction} corresponding to the so-called `solution A' of that table. 
        Using Eq.~\eqref{eq:yoshida_composition_formula_general_num_6_order} with the $SABA_2$~\eqref{eq:saba2} SI in the place of $S_2(\tau)$, we construct the order 6 SI $SABA_2Y6$ having $29$ steps.

    \item[] {\bf The $ABA864Y6$ scheme.}
        Implementing the composition method~\eqref{eq:yoshida_composition_formula_general_num} for $n = 4$, with the $ABA864$~\eqref{eq:abah864} in the role of $S_4(\tau)$, we construct the order $6$ integrator $ABA864Y6$, possessing $43$ steps.
        This integrator was used in~\citep{senyange2018computational}.
    \item[] {\bf The $s9SABA_26$ integrator.}
        We also consider the composition technique~\citep{kahan1997composition}
        \begin{equation}
            s9odr6b(\tau ) = S_2(\delta _1 \tau)S_2(\delta _2 \tau)S_2(\delta _3 \tau)S_2(\delta _4 \tau)S_2(\delta _5 \tau)S_2(\delta _4 \tau)S_2(\delta _3 \tau)S_2(\delta _2 \tau)S_2(\delta _1 \tau),
            \label{eq:s9odr6b_general}
        \end{equation}
        with the values of the coefficients $\delta_i$, $i=1,2,3,4,5$ given in the Appendix A of~\citep{kahan1997composition}.
        Implementing the composition method~\eqref{eq:s9odr6b_general} using the SI $SABA_2$~\eqref{eq:saba2} in place of $S_2(\tau)$, we build the $s9SABA_26$ order 6 integrator, having $37$ steps.

\end{enumerate}

\subsection{\label{subsec:symp_3ps_num}Three-part split symplectic integrators}
For some models such as the 1D and 2D DDNLS systems~\citep{senyange2018characteristics,manda2020chaotic,sun2020anomalous} studied here, it is preferable (or even necessary) to split the Hamiltonian in more than two integrable parts~\citep{skokos2014high,gerlach2016symplectic,laskar2019dedicated,casas2020composition}. 
Here, we construct some basic SIs for such systems, i.e.~for Hamiltonian functions which split in three integrable parts as 
\begin{equation}
    \mathcal{H} = \mathcal{A} + \mathcal{B} + \mathcal{C}.
    \label{eq:general_three_part_abc}
\end{equation}
In this case, we are looking for appropriate coefficients $c_i,~d_i$ and $e_i$ in order to obtain the approximation
\begin{equation}
    e^{\tau (L_{\mathcal{A}_V} + L_{\mathcal{B}_V} + L_{\mathcal{C}_V})} = \prod _{i=1}^{k} e^{\tau a_i L_{\mathcal{A}_V}}e^{\tau b_i L_{\mathcal{B}_V}}e^{\tau c_i L_{\mathcal{C}_V}} + \mathcal{O} (\tau ^{n + 1}).
    \label{eq:elh_elabc_general}
\end{equation}
Before proceeding, it is worth mentioning that some additional well-performing three-part split schemes were proposed recently in~\citep{casas2020composition} which were also implemented for the 1D DDNLS model.

\subsubsection{\label{subsubsec:3ps_si_order_2}Symplectic integrators of order two.}
\begin{enumerate}
    \item[] {\bf The $\mathcal{A}\mathcal{B}\mathcal{C}2$ integrator.}
        The simplest possible scheme of this type is 
        \begin{equation}
            \mathcal{ABC}2 (\tau ) = e^{a_1 \tau \bm{L}_{\mathcal{A}_V}}e^{b_1 \tau \bm{L}_{\mathcal{B}_V}}e^{c_1 \tau \bm{L}_{\mathcal{C}_V}}e^{b_1 \tau \bm{L}_{\mathcal{B}_V}}e^{a_1 \tau \bm{L}_{\mathcal{A}_V}},
            \label{eq:abc2}
        \end{equation}
        with $a_1 = b_1 = 1/2$ and $c_1 = 1$, which is in some sense the equivalent to the LF~\eqref{eq:leapfrog} we saw in the case of two-part split.
        This integrator has $5$ steps and is of order $2$.
        Equation~\eqref{eq:abc2} can be considered as the outcome of two successive applications of the LF~\eqref{eq:leapfrog}, first to the Hamiltonian $\mathcal{H} = \mathcal{A} + B$ and then to the Hamiltonian $B = \mathcal{B}+ \mathcal{C}$, considering it to be split into two parts.   
        This scheme has already been used in several studies (see for example~\citep{koseleff1996exhaustive,chambers1999hybrid, gozdziewski2008long,chambers2010n,quinn2010symplectic,sun2020anomalous}).
        As in the case of two-part split SIs we can construct higher order schemes based on composition methods.
        Some integrators constructed this way are reported in the rest of this section.
\end{enumerate}

\subsubsection{\label{subsub:3ps_si_order_4}Symplectic integrators of order four.}
\begin{enumerate}
    \item[] {\bf The $\mathcal{ABC}Y4$ scheme.}
        Implementing the Yoshida composition technique~\eqref{eq:yoshida_composition_formula_general_num} for $n = 2$ with $\mathcal{ABC}2$~\eqref{eq:abc2} in the place of the second order SI $S_2 (\tau)$, we obtain the SI $\mathcal{ABC}Y4\left(\tau \right)$~\citep{koseleff1996exhaustive} having $13$ steps.
        This integrator has already been used by~\citep{gerlach2016symplectic} for studying multidimensional Hamiltonian systems. In that work, the integrator was named $\mathcal{ABC}^{4}_{[Y]}$.
    
    \item[] {\bf The $\mathcal{ABC}S4$ integrator.}
        Implementing the composition technique proposed in~\citep{suzuki1990fractal}, we build the fourth order SI
        \begin{equation}
            \mathcal{ABC}S4\left(\tau \right) = \mathcal{ABC}2\left(p_2\tau\right) \mathcal{ABC}2\left(p_2\tau\right))\mathcal{ABC}2\left((1- 4p_2)\tau\right)\mathcal{ABC}2\left(p_2\tau\right) \mathcal{ABC}2\left(p_2\tau\right),
            \label{eq:abcs4}
        \end{equation}
        with $p_2 = \frac{1}{4 -4^{1/3}}$ and $1 - 4p_2 = \frac{-4^{1/3}}{4 - 4^{1/3}}$.
        This integrator has $21$ individual steps. 
        The $\mathcal{ABC}S4$ SI was also considered in~\citep{kahan1997composition} where it was denoted as the $s5odr4$ method.
        In addition, it was also used in~\citep{gerlach2016symplectic} where it was called $\mathcal{ABC}^{4}_{[S]}$.
     
    \item[] {\bf The $SS864S$ scheme.}
        As we mentioned in Sec.~\ref{subsec:symp_2ps_num}, the $ABAH864$~\eqref{eq:abah864} SI was initially designed for Hamiltonians which can be split in three integrable parts and for which the numerical evolution of two of these parts (say $\mathcal{B} + \mathcal{C}$ in~\eqref{eq:general_three_part_abc}) is approximated by the application of a second order two-part split SI. 
        Using the $SABA_2$~\eqref{eq:saba2} SI for the integration of the $\mathcal{B} + \mathcal{C}$~\eqref{eq:general_three_part_abc} part, the application of the $ABAH864$~\eqref{eq:abah864} SI results in a method having $49$ steps which we name $SS864S$.
        This integrator was called $SS_{864}^4$ in~\citep{skokos2014high}.
\end{enumerate}

\subsubsection{\label{subsubsec:3ps_si_order_6}Symplectic integrators of order six.}
\begin{enumerate}
    \item[] {\bf The $\mathcal{ABC}Y4Y6/\mathcal{ABC}S4Y6$ integrators.}
    Implementing the Yoshida composition technique~\eqref{eq:yoshida_composition_formula_general_num} for $n=4$, with the $\mathcal{ABC}Y4$ and $\mathcal{ABC}S4$~\eqref{eq:abcs4} methods in the place of $S_4 (\tau)$, we respectively construct the $\mathcal{ABC}Y4Y6$ and $\mathcal{ABC}S4Y6$ schemes having $37$ and $49$ steps.

    \item[] {\bf The $\mathcal{ABC}Y6\_A$ scheme.}
        Applying the composition technique~\eqref{eq:yoshida_composition_formula_general_num_6_order} with $\mathcal{A}\mathcal{B}\mathcal{C}2$~\eqref{eq:abc2} in the role of $S_2(\tau)$, we create the $\mathcal{ABC}Y6\_A (\tau)$ integrator
        with $29$ steps.
        This SI has already been implemented in~\citep{skokos2014high,gerlach2016symplectic}, where it was refereed to as $\mathcal{ABC}^{6}_{[Y]}$.
        
    \item[] {\bf The $s9\mathcal{ABC}6$ integrator.} 
        Implementing the composition technique~\eqref{eq:s9odr6b_general}, with $\mathcal{ABC}2$~\eqref{eq:abc2} in the place of $S_2(\tau)$, we build the SI $s9\mathcal{ABC}6(\tau)$
        which has $37$ steps.
        This integrator was also studied in~\citep{skokos2014high,gerlach2016symplectic} where it was named $\mathcal{ABC}^{6}_{[KL]}$.
    
    \item[] {\bf The $s11\mathcal{ABC}6$ scheme.} 
        The application of the composition technique~\citep{sofroniou2005derivation}
        \begin{equation}
            \begin{split}
                s11odr6(\tau) = S_2 (\gamma _1\tau)S_2 (\gamma _2\tau) &  S_2 (\gamma _3\tau)S_2 (\gamma _4\tau) S_2 (\gamma _5\tau)S_2 (\gamma _6\tau)S_2 (\gamma _5\tau) \\
                        & \times S_2 (\gamma _4\tau)S_2 (\gamma _3\tau)S_2 (\gamma _2\tau)S_2 (\gamma _1\tau),
            \end{split}
            \label{eq:s11odr6_general}
        \end{equation}
        with $\mathcal{ABC}2$~\eqref{eq:abc2} in the role of $S_2(\tau)$, leads to the scheme $s11\mathcal{ABC}6(\tau)$
        which has $45$ steps. 
        The values of the coefficients $\gamma_i$, $i=1, 2,\ldots, 6$ of~\eqref{eq:s11odr6_general} are given in section 4.2 of~\citep{sofroniou2005derivation}.
        The $s11\mathcal{ABC}6$ SI was called $\mathcal{ABC}^{6}_{[SS]}$ in~\citep{skokos2014high,gerlach2016symplectic}.
    
\end{enumerate}

\subsubsection{\label{subsubsec:3ps_si_order_8}Symplectic integrators of order eight.}
\begin{enumerate}
    \item[] {\bf The $\mathcal{ABC}Y8\_A/\mathcal{ABC}Y8\_D$ schemes.} 
        Implementing the composition technique~\citep{yoshida1990construction} 
        \begin{align}
            \nonumber
            S_8(\tau)=S_2(w_7 \tau)S_2(w_6 \tau)S_2(w_5 \tau) & S_2(w_4 \tau)S_2(w_3 \tau)S_2(w_2 \tau)S_2(w_1 \tau)S_2(w_0 \tau)S_2(w_1 \tau)S_2(w_2 \tau) \\
                & \times S_2(w_3 \tau)S_4(w_4 \tau)S_2(w_5 \tau)S_2(w_6 \tau)S_2(w_7\tau),
            \label{eq:yoshida_composition_formula_general_num_8_order}
        \end{align}
        with the $\mathcal{ABC}2$~\eqref{eq:abc2} in the role of $S_2(\tau)$, we build an order $8$ SI $\mathcal{ABC}Y8$ which has $61$ steps.
        For this scheme, we use two sets of coefficient values $w_i$, $i = 0, \ldots, 7$ corresponding to the so-called `solution A' and `solution D' in~\citep{yoshida1990construction}.
        The particular sets of these values can be found in Table 2 of~\citep{yoshida1990construction}.
        Here, we generate the SI $\mathcal{ABC}Y8\_A(\tau)$ [$\mathcal{ABC}Y8\_D(\tau)$] using the coefficient values of `solution A' [`solution D'].
    \item[] {\bf The $s17\mathcal{ABC}8$ integrator.} 
        Implementing the composition technique~\citep{kahan1997composition}
        \begin{align}
            s17odr8b (\tau) = S_2 (\delta _1\tau)S_2 (\delta _2\tau) \times \dots \times S_2 (\delta _8\tau)S_2 (\delta _9\tau)S_2 (\delta _8\tau)  \dots \times S_2 (\delta _2\tau)S_2(\delta _1\tau),
            \label{eq: abc8kl}
        \end{align}
        with the $\mathcal{ABC}2$~\eqref{eq:abc2} SI playing the role of the second order SI $S_2(\tau)$, we create the scheme $s17\mathcal{ABC}8$ having $69$ steps.
        The coefficient values $\delta_i$, $i = 1, 2, \ldots, 9$ are given in Appendix A of~\citep{kahan1997composition}.
        The $s17\mathcal{ABC}8$ integrator was named $\mathcal{ABC}^{8}_{[KL]}$ in~\citep{skokos2014high,gerlach2016symplectic}.
        
    \item[] {\bf The $s19\mathcal{ABC}8$ scheme.} 
        Implementing the composition technique~\citep{sofroniou2005derivation}
        \begin{align}
            s19odr8b(\tau) = S_2(\gamma _1\tau)\mathcal{ABC}2(\gamma _2\tau)  \times \dots \times S_2(\gamma _8\tau)S_2(\gamma _9\tau)S_2(\gamma _8\tau)\dots \times S_2(\gamma _2\tau)S_2(\gamma _1\tau),
            \label{eq:abc8ss}
        \end{align}
        with the $\mathcal{ABC}2$~\eqref{eq:abc2} integrator being the required second order SI $S_2(\tau)$, we obtain the scheme $s19\mathcal{ABC}8$ having $72$ steps.
        In~\citep{sofroniou2005derivation}, the values of the coefficients $\gamma_i$, $i = 1, 2, \ldots, 9$  are listed in equation ($13$).
        The $s19\mathcal{ABC}8$ SI was called $\mathcal{ABC}^{8}_{[SS]}$ in~\citep{skokos2014high,gerlach2016symplectic}.
\end{enumerate}

\section{\label{sec:num_comparison_num}Efficient integration of the 1D and 2D DDNLS systems}
We will compare the numerical efficiency of several of the integration schemes presented in Secs.~\ref{subsec:symp_2ps_num} and~\ref{subsec:symp_3ps_num} for studying the DDNLS model in one~\eqref{eq:Hamilton_complex_dnls_1d} and two~\eqref{eq:hamilton_complex_ddnls_2d} spatial dimensions.
In particular, we will compute the time evolution of initial excitations by solving the system's equations of motion and variational equations through numerical integration.
The efficiency of the integration methods will be judged by their ability to correctly reproduce the system's dynamical behaviors in the least possible needed central processing unit (CPU) time.
Since our focus is on the implementation of SIs for these tasks, our results will be presented mainly in relation to such integration techniques.

\subsection{\label{subsec:two_part_split}Two-part split symplectic integrators}
Two-part split SIs can be used for the Hamiltonian functions $\mathcal{H}_{1D}$~\eqref{eq:Hamilton_complex_dnls_1d} and $\mathcal{H}_{2D}$~\eqref{eq:hamilton_complex_ddnls_2d} of the 1D and 2D DDNLS models, although as we will disucss below this is not the best approach in integrating such systems if the number of sites is large.
Let us discuss in more detail the case of the 1D DDNLS model~\eqref{eq:Hamilton_complex_dnls_1d}. 
Its Hamiltonian function splits into two integrable parts $\mathcal{H}_{1D} = A + B$ as 
\begin{equation}
    A = \sum _{l} \epsilon _l \lvert \psi _l \rvert ^2 + \frac{\beta}{2} \lvert \psi _l \rvert ^4,
    \label{eq:a_2ps_1d_ddnls}
\end{equation}
possessing $N$ integrals of motion, namely the quantities $s_l = \lvert \psi_l \rvert ^2$, and 
\begin{equation}
    B = -\sum _{l} \psi _{l + 1}^\star \psi _{l} + \psi _{l + 1} \psi _{l}^\star,
    \label{eq:b_2ps_1d_ddnls}
\end{equation}
where $\psi_l$ is a complex valued quantity representing the wave packet's state at site $l$.
Following~\citep{skokos2014high,gerlach2016symplectic}, we introduce real phase space variables 
\begin{equation}
    \psi _l = \frac{1}{\sqrt{2}} \left(q_l + ip_l\right), \quad \psi_l ^\star = \frac{1}{\sqrt{2}} \left(q_l - ip_l\right),
    \label{eq:transformation_complex_real_1d_num}
\end{equation}
with $q_l$ and $p_l$ respectively being the generalized position and momentum of the system at site $l$.
This transformation brings the A~\eqref{eq:a_2ps_1d_ddnls} and B~\eqref{eq:b_2ps_1d_ddnls} parts of the $\mathcal{H}_{1D}$~\eqref{eq:Hamilton_complex_dnls_1d} Hamiltonian to
\begin{equation}
    A = \sum _{l} \frac{\epsilon _l }{2}\left(q_l ^2 + p_l ^2 \right) + \frac{\beta}{8}\left(q_l ^2 + p_l ^2 \right)^4, \quad
    B =  \sum _{l} - p_{l + 1}p_l - q_{l + 1} q_l.
    \label{eq:b_2ps_1d_ddnls_real}
\end{equation}
In~\citep{gerlach2016symplectic}, the efficiency of several two-part split SIs (in particular the $LF$~\eqref{eq:leapfrog}, $SABA_2$~\eqref{eq:saba2}, $ABA864$~\eqref{eq:aba864} and $SABA_2Y6$~\eqref{eq:yoshida_composition_formula_general_num_6_order}) were tested mainly in comparison with three-part split schemes.
In that work, the ability of the integrator to accurately reproduce the correct system's dynamics in the least CPU time was investigated primarily through the computation of the wave packet's second moment~\eqref{eq:second_moment_distr_num} and participation number~\eqref{eq:participation_ratio_num}.
As the 1D DDNLS possesses two integrals of motion, namely the total norm $\mathcal{S}_{1D}$~\eqref{eq:norm_complex_dnls_1d} and energy $\mathcal{H}_{1D}$~\eqref{eq:Hamilton_complex_dnls_1d}, the accuracy of the integration of the equations of motion was checked via the computation of the {\it relative energy error}
\begin{equation}
    E_r (t) = \left\lvert \frac{\mathcal{H}_{1D} (t) - \mathcal{H}_{1D} (0)}{\mathcal{H}_{1D} (0)} \right\rvert,
    \label{eq:rel_energy_error_1dDDNLS}
\end{equation}
and {\it  relative norm error}
\begin{equation}
    S_r (t) = \left\lvert \frac{\mathcal{S}_{1D} (t) - \mathcal{S}_{1D} (0)}{\mathcal{S}_{1D} (0)} \right\rvert.
    \label{eq:rel_norm_error_1dDDNLS}
\end{equation}

The main outcome of that study was that for large 1D DDNLS systems having $N \gtrsim  70$ lattice sites, three-part split SIs (whose efficiency will be discussed below) have to be preferred over two-part split schemes.
Although the investigation in~\citep{gerlach2016symplectic} was limited to the numerical integration of the system's equations of motion it is reasonable to assume that the simultaneous integration of the variational equations (needed for the computation of chaos indicators like the MLE) will not change the superiority of three-part split methods for large systems.
Since in our work, we always consider systems with $N > 70$ sites, typically $N$ is of the order of several hundreds, we will not use two-part split SIs.
Nevertheless, for the sake of completeness, we present in Sec.~\ref{sec:two-part_split_app} of Appendix~\ref{app:diff_op_ddnls} the explicit expression of the operators appearing in SI implementations for solving both the equations of motion and the variational equations of the 1D DDNLS system~\eqref{eq:Hamilton_complex_dnls_1d}.
Due to the inefficiency of two-part split SIs for the 1D DDNLS system, such SI were not also considered for the 2D DDNLS model. 

As a final remark, let us note that the two-part split SIs are adequate for computing the dynamics (integration of the equations of motion) and tangent dynamics (integration of the variational equations) for several many body systems like the 1D~\eqref{eq:hamilton_1d_dkg} and 2D~\eqref{eq:hamilton_2d_dkg} DKG models (see~\citep{skokos2013nonequilibrium,danieli2017intermittent}), the FPUT model and the XY model of Josephson junction array~\citep{danieli2017intermittent,mithun2019dynamical}.
The performance of several such schemes were studied in detail in~\citep{senyange2018computational,danieli2019computational}.
There, it was found that the integrators $ABA864$~\eqref{eq:aba864}, $ABAH864$~\eqref{eq:abah864} of order $4$ and the $SABA_2Y6$~\eqref{eq:yoshida_composition_formula_general_num_6_order}, $s9SABA_26$~\eqref{eq:s9odr6b_general} and $ABA864Y6$~\eqref{eq:yoshida_composition_formula_general_num} SIs of order $6$ exhibited the best performance when a moderate energy conservation ($E_r\approx 10^{-5})$ was required.

\subsection{\label{subsec:three_part_plit}Three-part split symplectic integrators}
Let us now consider the case of SIs for Hamiltonian systems which can be split in three integrable parts.
Example cases of such Hamiltonian functions include the 1D~\eqref{eq:Hamilton_complex_dnls_1d} and 2D~\eqref{eq:hamilton_complex_ddnls_2d} DDNLS models~\citep{skokos2014high,gerlach2016symplectic,danieli2019computational}, system's describing the motion of a rotating body~\citep{laskar2019dedicated} and of charged particles under Lorentz force fields~\citep{he2015volume,he2016higher,casas2020composition}.

The 1D DDNLS Hamiltonian~\eqref{eq:Hamilton_complex_dnls_1d} can be split in three integrable parts
\begin{equation}
    \mathcal{A} = \sum_l \frac{\epsilon_l}{2}\left(q_l^2 + p_l^2\right) + \frac{\beta}{8}\left(q_l^2 + p_l^2\right)^2, \quad \mathcal{B} = -\sum _l  p_{l + 1} p_l, \quad \mathcal{C} = -\sum_l q_{l + 1} q_{l},
    \label{eq:bc_3ps_ddnls_real}
\end{equation}
where obviously $\mathcal{A} = A$~\eqref{eq:a_2ps_1d_ddnls}.
Similarly, the Hamiltonian function of the 2D DDNLS system~\eqref{eq:hamilton_complex_ddnls_2d} can be split as 
\begin{equation}
    \begin{split}
    \mathsf{A} = \sum_{l,m} \frac{\epsilon_{l,m}}{2} \left(q_{l,m}^2 + p_{l,m}^2\right) + \frac{\beta}{8}& \left(q_{l,m}^2 + p_{l,m}^2\right)^2, \quad \mathsf{B} = -\sum_{l,m} p_{l+1, m}p_{l,m} + p_{l, m+1}p_{l,m},  \\ 
            &\mathsf{C} =-\sum_{l,m} q_{l+1, m}q_{l,m} + q_{l, m+1}q_{l,m}.
    \end{split}
    \label{eq:bc_3ps_ddnls_real_2d}
\end{equation} 
The explicit expressions of the operators needed for the application of three-part split SIs for these two systems are respectively given in sections~\ref{subsec:1D_DDNLS_thps_app} and~\ref{susec:3ps_2d_ddnsl_app} of Appendix~\ref{app:diff_op_ddnls}.

Let us first focus on the 1D DDNLS system~\eqref{eq:bc_3ps_ddnls_real} and check the ability of the various integration schemes to accurately reproduce the system's evolution in its phase space, as well as its tangent dynamics.
To that end, we monitor the time evolution of the energy $E_r(t)$~\eqref{eq:rel_energy_error_1dDDNLS} and norm $S_r(t)$~\eqref{eq:rel_norm_error_1dDDNLS} errors along with a basic observable like the wave packet's second moment $m_2(t)$~\eqref{eq:second_moment_distr_num}. 
For monitoring the accurate description of the system's the tangent dynamics we compute the motion's ftMLE~\eqref{eq:finite_mle}.
\begin{figure}[!hhtb]
    \centering
    \includegraphics[scale=0.55]{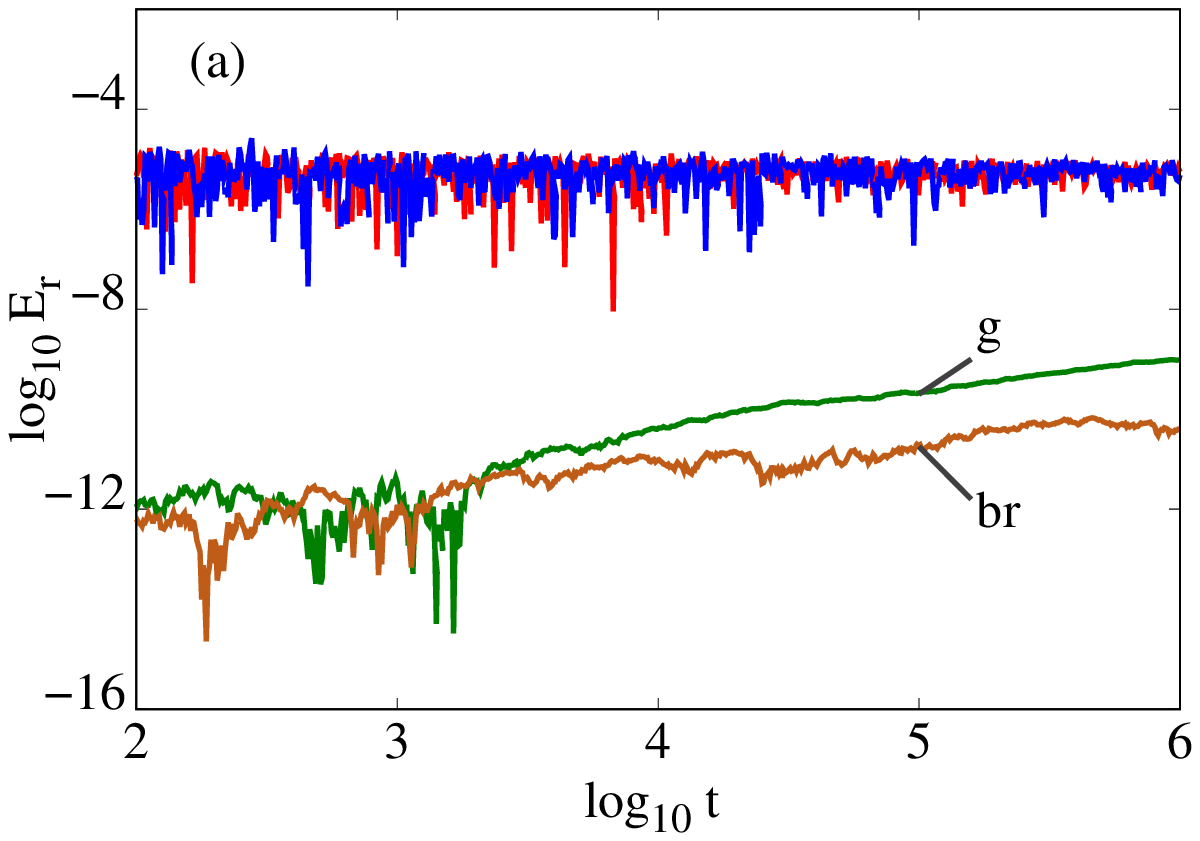}
    \includegraphics[scale=0.55]{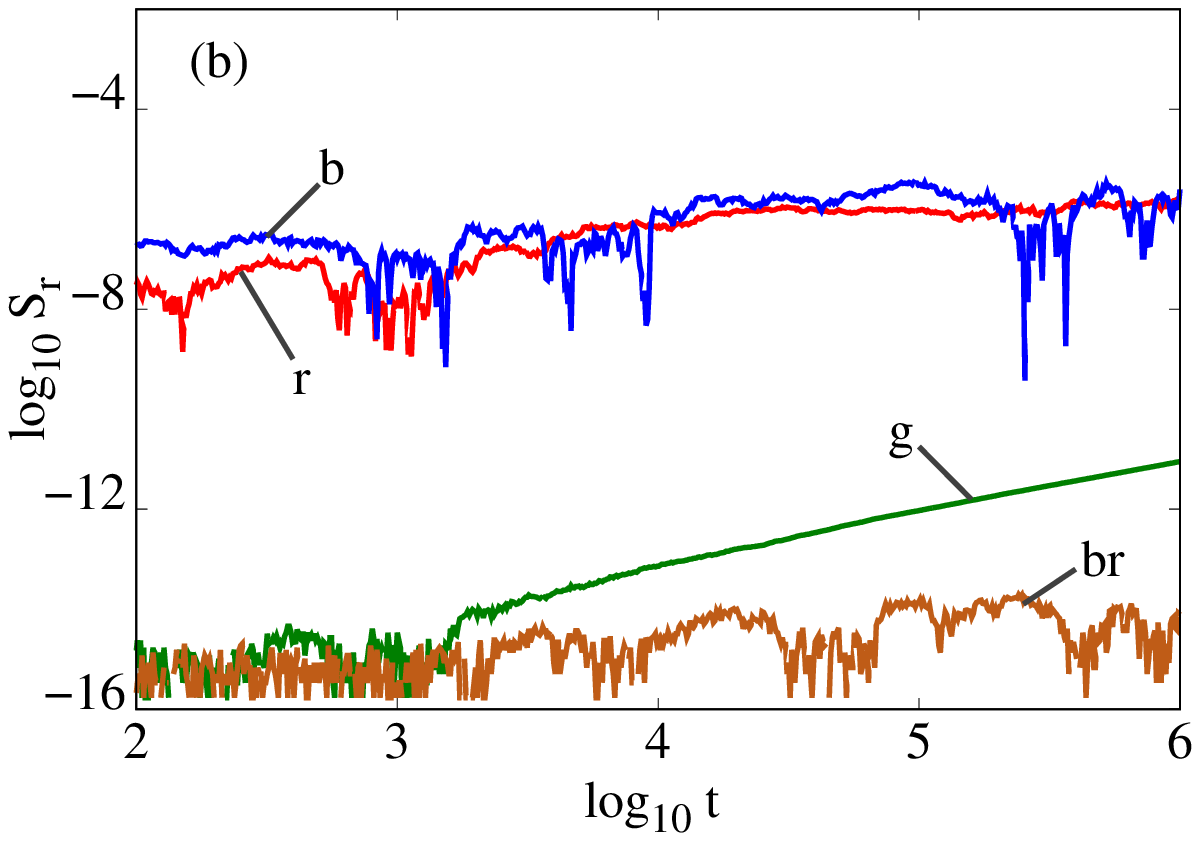}
    \includegraphics[scale=0.55]{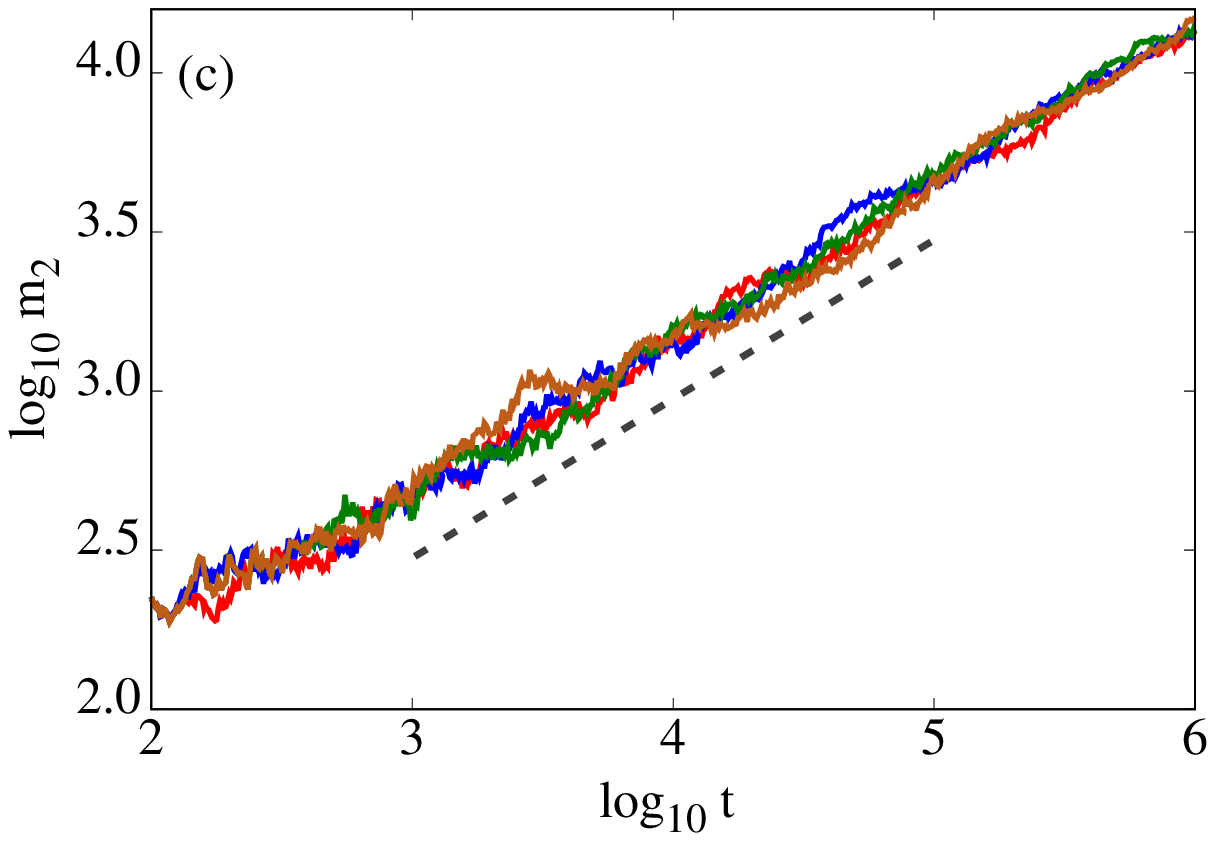}
    \includegraphics[scale=0.55]{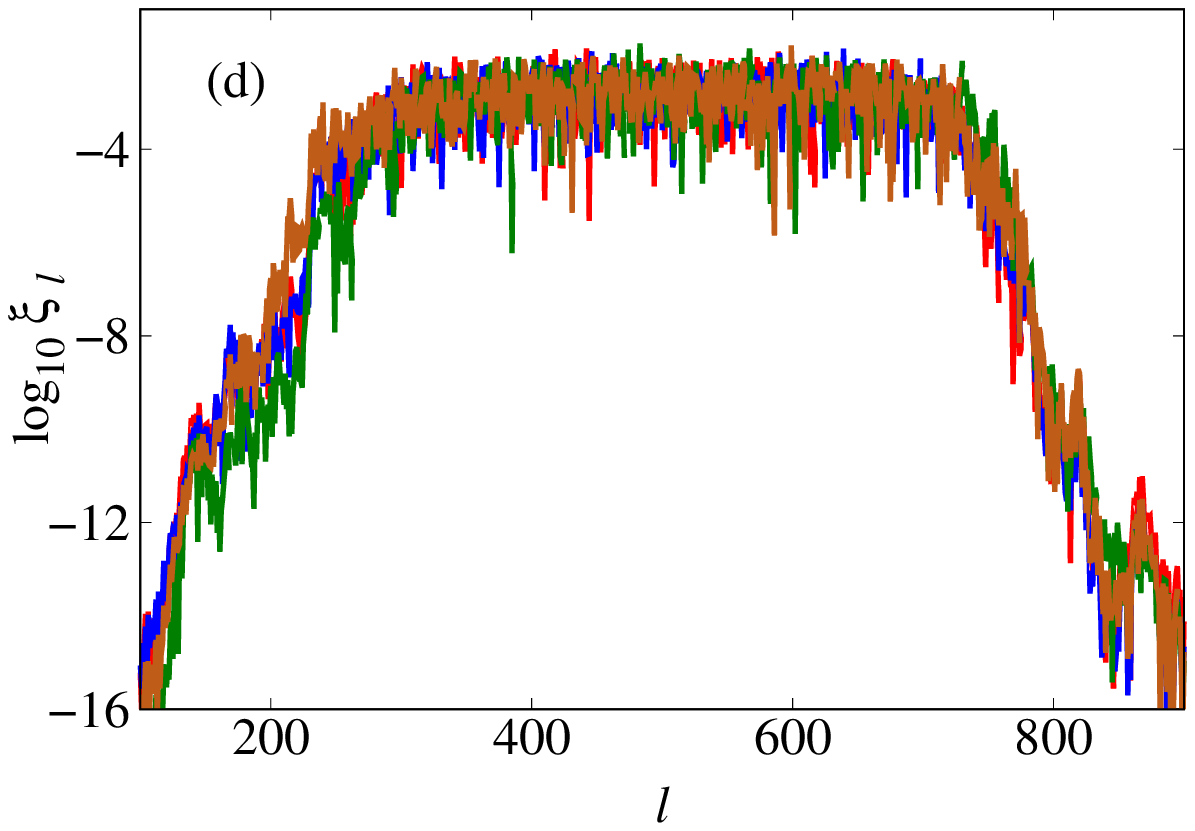}
    \caption{Results for the numerical integration of case $\text{A}1$ (see text for details) of the 1D DDNLS system~\eqref{eq:Hamilton_complex_dnls_1d} by the second order SI
        $\mathcal{ABC}2$~\eqref{eq:abc2} for $\tau = 0.0002$ [(r)ed curves], the fourth order SI $\mathcal{ABC}Y4$ (\ref{subsec:symp_3ps_num}) for $\tau = 0.0125$ [(b)lue curves] and the non-symplectic schemes $DOP853$ (Sec.~\ref{subsec:rk_family_num}) [(g)reen curves] and $TIDES$ (Sec.~\ref{subsec:taylor_series}) [(br)own curves].
        The time evolution of (a) the  relative energy error $E_r(t)$~\eqref{eq:rel_energy_error_1dDDNLS}, (b) the  relative norm error $S_r(t)$~\eqref{eq:rel_norm_error_1dDDNLS} and (c) the second moment $m_2(t)$~\eqref{eq:second_moment_distr_num} of the wave packet. 
        In (d) the norm density distribution $\xi_l$~\eqref{eq:norm_density_distribution_num} at time $t_f=10^{6}$ is shown. 
        The dashed line in (c) guides the eye for slope $1/2$ (from~\citep{danieli2019computational}).
    }
    \label{fig:figE9S19S11DOPTIDES}
\end{figure}
We consider two cases belonging respectively to the strong and weak chaos spreading regimes, which were also investigated in~\citep{senyange2018characteristics}:
\begin{enumerate}
    \item[$\bullet$] {\bf Case $\text{A}1$.} We excite $L = 21$ central sites, with a constant norm per site $s=1$. 
        In addition, the parameters $W = 3.5$ and $\beta = 0.62$ were chosen for the $\mathcal{H}_{1D}$~\eqref{eq:Hamilton_complex_dnls_1d} such that its value is approximately $0.0212$ for the disorder realization we are using. 
    \item[$\bullet$] {\bf Case $\text{A}2$.} Here, we consider a similar setup as in case $\text{A}1$, but the parameters of the $\mathcal{H}_{1D}$~\eqref{eq:Hamilton_complex_dnls_1d} are $W = 3$ and $\beta = 0.03$, giving for the disorder realization we selected $\mathcal{H}_{1D} \approx 3.444$. 
\end{enumerate}
In both cases, the total size of the lattice is $N = 1024$ sites and the initial DV needed for the computation of the MLE is chosen such that only one site at the center of the lattice has non zero $\delta q_l$ and $\delta p_l$ values, while its other components are set to zero.
As usual, the initial DV was normalized so that $\lVert \bm{w} (0)\rVert = 1$.

Some of the result of our analysis for the case $\text{A}1$ are presented in Fig.~\ref{fig:figE9S19S11DOPTIDES}.
In particular, we depict the evolution in time of the energy error $E_r (t)$~\eqref{eq:rel_energy_error_1dDDNLS} in Fig.~\ref{fig:figE9S19S11DOPTIDES}(a), the norm error $S_r (t)$~\eqref{eq:rel_norm_error_1dDDNLS} in Fig.~\ref{fig:figE9S19S11DOPTIDES}(b), and the $m_2 (t)$~\eqref{eq:second_moment_distr_num} in Fig.~\ref{fig:figE9S19S11DOPTIDES}(c), while in Fig.~\ref{fig:figE9S19S11DOPTIDES}(d) we plot the norm profile at the end of our integration.
We see that the used SIs: namely the $\mathcal{ABC}2$ [(r)ed curves] and $\mathcal{ABC}Y4$ [(b)lue curves] schemes conserve quite well the value of the computed energy as the relevant error is bounded from above by $E_r \approx 10^{-5}$ [Fig.~\ref{fig:figE9S19S11DOPTIDES}(a)].
On the other hand, the general purpose integrators (non-symplectic schemes): $DOP853$ [(g)reen curves] and $TIDES$ [(br)own curves] integrators although having $E_r(t)$ values much smaller than that of the SIs, their errors are increasing in time.
From the results of Fig.~\ref{fig:figE9S19S11DOPTIDES} we see that the value of $S_r (t)$ grows in time for all the used integrators, with $TIDES$ showing the best behavior as its $S_r$ values are the smallest and show a mild increase up to the maximum considered time $t_f = 10^{6}$.
We see that the SIs do not conserve the second integral of motion i.e.~the system's norm, although the increase of the related error $S_r(t)$ is in general slower than the ones observed for the non-symplectic methods. 
We emphasize that all the used schemes accurately reproduce the system's dynamics as the evolution of the wave packet's $m_2$ [Fig.~\ref{fig:figE9S19S11DOPTIDES}(c)] and its profile [Fig.~\ref{fig:figE9S19S11DOPTIDES}(d)] are correctly captured by all methods.
We also note that similar behaviors are seen for case A$2$, although we do not report these results here in order to avoid repetition of similar outcomes.

\begin{figure}[!hhtb]
    \centering
    \includegraphics[scale=0.55]{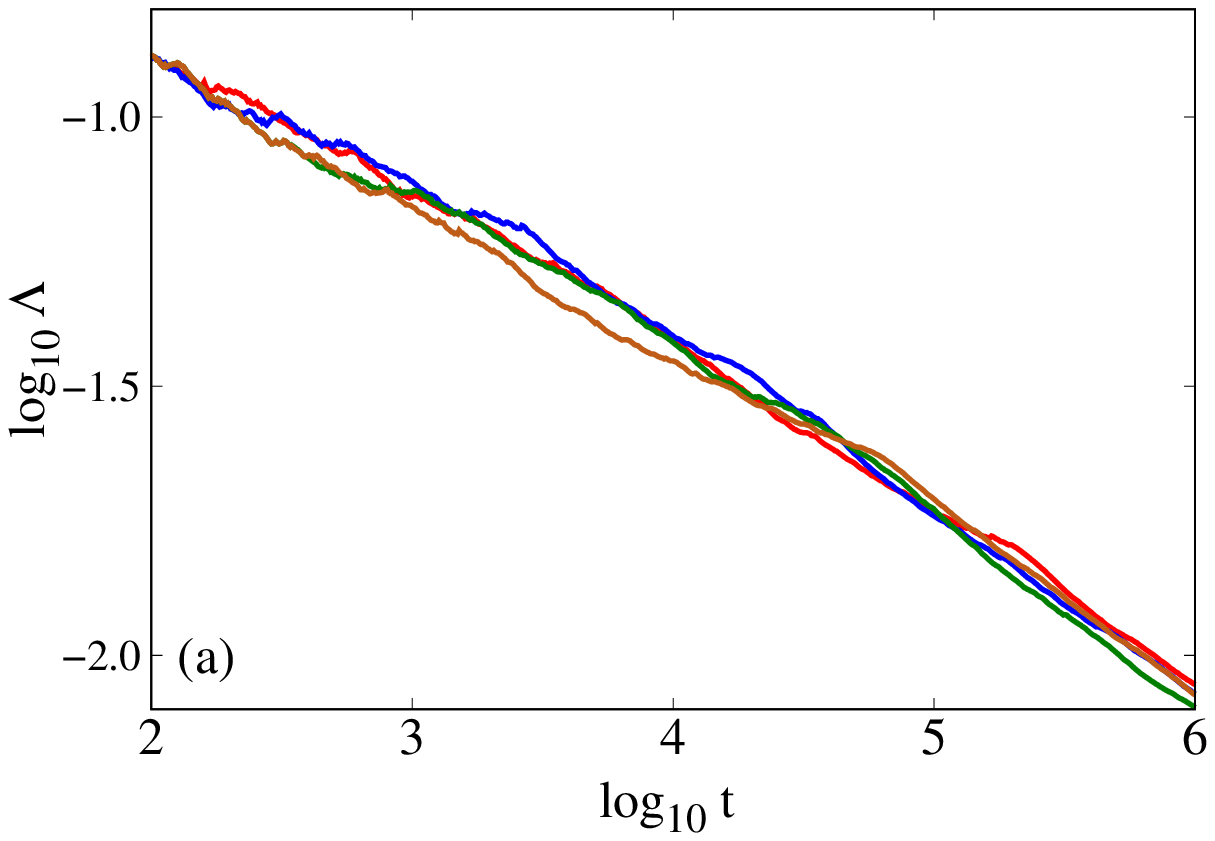}
    \includegraphics[scale=0.55]{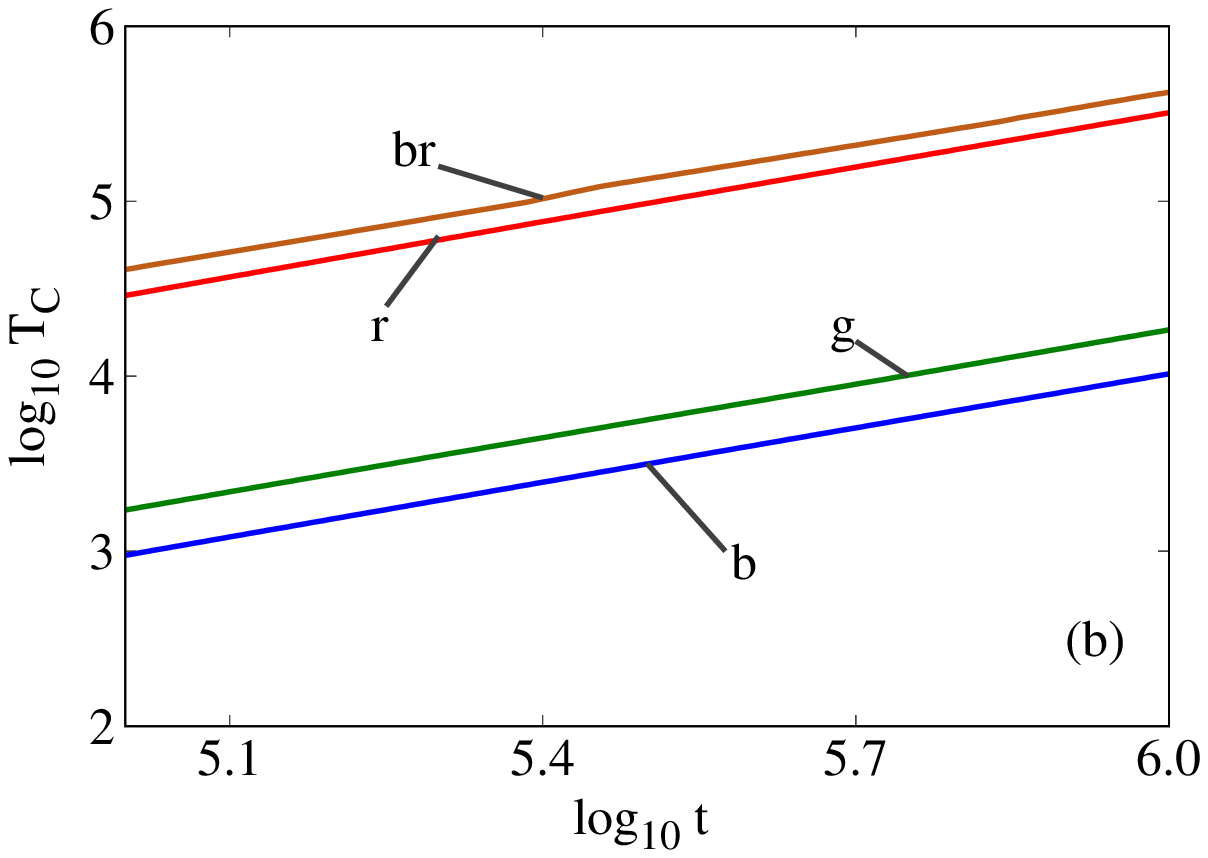}
    \includegraphics[scale=0.55]{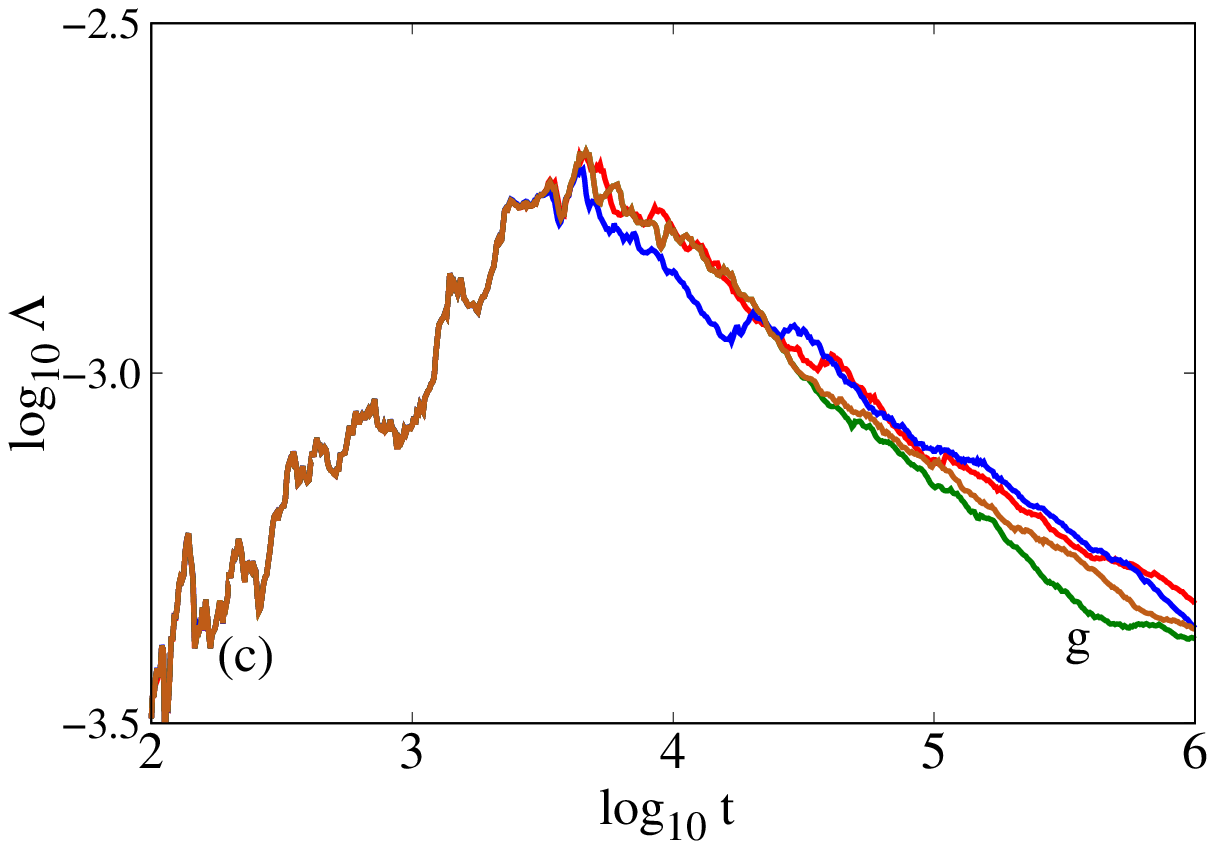}
    \includegraphics[scale=0.55]{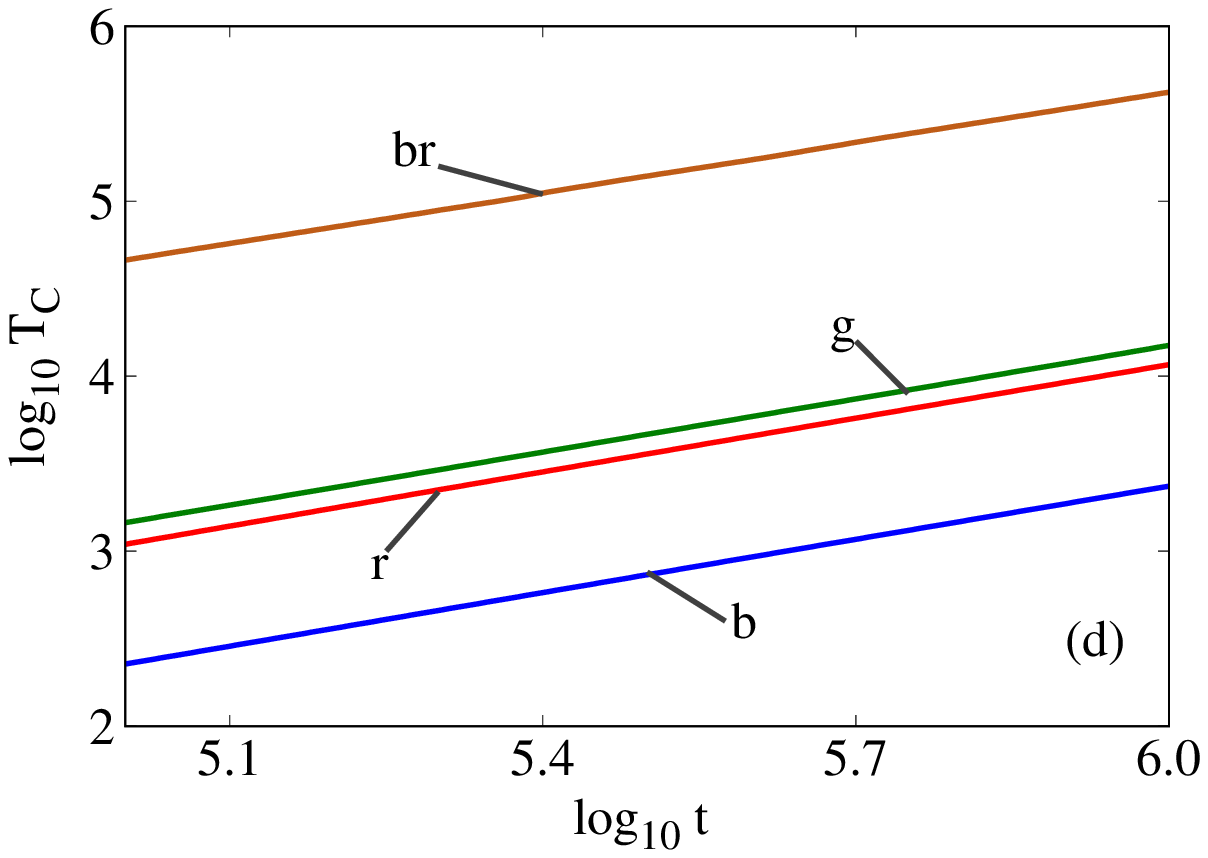}
    \caption{Results obtained by the numerical integration of the variational equations of the 1D DDNLS Hamiltonian~\eqref{eq:Hamilton_complex_dnls_1d} for the initial conditions described in cases $\text{A}1$ [panels (a) and (b)] and $\text{A}2$ [panels (c) and (d)] (see text for details).
            The time evolution of, (a) and (c) the ftMLE $\Lambda(t)$~\eqref{eq:finite_mle}, and (b) and (d) the required CPU time $T_C$ in seconds. 
            The integrators and the curves' colors are the ones used in Fig.~\ref{fig:figE9S19S11DOPTIDES}.
    }
    \label{fig:figE5S19S11DOPTIDES}
\end{figure}

The performance of seven three-part split SIs along with non-symplectic schemes for the integration of the equations of motion of the 1D DDNLS system was also investigated in~\citep{skokos2014high,gerlach2016symplectic}.
Here, we want to go one step further and investigate the ability of three-part split SIs to efficiently integrate the system's variational equations through the application of the `tangent map' method.
In Figs.~\ref{fig:figE5S19S11DOPTIDES}(a) and (c), we present the evolution of the ftMLEs, computed through the numerical solution of the variational equations  and see that for both sets of conditions considered here, all schemes practically reproduce similar results.
In Fig.~\ref{fig:figE5S19S11DOPTIDES}(b) and (d) we show the required CPU time $T_C$ for these computations and see that the best performing integrator among the ones presented in Fig.~\ref{fig:figE9S19S11DOPTIDES} is the $ABCY4$ scheme.

In Figs.~\ref{fig:figE9S19S11DOPTIDES} and~\ref{fig:figE5S19S11DOPTIDES}, we showed some representative results of only four integration schemes, although we performed a more detailed study by repeating the same simulations with several integrators.
The obtained results for cases A$1$ and A$2$ are respectively presented in Tables~\ref{table:performance1DDNLSstrong} and~\ref{table:performance1DDNLSweak}.
In these Tables accurate simulations with $E_r\lesssim 10^{-9}$ are also considered.
We see that overall, both tables give qualitatively the same information.
The best performing integrators for moderate accuracy ($E_r \lesssim 10^{-5}$) are the $s11\mathcal{ABC}6$ and $s9\mathcal{ABC}6$ SIs, while for higher accuracies ($E_r \lesssim  10^{-9}$), the $s19\mathcal{ABC}8$ and $s17\mathcal{ABC}8$ schemes are behaving better.
Consequently, we conclude that the $s11\mathcal{ABC}6$ and $s9\mathcal{ABC}6$ SIs are particularly well suited for the integration of the 1D DDNLS for moderate accuracies that are typically required in dynamical studies of this model~\citep{skokos2009delocalization,bodyfelt2011nonlinear,danieli2017intermittent,danieli2019dynamical,senyange2018characteristics,manda2020chaotic}.
\begin{table}[!hhhtb]
    \centering
    \caption{Information on the performance of the numerical shcmes used for the integration of the equations of motion and the variational equations of the 1D DDNLS system~\eqref{eq:Hamilton_complex_dnls_1d} up to final time of $t_f = 10^{6}$ time units for the case A$1$ (see text for details).
        The order $n$, the number of `Steps' of each SI along with the required CPU time $T_C$ in seconds are reported.
        The integration time step $\tau$ used for each integrator fixes the relative energy error $E_r(t)\lesssim 10^{-5}$ and $E_r(t) \lesssim 10^{-9}$.  
        All simulations were performed on a workstation using $3.00$ GHz Intel Xeon E$5-2623$ processors.
        All codes were written in Fortran90 language and were compiled by using the gfortran compiler (\url{https://gcc.gnu.org/}) with $-$O$3$ optimization flag. 
        No advanced vectorization mode has been implemented.
    }
    \begin{tabular}{lcclr|lcclr}
        \toprule
        \multicolumn{5}{c}{$E_r \approx 10^{-5}$} & \multicolumn{5}{c}{$E_r \approx 10^{-9}$} \\
        \toprule
        Integrator           & $n$ & Steps & $\tau$    & $T_{\text{C}}$ & Integrator           & $n$ & Steps               & $\tau$  & $T_{\text{C}}$ \\
        \midrule
        $s11\mathcal{ABC}6$  & $6$ & $45$  & $0.115$   & $3395$         & $s19\mathcal{ABC}8$  & $8$ & $77$                & $0.09$  & $7242$         \\
        $s9\mathcal{ABC}6$   & $6$ & $37$  & $0.095$   & $3425$         & $s17\mathcal{ABC}8$  & $8$ & $69$                & $0.08$  & $7301$         \\
        $\mathcal{ABC}Y6\_A$ & $6$ & $29$  & $0.07$    & $3720$         & $s11\mathcal{ABC}6$  & $6$ & $45$                & $0.025$ & $15692$        \\
            $SS864S$             & $4$ & $17$  &  $0.05$    & $6432$         & $s9\mathcal{ABC}6$   & $6$ & $37$                & $0.02$  & $16098$        \\
        $\mathcal{ABC}Y4$    & $4$ & $13$  & $0.0125$  & $10317$        & $DOP853$             & $8$ & $\delta_a=10^{-16}$   & $0.05$  & $18408$        \\
        $\mathcal{ABC}S4Y6$  & $6$ & $49$  & $0.015$   & $35417$        & $\mathcal{ABC}Y8\_D$ & $8$ & $61$                & $0.002$ & $258891$       \\
        $\mathcal{ABC}Y4Y6$  & $6$ & $37$  & $0.008$   & $40109$        & $TIDES$              & $-$ & $\delta_a = 10^{-16}$ & $0.05$  & $419958$       \\
        $\mathcal{ABC}S4$    & $4$ & $21$  & $0.00085$ & $267911$       &                      &     &                     &         &                \\
        $\mathcal{ABC}2$     & $2$ & $5$   & $0.0002$  & $320581$       &                      &     &                     &         &                \\
        \bottomrule
    \end{tabular}
    \label{table:performance1DDNLSstrong}
\end{table}
\begin{table}[!hhhtb]
    \centering
    \caption{Similar to Table \ref{table:performance1DDNLSstrong} but for  case $\text{A}2$ (see text for details) of the 1D DDNLS system~\eqref{eq:Hamilton_complex_dnls_1d}}
    \begin{tabular}{lcclr|lcclr}
        \toprule
        \multicolumn{5}{c}{$E_r \approx 10^{-5}$} &
        \multicolumn{5}{c}{$E_r \approx 10^{-9}$} \\
        \toprule
        Integrator           & $n$ & Steps & $\tau$  & $T_{\text{C}}$ & Integrator           & $n$ & Steps               & $\tau$  & $T_{\text{C}}$ \\
        \midrule
        $s11\mathcal{ABC}6$  & $6$ & $45$  & $0.4$   & $1132$         & $s19\mathcal{ABC}8$  & $8$ & $77$                & $0.3$   & $2184$         \\
        $s9\mathcal{ABC}6$   & $6$ & $37$  & $0.285$ & $1147$         & $s17\mathcal{ABC}8$  & $8$ & $69$                & $0.225$ & $2632$         \\
        $\mathcal{ABC}Y6\_A$ & $6$ & $29$  & $0.2$   & $1308$         & $s11\mathcal{ABC}6$  & $6$ & $45$                & $0.1$   & $4137$         \\
        $SS864S$             & $4$ & $17$  & $0.265$ & $1365$         & $s9\mathcal{ABC}6$   & $6$ & $37$                & $0.075$ & $4462$         \\
        $\mathcal{ABC}Y4$    & $4$ & $13$  & $0.055$ & $2354$         & $\mathcal{ABC}Y8\_D$ & $8$ & $61$                & $0.065$ & $8528$         \\
        $\mathcal{ABC}S4Y6$  & $6$ & $49$  & $0.105$ & $4965$         & $DOP853$             & $8$ & $\delta_a=10^{-16}$   & $0.05$  & $14998$        \\
        $\mathcal{ABC}Y4Y6$  & $6$ & $37$  & $0.04$  & $8091$         & $TIDES$              & $-$ & $\delta_a = 10^{-16}$ & $0.05$  & $420050$       \\
        $\mathcal{ABC}S4$&$4$&$21$&$0.02$&$9774$&  & & &	\\
        $\mathcal{ABC}2$&$2$&$5$&$0.0055$&$11700$&  & & &	\\
        \bottomrule
    \end{tabular}
    \label{table:performance1DDNLSweak}
\end{table}

Let us now perform similar computations for the 2D DDNLS system~\eqref{eq:hamilton_complex_ddnls_2d} and in particular for two typical particular setups.
\begin{enumerate}
    \item[$\bullet$] {\bf Case $\text{B}1$.} The initial wave packet is located at $L \times L = 7\times 7$ central sites having a uniformly distributed norm per site $s_{l, m} = 1/6$.
        The disorder strength is set at $W=15$, the nonlinear coefficient $\beta = 6$ and the specific disorder realization gives $\mathcal{H}_{2D} \approx 1.96$~\eqref{eq:hamilton_complex_ddnls_2d}.
    \item[$\bullet$] {\bf Case $\text{B}2$.} An initial wave packet of size $L\times L = 1$ at the central site with a unit norm i.e.~$s_{l, m} = 1$ per site for $W=16$ and $\beta = 1.25$.
        The specific realization of disorder parameters sets up $\mathcal{H}_{2D} \approx 0.625$.
\end{enumerate}
In both cases, a lattice of $200 \times 200$ sites was considered and the initial DV needed for the variational equations has non zero components chosen from a random uniform distribution for only the initially excited sites and its norm was set to unity i.e.~$\lVert \bm{w}(0)\rVert = 1$.
The obtained classifications for the performance of several integration schemes are shown in Tables~\ref{table:performance2DDNLSstrong_1new} and~\ref{table:performance2DDNLSweak_2new} respectively for cases B$1$ and B$2$.
From these tables, we see again that for moderate energy accuracy ($E_r \lesssim 10^{-5}$) the $s9\mathcal{ABC}6$ and $s11\mathcal{ABC}6$ SIs have the best performance, while at higher energy accuracy ($E_r \lesssim 10^{-9}$) both the $s19\mathcal{ABC}8$ and $s17\mathcal{ABC}8$ SIs exhibit the best performance.
\begin{table}[!hhhtb]
    \centering
    \caption{Similar to Table \ref{table:performance1DDNLSstrong} but for the case B$1$ of the 2D DDNLS model~\eqref{eq:hamilton_complex_ddnls_2d}.}
    \begin{tabular}{lcclr|lcclr}
        \toprule
        \multicolumn{5}{c}{$E_r \approx 10^{-5}$} &
        \multicolumn{5}{c}{$E_r \approx 10^{-9}$} \\
        \toprule
        Integrator           & $n$ & Steps & $\tau$  & $T_{\text{C}}$ & Integrator           & $n$ & Steps             & $\tau$   & $T_{\text{C}}$ \\
        \midrule
        $s9\mathcal{ABC}6$   & $6$ & $45$  & $0.105$ & $13914$        & $s17\mathcal{ABC}8$  & $8$ & $77$              & $0.075$  & $36528$        \\
        $s11\mathcal{ABC}6$  & $6$ & $37$  & $0.125$ & $14000$        & $s19\mathcal{ABC}8$  & $8$ & $69$              & $0.08$   & $38270$        \\
        $\mathcal{ABC}Y6\_A$ & $6$ & $29$  & $0.08$  & $15344$        & $s9\mathcal{ABC}6$   & $6$ & $45$              & $0.0235$ & $65287$        \\
        $\mathcal{ABC}Y4$    & $4$ & $13$  & $0.025$ & $23030$        & $s11\mathcal{ABC}6$  & $6$ & $37$              & $0.0275$ & $67314$        \\
        $SS864S$             & $4$ & $17$  & $0.085$ & $23887$       & $\mathcal{ABC}Y8\_D$ & $8$ & $61$              & $0.008$  & $140506$       \\
        $\mathcal{ABC}S4Y6$  & $6$ & $49$  & $0.03$  & $77424$        & $DOP853$             & $8$ & $\delta_a=10^{-16}$ & $0.05$   & $218704$       \\
        $\mathcal{ABC}Y4Y6$&$6$&$37$&$0.0165$&$87902$&  & & & \\
        $\mathcal{ABC}S4$&$4$&$21$&$0.0065$&$132713$& & & & \\
        $\mathcal{ABC}2$&$2$&$5$&$0.005$&$157694$&  & & &			\\
        \\
        \bottomrule
    \end{tabular}
    \label{table:performance2DDNLSstrong_1new}
\end{table}
\begin{table}[!hhhtb]
    \centering
    \caption{Similar to Table \ref{table:performance1DDNLSstrong} but for the case B$2$ of the 2D DDNLS model~\eqref{eq:hamilton_complex_ddnls_2d}}
    \begin{tabular}{lcclr|lcclr}
        \toprule
        \multicolumn{5}{c}{$E_r \approx 10^{-5}$} &
        \multicolumn{5}{c}{$E_r \approx 10^{-9}$} \\
        \toprule
        Integrator           & $n$ & Steps & $\tau$    & $T_{\text{C}}$ & Integrator           & $n$ & Steps             & $\tau$   & $T_{\text{C}}$ \\
        \midrule
        $s11\mathcal{ABC}6$  & $6$ & $45$  & $0.1515$  & $11443$        & $s19\mathcal{ABC}8$  & $8$ & $77$              & $0.135$  & $20952$        \\
        $s9\mathcal{ABC}6$   & $6$ & $37$  & $0.11$    & $13408$        & $s17\mathcal{ABC}8$  & $8$ & $69$              & $0.0875$ & $28966$        \\
        $\mathcal{ABC}Y6\_A$ & $6$ & $29$  & $0.0775$  & $14607$        & $s11\mathcal{ABC}6$  & $6$ & $45$              & $0.0335$ & $50301$        \\
        $SS864S$             & $4$ & $17$  & $0.0915$  & $15564$        & $s9\mathcal{ABC}6$   & $6$ & $37$              & $0.024$  & $58187$        \\
        $\mathcal{ABC}Y4$    & $4$ & $13$  & $0.0215$  & $25898$        & $\mathcal{ABC}Y8\_D$ & $8$ & $61$              & $0.009$  & $150045$       \\
        $\mathcal{ABC}S4Y6$  & $6$ & $49$  & $0.035$   & $64423$        & $DOP853$             & $8$ & $\delta_a=10^{-16}$ & $0.05$   & $166145$       \\
        $\mathcal{ABC}Y4Y6$  & $6$ & $37$  & $0.01375$ & $102580$       &                      &     &                   &          &                \\
        $\mathcal{ABC}S4$    & $4$ & $21$  & $0.005$   & $185615$       &                      &     &                   &          &                \\
        $\mathcal{ABC}2$     & $2$ & $5$   & $0.00155$ & $198534$       &                      &     &                   &          &                \\
        \bottomrule
    \end{tabular}
    \label{table:performance2DDNLSweak_2new}
\end{table}

Let us also stress a point which we consider important.
Although the results presented in this section suggest that the $DOP853$ and $TIDES$ schemes are not well suited for very long time integrations as they require a lot of CPU time, they are nevertheless extremely versatile and perform quite well for several tasks we expect an integrator to do.
This fact is emphasized in Fig.~4.2 of~\citep{hairer1993solving} and in the related discussion where the $DOP853$ scheme proved to be the best performing integrator among the ones considered there.
In the same spirit, the advantages of the $TIDES$ package were discussed in~\citep{barrio2005performance,abad2012algorithm}.

\section{\label{sec:conclusion_num}Summary}
In this chapter, we presented the numerical methods we use in our study.
Particular emphasis was given to a plethora of SIs developed for the integration of the equations of motion and the variational equations of the 1D and 2D DDNLS systems.
For the 1D DDNLS model, we considered symplectic schemes based on splitting the system's Hamiltonian into two and three integrable parts, while for the 2D DDNLS model, only the three-part split of the Hamiltonian function was explored.
We studied the performance of the obtained numerical SIs by comparing their efficiency to some commonly used integrators: The RK scheme of order $8$ based on the Dormand-Prince method $DOP853$ and the $TIDES$ package which is extracted from Taylors/Lie series expansions of the system's vector field.
Our results suggest that SIs should be preferred for very long time integrations as they need less CPU time to achieve a certain level of accuracy.
In the case of three-part split  methods, the $s11\mathcal{ABC}6$ and $s9\mathcal{ABC}6$ as well as the $s19\mathcal{ABC}8$ and $s17\mathcal{ABC}8$ SIs proved to be the most efficient ones for moderate and high accuracy respectively.
In addition, the three-part split approach leads to SIs which are faster than those produced by the two-part split method for large system sizes.
This result justifies our choice of the former approach as our used integration techniques.

The proper selection of initial excitations of the DDNLS models has been discussed in~\citep{kati2020density} as one has to take into account the values of both integrals of motion i.e.~the total energy and total norm.
Therefore, numerical approaches for designing initial conditions for these systems in order to obtain specific energy and norm values were also presented in detail.
We further outlined other numerical techniques we use in our study, such as the determination of the power law exponents from non smooth numerical data sets, the numerical characterization of the wave packet as well as the numerical calculations of its extent through definitions of the first and second moments, and the participation number.
Furthermore, we introduced the notion of tangent space of a trajectory in the phase space and presented numerical techniques for the computation of observables of the orbit's dynamics like the MLE and the DVD. 

So far we have discussed and presented the analytical approaches, as well as numerical techniques we are using in our work.
The two next chapters will be devoted on the application of these techniques in order to investigate in detail the characteristics of chaos of the 1D and 2D DDNLS systems.

\chapter{\label{chap:chaotic_one_dimensional}Chaotic dynamics of wave packet spreading in the one-dimensional disordered nonlinear Schr\"odinger equation}

\section{\label{sec:introduction_1d_spreading}Introduction}
In this chapter we turn the focus to one of the main objectives of this thesis, namely the investigation of the chaotic dynamics of wave packet spreading in the 1D DDNLS model~\eqref{eq:Hamilton_complex_dnls_1d}.
Although this subject has been considered in the past~\citep[see e.g.][]{mulansky2009dynamical,skokos2013nonequilibrium,antonopoulos2014complex,antonopoulos2016coupled,antonopoulos2017analyzing}, the detailed investigation of the asymptotic behavior of chaos for the different dynamical regimes in the 1D DDNLS system is still incomplete.
Due to both the presence of disorder in the equations of motion~\eqref{eq:eq_mot_1d_ddnls} and the stochastic nature of the tangent dynamics, visible for example if we consider a small perturbation in the reduced nonlinear stochastic ODE~\eqref{eq:evol_transmission_flach} approximating the system's dynamics in the limit of small amplitude wave packet spreading, the analytical study of the system chaotic dynamics is very difficult.
Thus, we rely on numerical simulations for our investigations.

The chapter is organized as follows:
In Sec.~\ref{sec:other_num_considerations}, we discuss some of the computational issues mainly related to the integration of the model's variational equations.
Section~\ref{sec:mle_1d_spreading} is devoted to presenting our computations of the MLE for the AL, weak chaos, and strong chaos dynamical regimes.
In Sec.~\ref{sec:dvd_1d_spreading} we analyse the local characteristics of chaos.
We investigate how the DVD behaves in the linear regime and discuss the idea of the presence of chaotic hotspots within the wave packet interior as a mechanism for promoting spreading in the weak and strong chaos regimes.
In Sec.~\ref{sec:discussion_1d_spreading} we discuss our findings. More specifically, we discuss the implication of the wave packet chaoticity for its spreading dynamics in the weak and strong chaos cases. 
The results of this chapter are heavily based on findings in~\citep{senyange2018characteristics}.

\section{\label{sec:other_num_considerations}Some aspects related to our numerical simulations}
\subsection{\label{subsec:choice_of_init_dvd}Choice of the initial deviation vector}
Let us now look more closely at the numerical aspect of the DV in our simulations.
In general the choice of the DVs is of practical importance for the computation of the Lyapunov exponents~\citep{skokos2010lyapunov}.
The most commonly suggested option being initial DVs having randomly selected elements in order to avoid any linear dependence between the coordinates of the DVs.
As the computation of the MLE only requires a single DV the practice above is not absolutely necessary.
However, most papers suggest it as good choice of DV~\citep[see~e.g.][]{skokos2010lyapunov}.
For low degrees of freedom Hamiltonian models such as the ones seen in pioneering fields of chaos theory like celestial mechanics~\citep{henon1964applicability,froeschle1972numerical,skokos2001alignment} or plasma physics~\citep{chirikov1971research,wingen2005stochastic,kominis2008explicit,white2013theory}, the dynamics of the system takes place in a rather `small' phase space such that if the system's orbit is chaotic, all choices of DVs for the computation of the ftMLE converge to the MLE practically at the same rate.
In systems with many more degrees of freedom, especially the ones we are working in this thesis which have in theory an infinite phase space dimension, the system's dynamics in the phase space take place in `tiny' regions compared to the overall size of the system's phase space.
Thus, one has to carefully select the type of DVs in order to correctly capture the dynamics.

\begin{figure}[!hhtb]
    \centering
    \includegraphics[width=0.7\textwidth]{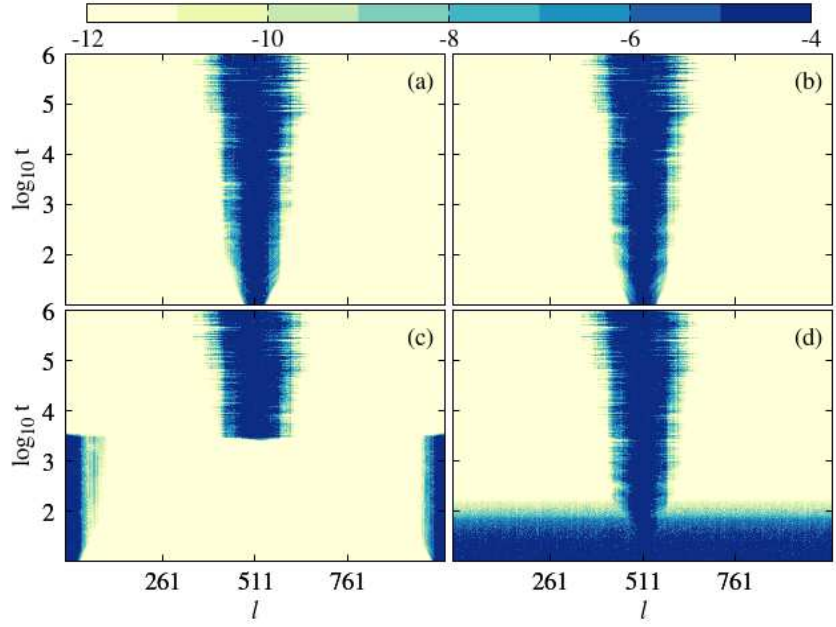}
    \caption{Spatiotemporal evolution of the DVD $\xi_l^D$~\eqref{eq:DVD_definition_num} profile for different initial DVs.
        The panel (a) corresponds to DV$_1$, the panel (b) to DV$_2$, the panel (c) to DV$_3$ and the panel (d) to DV$_4$ (see text for details).
        The color scales at the top of the figure are used for coloring the lattice sites according to their $\log_{10}\xi_l^D$ values.
    }
    \label{fig:choice_of_dvd_dvd_profiles_chaos_1d}
\end{figure}

\begin{figure}[!hhtb]
    \centering
    \includegraphics[width=0.49\textwidth]{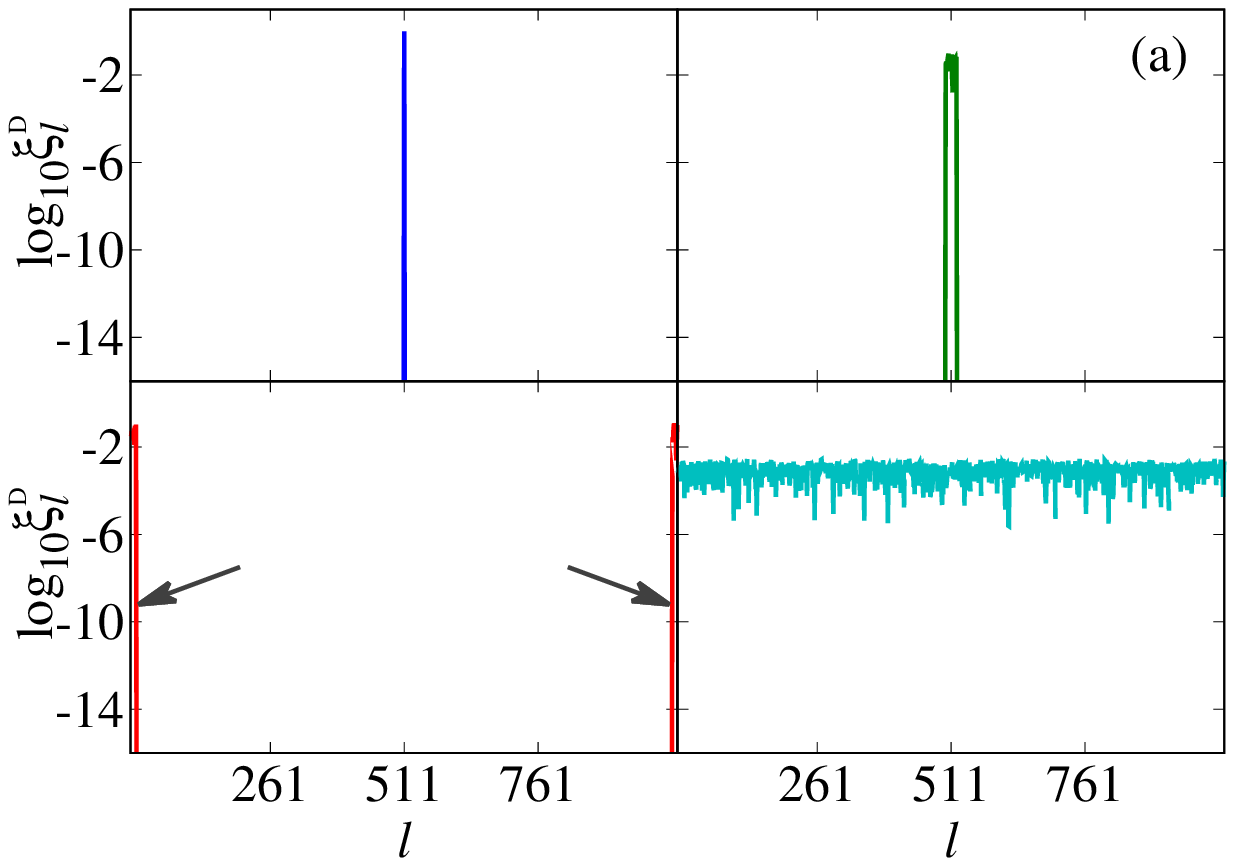}
    \includegraphics[width=0.49\textwidth]{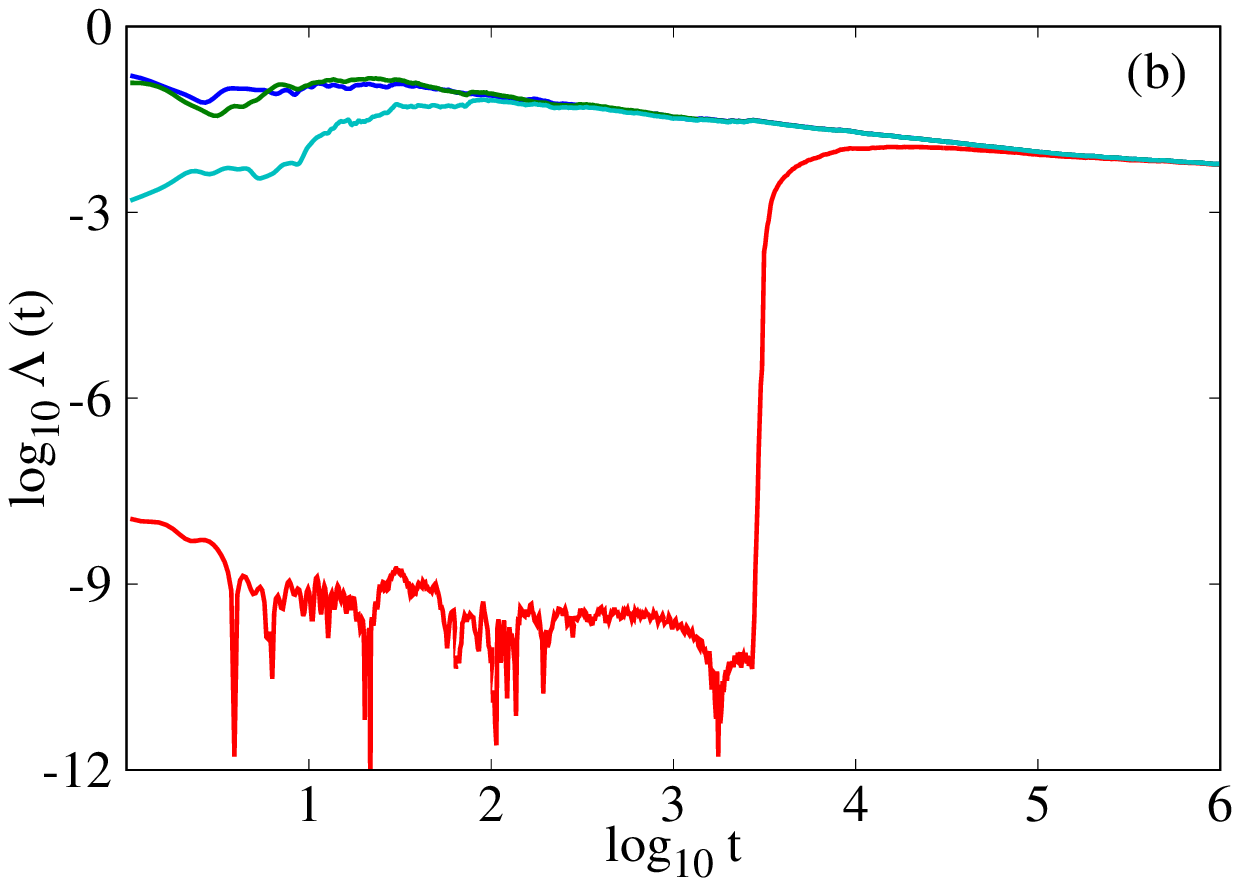}
    \caption{(a) Initial profiles $\xi_l^D$~\eqref{eq:norm_density_distribution_num} of the DVDs used for obtaining the results of Fig.~\ref{fig:choice_of_dvd_dvd_profiles_chaos_1d}.
            The positions of each panel corresponds to the arrangement seen in Fig.~\ref{fig:choice_of_dvd_dvd_profiles_chaos_1d}.
            (b) Time evolution of the ftMLE, $\Lambda (t)$~\eqref{eq:finite_mle} for the DVs of panel (a).
            The colors of the curves correspond to the ones in panel (a).}
    \label{fig:choice_of_dvd_mle_chaos_1d_2}
\end{figure}

To better understand this let us consider the example in Figs.~\ref{fig:choice_of_dvd_dvd_profiles_chaos_1d} and~\ref{fig:choice_of_dvd_mle_chaos_1d_2}(a), where we initially excite in a lattice of $N=1021$ sites the $q_l$ and $p_l$ coordinates of $L = 21$ central sites with a uniformly distributed norm per site $s = 0.3$, such that the total norm~\eqref{eq:norm_complex_dnls_1d} is $\mathcal{S}_{1D} = 6.3$.
Moreover, the nonlinear coefficient $\beta = 1$, the disordered strength $W=4$ and the random values $\epsilon_l$ are chosen such that the total energy~\eqref{eq:Hamilton_complex_dnls_1d} $\mathcal{H}_{1D} \approx 1.05$ is uniformly distributed among all the excited sites with an energy per site $h\approx 0.05$.
This initial wave packet belongs to the weak chaos regime of the 1D DDNLS model~\eqref{eq:Hamilton_complex_dnls_1d} according to the parameter spaces of Figs.~\ref{fig:parameter_space_dnls_1d_delta_W} and~\ref{fig:x_y_thermal_selftrapping_phases}.
We integrate the system's equations of motion and variational equations using the $s11\mathcal{ABC}6$ scheme~\eqref{eq:s11odr6_general} up to $t_f = 10^6$, fixing the integration time step $\tau = 0.15$ which leads to a maximal relative energy error~\eqref{eq:rel_energy_error_1dDDNLS} $E_r (t) \lesssim 10^{-6}$ and relative norm error~\eqref{eq:rel_energy_error_1dDDNLS} $S_r(t) \lesssim 10^{-4}$.
For the initial DV, we use four cases:
\begin{enumerate}
    \item[]{\bf Case DV$_1$.} 
        We give random values to $\delta q_l$ and $\delta p_l$ of only the central site of the lattice.
        This leads to the spatial distribution of the DVD $\xi_l^{D}$~\eqref{eq:DVD_definition_num} as shown in upper left panel of Fig.~\ref{fig:choice_of_dvd_mle_chaos_1d_2}(a). 
    \item[]{\bf Case DV$_2$.}
        We randomly excite the $\delta q_l$ and $\delta p_l$ coordinates of $L=21$ central sites i.e.~the DVD $\xi_l^D$~\eqref{eq:DVD_definition_num} has non-zero elements only in the lattice's initially excited region as shown in the upper right panel of Fig.~\ref{fig:choice_of_dvd_mle_chaos_1d_2}(a).  
    \item[]{\bf Case DV$_3$.} 
        We give to the $\delta q_l$ and $\delta p_l$ coordinates of the $L=21$ sites at either end of the lattice, randomly chosen values, which corresponds to the DVD $\xi_l^D$~\eqref{eq:DVD_definition_num} depicted in the lower left panel of Fig.~\ref{fig:choice_of_dvd_mle_chaos_1d_2}(a)
    \item[]{\bf Case DV$_4$.} 
        We excite all the elements of the DV with values drawn from a random uniform distribution.
        This results to a DVD $\xi_l^D$~\eqref{eq:DVD_definition_num} having non-zero elements at all lattice sites as shown in the lower right panel of Fig.~\ref{fig:choice_of_dvd_mle_chaos_1d_2}(a). 
\end{enumerate}
In Fig.~\ref{fig:choice_of_dvd_dvd_profiles_chaos_1d} we present the spatiotemporal evolution of the DVD, $\xi_l^{D}$~\eqref{eq:DVD_definition_num}, for the cases DV$_1$ in Fig.~\ref{fig:choice_of_dvd_dvd_profiles_chaos_1d}(a), DV$_2$ in Fig.\ref{fig:choice_of_dvd_dvd_profiles_chaos_1d}(b), DV$_3$ in Fig.~\ref{fig:choice_of_dvd_dvd_profiles_chaos_1d}(c) and DV$_4$ in Fig.~\ref{fig:choice_of_dvd_dvd_profiles_chaos_1d}(d).
A transient time where the dynamical behaviors of each DVD differs is clearly visible. 
For DV$_1$ and DV$_2$, a timid expansion of the DVD is observed which do not spread over the whole lattice, but always remain within the area of the wave packet extent where the dynamics of the system take place.   
This is further supported by the observations of the spatiotemporal evolution of DV$_3$ and DV$_4$, where the tangent dynamics has a tendency to enhance the magnitude of the coordinates of the DV inside the region where the motion of the wave packet occurs and the values of the coordinates of the DV outside the wave packet die out.
In the case of DV$_3$, this transition happens around $\log_{10}t \approx 3.5$ [see Fig.~\ref{fig:choice_of_dvd_dvd_profiles_chaos_1d}(c)], while for the case DV$_4$ it happens at $\log_{10}t \approx 2$ [see Fig.~\ref{fig:choice_of_dvd_dvd_profiles_chaos_1d}(d)].   

During these transient times, the ftMLEs in each DV case show different evolutions as seen in Fig.~\ref{fig:choice_of_dvd_mle_chaos_1d_2}(b).
Indeed, the ftMLEs obtained with DV$_1$ and DV$_2$ display larger values compared to the other DVs (DV$_3$ and DV$_4$).
This is due to the fact that these DVs evolve within the active domain, whose evolution is controlled by nonlinear deterministic equations~\eqref{eq:eq_mot_1d_ddnls}.
Consequently, a deterministic chaotic dynamics is visible from the beginning of their evolutions which explain the large values of the ftMLEs.
Next is the ftMLE of DV$_4$ which is at its peak two order of magnitudes smaller than the values observed for the ftMLEs of DV$_1$ and DV$_2$. 
The reason for such a measurement is the following: For time $t$ such that $\log_{10} t \lessapprox 2$ in the early evolution of DV$_4$, a large part of the DV coordinates evolve in the regular background outside the system's excited region.
Thus the contribution of the regular part is non-negligible compared to the contribution of the excited region of the lattice which is chaotic.
This leads to the smaller value of the ftMLE observed in Fig.~\ref{fig:choice_of_dvd_mle_chaos_1d_2}(b) compared to the ones obtained from DV$_1$ and DV$2$.
In the case of DV$_3$, when $\log_{10}t \lessapprox 3.5$, all the non-zero coordinates of the DV evolve only into the regular part of the lattice which therefore solely contribute to the value of the ftMLE which indicates $\Lambda (t)\approx 10^{-10}$ at its lowest pick i.e.~an almost zero ftMLE.  

For longer times, well above the largest transient time of DV$_3$, all the DVs converge to a consistent behavior as all the asymptotic spatiotemporal evolution of the DVDs $\xi_l^D$~\eqref{eq:DVD_definition_num} are practically the same in Fig.~\ref{fig:choice_of_dvd_mle_chaos_1d_2}.   
According to the theory of chaos~\citep[see~e.g.][]{skokos2010lyapunov,skokos2016smaller}, this asymptotic behavior of the DVs in Fig.~\ref{fig:choice_of_dvd_dvd_profiles_chaos_1d} corresponds to the spatial alignment of the tangent dynamics along the most chaotic direction of the system's reference orbit.
Thus, the average exponential rate of growth (or shrinking) of the magnitude of this DV measures the system's MLE.
This is what is observed in Fig.~\ref{fig:choice_of_dvd_mle_chaos_1d_2}(b), where for time $t$ above the transient time, all the ftMLEs practically indicate the same value. 

Based on this experiment, we can decide which type of DVs is the best from a practical point of view.
Indeed, our objective is to be able to retrieve the asymptotic behavior of the tangent dynamics as fast as possible since we are constrained by computational time limitations. 
Thus, DV$_3$ and DV$_4$, which possess the largest transient times are bad choices for our purpose.
This is worsened by the fact that for lattice size $N> 1021$ (with the size of the excited region $L$ almost constant), the regular background outside the excited central part of the lattice becomes larger, therefore longer transient times are expected for the DVs similar to DV$_3$ and DV$_4$ as the elements of the tangent dynamics outside the wave packet evolution take larger times to vanish (only the elements of the DV inside the lattice's excited part contribute to the MLE). 
On the other hand the cases DV$_1$ and DV$_2$ are good choices since they are located inside the central excited part of the lattice where the dynamics take place.
Consequently, they take less time to converge to the most chaotic direction whose non zero elements of the DV are always located inside the wave packet as indicated by the evolution of the DVDs in Fig.~\ref{fig:choice_of_dvd_dvd_profiles_chaos_1d}.
For the rest of this chapter, unless otherwise stated, our default initial DV is the DV$_1$.

\subsection{\label{subsec:dynamic_lattice_chaotic_dynamics_1d}The dynamic lattice expansion}
Because the initially excited sites are located at the central part of a lattice of large size, in our simulations we can consider a smaller subset of the lattice at the beginning of our simulations and expand it when ever it is necessary.
We refer to this approach as the {\it dynamic lattice simulation} which has been successfully implemented for both the 1D~\eqref{eq:Hamilton_complex_dnls_1d} and 2D~\eqref{eq:hamilton_complex_ddnls_2d} DDNLS models and DKG systems (Eqs.~\eqref{eq:hamilton_1d_dkg} and~\eqref{eq:hamilton_2d_dkg}) in~\citep{senyange2018characteristics,manda2020chaotic} in contrast to the {\it fixed lattice method} used in several studies on the subject~\citep{flach2009universal,skokos2010spreading,laptyeva2010crossover,laptyeva2012subdiffusion,kati2020density}.  
Let us note that this is not particularly unusual as even in the theory (see Chap.~\ref{chap:spreading} and references therein), the infinite lattice size i.e.~$N\rightarrow \infty$ in $\mathcal{H}_{1D}$~\eqref{eq:Hamilton_complex_dnls_1d} is often substituted by a large but finite value of $N$.
Then, we have to define a condition which controls our lattice expansion.
For the 1D DDNLS model~\eqref{eq:Hamilton_complex_dnls_1d}, we monitor the value of the average norm per site $\sum _{l\in \mathcal{D}}s_l/\mathcal{D}$ of the last $\mathcal{D}$ sites at both edges of the `active' (dynamic) lattice with size $N_{\mathcal{D}, \delta_{\mathcal{D}}}$ sites.
Then, whenever
\begin{equation}
    \frac{1}{\mathcal{D}_L}\sum _{l\in \mathcal{D}_L} s_l > \delta_{\mathcal{D}_L}, \mbox{ or } \frac{1}{\mathcal{D}_R}\sum _{l\in \mathcal{D}_R} s_l > \delta_{\mathcal{D}_R},
    \label{eq:condition_dynamic_latt_chaos_1d}
\end{equation}
the lattice is expanded at both ends (in order to maintain the symmetry with respect to the initial central excitation) by $\mathcal{N}$ sites.
Note that in Eq.~\eqref{eq:condition_dynamic_latt_chaos_1d}, $\mathcal{D}_{L}$ and $\mathcal{D}_{R}$ describe the number of sites at the left and right edges respectively (we often assume $\mathcal{D}_{L} = \mathcal{D}_R = \mathcal{D}$). 
In addition, $\delta_{\mathcal{D}_L} = \delta_{\mathcal{D}_R} = \delta_{\mathcal{D}}$ is the threshold above which we expand our current dynamic lattice at the edges.
It is worth mentioning that $\mathcal{N} > \mathcal{D}$.
\begin{figure}[!htb]
    \centering
    \includegraphics[width=0.8\textwidth]{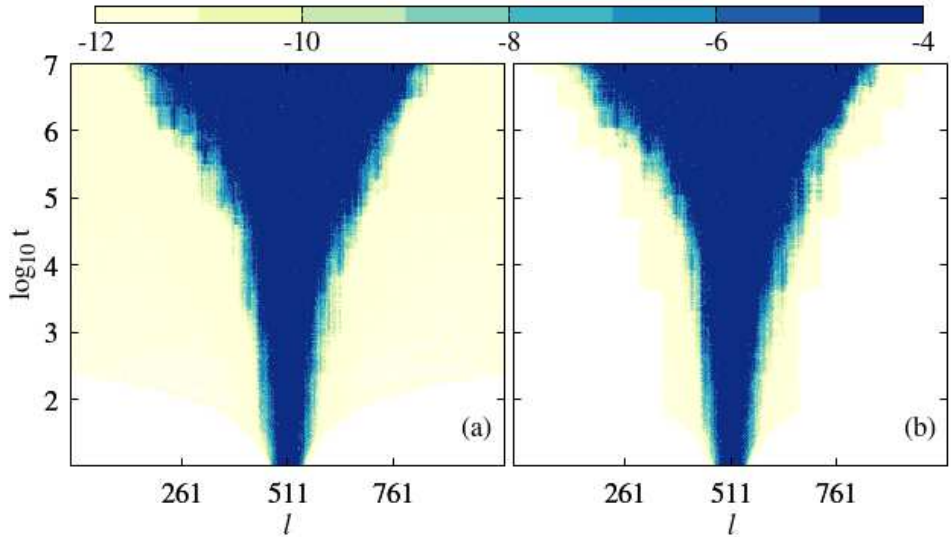}
    \caption{Heatmaps of the spatiotemporal evolutions of the normalized norm distribution $\xi_l$~\eqref{eq:DVD_definition_num} for an initial condition belonging to the strong chaos spreading regime (see text for details).
    We excited $L = 21$ central sites on a lattice of $N = 1021$ sites with the same norm per site $s = 1$ so that the total norm~\eqref{eq:norm_complex_dnls_1d} $\mathcal{S}_{1D} = 21$.
    The random values $\epsilon_l$, $1 \leq l \leq N$, give a total energy~\eqref{eq:Hamilton_complex_dnls_1d} $\mathcal{H}_{1D} \approx 3.21$, using the parameters $W=4$ and $\beta=0.72$.
    The panel (a) corresponds to the fixed lattice size as we see the residual norm i.e.~$s \sim 10^{-12}$ (light yellow in the heatmap) spreads over the entire lattice. 
    The panel (b) corresponds to the lattice expansion approach as the residual norm is concentrated around the wave packet (darker region in the heatmap).
    The white area, corresponds to (empty) lattice sites with no norm $s$.
    The color scales at the top of the figure are used for coloring the lattice sites according to the magnitude of their $\log_{10}\xi_l$ values. 
    }
    \label{fig:dynamic_latt_chaos_1d_00}
\end{figure}

To be more specific, in Fig.~\ref{fig:dynamic_latt_chaos_1d_00}, we plot the evolution of the normalized norm distribution $\xi_l$~\eqref{eq:norm_density_distribution_num} on a lattice of $N = 1021$ sites using at time $t = 0$, $L = 21$ centrally excited sites such that each site receives the same constant norm per site $s = 1$, leading to the total norm~\eqref{eq:norm_complex_dnls_1d} $\mathcal{S}_{1D} = 21$.
Further, the values of $\epsilon_l$ are chosen in a way that the Hamiltonian~\eqref{eq:Hamilton_complex_dnls_1d} $\mathcal{H}_{1D} \approx 3.21$, with parameters $W = 4$ and $\beta = 0.72$.
This initial condition belongs to the strong chaos spreading regime according to Figs.~\ref{fig:parameter_space_dnls_1d_delta_W} and~\ref{fig:x_y_thermal_selftrapping_phases} of Chap.~\ref{chap:spreading}.
We integrate the system's phase space and tangent dynamics using the $s11\mathcal{ABC}6$ SI~\eqref{eq:s11odr6_general}, with time step $\tau = 0.2$ up to the final integration time $t_f = 10^{7}$.
In doing so, we obtain a relative energy error~\eqref{eq:rel_energy_error_1dDDNLS} $E_r(t) \lesssim 10^{-5}$ and a relative norm error~\eqref{eq:rel_norm_error_1dDDNLS} $S_r(t) \lesssim 10^{-3}$.
In Fig.~\ref{fig:dynamic_latt_chaos_1d_00}(a), we plot the spatiotemporal wave packet evolution $\xi_l(t)$ for the fixed lattice size system, along with the one of the dynamic lattice implementation [see Fig.~\ref{fig:dynamic_latt_chaos_1d_00}(b)], starting from $N_{\mathcal{D}, \delta_{\mathcal{D}}} = 321$ at time $t = 0$.
In addition, we used as threshold $\delta_{\mathcal{D}} = 10^{-12}$.
We monitor $\mathcal{D} = 10$ points at the edges and increase each edges by $\mathcal{N} = 50$ sites each time that the wave packet has reached the borders of the lattice i.e.~whether the condition~\eqref{eq:condition_dynamic_latt_chaos_1d} is verified.

In Fig.~\ref{fig:dynamic_latt_chaos_1d_00}(b), the dynamic lattice size $N_{\mathcal{D}, \delta_{\mathcal{D}}} = 321$ sites at time $t = 0$, was increased by $2\mathcal{N} = 100$ at time $\log _{10} t \approx 3.64$ to $N_{\mathcal{D}, \delta_{\mathcal{D}}} = 421$ sites, at $\log_{10} t \approx 4.72$ to $N_{\mathcal{D}, \delta_{\mathcal{D}}}= 521$ sites, at $\log_{10} t \approx 5.57$ to $N_{\mathcal{D}, \delta_{\mathcal{D}}}= 621$ sites, at $\log_{10} t \approx 5.75$ to $N_{\mathcal{D}, \delta_{\mathcal{D}}}= 721$ sites, at $\log_{10} t \approx 6.34$ to $N_{\mathcal{D}, \delta_{\mathcal{D}}}= 821$ sites, and at time $\log_{10} t \approx 6.86$ reaching a total lattice size at the final time $t_f = 10^{7}$ of the integration $N_{\mathcal{D}, \delta_{\mathcal{D}}}= 921$ sites.
Consequently, after increasing the dynamic lattice $6$ times, we obtain the results shown in Fig.~\ref{fig:dynamic_latt_chaos_1d_00}(b) where the white areas at the edges indicate empty space i.e. lattice sites with norm per site $s = 0$.
Further, it clearly appears from Figs.~\ref{fig:dynamic_latt_chaos_1d_00}(a) and (b) that the spatiotemporal evolutions of the normalized norm distribution $\xi_l (t)$~\eqref{eq:norm_density_distribution_num} are practically the same.
It is worth mentioning that the initial lattice size $N_{\mathcal{D}, \delta_{\mathcal{D}}}=321$ is much larger than the NM average localization volume~\eqref{eq:loc_vol_loc_length_2d_ddnls_estimation} $\overline{V}\approx 21$ sites in this context.

\begin{figure}[!htbp]
    \centering
    \includegraphics[width=0.4\textwidth, height=0.4\linewidth]{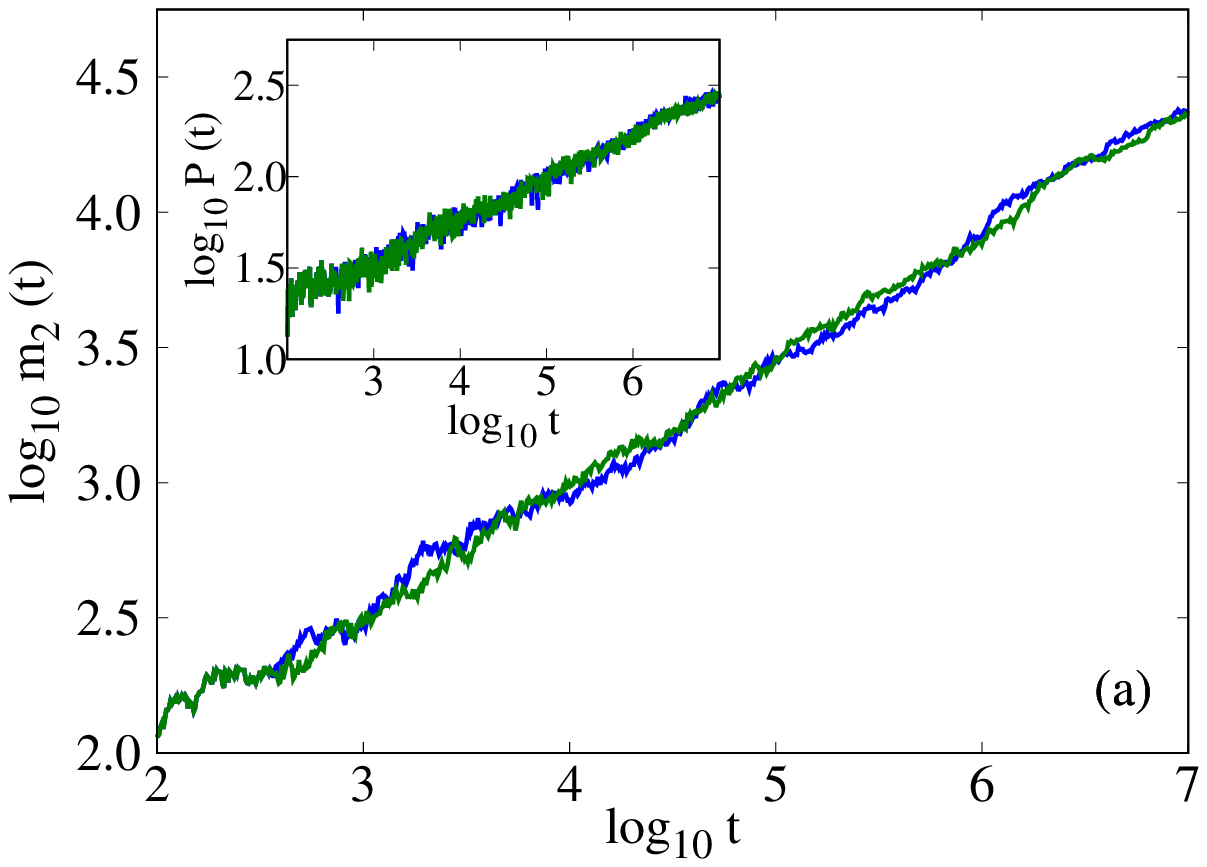}
    \includegraphics[width=0.4\textwidth, height=0.4\linewidth]{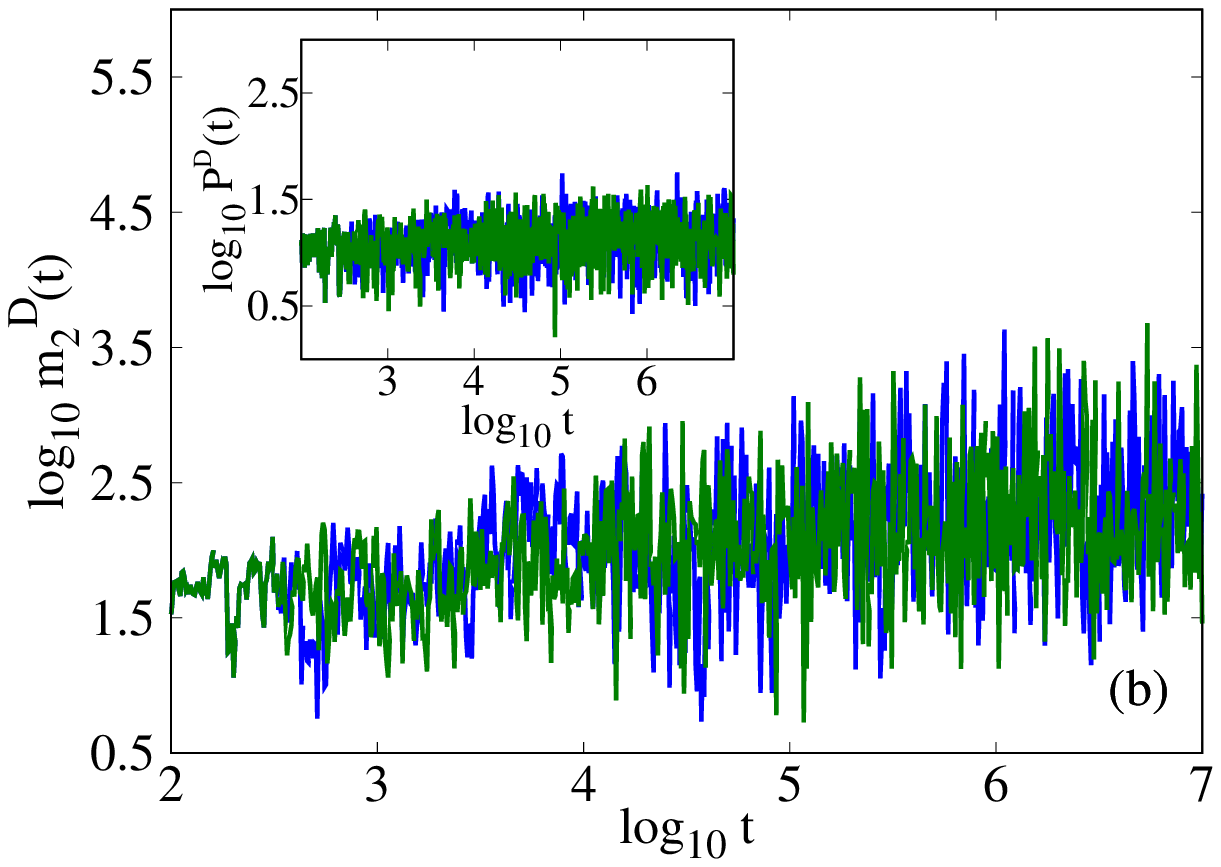}
    \includegraphics[width=0.4\textwidth, height=0.4\linewidth]{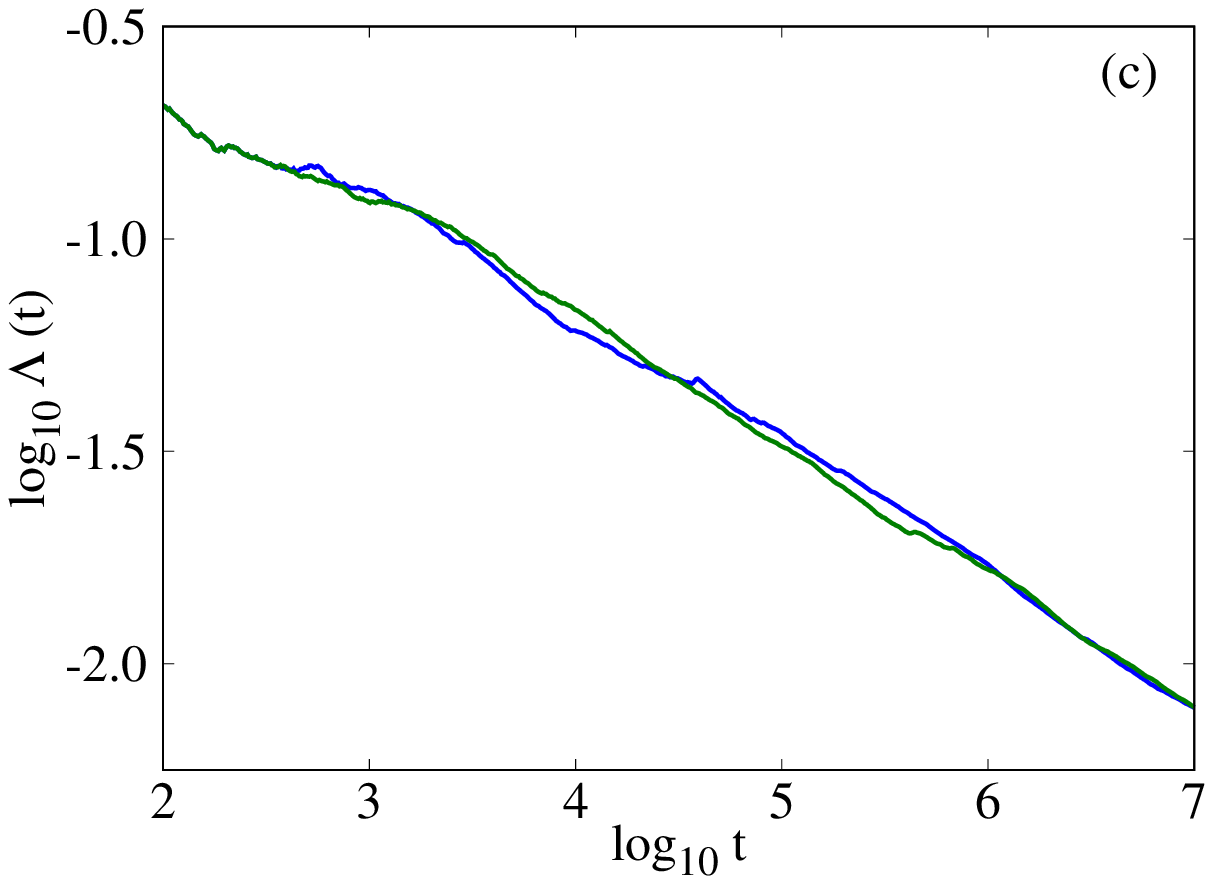}
    \includegraphics[width=0.4\textwidth, height=0.4\linewidth]{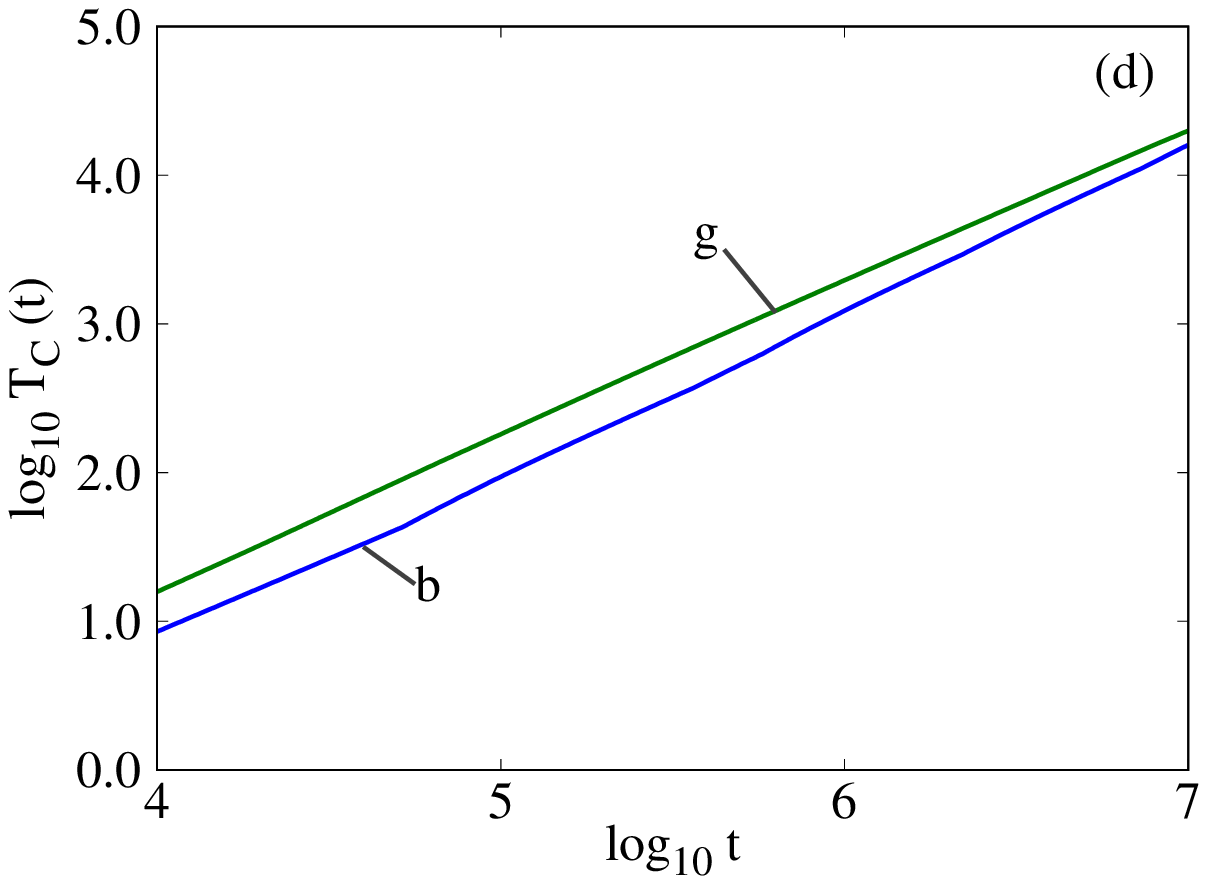}
    \caption{Results of the time evolution of the observables of the phase space and tangent dynamics of the case used in Fig.~\ref{fig:dynamic_latt_chaos_1d_00}.
     The panel (a) presents the second moment $m_2(t)$~\eqref{eq:second_moment_distr_num} and the participation number $P(t)$~\eqref{eq:participation_ratio_num} [inset of panel (a)] of the norm distribution $\xi_l$~\eqref{eq:norm_density_distribution_num}.
        The panel (b) shows the second moment $m_2^D(t)$ and participation number $P^D(t)$ [inset of panel (b)] of the DVD $\xi_l^D$~\eqref{eq:DVD_definition_num}.
        In panel (c) we see the ftMLEs $\Lambda (t)$~\eqref{eq:finite_mle} and in panel (d) the required CPU time.
        The (g)reen curves are for the fixed and the (b)lue curves the dynamic lattice simulations.
    }
    \label{fig:time_evol_m2_P_mD_PD_MLE_CPU_1d_spreading}
\end{figure}

In order to quantify the qualitative observations of the evolution of the norm distribution $\xi_l$~\eqref{eq:norm_density_distribution_num} in Fig.~\ref{fig:dynamic_latt_chaos_1d_00}, we plot the time evolution of the $m_2(t)$~\eqref{eq:second_moment_distr_num} [panel (a) of Fig.~\ref{fig:time_evol_m2_P_mD_PD_MLE_CPU_1d_spreading}] and $P(t)$~\eqref{eq:participation_ratio_num} [inset of panel (a) in Fig.~\ref{fig:time_evol_m2_P_mD_PD_MLE_CPU_1d_spreading}] of $\xi_l $~\eqref{eq:norm_density_distribution_num} for the fixed [(g)reen curve] and dynamic [(b)lue curves] lattice simulations.
Both the $m_2$ and $P$ of $\xi_l$ are practically overlapping up to the largest simulation time $t_f = 10^{7}$.
As in this work we are also interested in the evolution of the variational equations, we depict the time evolution of the second moment $m_2^{D}(t)$~\eqref{eq:second_moment_distr_num} and participation number $P^D(t)$~\eqref{eq:participation_ratio_num} of the DVD $\xi_l^D$~\eqref{eq:DVD_definition_num} as well as the values of the ftMLE $\Lambda (t)$~\eqref{eq:finite_mle} in Figs.~\ref{fig:time_evol_m2_P_mD_PD_MLE_CPU_1d_spreading}(b-c) for both the fixed [(g)reen curves] and dynamic [(b)lue curves] lattices computations of Fig.~\ref{fig:dynamic_latt_chaos_1d_00}.
We notice a good correspondence in magnitude of the values of the $m_2^D$ and $P^D$ between the two lattice types as they are randomly oscillating in the same range of $m_2^D$ and $P^D$ values.
Further, the chaoticity of the reference orbit depicted in the initial condition used in Fig.~\ref{fig:dynamic_latt_chaos_1d_00} is well reported by the two type of simulations as the curves of the ftMLEs practically overlap. 
Finally, we analyze the CPU time $T_C$ necessary to reach the final time of integration $t_f=10^7$ [Fig.~\ref{fig:time_evol_m2_P_mD_PD_MLE_CPU_1d_spreading}(d)].
Clearly, the dynamic lattice appears advantageous as the final CPU time is noticeably smaller on the logarithmic scale.
More specifically, we calculate the gain factor 
\begin{equation}
    g = \frac{T_C^{(F)} - T_C^{(D)}}{T_C^{(F)}},
    \label{eq:speedup_factor}
\end{equation}
with  $T_C^{(F)}$ and $T_C^{(D)}$ being respectively the CPU time needed for the fixed and dynamics lattice simulations.
In the example above, we obtain an improvement of approximately $20\%$, using $T_C^{(F)} \approx 5.5~\text{hours}$ and $T_C^{(D)} \approx 4.43~\text{hours}$.

It is worth commenting a bit more on the bent nature of the curve of the CPU time for the dynamic lattice simulation.
For the fixed lattice simulation, a fairly straight line is seen on the dependence of $\log_{10} T_C$ on the logarithmic time i.e. $\log_{10}t$.
This is due to the fact that the lattice size remains fixed along the integration of the system's orbit.
Thus the number of basic computer operations (addition, multiplication, and so on) necessary to evolve the equations of motion and the variational equations, as well as to compute all the observables of the phase space and tangent dynamics, remains constant for each integration time step $\tau$.
So the constant CPU time necessary for a single integration time step $\tau$ evolution adds up to the cumulative CPU time $\log_{10} T_C$ in the constant rate (or slope) seen on the (g)reen curve of Fig.~\ref{fig:time_evol_m2_P_mD_PD_MLE_CPU_1d_spreading}(d). 
This is not the case for the dynamic lattice simulation.
Indeed, the number of lattice sites is constantly increased, leading to an augmentation of the number of basic computer operations for the time evolution of the equations of motion and variational equations along with the calculation of observables of the phase space and tangent dynamics.  
Consequently, the CPU time needed for a single integration time step $\tau$ increases in time leading to a constant increase of the slope of the (b)lue curve in Fig.~\ref{fig:time_evol_m2_P_mD_PD_MLE_CPU_1d_spreading}(d) which create the bending effects of the $\log_{10} T_C$ curve.  
Ultimately, there exists a time $t$ where the CPU time of both lattice simulations overlap and no gain from using the dynamic lattice is visible.

\begin{figure}[!htbp]
    \centering
    \includegraphics[width=0.49\textwidth]{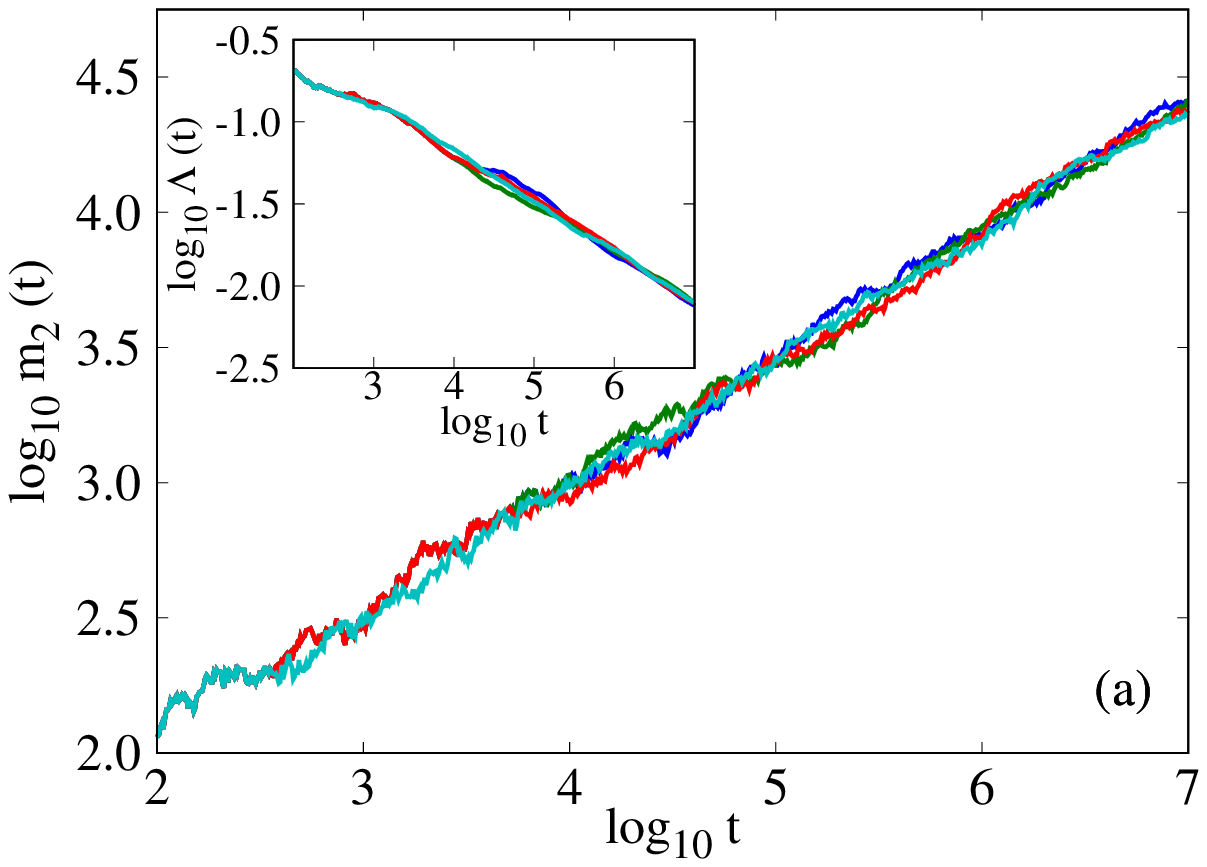}
    \includegraphics[width=0.49\textwidth]{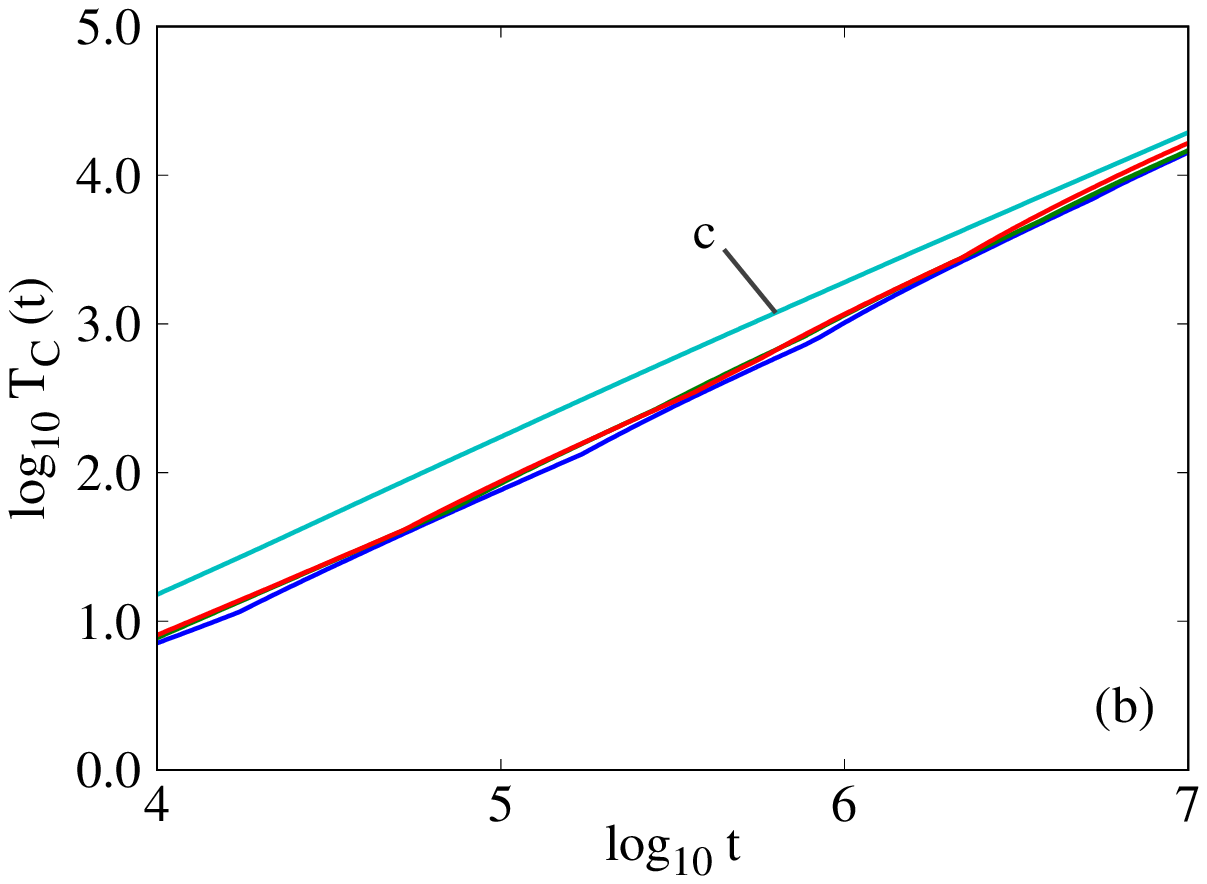}
    \caption{Similar to Fig.~\ref{fig:time_evol_m2_P_mD_PD_MLE_CPU_1d_spreading}, but only the $m_2(t)$~\eqref{eq:second_moment_distr_num} [panel (a)], the ftMLE $\Lambda(t)$~\eqref{eq:finite_mle} [inset of panel (a)] and the CPU time $T_C(t)$ [panel (b)] are reported. 
        The curve coloring corresponds to $\delta_{\mathcal{D}} = 10^{-8}$ [(b)lue], $\delta_{\mathcal{D}} = 10^{-10}$ [(g)reen], $\delta_{\mathcal{D}} = 10^{-12}$ [(r)ed] and the fixed lattice simulation [(c)yan].
        See text for details.    
    }
    \label{fig:time_evol_m2_MLE_CPU_various_threshold_1d_spreading}
\end{figure}
Let us now discuss what happens when we change the value of the threshold $\delta_\mathcal{D}$ in Eq.~\eqref{eq:condition_dynamic_latt_chaos_1d}.
In Fig.~\ref{fig:time_evol_m2_MLE_CPU_various_threshold_1d_spreading}, we perform simulations with the initial condition used in the example of Fig.~\ref{fig:dynamic_latt_chaos_1d_00}.
We use different parameter values of the norm per site threshold at the edge points $\delta_\mathcal{D}$ ranging from $10^{-8}$ to $10^{-12}$.
The result is depicted in Fig.~\ref{fig:time_evol_m2_MLE_CPU_various_threshold_1d_spreading}.
There, we plot the time evolution of $m_2(t)$~\eqref{eq:second_moment_distr_num} of $\xi_l$~\eqref{eq:norm_density_distribution_num} as observable of the evolution of the equations of motion [see Fig.~\ref{fig:time_evol_m2_MLE_CPU_various_threshold_1d_spreading}(a)] and the ftMLE $\Lambda (t)$~\eqref{eq:finite_mle} [see inset of Fig.~\ref{fig:time_evol_m2_MLE_CPU_various_threshold_1d_spreading}(a)] for of the tangent dynamics with $\delta _{\mathcal{D}} = 10^{-8}$ [(b)lue curves], $\delta _{\mathcal{D}} = 10^{-10}$ [(g)reen curves], $\delta _{\mathcal{D}} = 10^{-12}$ [(r)ed curves] and the fixed lattice computation [(c)yan curves]. 
The evolutions of the $m_2(t)$~\eqref{eq:second_moment_distr_num} and the $\Lambda(t)$~\eqref{eq:finite_mle} practically overlap in all used cases, indicating a tendency to follow the same power law behaviors. 
In Fig.~\ref{fig:time_evol_m2_MLE_CPU_various_threshold_1d_spreading} (b), we depict the time evolution of the CPU time $T_C$ needed for these simulations.
Once again the fixed lattice takes a longer time to complete the integration compared to the dynamic lattice.
In addition, dynamic lattice simulations with larger norm per site threshold $\delta_{\mathcal{D}}$ for the control of the endpoints take less time.
Overall, a gain of $g \approx 27\%$, $24\%$, $15\%$ respectively when $\delta_{\mathcal{D}} = 10^{-8}$, $10^{-10}$ and $10^{-12}$ are observed.   
The closeness of the gains in the case $\delta _{\mathcal{D}} = 10^{-8}$ and $\delta _{\mathcal{D}} = 10^{-10}$ is explained by the fact that the norm flows rather quickly at the boundaries, as the case used here belongs to the strong chaos spreading regime.
A larger difference appears between the different $\delta_\mathcal{D}$ values when initial conditions are taken from the weak chaos spreading regime. 

From what has been observed above, the overall idea of the dynamic lattices is to find the optimal lattice size for the numerical evolution of the Hamiltonian models (in the present context $\mathcal{H}_{1D}$~\eqref{eq:Hamilton_complex_dnls_1d} and $\mathcal{H}_{2D}$~\eqref{eq:hamilton_complex_ddnls_2d}) for a specific set of parameter values $W$, $\beta$, $\mathcal{H}_{1D}$~\eqref{eq:Hamilton_complex_dnls_1d}, and $\mathcal{S}_{1D}$~\eqref{eq:norm_density_distribution_num} realizing a specific level of accuracy $\delta_{\mathcal{D}}$ in capturing the evolution of a central initial wave packet excitation.
Of course, the smaller $\delta_{\mathcal{D}}$ is the larger lattice size we are going to need.
In practice, our choice of $\delta_{\mathcal{D}}$  depends on the final values of the average norm per site attained in our longest simulations. 
In the case of the 1D DDNLS model, we are reaching a wave packet width $P_{w} \propto 10^3$ sites.
Since the constant total norm $\mathcal{S}_{1D} \propto 10$ is shared among all the excited sites, the final avarage norm per site 
\begin{equation}
    s_f \propto \frac{\mathcal{S}_{1D}}{P_{w}}  \sim 10^{-2}-10^{-3},
\end{equation}
such that using 
\begin{equation}
    \delta _{\mathcal{D}} \sim 10^{-8},
\end{equation}
we capture the wave packet evolution with a rather good accuracy as shown in Fig.~\eqref{fig:time_evol_m2_P_mD_PD_MLE_CPU_1d_spreading} up to our largest simulation time $t_f = 10^{8}$.

In order to avoid repetition, in all our numerical simulations, we consider for the 1D DDNLS system~\eqref{eq:Hamilton_complex_dnls_1d} initially localized wave packet excitations of size $L$ at the center of a lattice whose maximal size reach in some cases $N \approx 5000$ sites in order to exclude effects from the lattice edges (i.e.~in order to secure that the evolved wave packet does not reach the fixed edges). 
We give to each excited site a single oscillator frequency shift (or rescaled norm per site) $x = \beta s$~\eqref{eq:rescaled_densities_stat}, where $s$ is the constant norm at each site so that the total norm~\eqref{eq:norm_complex_dnls_1d} is $\mathcal{S}_{1D} = \sum _{l\in L}s = Ls$ (or $\mathcal{S}_{1D} = Lx/\beta$, $\beta$ being the nonlinear coefficient of the Hamiltonian $\mathcal{H}_{1D}$~\eqref{eq:Hamilton_complex_dnls_1d}). 
As mentioned in Sec.~\eqref{subsec:choice_of_init_dvd}, for the initial DV $\bm{w}(0)$, we choose to excite only the $\delta q_l$ and $\delta p_l$ values of the site at the centre the lattice (it corresponds to DV$_{1}$ in Sec.~\ref{subsec:choice_of_init_dvd}) and ensure that the norm $\lVert \bm{w}(0) \rVert = 1$. 
Further we integrate the equations of motion and the variational equations using the SI $s11\mathcal{ABC}6$~\eqref{eq:s11odr6_general} of order $6$ with an integration time step $\tau \approx 0.15-0.25$, always ensuring that the maximal relative energy~\eqref{eq:rel_energy_error_1dDDNLS} $E_r$ and norm~\eqref{eq:norm_density_distribution_num} $S_r$ errors remain below $10^{-3}$.
Our typical final integration time is $t_f \approx 10^7-10^{8}$ time units.
For all the considered realizations with $\beta > 0$ in the $\mathcal{H}_{1D}$~\eqref{eq:Hamilton_complex_dnls_1d}, we set the same random parameter $\epsilon _l = 0$ at the lattice's centre.
This ensures that all simulations of single (the central) site excitation ($L=1$) results in the same energy value for the same initial norm per site $s_l$.
On the other hand, in case of multi-site ($L > 1$) initial excitation, all the active sites receive the same value of the norm per site $s_l$~\eqref{eq:ham_norm_inf_temp} and we take particular care so the value of the corresponding $\mathcal{H}_{1D}$~\eqref{eq:Hamilton_complex_dnls_1d} is in a region belonging to the system's thermal phase~\citep{flach2016spreading,kati2020density} (see also Sec.~\ref{sec:initialization_wave}).

\section{\label{sec:mle_1d_spreading} The maximum Lyapunov exponent}

\subsection{\label{subsec:linear_regime_1d_mle}The linear regime}
Here, we look at the linear regime in the 1D DDNLS model~\eqref{eq:Hamilton_complex_dnls_1d}.
Its Hamiltonian corresponds to $\mathcal{H}_{L}$~\eqref{eq:hamilton_anderson_model}, or equivalently $\mathcal{H}_{1D}$~\eqref{eq:Hamilton_complex_dnls_1d} with the nonlinear coefficient $\beta = 0$.
The theory in Sec.~\ref{sec:anderson_localization} tells us that no diffusion of initially localized excitations is present as AL prevails.
In this situation, a saturation of the $m_2$ and $P$ is observed once the wave packet extent reaches the border of the localization volume $\overline{V}$~\eqref{eq:localization_volume_length_chap_anderson}.
This corresponds to equilibrium non-ergodic (regular) dynamics in the system's phase space where the trajectories belong to a torus associated with energy level $E$ of the periodic motion $\psi _l = A_l e^{-i E t}$ at each site $l$~\eqref{eq:stationary_states_anderson} of the chain of oscillators. 
Thus no chaos is present and one expects the ftMLE $\Lambda(t) \propto t^{-1}$.
\begin{figure}[!htbp]
    \centering
    \includegraphics[width=0.49\textwidth]{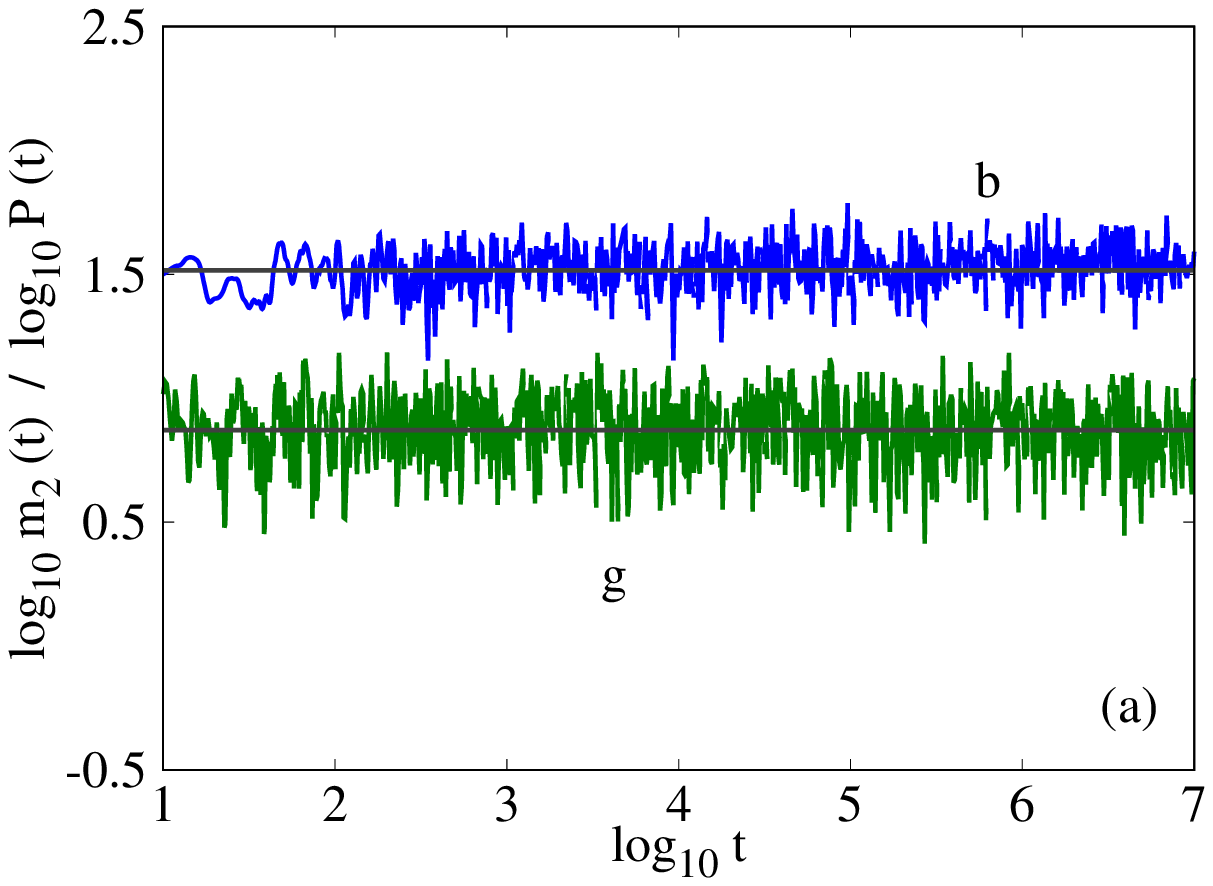}
    \includegraphics[width=0.49\textwidth]{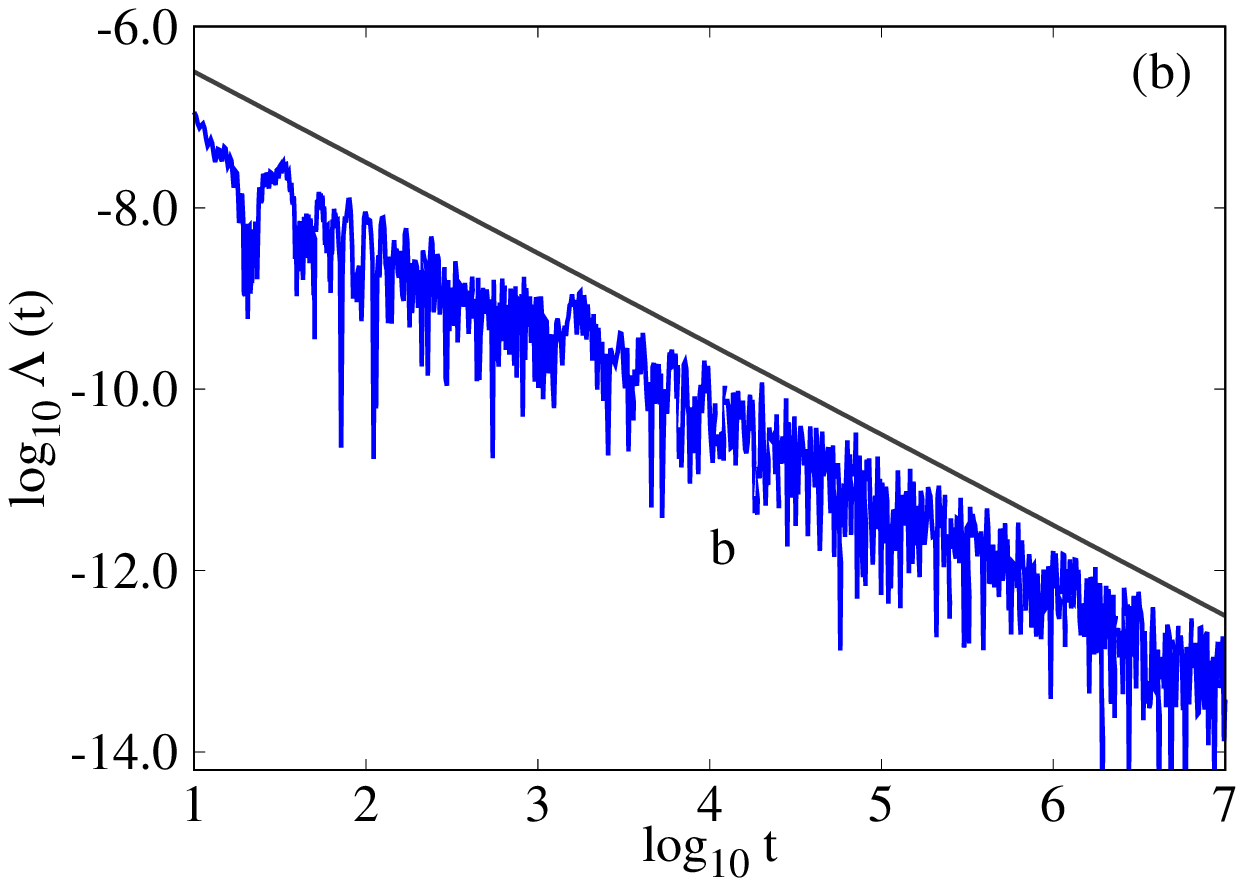}
    \caption{The time evolution of the second moment $m_2(t)$~\eqref{eq:second_moment_distr_num} [(b)lue curve in panel (a)], the participation number $P(t)$~\eqref{eq:participation_ratio_num} [(g)reen curve in panel (a)] of the norm density distribution $\xi_l$~\eqref{eq:norm_density_distribution_num}.
        The related ftMLE $\Lambda(t)$ is shown in panel (b) for a representative of case A$1_{1D}$ (see text for details). 
        The black lines in panel (a) guide the eye for zero slopes along the observable's mean values $\approx 32$ for the $m_2$ [upper black line] and $ \approx 8$ [lower black line] for $P$ respectively.
        In addition, the one in panel (b) guides the eye for slope $-1$.
    }
    \label{fig:linear_case_m2_P_mle_1d_spreading}
\end{figure}

In Fig.~\ref{fig:linear_case_m2_P_mle_1d_spreading}, we plot the results of the time evolution of the $m_2(t)$~\eqref{eq:second_moment_distr_num} and $P(t)$~\eqref{eq:participation_ratio_num} of the norm distribution $\xi_l$~\eqref{eq:norm_density_distribution_num} as observables of the evolution of the equations of motion, and the ftMLE $\Lambda(t)$~\eqref{eq:finite_mle} for the variational equations for a single site $L=1$ excitation (we label this as case A$1_{1D}$ for future reference) with a unit norm per site i.e.~$s=1$ at the center of a lattice of size $N=1021$ sites.
The energy $\mathcal{H}_{1D} = 1$ is obtained from the disorder realization we used, with parameters $W = 4$ and $\beta = 0$. 
In panel (a) of Fig.~\ref{fig:linear_case_m2_P_mle_1d_spreading}, we depict the obtained evolution of $m_2(t)$~\eqref{eq:second_moment_distr_num} [(blue) curve], and $P(t)$~\eqref{eq:participation_ratio_num} [(g)reen curve].
We see that the $m_2(t)$ and $P(t)$ saturate as their values randomly oscillate around their the means $\overline{m}_2 \approx 32$ [upper black line in Fig.~\ref{fig:linear_case_m2_P_mle_1d_spreading}(a)] and $\overline{P} \approx 8$ [lower black line in Fig.~\ref{fig:linear_case_m2_P_mle_1d_spreading}(a)] rather quickly ($\log_{10}t \approx 1$).
In panel (b) of Fig.~\ref{fig:linear_case_m2_P_mle_1d_spreading}, we show the evolution of the ftMLE $\Lambda (t)$~\eqref{eq:finite_mle}.
A clear behavior $\Lambda(t) \propto t^{-1}$ is visible as the (b)lue curve in Fig.~\ref{fig:linear_case_m2_P_mle_1d_spreading}(b) has the tendency to lay along the black straight line whose slope is $-1$. 
This was also reported for the DKG of harmonic oscillators~\citep[see~e.~g.][]{skokos2013nonequilibrium}.
Next, we inch up the nonlinearity $\beta$ in $\mathcal{H}_{1D}$~\eqref{eq:Hamilton_complex_dnls_1d} and numerically investigate the fate of the wave packet's localization.

\subsection{\label{subsec:weak_chaos_1d}The weak chaos regime}
So far, the results predicted in Chap.~\ref{chap:spreading} regarding the spreading of a nonlinear localized excitation in the 1D DDNLS model~\eqref{eq:Hamilton_complex_dnls_1d} appear to be accurately reproduced by numerical simulations.
Indeed, the first numerical experiments performed in~\citep{flach2009universal} were a complete success in confirming these predictions.
Furthermore, the observation of two distinct power laws of wave packet expansion during the spreading was extensively studied in~\citep{skokos2009delocalization,flach2010spreading,laptyeva2010crossover,bodyfelt2011nonlinear,bodyfelt2011wave,laptyeva2013nonlinear} with the crossover from the strong to weak chaos spreading characteristics also reported in~\citep{laptyeva2010crossover}.
Here we present the results of the numerical investigation of the chaotic dynamics of the wave packet spreading within the weak chaos regime.
In particular we focus on the following cases:
\begin{enumerate}
    \item[$\bullet$]{\bf Case $\text{W}1_\text{1D}$.} $W = 3$, $\beta = 0.03$, $L = 21$ and $x = 0.03$.
    \item[$\bullet$]{\bf Case $\text{W}2_\text{1D}$.} $W = 3$, $\beta = 0.6$, $L = 1$ and $x= 0.6$.
    \item[$\bullet$]{\bf Case $\text{W}3_\text{1D}$.} $W = 4$, $\beta = 1$, $L = 1$ and $x = 1$.
    \item[$\bullet$]{\bf Case $\text{W}4_\text{1D}$.} $W = 4$, $\beta = 0.04$, $L = 21$ and $x = 0.04$.
\end{enumerate} 
We can justify the choice of these parameters based on the first condition of Eq.~\eqref{eq:spreading_regime_1d_ddnls_1}.
Indeed, a multisite excitation is expected to evolve in the the weak chaos regime whenever 
\begin{equation}
    x \frac{L}{\overline{V}} < d.
    \label{eq:condition_weak_chaos_chaotic_1d}
\end{equation} 
where $x = \beta s$~\eqref{eq:rescaled_densities_stat} is the rescaled average norm per site, $\overline{V}$~\eqref{eq:localization_volume_length_chap_anderson} is the average localization volume and $d$~\eqref{eq:average_spacing_between_freq_anderson_model} is the average spacing between frequencies in a NM volume.
In Table~\ref{tab:localization_volume_and_av_spacing_spreading}, we give the various values of $d$~\eqref{eq:average_spacing_between_freq_anderson_model}, $\overline{V}$~\eqref{eq:localization_volume_length_chap_anderson}, and $\Delta$~\eqref{eq:characteristics_eigfreq_anderson_model} the frequency bandwidth of NMs for the different disorder strength $W$ values of the 1D DDNLS Hamiltonian model $\mathcal{H}_{1D}$~\eqref{eq:Hamilton_complex_dnls_1d} we are using in this chapter.
Applying the condition~\eqref{eq:condition_weak_chaos_chaotic_1d} in case W$1_{1D}$, we see that $x L/\overline{V} = 0.03\cdot 21/37 \approx 0.017$ which is one order of magnitude smaller than the value of $d = 0.2$ indicated in Table~\ref{tab:localization_volume_and_av_spacing_spreading}.
For the case W$4_{1D}$, the same calculation leads to the number $0.04\cdot 21/21$ which is again much smaller than $d= 0.39$ of $W = 4$ in Table~\ref{tab:localization_volume_and_av_spacing_spreading}.
In the case of single excitation wave packet, it is the first condition of inequality~\eqref{eq:spreading_regime_1d_ddnls_3}
\begin{equation}
    x < \Delta,
    \label{eq:condition_weak_chaos_chaotic_1d_02}
\end{equation} 
which governs the weak chaos spreading dynamics appearance.
For the case W$2_{1D}$ the inequality~\eqref{eq:condition_weak_chaos_chaotic_1d_02} gives $0.6<7$, while in the case W$3_{1D}$ we obtain $1< 8$, both of which are true.
It is worth mentioning that the cases $\text{W}1_\text{1D}$, $\text{W}2_\text{1D}$, $\text{W}3_\text{1D}$ and $\text{W}4_\text{1D}$ have already been used in several works~\citep{laptyeva2010crossover,bodyfelt2011nonlinear,senyange2018computational,senyange2018characteristics}.
In all the cases, we are averaging our observables of the phase space and tangent dynamics over $100$ disorder realizations in order to obtain a robust representation of the system's dynamical behavior which takes into account several trajectories of the same total norm $\mathcal{S}_{1D}$~\eqref{eq:norm_density_distribution_num}.
\begin{table}[!htbp]
    \centering 
    \begin{tabular}{c|rrr}
         NM properties & $W = 3$ & $W = 3.5$ & $W = 4$ \\
        \toprule
        $\overline{V}$ & $37$ & $27$ & $21$\\
        \midrule 
        $\Delta$ & $7$ & $7.5$ & $8$ \\
        \midrule
        $d = \Delta/\overline{V}$ & $0.2$ & $0.28$ & $0.39$\\
        \bottomrule
    \end{tabular}
    \caption{Average localization volume $\overline{V}$~\eqref{eq:localization_volume_length_chap_anderson}, frequency bandwidth $\Delta$~\eqref{eq:characteristics_eigfreq_anderson_model} and average spacing between NM $d$~\eqref{eq:average_spacing_between_freq_anderson_model} in case of a multisite excitation of width $L = \overline{V}$ sites at the center of the 1D DDNLS lattice system~\eqref{eq:Hamilton_complex_dnls_1d}.}
    \label{tab:localization_volume_and_av_spacing_spreading}
\end{table}
\begin{figure}[!htb]
    \centering 
    \includegraphics[width=0.4\textwidth, height=0.4\linewidth]{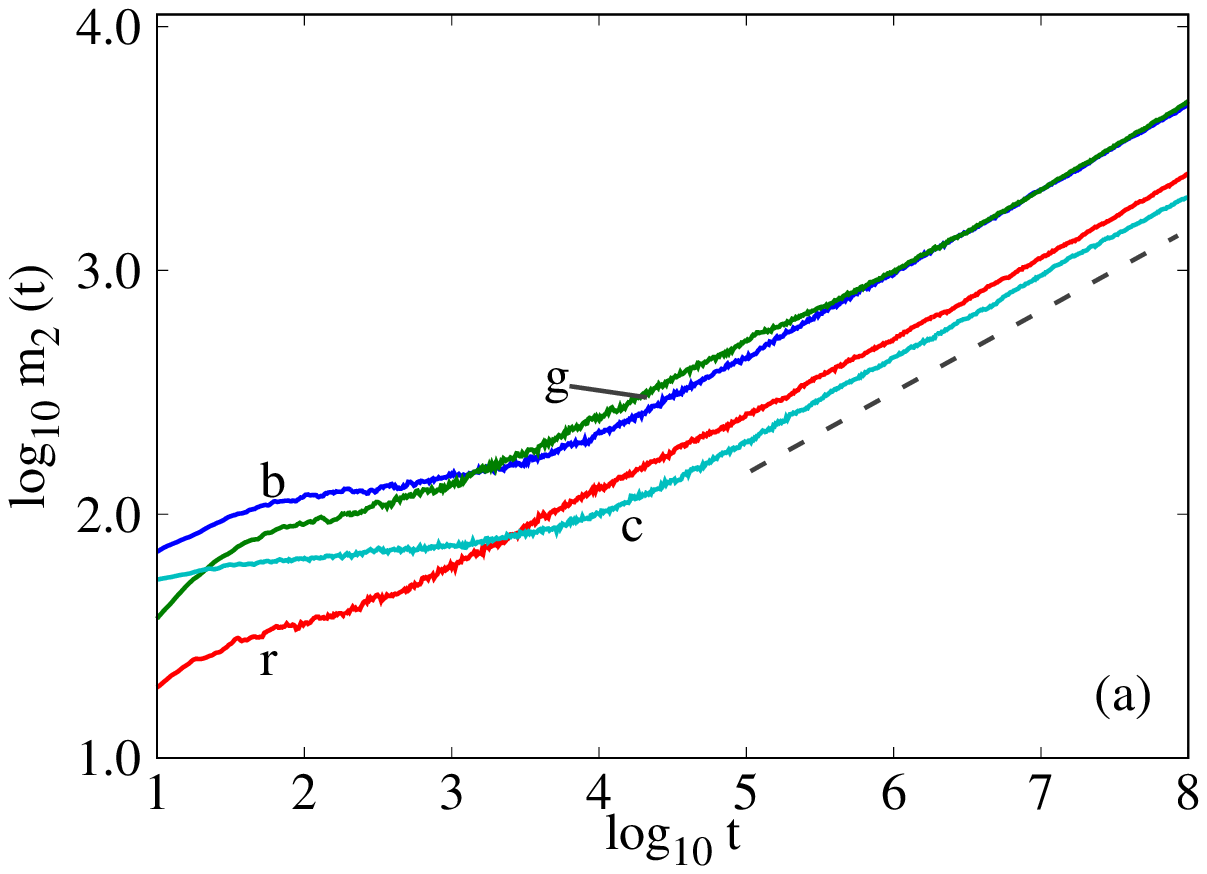}
    \includegraphics[width=0.4\textwidth, height=0.4\linewidth]{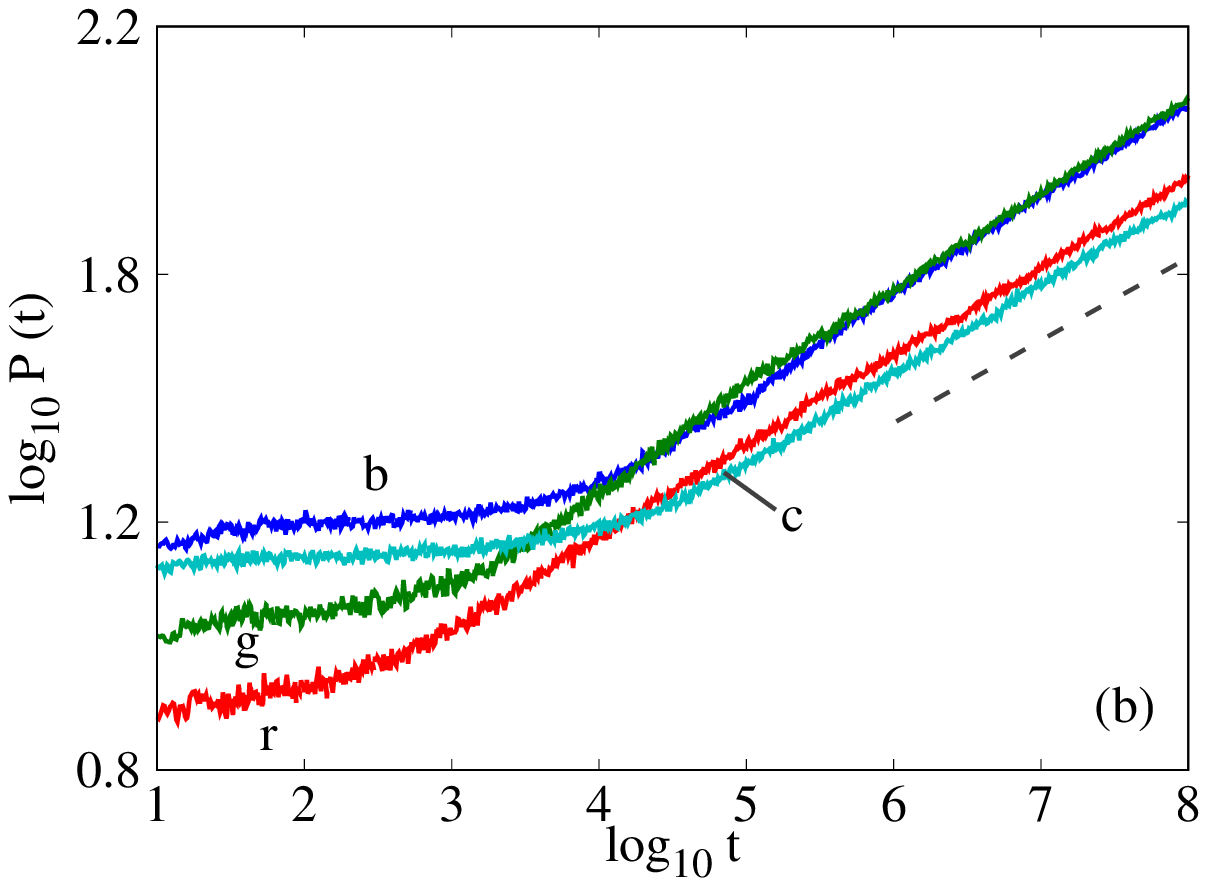}
    \includegraphics[width=0.4\textwidth, height=0.4\linewidth]{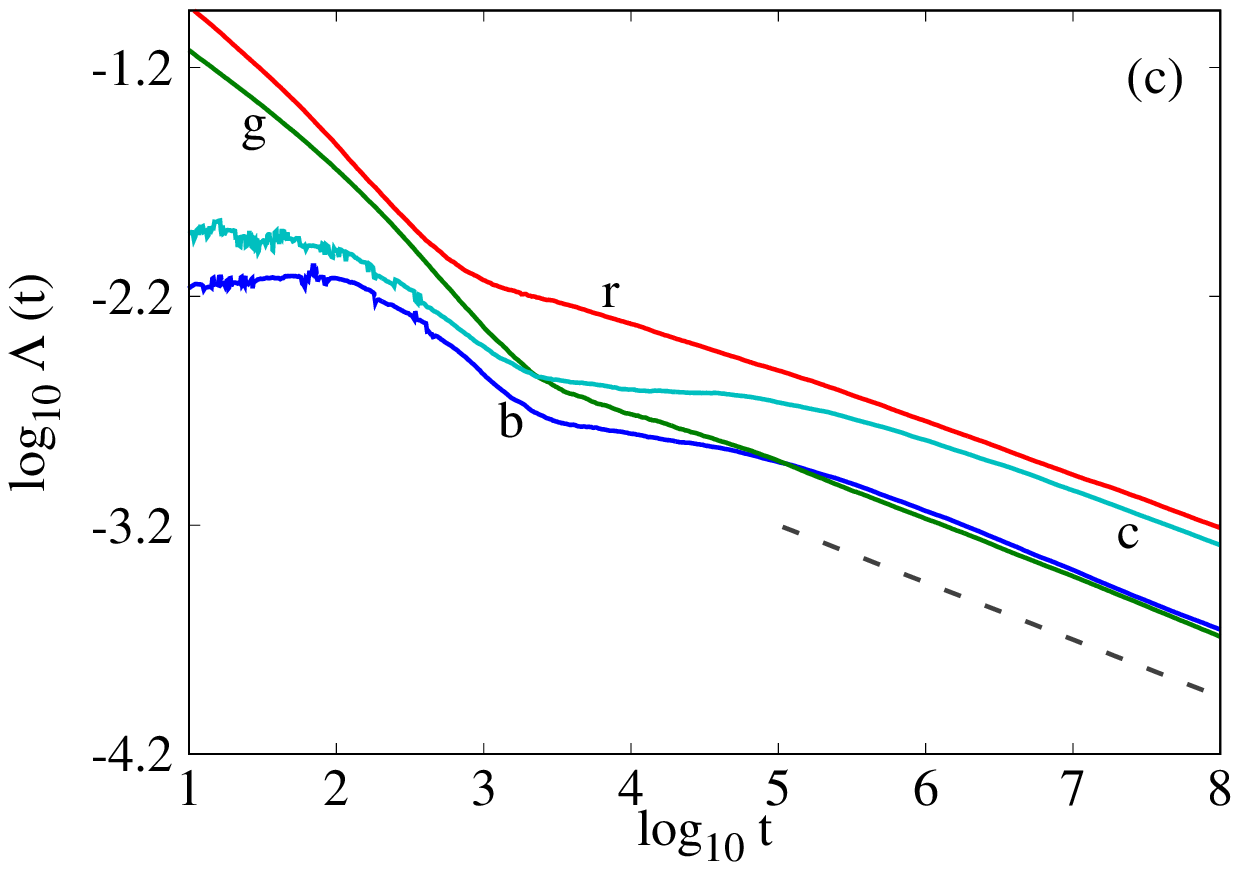}
    \includegraphics[width=0.4\textwidth, height=0.4\linewidth]{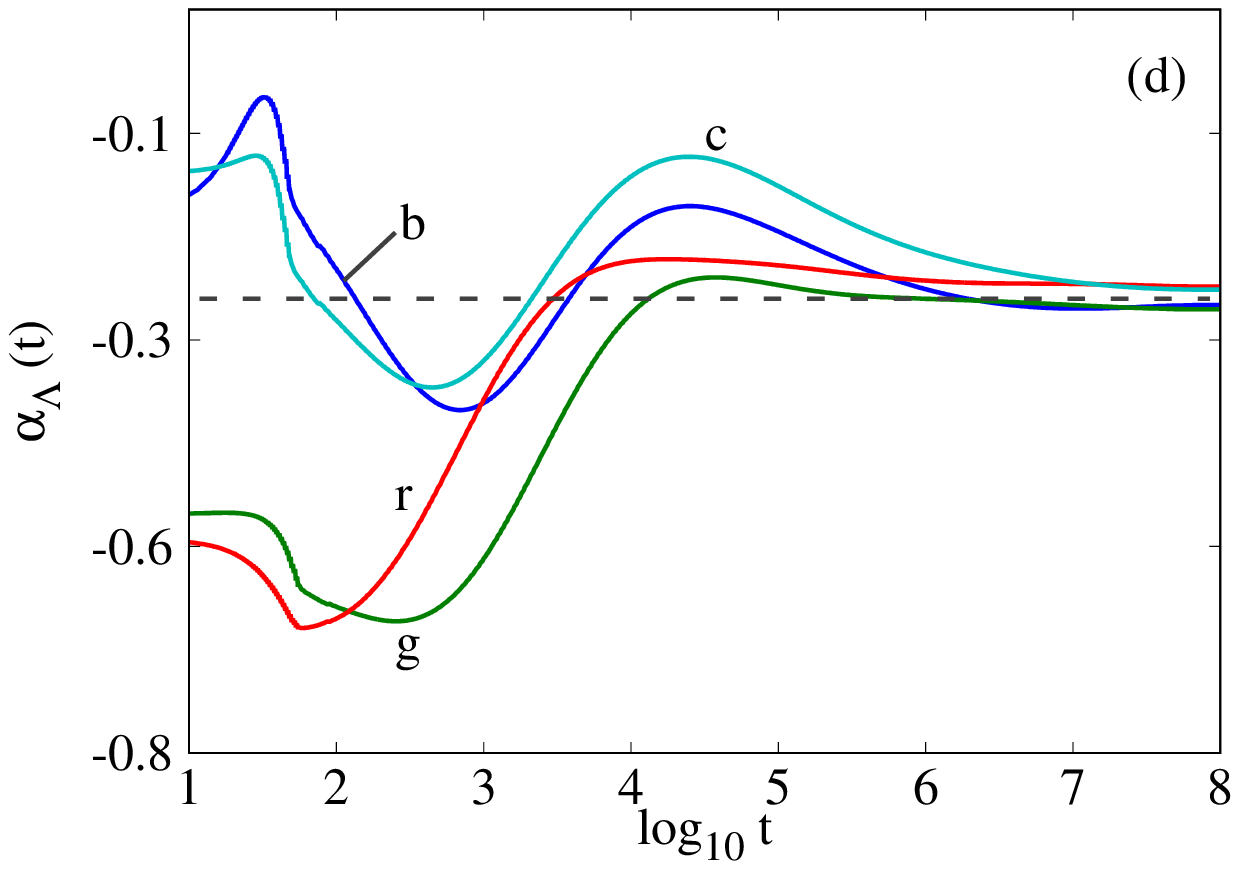}
    \caption{Time dependence of the second moment $m_2(t)$~\eqref{eq:second_moment_distr_num} [panel (a)], the participation number $P(t)$~\eqref{eq:participation_ratio_num} [panel (b)] of $\xi_l$~\eqref{eq:norm_density_distribution_num}, the ftMLE $\Lambda(t)$~\eqref{eq:finite_mle} (c) and $\alpha_\Lambda(t)$~\eqref{eq:slope_theoretical_num} [panel (d)] for the weak chaos cases W$1_{1D}$ [(b)lue], W$2_{1D}$ [(g)reen], W$3_{1D}$ [(r)ed] and W$4_{1D}$ [(c)yan] (see text for details).
        The black dashed lines guide the eye for slopes $1/3$ [panel (a)], $1/6$ [panel (b)] and $-0.25$ [panel (c)]. 
        The horizontal dashed black line in panel (d) indicates the value $\alpha_\Lambda = -0.25$.
        }
    \label{fig:m2P_weak_chaos_1d}
\end{figure}

The fact that all these cases belong to weak chaos becomes evident in Fig.~\ref{fig:m2P_weak_chaos_1d}(a) and (b), where we plot the time evolution of the second moment $m_2(t)$~\eqref{eq:second_moment_distr_num} and the participation number $P(t)$~\eqref{eq:participation_ratio_num} of the norm density distribution $\xi_l$~\eqref{eq:norm_density_distribution_num} respectively.
For both quantities, a subdiffusive spreading is noticeably well fitted by the monotonic asymptotic power laws $m_2(t) \propto t^{\alpha_m}$ [see black line in Fig.~\ref{fig:m2P_weak_chaos_1d}(a)] with $\alpha_m = \frac{1}{3}$ and $P(t) \propto t^{\alpha_P}$ [see black line in Fig.~\ref{fig:m2P_weak_chaos_1d}(b)] where $\alpha_P = \frac{1}{6}$ in agreement with previous works~\citep{flach2009universal,skokos2009delocalization,laptyeva2010crossover,flach2010spreading,bodyfelt2011nonlinear}.
Further, in Fig.~\ref{fig:m2P_weak_chaos_1d}(c) we present the time evolution of the ftMLE, $\Lambda(t)$~\eqref{eq:finite_mle} which is the primary quantity of our interest.
An asymptotic decay following the power law $\Lambda \propto t^{\alpha_{\Lambda}}$ ($\alpha _\Lambda < 0$) is observed for all cases.
In Fig.~\ref{fig:m2P_weak_chaos_1d}(d), we present the time evolution of the related power law exponent $\alpha_\Lambda (t)$ computed following the techniques presented in Sec.~\ref{sec:slope_determination_num}.
The dependence on time of the power law exponent values $\alpha_\Lambda(t)$ shows a clear tendency to saturate at $\alpha _{\Lambda} \approx -0.25$ in agreement with~\citep{skokos2013nonequilibrium,senyange2018characteristics}. 
This number is retrieved in this work taking the average value of all the cases in the last decade of the evolution of $\alpha _\Lambda(t)$.  
These results are not only specific to the 1D DDNLS model, as they have also been found for the 1D DKG model of anharmonic oscillators~\eqref{eq:hamilton_1d_dkg} in~\citep{skokos2013nonequilibrium,senyange2018characteristics}.

\subsection{\label{subsec:strong_chaos_1d_spreading}The strong chaos regime}
We also examine the strong chaos regime of the 1D DDNLS system~\eqref{eq:Hamilton_complex_dnls_1d}.
We use the following initial localized wave packet excitations:
\begin{itemize}
    \item[$\bullet$]{\bf Case $\text{S}1_{\text{1D}}$.} $W= 3$, $\beta = 1$, $L = 37$ and $x = 0.3$.
    \item[$\bullet$]{\bf Case $\text{S}2_{\text{1D}}$.} $W= 3.5$, $\beta = 0.62$, $L = 21$ and $x = 0.62$.
    \item[$\bullet$]{\bf Case $\text{S}3_{\text{1D}}$.} $W= 3.5$, $\beta = 0.72$, $L = 21$ and $x = 0.72$.
    \item[$\bullet$]{\bf Case $\text{S}4_{\text{1D}}$.} $W= 3.5$, $\beta = 1$, $L = 27$ and $x = 0.525$.   
\end{itemize}
Again, we justify the clustering of these initial parameter in the strong chaos spreading regime applying the second condition in Eq.~\eqref{eq:spreading_regime_1d_ddnls_1}.
It states that if   
\begin{equation}
    x \frac{L}{\overline{V}} > d,
    \label{eq:condition_strong_chaos_chaotic_1d}
\end{equation}
the initial wave packet excitation will evolve for a finite amont of time within the strong chaos region of Fig.~\ref{fig:x_y_thermal_selftrapping_phases}.
As we did in the case of weak chaos, we numerically check the condition~\eqref{eq:condition_strong_chaos_chaotic_1d} on the parameter values above.
For the case S$1_{1D}$ the left hand side of inequality~\eqref{eq:condition_strong_chaos_chaotic_1d} results in $0.3\cdot 37/37 = 0.3$, which is larger than $d = 0.2$ of $W = 3$.
For the cases S$2_{1D}$, S$3_{1D}$ and S$4_{1D}$ which all use the value $W = 3.5$, we calculate the left hand side of Eq.~\eqref{eq:condition_strong_chaos_chaotic_1d} as $0.62\cdot 21/27 \approx 0.482$, $0.72 \cdot 21/27 = 0.56$ and $0.525\cdot 27/27 = 0.525$ respectively which are all greater than $d = 0.28$ in Table~\ref{tab:localization_volume_and_av_spacing_spreading}.
Thus all the cases above belong to the strong chaos regimes according to the theory of Chap.~\ref{chap:spreading}.
Note that the cases S$2_{1D}$ and S$3_{1D}$ were used in~\citep{senyange2018characteristics}, while we introduce the cases S$1_{1D}$ and S$4_{1D}$ in order to investigate other initial conditions than the ones already presented in the literature~\citep{laptyeva2010crossover,bodyfelt2011nonlinear,senyange2018characteristics}. 
Like for the weak chaos regime, we average our observables of the phases space and tangent dynamics over $100$ disorder realizations in order to ensure the mixing of several different trajectories with constant $\mathcal{S}_{1D}$ values in the system's phase space in order to ensure the robustness of our computations.
\begin{figure}[!htb]
    \centering 
    \includegraphics[width=0.4\textwidth, height=0.4\linewidth]{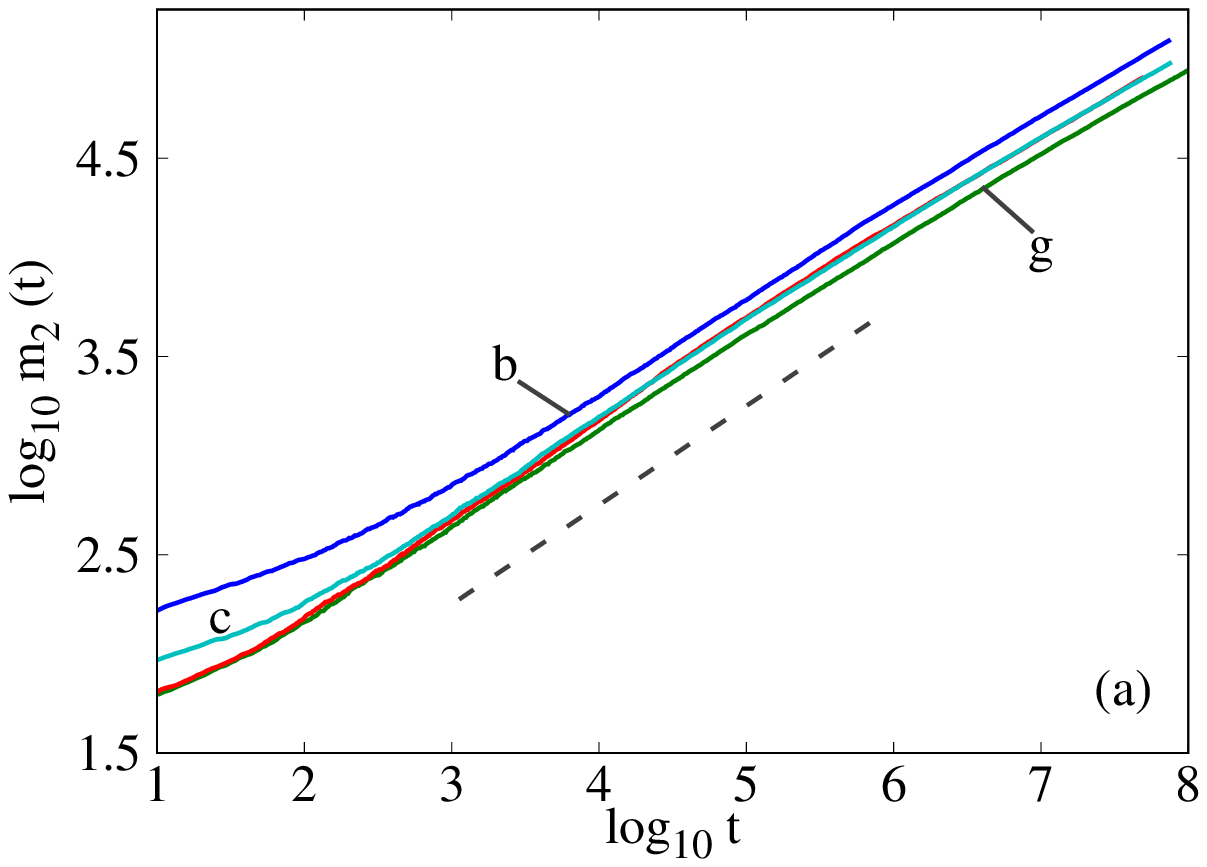}
    \includegraphics[width=0.4\textwidth, height=0.4\linewidth]{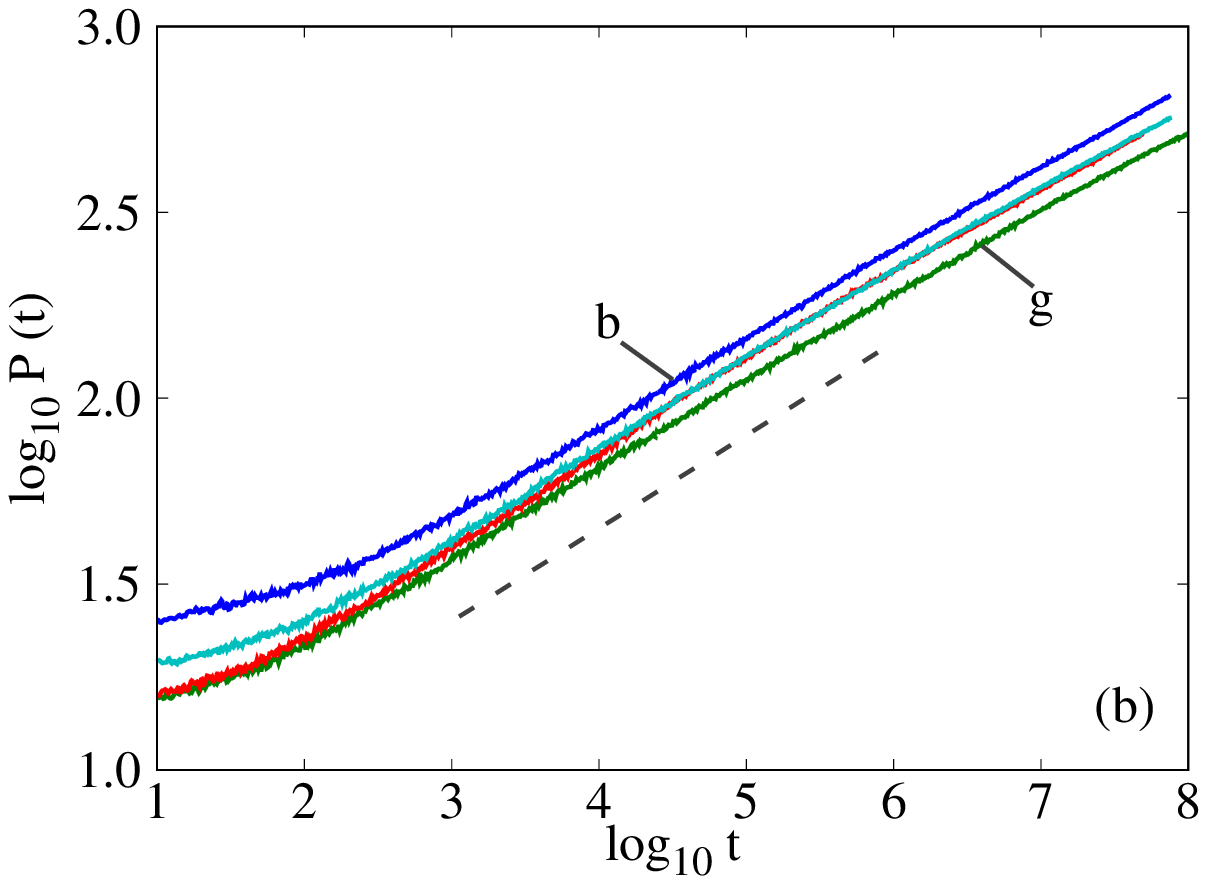}
    \includegraphics[width=0.4\textwidth, height=0.4\linewidth]{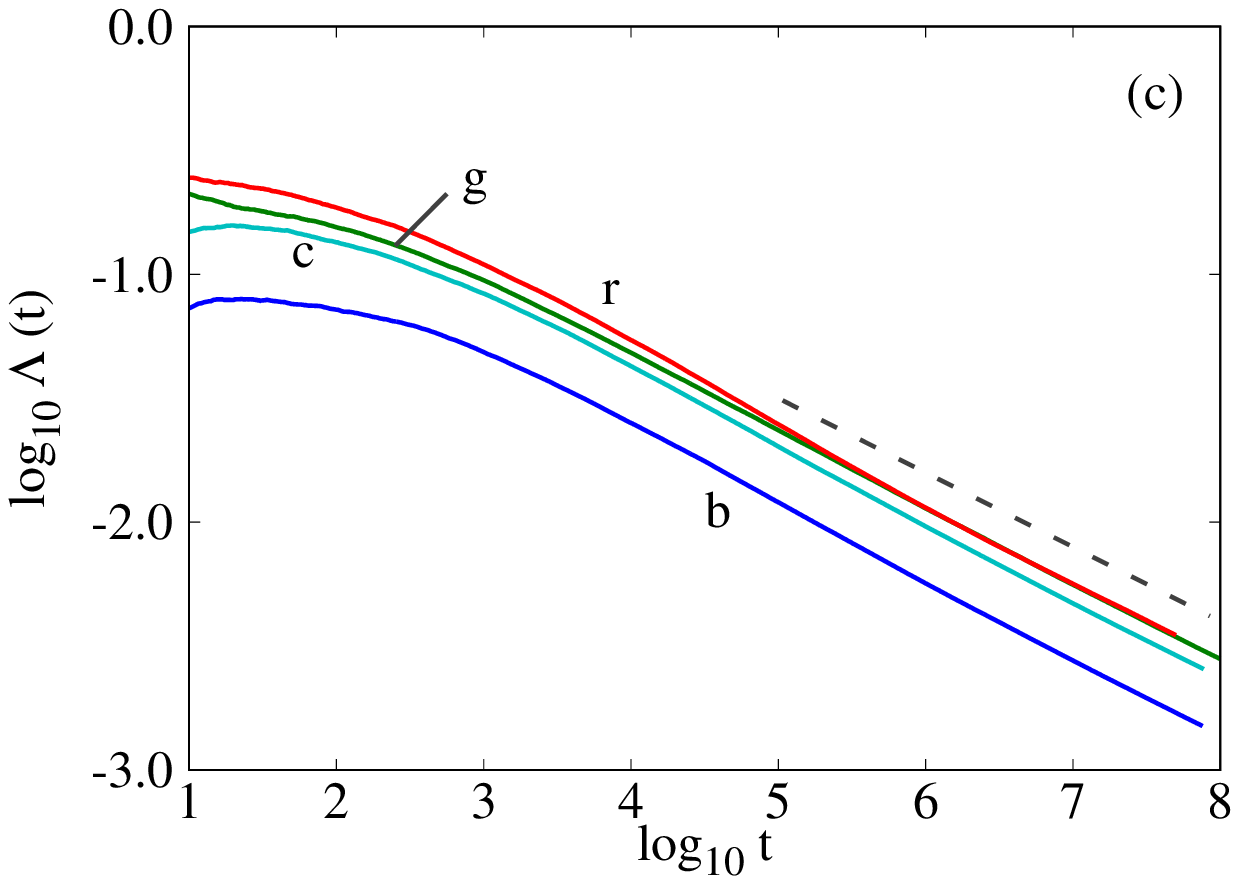}
    \includegraphics[width=0.4\textwidth, height=0.4\linewidth]{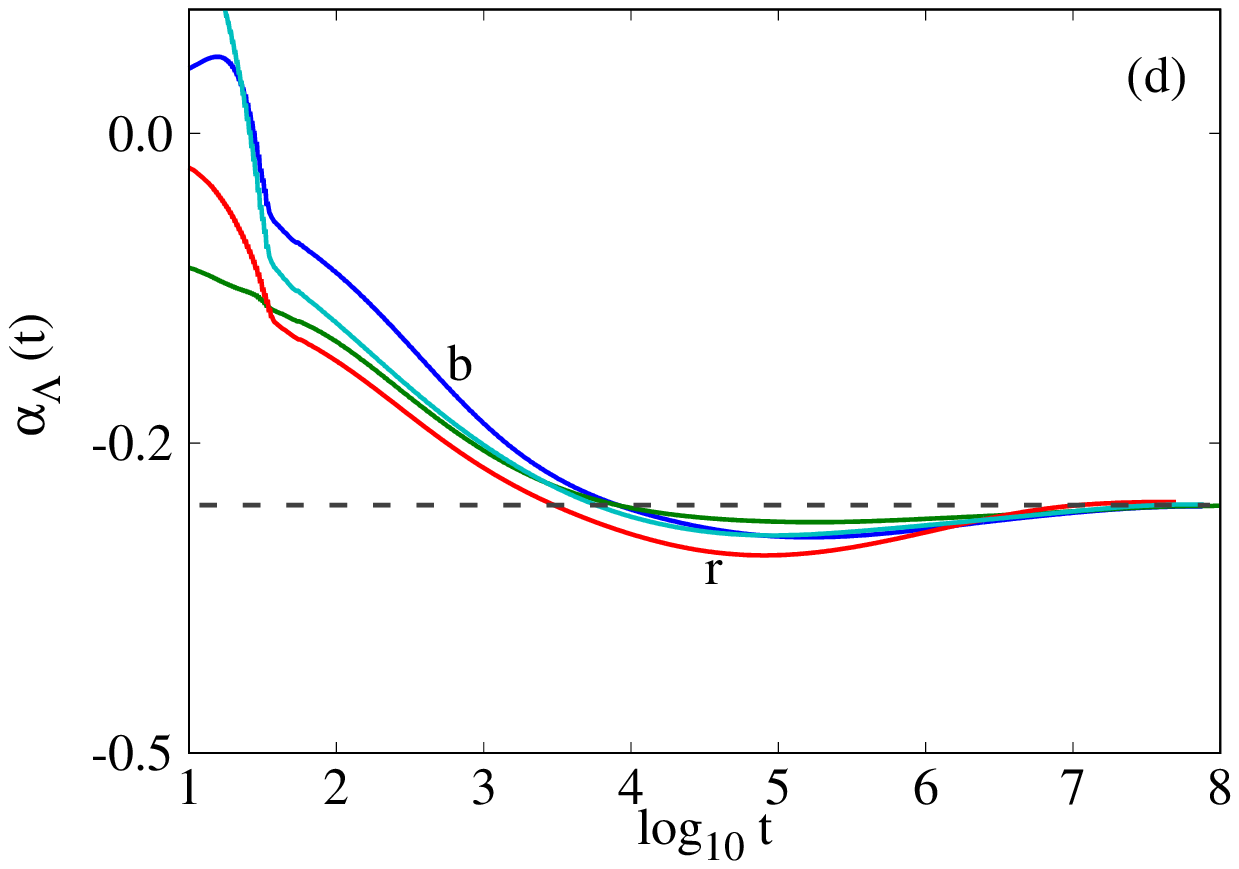}
    \caption{Similar to Fig.~\ref{fig:m2P_weak_chaos_1d} but for strong chaos regime.
            The coloring corresponds to the cases S$1_{1D}$ [(b)lue], S$2_{1D}$ [(g)reen], S$3_{1D}$ [(r)ed] and S$4_{1D}$ [(c)yan] (see text for details).
            The straight lines guide the eye for slopes $\frac{1}{2}$ [panel (a)], $\frac{1}{4}$ [panel (b)] and $-0.3$ [panel (c)].
            The horizontal dashed line in panel (d) indicates the value $\alpha_\Lambda = -0.3$.
        }
    \label{fig:mle_slopes_strong_chaos_1d}
\end{figure}

The panels of Figs.~\ref{fig:mle_slopes_strong_chaos_1d}(a) and (b) depict the evolution of the wave packet second moment $m_2(t)$~\eqref{eq:second_moment_distr_num} and participation number $P(t)$~\eqref{eq:participation_ratio_num} respectively.
Again, a subdiffusive spreading of the wave packet initial excitation following the power laws $m_2(t) \propto t^{\alpha_m}$ and $P (t) \propto t^{\alpha_P}$ with $\alpha_m, \alpha _P< 1$ is clearly visible.
This time the growth rates are well fitted by $\alpha_m = \frac{1}{2}$ [see black curve in Fig.~\ref{fig:mle_slopes_strong_chaos_1d}(a)] and $\alpha _P  = \frac{1}{4}$ [see black curve in Fig.~\ref{fig:mle_slopes_strong_chaos_1d}(b)] in a region of around the early stages of the time evolution i.e. $\log_{10} t \approx 10^{3} - 10^{6}$.
This numerically confirms our choice of these cases as strong chaos ones.
We can therefore compute the  ftMLEs of all these strong chaos cases in Fig.~\ref{fig:mle_slopes_strong_chaos_1d}(c).
There, a similar tendency $\Lambda(t) \propto t^{\alpha_\Lambda}$ ($\alpha_\Lambda < 0$) to the one noticed for the weak chaos cases is seen for all the used cases. 
Nevertheless, the $\alpha_{\Lambda}(t)$ values tend to saturate at $\alpha_\Lambda \approx -0.3$ as shown in Fig.~\ref{fig:mle_slopes_strong_chaos_1d}(d).
Again, this value is obtained averaging the data of the time evolution of $\alpha_\Lambda(t)$ in the last decade of the evolution where the saturation is observed.
It is worth mentioning that once again the same results have also been found for the 1D DKG model~\eqref{eq:hamilton_1d_dkg}~\citep{senyange2018characteristics} underlining the generality of this power exponent.

\section{\label{sec:dvd_1d_spreading}The deviation vector distribution}
There is a need of a second ingredient, a non-correlated spatiotemporal force $f(t)$ acting on the chaotic fraction of the system's NMs evolution as seen in Eq.~\eqref{eq:evol_transmission_flach} in order to obtain the chaotic spreading of the wave packet.
These random fluctuations of chaotic seeds are known to support the subdiffusive spreading of the lattice's excited part~\citep{flach2009universal,skokos2013nonequilibrium}.
We have been stressing in Sec.~\ref{subsec:dvd_num} and in the first part of this chapter that the DV used for the computation of the MLE aligns with the most chaotic direction intrinsic to a system's orbit in its phase space.
Therefore, the DVD~\eqref{eq:DVD_definition_num}
\begin{equation}
    \xi_l^D(t) = \frac{\delta q_l^2(t) + \delta p_l^2(t)}{\sum _l \delta q_l^2(t) + \delta p_l^2(t)},
    \label{eq:dvd_1d_chaotic}
\end{equation}
can be used to represent sites which are more sensitive to perturbation at time $t>0$.
Consequently, it can be used to visualize the motion of chaotic seeds~\citep{oganesyan2009energy,basko2011weak,basko2012local,mulansky2014scaling,skokos2013nonequilibrium,senyange2018characteristics,manda2020chaotic}.
In this section, we analyse in depth the dynamical behavior of the DVD in the AL, weak chaos and strong chaos regimes of the initially localized wave packet excitation in the 1D DDNLS lattice system~\eqref{eq:Hamilton_complex_dnls_1d}.

\subsection{\label{subsec:dvd_linear_1d_spreading}The linear regime}
\begin{figure}[!hbp]
    \centering 
    \includegraphics[width=0.7\textwidth]{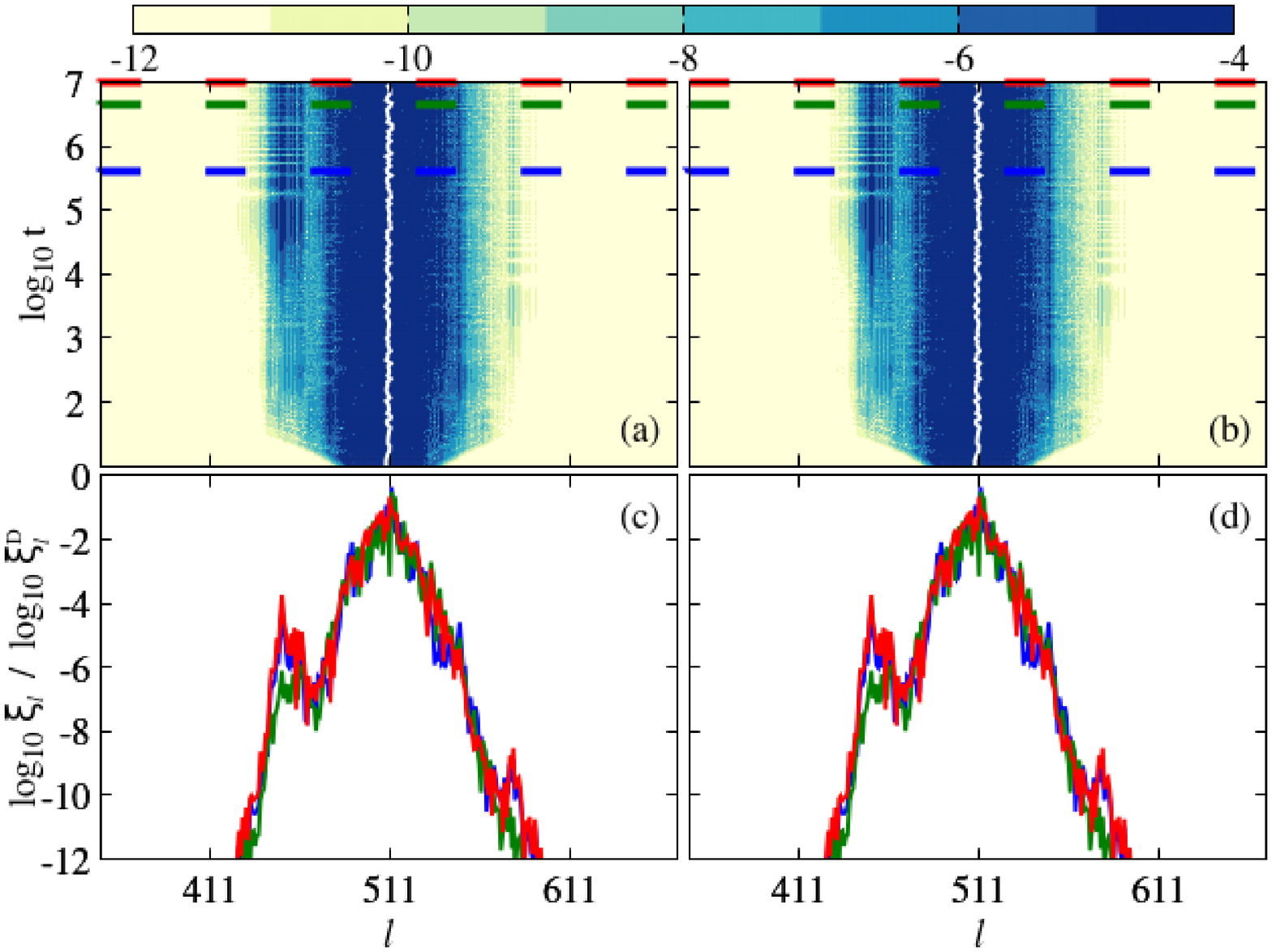}
    \caption{Spatiotemporal dependence of the norm density distribution $\xi_l$~\eqref{eq:norm_density_distribution_num} and DVD $\xi_l^D$~\eqref{eq:dvd_1d_chaotic} for a representative case A$1_{1D}$ (see text for details) of the linear regime.
        The color scales at the top of the figure are used for coloring the lattice sites according to their $\log _{10} \xi_l$ [panel (a)] and $\log_{10}\xi_l^D$ [panel (b)] values.     
        In both panels, the white curves represent the motion of the distribution's centers.
        The panels (c) and (d) show snapshots of the spatial dependence of the distributions $\xi_l$ and $\xi_l^D$ respectively at times $\log_{10} t \approx 5.6$ [(b)lue curves], $\log_{10} t \approx 6.65$ [(g)reen curves] and $\log_{10} t \approx 7$ [(r)ed curves]. 
        These times are also indicated using the same coloring for the dashed horizontal lines in panels (a) and (b).
    }
    \label{fig:linear_case_norm_dvd_and_snapshots_1d_spreading}
\end{figure}

Let us first look more closely to the behavior of the norm density distribution $\xi_l(t) = s_l(t)/\sum_l s_l(t)$~\eqref{eq:norm_density_distribution_num} with $s_l = (q_l^2 + p_l^2)/2$ and the DVD $\xi_l^D$~\eqref{eq:DVD_definition_num} generated during the computation of the ftMLE $\Lambda (t)$~\eqref{eq:mle} when nonlinearity is absent from the system's dynamical equations i.e. $\beta =0$ in $\mathcal{H}_{1D}$~\eqref{eq:Hamilton_complex_dnls_1d}.
We use the case A$1_{1D}$ of Fig.~\ref{fig:linear_case_m2_P_mle_1d_spreading}.
For this case, we compute the spatiotemporal evolution of the norm distribution $\xi_l (t)$ and the DVD $\xi_l^D (t)$ in Figs.~\ref{fig:linear_case_norm_dvd_and_snapshots_1d_spreading}(a) and (b) respectively.
For both distributions, snapshots at specific times $\log_{10}t \approx 5.6$ [(b)lue dashed line in Figs.~\ref{fig:linear_case_norm_dvd_and_snapshots_1d_spreading}(a) and (b)], $\log_{10}t \approx 6.65$ [(g)reen dashed line in Figs.~\ref{fig:linear_case_norm_dvd_and_snapshots_1d_spreading}(a) and (b)] and $\log_{10}t \approx 7$ [(r)ed dashed line in Figs.~\ref{fig:linear_case_norm_dvd_and_snapshots_1d_spreading}(a) and (b)] are displayed in Figs.~\ref{fig:linear_case_norm_dvd_and_snapshots_1d_spreading}(c) and (d) respectively for the $\xi_l$ and $\xi_l^D$.

With $\xi_l(t)$~\eqref{eq:norm_density_distribution_num}, a clear spatial saturation of the wave packet extents appears in Figs.~\ref{fig:linear_case_norm_dvd_and_snapshots_1d_spreading}(a) and (c), in agreement with the saturation of the wave packet $m_2(t)$ and $P(t)$ noticed in Fig.~\ref{fig:linear_case_m2_P_mle_1d_spreading}(a).
For $\xi_l^D$~\eqref{eq:DVD_definition_num} in Figs.~\ref{fig:linear_case_norm_dvd_and_snapshots_1d_spreading}(b) and (d) a saturation of its spatial evolution is also present.
Further, a striking resemblance between the distributions $\xi_l$ and $\xi_l^D$ in Fig.~\ref{fig:linear_case_norm_dvd_and_snapshots_1d_spreading} is visible.
This is explained by the fact that if one computes the variational equations of the 1D DDLS model~\eqref{eq:hamilton_anderson_model}, we obtain 
\begin{equation}
    i \dot{\delta \psi} _l = \epsilon _l \delta \psi _l - \left( \delta \psi _{l - 1} + \delta \psi _{l + 1} \right),
    \label{eq:var_eq_anderson}
\end{equation}  
whose solutions are oscillatory motion of the form $\delta \psi_l = \delta A_l \exp{iEt}$ which oscillates in the same manner as the system's reference orbit $\psi_l = A_l \exp iEt$ which resolves the equations of motion of $\mathcal{H}_L$~\eqref{eq:hamilton_anderson_model} having the same form to~\eqref{eq:var_eq_anderson}.
Consequently, both the $\xi_l(t)$ and $\xi_l^D(t)$, are expected to depict qualitatively the same spatiotemporal dynamics.
These properties are expressions of the regularity (non-ergodicity) of the system's phase space in the linear regime and tell us that an initial condition taken a bit further away from the system's reference orbit will also experience AL as seen in Sec.~\ref{sec:anderson_localization}.  
\begin{figure}[!htb]
    \centering 
    \includegraphics[width=0.39\textwidth, height=0.4\linewidth]{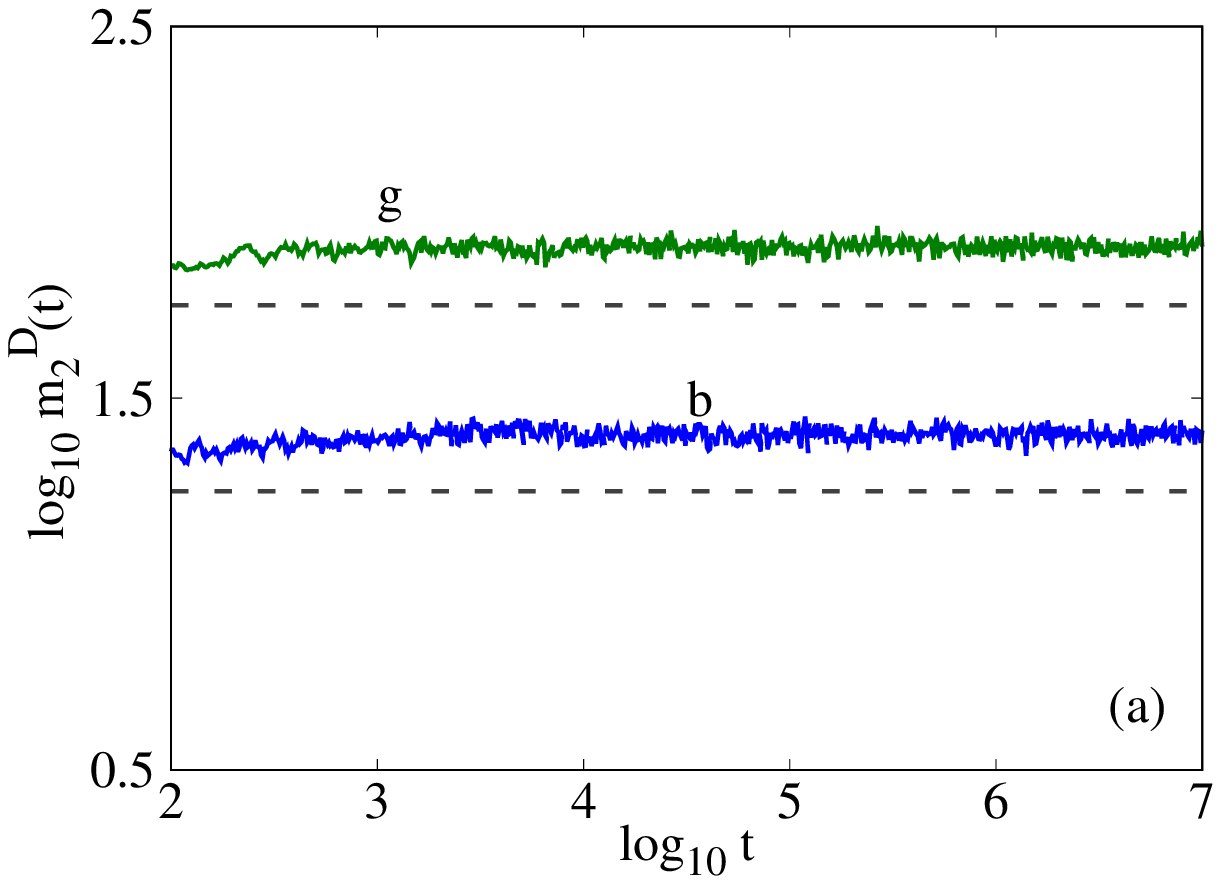}
    \includegraphics[width=0.39\textwidth, height=0.4\linewidth]{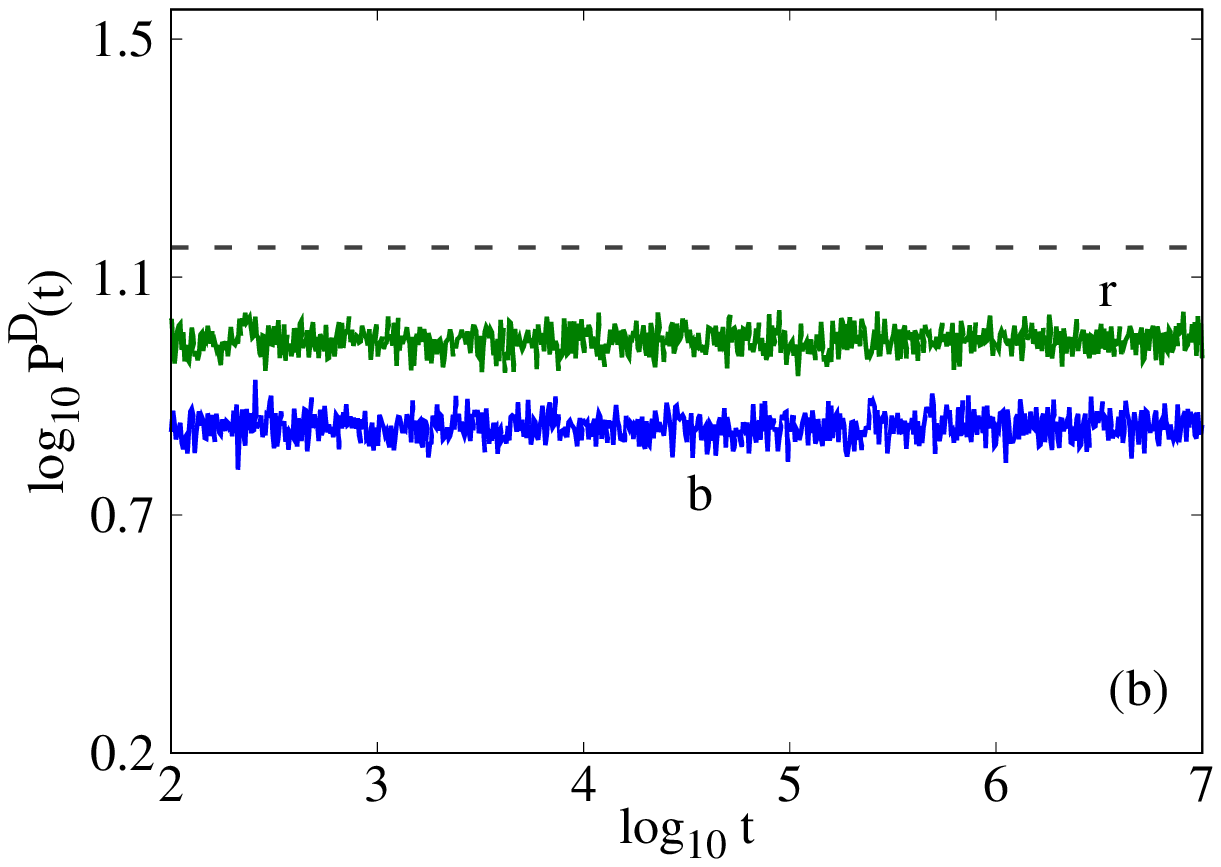}
    \includegraphics[width=0.39\textwidth, height=0.4\linewidth]{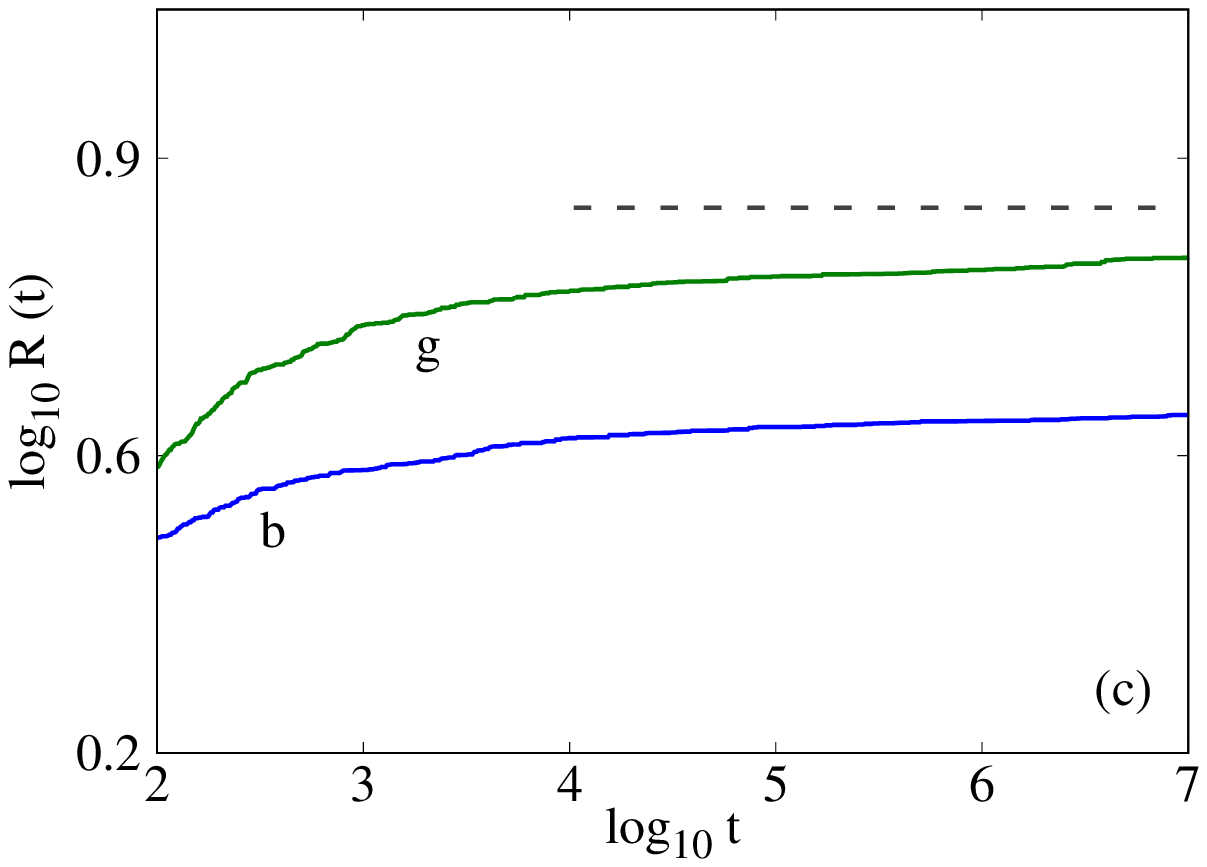}
    \includegraphics[width=0.39\textwidth, height=0.4\linewidth]{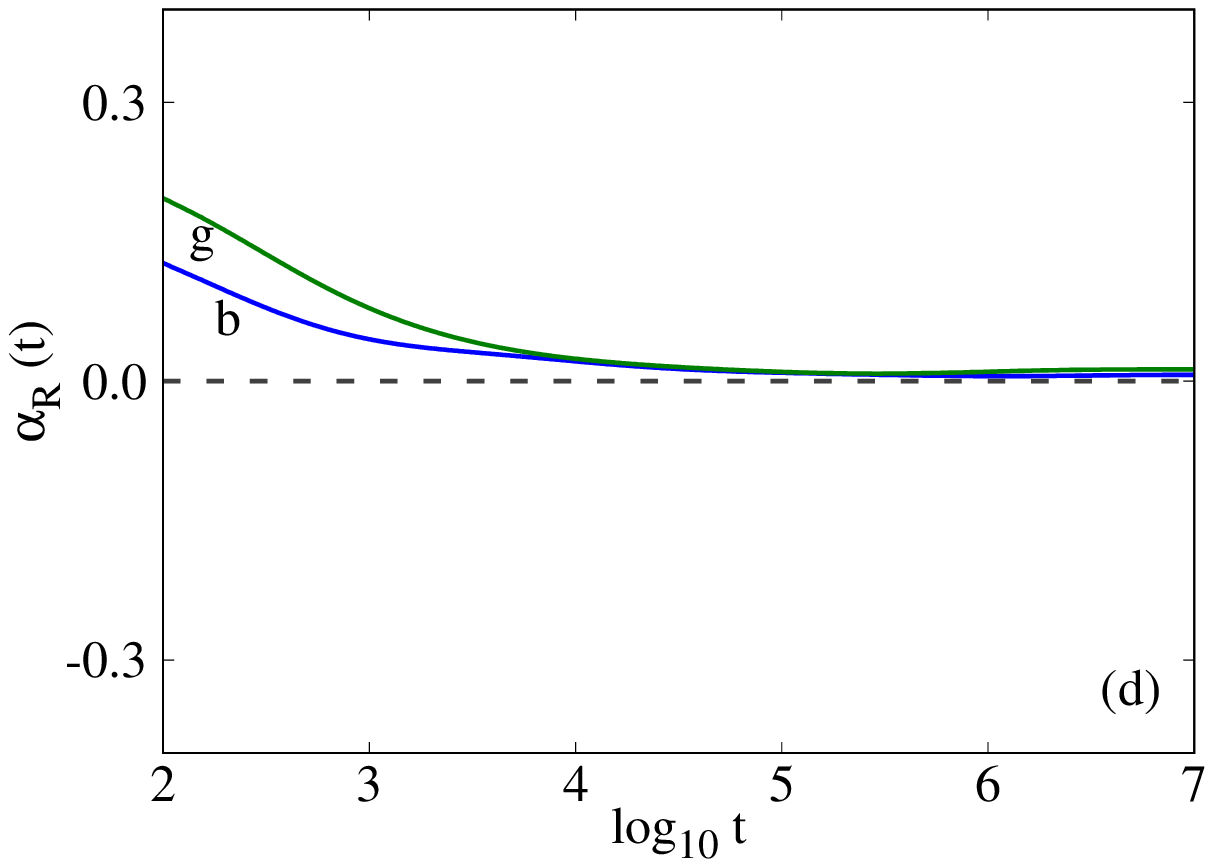}
    \caption{Time dependence of the second moment $m_2^D(t)$~\eqref{eq:second_moment_distr_num} in panel (a), participation number $P^D(t)$~\eqref{eq:participation_ratio_num} in panel (b) of $\xi_l^D$~\eqref{eq:DVD_definition_num}.
        In panel (c) we plot the length covered by the distribution's center $R(t)$~\eqref{eq:range_covered_by_DVD}, while in panel (d) we show the related power law exponent $\alpha _R$ when $R$ is fitted by $R\propto t^{\alpha_R}$ at every point of its evolution.
        The (b)lue curves correspond to case A$1_{1D}$ and the (g)reen ones to case A$2_{1D}$ of the linear regime (see text for details).
        The horizontal lines in panels (a), (b) and (c) guide the eye for slope $0$.
        Similarly, the horizontal dashed line in panel (d) indicates the value $\alpha_R = 0$.
        The results of panels (a), (b) and (c) are averaged over $48$ disorder realizations.
    }
    \label{fig:linear_case_m2_P_R_and_aR_1d_spreading}
\end{figure}

Our observations, on the spatial evolution of the DVD~\eqref{eq:DVD_definition_num} are confirmed in Fig.~\ref{fig:linear_case_m2_P_R_and_aR_1d_spreading}(a) and (b) where we plot the time evolution of the DVD's second moment $m_2^D(t)$ [see Fig.~\ref{fig:linear_case_m2_P_R_and_aR_1d_spreading}(a)] and participation number $P^D(t)$ [see Fig.~\ref{fig:linear_case_m2_P_R_and_aR_1d_spreading}(b)] of $\xi_l^D$~\eqref{eq:DVD_definition_num}.
Indeed, a clear saturation of the $m_2^D$ [panel (a) of Fig.~\ref{fig:linear_case_m2_P_R_and_aR_1d_spreading}] and the $P^D$ [panel (b) of Fig.~\ref{fig:linear_case_m2_P_R_and_aR_1d_spreading}] is seen, indicative of a halt of the DVD $\xi_l^D(t)$~\eqref{eq:dvd_1d_chaotic}.
Note that in Fig.~\ref{fig:linear_case_m2_P_R_and_aR_1d_spreading} the (b)lue curves corresponds to the case A$1_{1D}$ of Figs.~\ref{fig:linear_case_m2_P_mle_1d_spreading} and~\ref{fig:linear_case_norm_dvd_and_snapshots_1d_spreading}, while the (g)reen one (Case A$2_{1D}$) is the same condition but this time obtained with $W = 3$ in order to stress the general nature of these findings.
The curves represent averages over $48$ disorder realizations in order to obtain a robust representation of the behavior of the tangent dynamics in case of regular motion (we do this because of the stochasticity of the random disorder parameters).

In~\citep{skokos2013nonequilibrium}, it was shown that the DVD oscillates inside the wave packet extent with increasing amplitudes.
In order to capture this dynamical behavior, in this work we also investigate the spatiotemporal evolution of the center $\overline{l}^D(t) = \sum_{l}l\xi_l^D(t)$ of the DVD~\eqref{eq:DVD_definition_num} by computing
\begin{equation}
    R(t) = \max _{[0, t]} \left\{\overline{l}^D(t)\right\} - \min _{[0, t]} \left\{\overline{l}^D(t)\right\},
    \label{eq:range_covered_by_DVD}
\end{equation}
which measures the distance between the two extreme positions reached by $\overline{l}^D(t)$ at time $t>0$.
Thus, $R(t)$~\eqref{eq:range_covered_by_DVD} measures the total size of the region visited by the DVD's center along its evolution.
If the $\overline{l}^D$ is fixed in time then $\max_{\left[0, t\right]} \left\{\overline{l}^D(t)\right\} = \min_{\left[0, t\right]} \left\{\overline{l}^D(t)\right\}$, so that 
\begin{equation}
    R(t) = 0.
    \label{eq:fix_dvd_center_01}
\end{equation}
This situation is highly improbable as small oscillations of the DVD center can appear due to numerical fluctuations.
In the present case, we say that whether the $R(t)$ has a tendency to saturate to values smaller or comparable to the localization volume $\overline{V}$~\eqref{eq:localization_volume_length_chap_anderson} i.e.
\begin{equation}
    R \lesssim \overline{V}, \quad R(t) \sim t^{0},
    \label{eq:R_fix_1d_spreading}
\end{equation}  
we conclude that the DVD center $\overline{l}^D(t)$ is practically fixed in time.
Notice that Eq.~\eqref{eq:R_fix_1d_spreading} implies that no motion of the DVD center outside the initially excited region is observed, leading to no thermalization of more lattice's sites (or modes). 
On the other hand, if
\begin{equation}
    R(t) > \overline{V},
\end{equation}
after a certain time $t$, then we can distinguish between two cases.
(i) A saturation of the values of $R$ take place i.e. $R(t) \sim t^{0}$, or (ii) we observe a consistent monotonic increasing tendency of $R(t)$ such that after a certain time $R\gg \overline{V}$.
In the case (i) the DVD visited the neighboring modes of the initially excited one, therefore a wave packet expansion takes place in one of the regimes mentionned above and then slow down up to a complete halt after a certain amount of time, restoring the AL into the system.
This situation, corresponds to the predictions of~\citep{johansson2010kam,aubry2011kam,mulansky2011scaling,pikovsky2011scaling}.
On another hand, the case (ii) means that the DVD's center is visiting sites far from the initially excited ones.
This is the indication of an infinite spreading of the wave packet into the system and corresponds to situations metioned in~\citep{shepelyansky1993delocalization,pikovsky2008destruction,flach2010spreading,laptyeva2010crossover,laptyeva2012subdiffusion}.

Figure~\ref{fig:linear_case_m2_P_R_and_aR_1d_spreading}(c) shows the evolution of the $R(t)$~\eqref{eq:range_covered_by_DVD} of the center of the DVD $\overline{l}^D$ for the initial conditions A$1_{1D}$ [(b)lue curve] and A$2_{1D}$ [(g)reen curve] used above.
A clear tendency to saturate a value $R(t) \approx 3$ for the case A$1_{1D}$ and $R(t) \approx 5$ in case of A$2_{1D}$ is seen.
These values are much smaller than the localization volume $\overline{V}= 21$ and $37$ respectively for the values $W=4$ and $W=3$.
Further, we assumed that the behavior of $R(t)$ can be fitted with a functional form 
\begin{equation}
    R(t) \propto t^{\alpha_R},
    \label{eq:functional_form_R_1d}    
\end{equation}
and plotted the time evolution of the power law exponent $\alpha_R(t)$ in Fig.~\ref{fig:linear_case_m2_P_R_and_aR_1d_spreading}(d), where a clear tendency to converge to values $\alpha_R \approx 0$ is seen, supporting the fact that the center of the DVD [see white curve in Fig.~\ref{fig:linear_case_norm_dvd_and_snapshots_1d_spreading}(b)] remains practically fixed in Figs.~\ref{fig:linear_case_norm_dvd_and_snapshots_1d_spreading}(b) and (d).
Therefore, no modes' chaotic heating take place as expected in the linear regime of disordered systems.
Note that the value $\alpha_R \approx 0$ was calculated averaging the slopes $\alpha _R (t)$ of the cases A$1_{1D}$ and A$2_{1D}$ over the last decade of evolution. 
\subsection{\label{subsec:dvd_weak_1d_spreading}The weak chaos regime}
\begin{figure}[!htb]
    \centering 
    \includegraphics[width=0.7\textwidth]{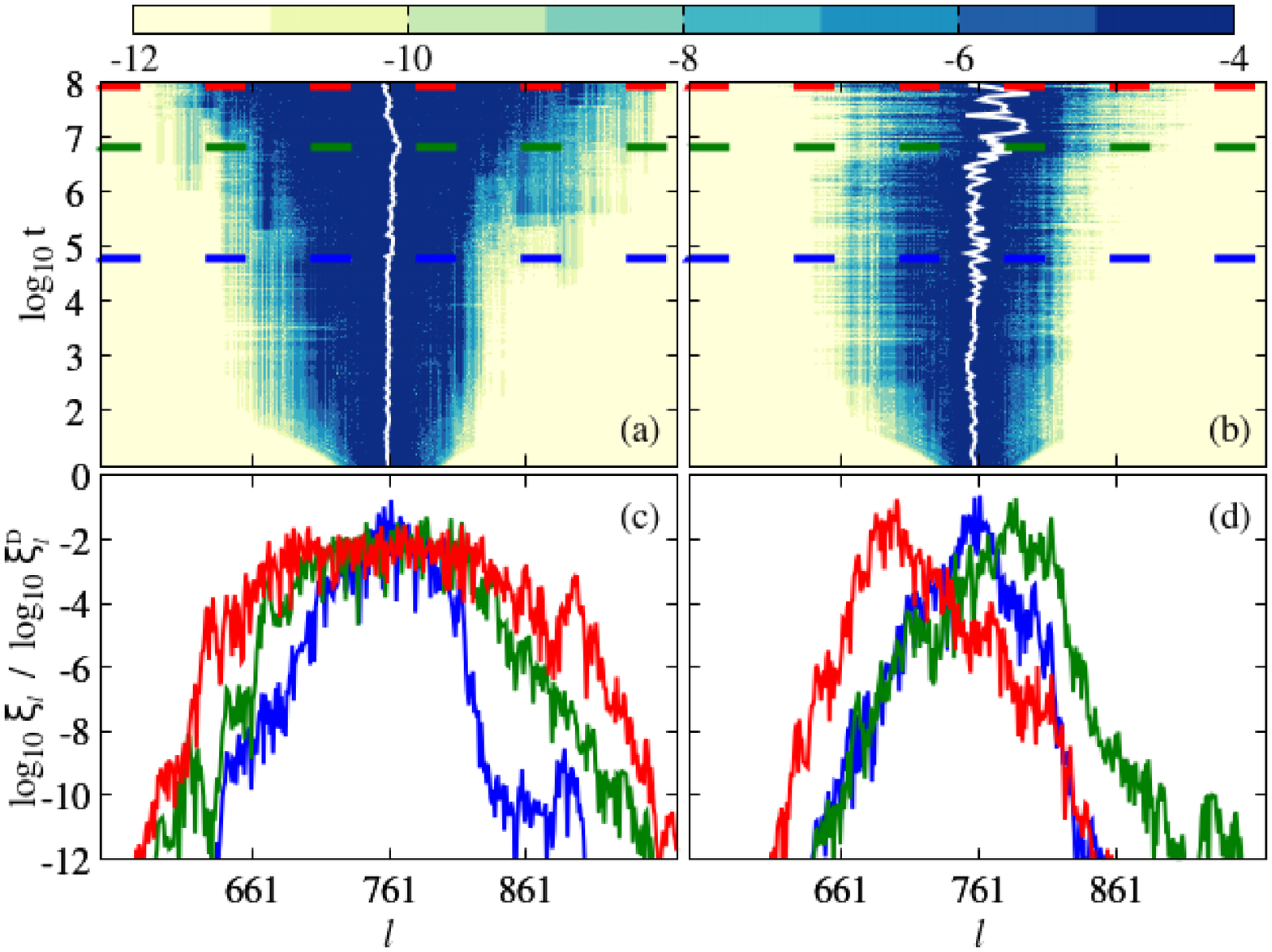}
    \caption{Spatiotemporal dependence of the norm density distribution $\xi_l$~\eqref{eq:norm_density_distribution_num} and DVD $\xi_l^D$~\eqref{eq:dvd_1d_chaotic} for the representative case W$4_{1D}$ (see text for details) of the weak chaos regime.
        The color scales at the top of the figure are used for coloring the lattice sites according to their $\log _{10} \xi_l$ [panel (a)] and $\log_{10}\xi_l^D$ [panel (b)] values.     
        In both panels, the white curves represent the motion of the distribution's center.
        The panels (c) and (d) show snapshots of the spatial dependence of the distributions $\xi_l$ and $\xi_l^D$ respectively at times $\log_{10} t \approx 4.8$ [(b)lue curves], $\log_{10} t \approx 6.82$ [(g)reen curves] and $\log_{10} t \approx 7.94$ [(r)ed curves]. 
        These times are also indicated using similar colors for the dashed horizontal lines in panels (a) and (b).
    }
    \label{fig:norm_and_dvd_distributoin_weak_1d}
\end{figure}

Let us now look at the dynamical behavior of the DVD~\eqref{eq:DVD_definition_num} in the case $\beta > 0$ for the 1D DDNLS model with Hamiltonian $\mathcal{H}_{1D}$~\eqref{eq:Hamilton_complex_dnls_1d}.
In Fig.~\ref{fig:norm_and_dvd_distributoin_weak_1d}(a), we sketch the time evolution of the norm density distribution $\xi _l$~\eqref{eq:norm_density_distribution_num} along with its related DVD $\xi_l^D$~\eqref{eq:dvd_1d_chaotic} in Fig.~\ref{fig:norm_and_dvd_distributoin_weak_1d}(b) for an individual realization of the weak chaos case $\text{W}4_\text{1D}$.
Snapshots of these distributions are presented at times $\log_{10}t \approx 4.8$ [(b)lue dashed horizontal line] $\log_{10} t \approx 6.82$ [(g)reen horizontal dashed line] and $\log_{10} t \approx 7.94$ [(r)ed dashed horizontal line] in Figs.~\ref{fig:norm_and_dvd_distributoin_weak_1d}(c) and~(d).

From Figs.~\ref{fig:norm_and_dvd_distributoin_weak_1d}(a) and (c), we see that the norm density $\xi _l$ is expanding smoothly and in a more or less symmetric way about the distribution center which remains (almost) fix at the position of the initial excitation [see white curve in Fig.~\ref{fig:normal_modes_localization_anderson_model}(a)].
On the other hand, the DVD $\xi_l^D$~\eqref{eq:dvd_1d_chaotic} in Figs.~\ref{fig:norm_and_dvd_distributoin_weak_1d}(b) and (d) retains a rather localized character, especially compared to the norm extent.
In addition, its centre $\overline{l}^D$ [see the white curve in Figs.~\ref{fig:norm_and_dvd_distributoin_weak_1d}(b) and (d)] is randomly wandering inside the lattice excited part as suggested by the fuzziness of the white curve in Fig.~\ref{fig:norm_and_dvd_distributoin_weak_1d}(b).
These random fluctuations increase in amplitude with time which becomes evident for $\log_{10} t \gtrsim 6$ where large excursions of the $\xi_l^D$ centre are clearly visible.

\begin{figure}[!htb]
    \centering 
    \includegraphics[width=0.4\textwidth, height=0.4\linewidth]{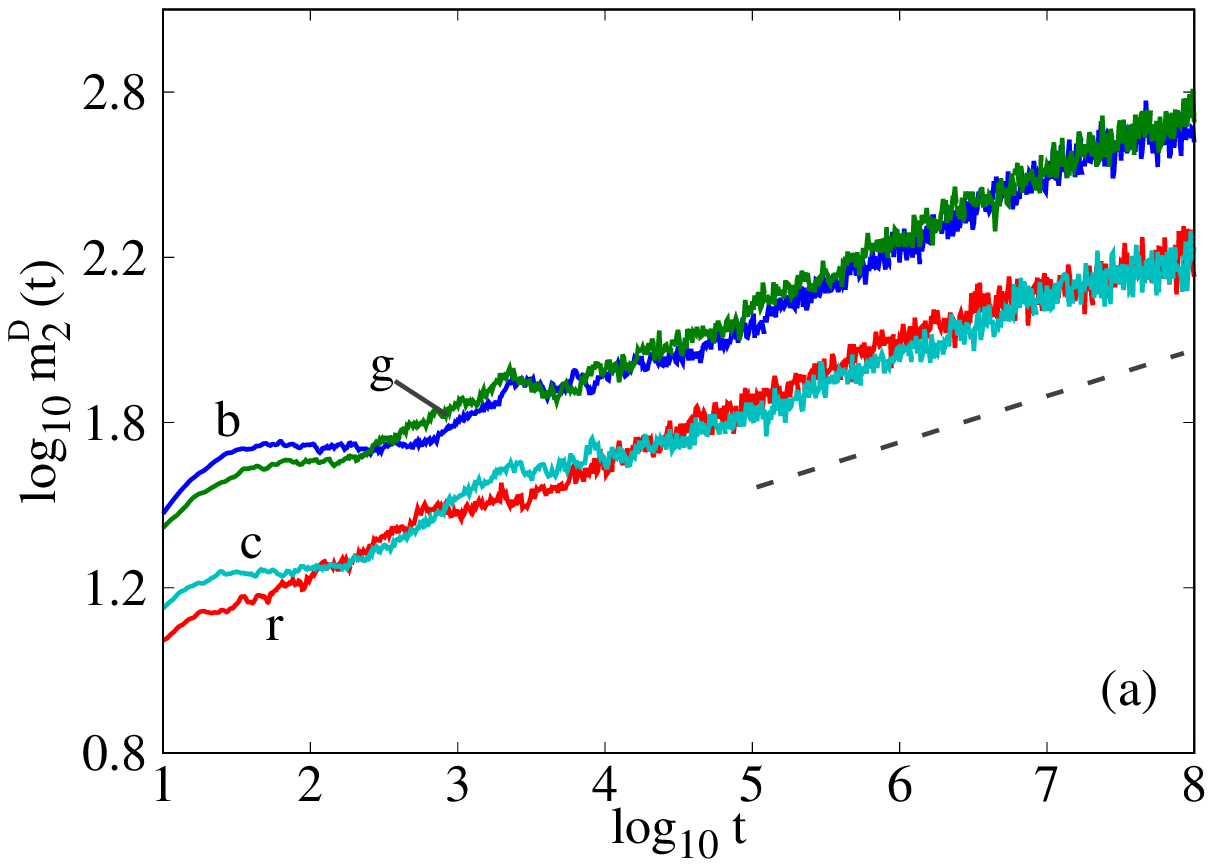}
    \includegraphics[width=0.4\textwidth, height=0.4\linewidth]{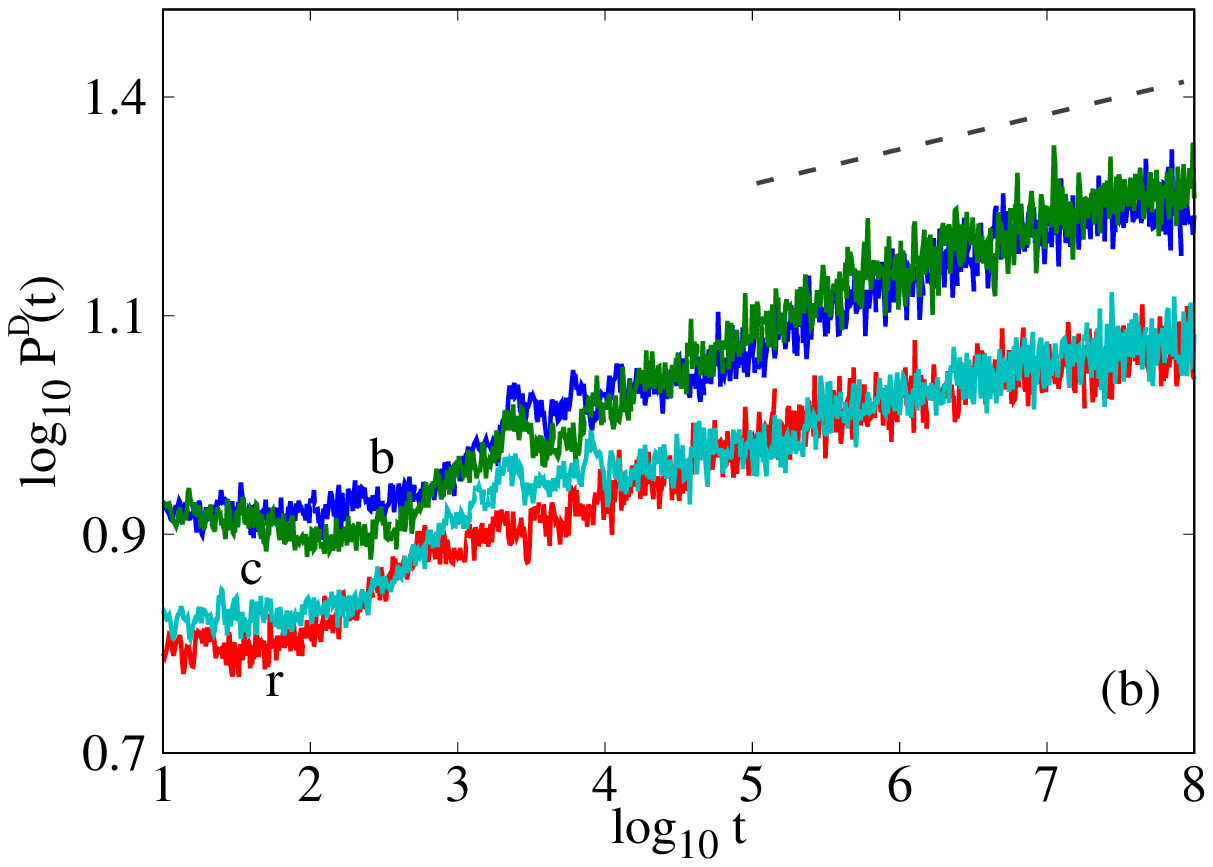}
    \includegraphics[width=0.4\textwidth, height=0.4\linewidth]{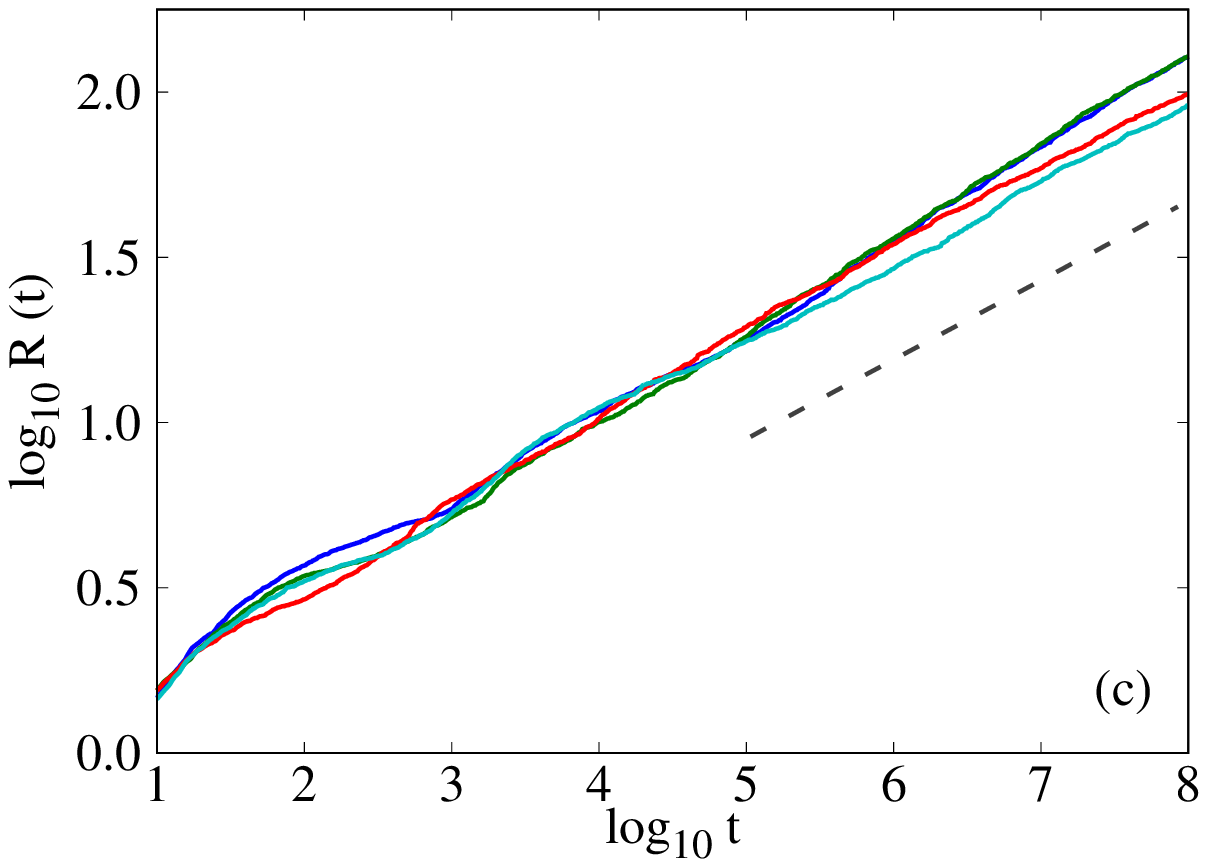}
    \includegraphics[width=0.4\textwidth, height=0.4\linewidth]{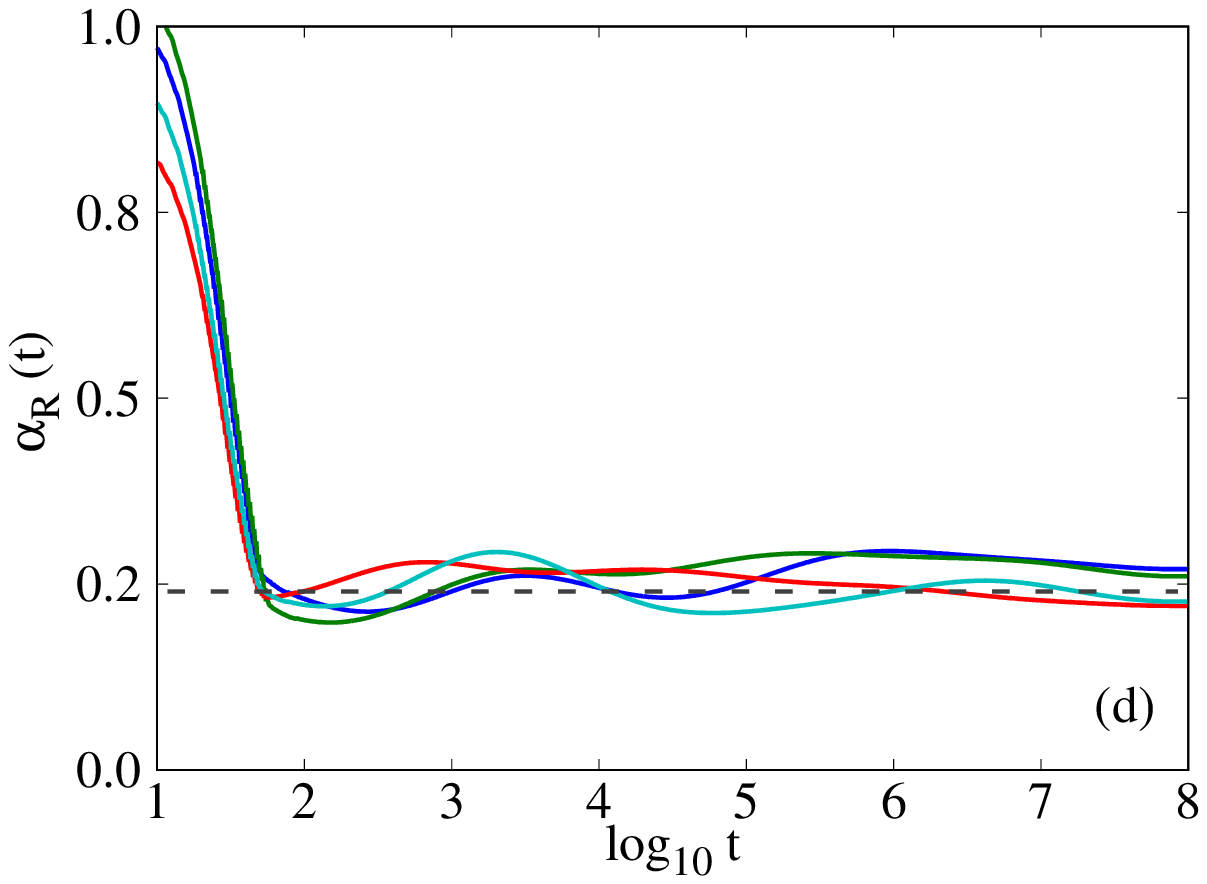}
    \caption{Results of the temporal dependence of the DVD's $\xi_l^D$~\eqref{eq:dvd_1d_chaotic} second moment $m_2^D$~\eqref{eq:second_moment_distr_num} in panel (a) and participation number $P^D$~\eqref{eq:participation_ratio_num} in panel (b).
        Panel (c) shows the region $R$~\eqref{eq:range_covered_by_DVD} visited by the distribution's center and in panel (d) we depict the power exponent $\alpha _R$ when the values of $R(t)$ are fitted with $t^{\alpha_R}$.
        All results corresponds to the weak chaos regime of the 1D DDNLS system~\eqref{eq:Hamilton_complex_dnls_1d}.
        The used initial wave packet parameters are for cases W$1_{1D}$ [(b)lue], W$2_{1D}$ [(g)reen], W$3_{1D}$ [(r)ed] and W$4_{1D}$ [(c)yan] (see text for details).
        The black lines in the panels (a), (b) and (c) guide the eye along the slopes $0.14$, $0.04$ and $0.24$.
        The horizontal line in panel (d) indicates the value $\alpha _R = 0.24$ obtained averaging the two last decades of the evolution of $\alpha _R(t)$ for all the cases.         
        The curves presented in panels (a), (b) and (c) show average results over $100$ disorder realizations.
    }
    \label{fig:dvd_m2_P_R_slopes_R_weak_chaos_1d}
\end{figure}

In order to validate our observations, we compute the evolution of the DVD's~\eqref{eq:DVD_definition_num} second moment $m_2^D(t)$~\eqref{eq:second_moment_distr_num} (see Fig.~\ref{fig:dvd_m2_P_R_slopes_R_weak_chaos_1d}(a)) and participation number $P^D(t)$~\eqref{eq:participation_ratio_num} (see Fig.~\ref{fig:dvd_m2_P_R_slopes_R_weak_chaos_1d}(b)) for all the weak chaos cases of Sec.~\ref{subsec:weak_chaos_1d}.
A growth of the $m_2^{D}(t)$ well fitted by the power law $m_2^D \propto t^{0.14}$ (see black line in Fig.~\ref{fig:dvd_m2_P_R_slopes_R_weak_chaos_1d}(a)) is observed smaller than the $m_2(t) \propto t^{1/3}$ in the case of the $\xi_l$~\eqref{eq:norm_density_distribution_num}.
In addition, the $P^D(t)$ slowly increases in all the studied cases following a timid power law $P^D \propto t^{0.04}$ in Fig.~\ref{fig:dvd_m2_P_R_slopes_R_weak_chaos_1d}(b) and reaching values similar to the mode's localization volume $\overline{V}$~\eqref{eq:localization_volume_length_chap_anderson} (in the cases W$1_{1D}$ [(b)lue] and W$2_{1D}$ [(g)reen] $P^D \approx 10^{1.3}\approx 20$) before showing a tendency to saturate at the last stages of the evolution in accordance to the localized character of the DVD observed in Figs.~\ref{fig:norm_and_dvd_distributoin_weak_1d}(b) and (d).

We also compute the size $R(t)$~\eqref{eq:range_covered_by_DVD} of the region visited by the centre of the DVD $\overline{l}^D(t) = \sum_l l \xi_l^D(t)$.
The result is depicted in Fig.~\ref{fig:dvd_m2_P_R_slopes_R_weak_chaos_1d}(c), where a monotonic increase of $R$ with time is seen which attains values $R(t)\approx 10^2$, much larger than the NM localization volume.
This increase is well fitted by an asymptotic power law $R \propto t^{\alpha_R}$ ($\alpha_R>0$) where the evolution of $\alpha_R (t)$ is shown in Fig.~\ref{fig:dvd_m2_P_R_slopes_R_weak_chaos_1d}(d).
There, all the curves of the power law exponent $\alpha_R(t)$ exhibit a tendency to saturate to an estimated exponent $\alpha _R \approx 0.24$ (we averaged the $\alpha_R(t)$ of all the studied cases in the two last decades of their evolutions to obtain this value).
This asymptotic value of the variations is reached at an early stage ($\log_{10}t \approx 2$) of the evolution and persists for the remaining $6$ decades of our numerical integration of the phase space and tangent space trajectories.
Consequently, the fluctuations of the chaotic seeds exist from near the beginning of the evolution and sustain their growth along with the wave packet expansion.
It is worth mentioning that the same results were found also for the 1D DKG model of anharmonic oscillators~\eqref{eq:hamilton_1d_dkg} in~\citep{senyange2018characteristics}.

\subsection{\label{subsec:dvd_strong_1d_spreading}The strong chaos regime}
\begin{figure}[!htb]
    \centering 
    \includegraphics[width=0.7\textwidth]{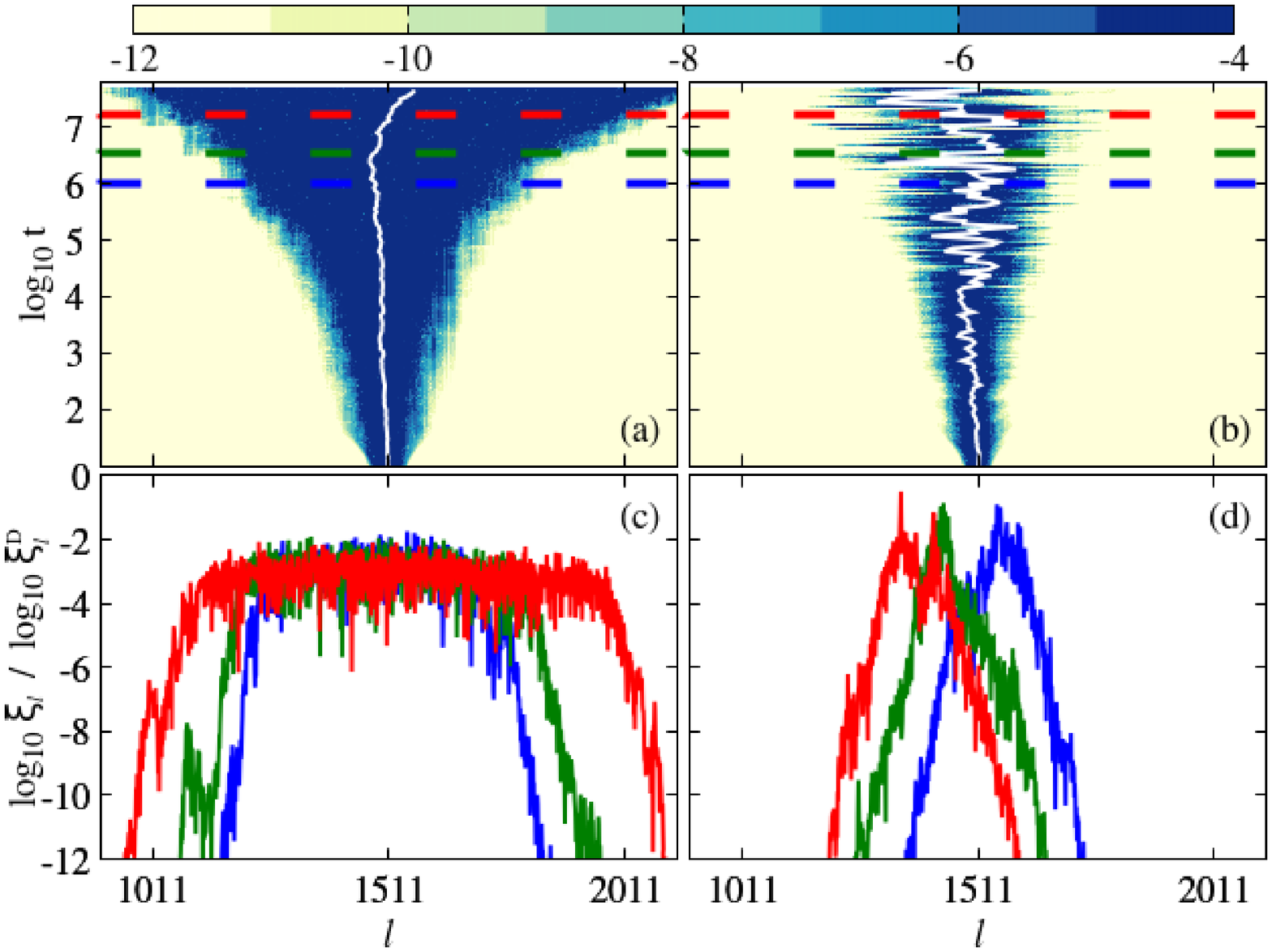}
    \caption{Spatiotemporal dependence of the norm density distribution $\xi_l$~\eqref{eq:norm_density_distribution_num} and DVD $\xi_l^D$~\eqref{eq:dvd_1d_chaotic} for the representative case S$3_{1D}$ (see text for details) of the strong chaos regime.
    The color scales at the top of the figure are used for coloring the lattice sites according to their $\log _{10} \xi_l$ [panel (a)] and $\log_{10}\xi_l^D$ [panel (b)] values.     
    In both panels, the white curves represent the motion of the center's distributions.
    The panels (c) and (d) show snapshots of the spatial dependence of the distributions $\xi_l$ and $\xi_l^D$ respectively at times $\log_{10} t \approx 6.0$ [(b)lue curves], $\log_{10} t \approx 6.53$ [(g)reen curves] and $\log_{10} t \approx 7.23$ [(r)ed curves]. 
    These times are also indicated using similar colors for the dashed horizontal lines in panels (a) and (b). 
    }
    \label{fig:norm_and_dvd_distributoin_strong_1d}
\end{figure}

We also investigate the dynamical behavior of the DVD $\xi_l^D$~\eqref{eq:dvd_1d_chaotic} in the strong chaos region of Fig.~\ref{fig:parameter_space_dnls_1d_delta_W} and~\ref{fig:x_y_thermal_selftrapping_phases}.
Here, we use a realization of the $\text{S}3_{\text{1D}}$ case as a representative of the strong chaos region dynamical behavior in Fig.~\ref{fig:norm_and_dvd_distributoin_strong_1d}.
We plot the evolution of the norm density $\xi _l$~\eqref{eq:norm_density_distribution_num} (see Fig.~\ref{fig:norm_and_dvd_distributoin_strong_1d}(a)), and its related DVD $\xi_l ^D$ (see Fig.~\ref{fig:norm_and_dvd_distributoin_strong_1d}(b)) along with snapshots of these distributions at specific times (see Figs.~\ref{fig:norm_and_dvd_distributoin_strong_1d}(c) and (d)) namely $\log_{10}t \approx 6$ [(b)lue horizontal lines], $\log_{10}t \approx 6.54$ [(g)reen horizontal lines] and $\log_{10}t \approx 7.24$ [(r)ed horizontal lines] in Figs.~\ref{fig:norm_and_dvd_distributoin_strong_1d}(a) and (b).
Similar to what was observed for the weak chaos case in Fig.~\ref{fig:norm_and_dvd_distributoin_weak_1d}, the norm distribution expands rather smoothly and symmetrically around its center [the white curve in Fig.~\ref{fig:norm_and_dvd_distributoin_strong_1d}(a)].
The $\xi _l$ spreads more widely compared to the weak chaos case in Fig.~\ref{fig:norm_and_dvd_distributoin_weak_1d}(a) as in this regime the wave packet spreads faster.
On the other hand, the $\xi _l^D$ exhibits a localized character [see Fig.~\ref{fig:norm_and_dvd_distributoin_strong_1d}(b)], with the distribution profile retaining a confined `pointy' shape [see Fig.~\ref{fig:norm_and_dvd_distributoin_strong_1d}(d)].
In addition, we see that the centre of $\xi_l^D$ randomly oscillates inside the wave packet width [see the fuzziness of the white curve of Fig.~\ref{fig:norm_and_dvd_distributoin_strong_1d}(b) and the profile positions in (d)] with increasing amplitudes, moving faster and wider than in the weak chaos regime [see Figs.~\ref{fig:norm_and_dvd_distributoin_weak_1d}(b) and (d)].

\begin{figure}[!htb]
    \centering 
    \includegraphics[width=0.4\textwidth, height=0.4\linewidth]{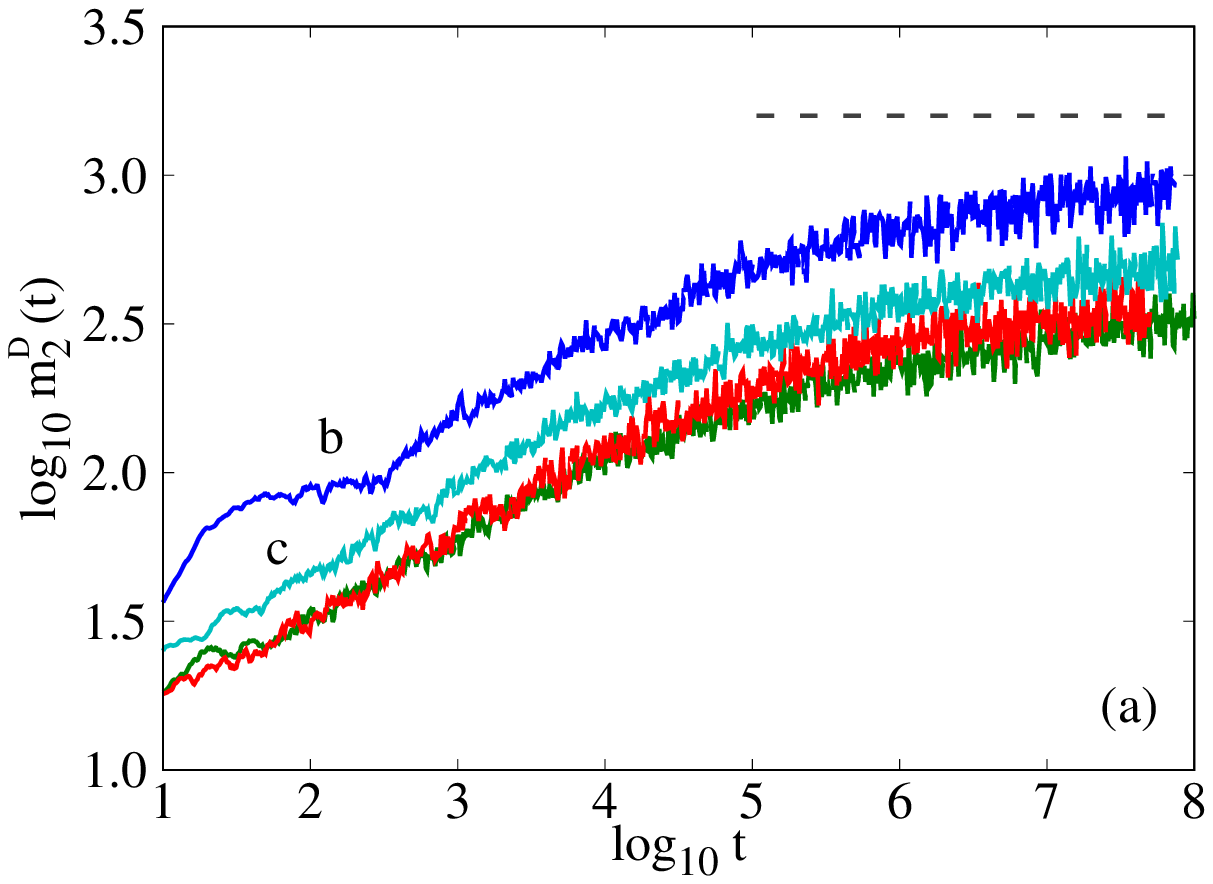}
    \includegraphics[width=0.4\textwidth, height=0.4\linewidth]{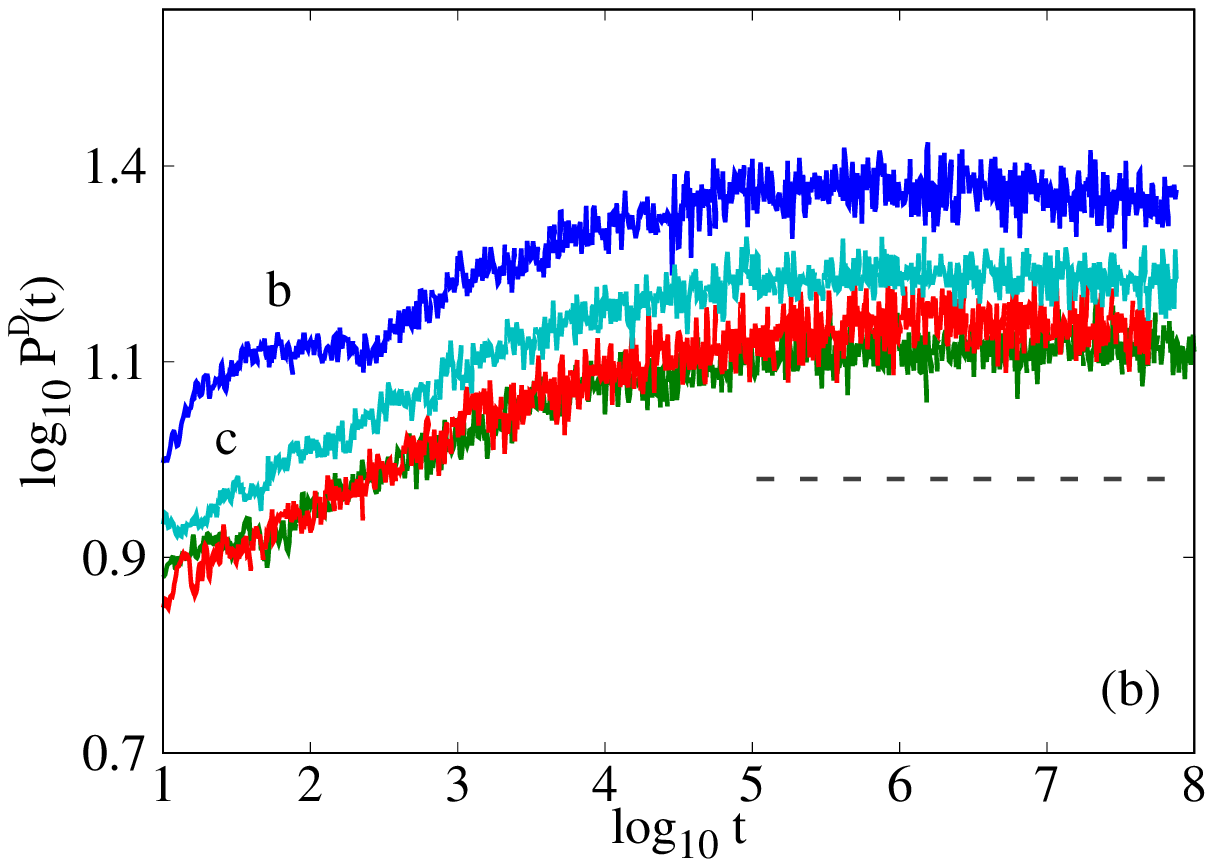}
    \includegraphics[width=0.4\textwidth, height=0.4\linewidth]{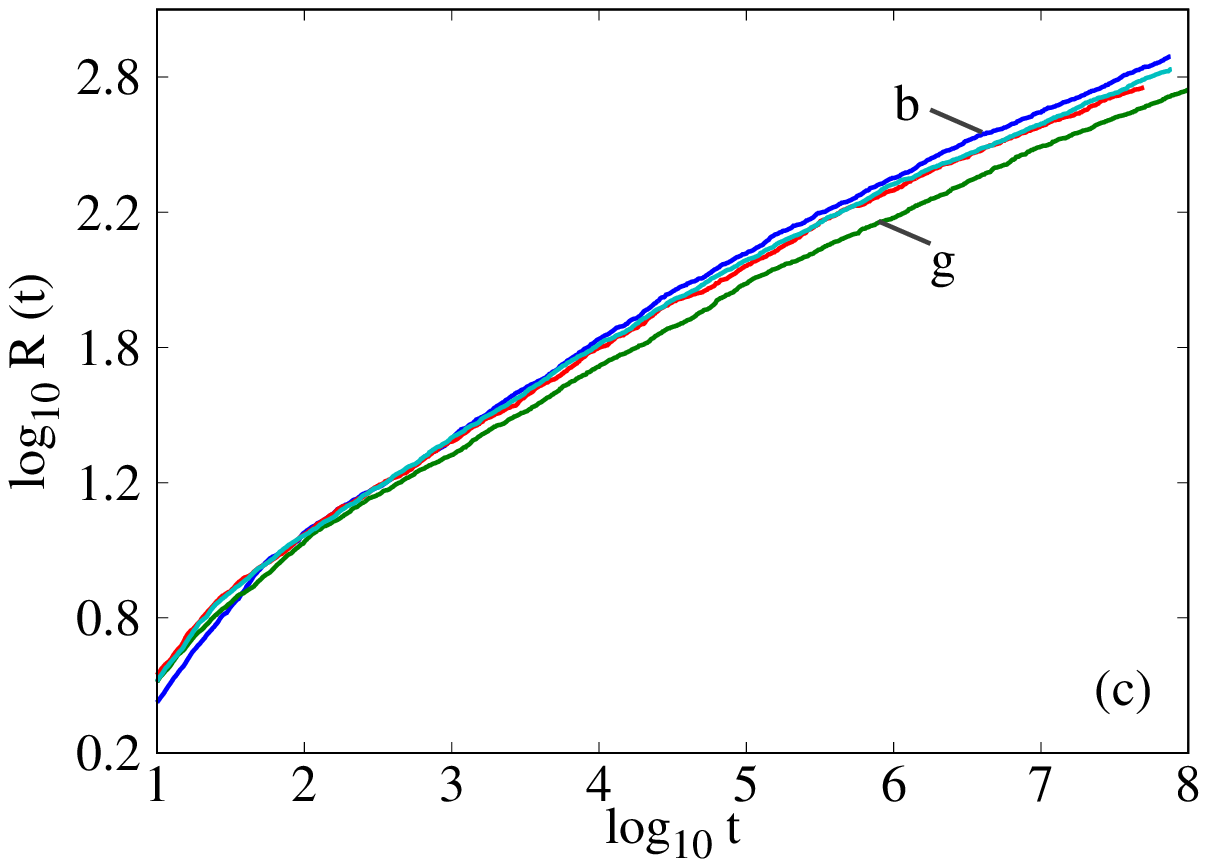}
    \includegraphics[width=0.4\textwidth, height=0.4\linewidth]{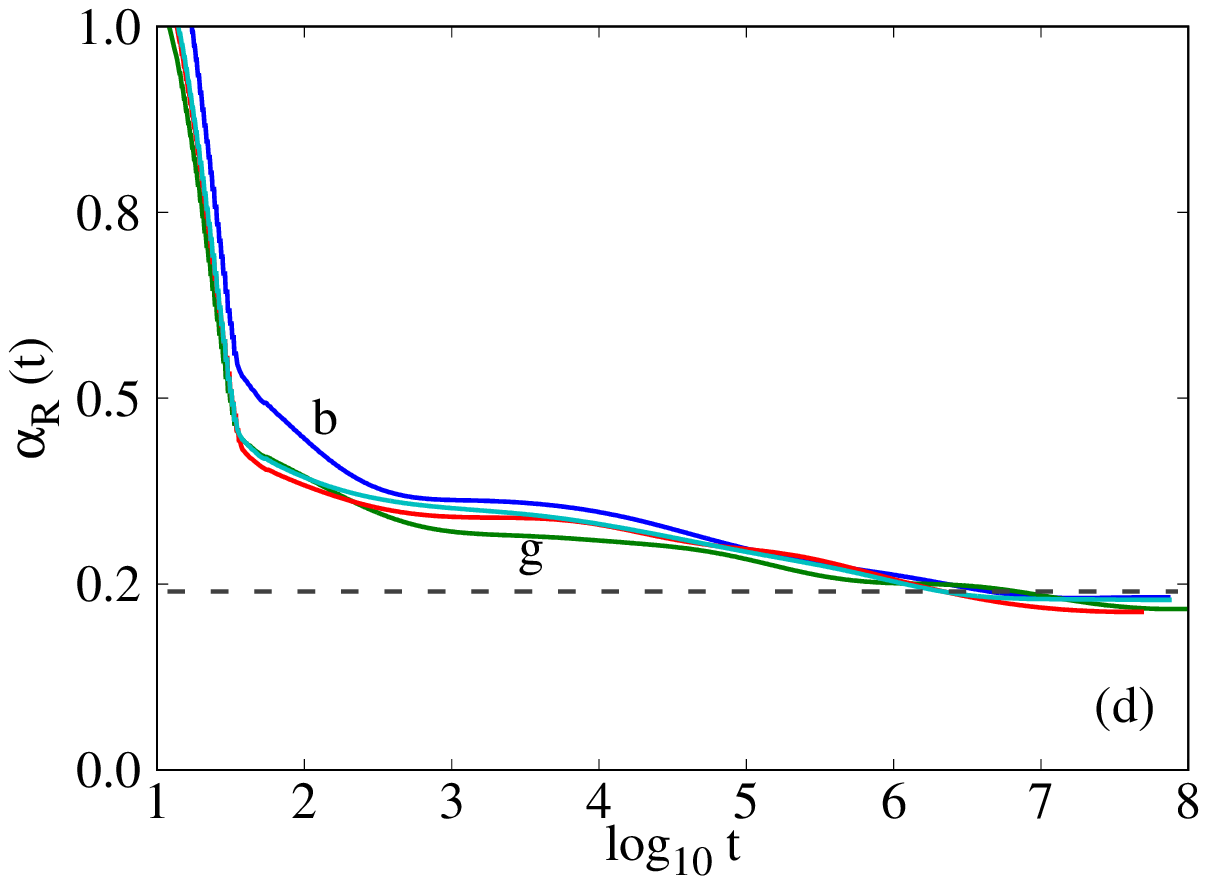}
    \caption{ Results of the time dependence of the second moment $m_2^D$~\eqref{eq:second_moment_distr_num} in panel (a) and the participation number $P^D$~\eqref{eq:participation_ratio_num} in panel (b) of the DVD $\xi_l^D$~\eqref{eq:dvd_1d_chaotic}.
        The region visited by the distribution's center $R$~\eqref{eq:range_covered_by_DVD} is in panel (c) and the power exponent $\alpha _R$ in panel (d) when the values of $R(t)$ are fitted with the relation $t^{\alpha_R}$ for the strong chaos regime.
        The used initial wave packet parameters are for cases S$1_{1D}$ [(b)lue], S$2_{1D}$ [(g)reen], S$3_{1D}$ [(r)ed] and S$4_{1D}$ [(c)yan] (see text for details).
        The black lines in the panels (a), (b) guide the eye for slope $0.0$.
        The horizontal dashed line in panel (d) indicates the value $\alpha _R = 0.24$.         
        The curves presented in panels (a), (b) and (c) show average results over $100$ disorder realizations.
    }
    \label{fig:dvd_m2_P_R_slopes_R_strong_chaos_1d}
\end{figure}

We quantify these observations by computing the DVD's $\xi_l^D$~\eqref{eq:dvd_1d_chaotic} second moment $m_2^D(t)$~\eqref{eq:second_moment_distr_num} [see panel (a) of Fig.~\ref{fig:dvd_m2_P_R_slopes_R_strong_chaos_1d}] and participation number $P^D(t)$ [see panel (b) of Fig.~\ref{fig:dvd_m2_P_R_slopes_R_strong_chaos_1d}] for all the strong chaos cases of Sec.~\ref{subsec:strong_chaos_1d_spreading}.
After a fast growth of the $m_2^D$ and $P^D$ values at the early stage of the evolution ($\log_{10} t \lesssim 5$), a tendency to saturate to a constant value is clearly visible for both quantities.
For instance, we have fitted the last decades of evolution of the $P^D(t)$ with a power law $P^D(t) \propto t^{a}$ and obtained $a \approx 0$. 
It is worth mentioning that the number of highly excited sites reached by the $\xi_l^D$ is $P^D \approx \overline{V}$ (for example W$1_{1D}$ gives $P^D \approx 10^{1.4} \approx 25$) which suggests that the DVD always remains confined in a single mode's localization volume at a time while moving around from site to site.
These results support the pointy and concentrated profile seen in Figs.\ref{fig:norm_and_dvd_distributoin_strong_1d}(b) and (d).

The evolution of the size $R(t)$~\eqref{eq:range_covered_by_DVD} of the region visited by the DVD center $\overline{l}^D$ is depicted in Fig.~\ref{fig:dvd_m2_P_R_slopes_R_strong_chaos_1d}(c). 
A clear monotonic increase is visible from the early stage of the evolution with $R(t)$ values reaching up to $R \approx 10^{2.8} \approx 700$ sites at the final time of the integration, several times larger than the average localization volume $\overline{V}$ of a NM. 
The dependence of $R(t)$ on time can be fitted with a power law $R \propto t^{\alpha_R}$ with $\alpha_R > 0$.
In Fig.~\ref{fig:dvd_m2_P_R_slopes_R_strong_chaos_1d}(d), the evolution of the power law exponent $\alpha _R(t)$ is shown.
We see that, in a large part of the dynamical evolution, the value of $\alpha _R(t)$ is always well above the ones observed for the weak chaos regime [black horizontal dashed line in Fig.~\ref{fig:dvd_m2_P_R_slopes_R_strong_chaos_1d} (d)].
Then, at later stages of the evolution ($\log_{10}t > 7$), the $\alpha_R(t)$ asymptotically converges to values $\alpha _R \approx 0.24$ seen in the weak chaos regime. 

This transitional behavior is well in agreement with previous works~\citep[see~e.g.][]{flach2010spreading,laptyeva2010crossover,skokos2010spreading,bodyfelt2011nonlinear}.
Indeed, the fact that the $R$ values reached in the weak chaos regime ($R\approx 10^2$) are smaller than the ones observed in the strong chaos case ($R\approx 10^3$) at the final time of our numerical integration is likely explained by the fact that the faster spreading in the strong chaos regime implies faster and wider wandering of chaotic seeds as a more rapid randomization and thermalization are required.
In addition, because the strong chaos is a transient regime, an asymptotic crossover of the wave packet dynamics to the weak chaos regime should take place.
This is what we see in Fig.~\ref{fig:dvd_m2_P_R_slopes_R_strong_chaos_1d}(d) for example since the rate of growth of $R(t) \propto t^{\alpha_R}$ i.e. $\alpha_R$ is always larger for the strong chaos regime and converge to the $\alpha_R$ values seen in the weak chaos dynamical behavior at later stages.
This numerically demonstrates that the nature of the chaotic dynamics cross over well before the wave packet dynamics itself for which up to now a complete crossover of a strong chaos initial condition has never been reported in the literature~\citep[see~e.g.][]{laptyeva2010crossover}.
It is worth noticing that the same behavior of the DVD in the strong chaos regime was retrieved also for the 1D DKG model in~\citep{senyange2018characteristics}. 

\section{\label{sec:discussion_1d_spreading} Discussion and summary}
Let us now discuss the main implications of our results in this chapter.
%
The outputs of the computation of the ftMLE  $\Lambda \propto t^{-0.25}$ for the weak and $\Lambda \propto t^{-0.3}$ for the strong chaos spreading regimes are quite interesting.
In studies of finite size multidimensional dynamical systems the chaotic dynamics show a plateau of the $\Lambda(t)$ values for chaotic trajectories (see e.g.~\citep{benettin1980lyapunov,mithun2018weakly,benettin2018fermi,hillebrand2019heterogeneity,mithun2018weakly,hillebrand2020chaotic}), while quasi-periodic trajectories reveal a $\Lambda(t) \propto t^{-1}$ tendency~\citep{benettin1980lyapunov,skokos2010lyapunov}.
The results we are obtaining here can be interpreted in the following way.
As the constant total norm $\mathcal{S}_{1D}$~\eqref{eq:norm_density_distribution_num} (energy $\mathcal{H}_{1K}$~\eqref{eq:hamilton_1d_dkg} for the 1D DKG model) is shared among more and more thermalized sites as additional degrees of freedom get activated, the average norm per site $s$~\eqref{eq:rescaled_densities_stat} of the excited part decreases in time.
Therefore both the single oscillator nonlinear frequency shift $\delta \propto \beta s$ [used for example in the the Chirikov criterion~\ref{eq:chirikov_criterion}] and the {\it effective} nonlinear strength $\propto \beta s^2$ [present for example in the equations of motion~\eqref{eq:eq_mot_1d_ddnls} and~\eqref{eq:evol_transmission_flach}] decay in time, taking the system closer to the integrable limit.
While one may expect a cross over to quasi-periodic (regular) motion as speculated in~\citep{johansson2010kam,aubry2011kam}, this is never the case, as the exponents of the power law $\Lambda \propto t^{\alpha_\Lambda}$ with $\alpha_\Lambda < 0$ are $\alpha _{\Lambda} = -0.25$ (weak chaos) and $\alpha _{\Lambda} = -0.3$ (strong chaos) which are different from the value $\alpha _\Lambda = -1$ seen in the case of regular motion [see for example Fig.~\ref{fig:linear_case_m2_P_mle_1d_spreading}(b)]. 
Thus, our results suggest that the system always remains chaotic although its chaoticity decreases in time.

Further, the fact that the exponent of the strong chaos regime $\alpha_\Lambda = -0.3$ is smaller than the one in the weak chaos case $\alpha_\Lambda = -0.25$ is explained by the fact that the wave packet spreads faster in the strong chaos regime where $m_2(t)\propto t^{1/2}$ is observed than in the weak chaos region where $m_2(t)\propto t^{1/3}$, so that the average norm per site $s$ within the wave packet excited part and the related effective nonlinearity $\beta s^2$ of the system decrease faster in the strong chaos case.
Thus  orbits in the parameter spaces of the 1D DDNLS model~\eqref{eq:Hamilton_complex_dnls_1d} in the strong chaos region move faster toward the integrable limit than in the weak chaos region.
This means that the value of the exponent in the power law $\Lambda \propto t^{\alpha_\Lambda}$ in the strong chaos case is closer to the $-1$ value observed in the integrable limit compared to the weak chaos case (indeed we see that $-1<-0.3<-0.25$).

Since the wave packet chaoticity is decreasing, an important aspect is to know whether the wave packet chaotic dynamics will always be able to support the subdiffusive spreading.
In order to analyze that point, we compare the system's timescales.
This was first done by~\citep{skokos2013nonequilibrium} for the weak chaos regime of the 1D DKG model~\eqref{eq:hamilton_1d_dkg}.
Here, we extend this analysis to the weak and strong chaos regimes of the 1D DDNLS system~\eqref{eq:Hamilton_complex_dnls_1d}.

A characteristic timescale for the system's chaotic dynamics is the Lyapunov time~\eqref{eq:lyapunov_time_chap_03} 
\begin{equation}
    T_\Lambda = \frac{1}{\Lambda},
    \label{eq:lyapunov_time_1d}   
\end{equation}
which gives an estimation of the time it takes for a dynamical system to become chaotic.
On the other hand, if we assume that the wave packet dynamics is mainly characterized by its second moment $m_2$~\eqref{eq:second_moment_distr_num}, then 
\begin{equation}
    T_M \sim \frac{1}{D},
    \label{eq:spreading_m2_time_1d}
\end{equation}
describes a timescale above which the subdiffusive spreading effectively starts.
Note that 
\begin{equation}
    D \sim \frac{dm_2}{dt},
\end{equation}
as the behavior $m_2 \sim Dt$ is assumed by definition of the diffusion coefficient $D$ (see also Sec.~\ref{subsec:spreading_mechamism_1d_sigma_2}).
Alternatively, if we assume that it is the participation number which mainly measures the wave packet subdiffusive spreading, we could define in a similar way the spreading timescale
\begin{equation}
    T_P \sim \frac{1}{\dot{P}},
    \label{eq:spreading_p_time_1d}
\end{equation}
where $\dot{P} = dP/dt$.
In the case of the spreading regime mentioned here, we have $m_2 \sim t^{\alpha _m}$ and $P\sim t^{\alpha _m/2}$.
We also saw that $\Lambda \sim t^{\alpha_\Lambda}$.
Consequently,
\begin{equation}
    \frac{T_M}{T_L} \sim t^{1 + \alpha_\Lambda - \alpha _m}, \qquad \frac{T_P}{T_L} \sim t^{1 + \alpha _\Lambda -\alpha _m/2},
\end{equation}
which reduce to 
\begin{equation}
    \frac{T_M}{T_L} \sim t^{5/12}, \qquad \frac{T_P}{T_L} \sim t^{7/12},
\end{equation}
for the weak chaos spreading regime in which $\alpha _m = 1/3$, $\alpha _\Lambda = -0.25$ was found, and to 
\begin{equation}
    \frac{T_M}{T_L} \sim t^{1/5}, \qquad \frac{T_P}{T_L} \sim t^{9/20}
\end{equation}
in the strong chaos case where we obtained $\alpha _m = 1/2$ and $\alpha_\Lambda = -0.3$.
Consequently the Lyapunov time $T_L$~\eqref{eq:lyapunov_time_1d} always remains smaller than the spreading timescales defined by $T_M$~\eqref{eq:spreading_m2_time_1d} and $T_P$~\eqref{eq:spreading_p_time_1d}, so that the wave packet chaotization is always faster than its spreading to new lattice's sites.

The fact that the DV needed for the computation of the MLE is aligned to the most chaotic direction is helpful to visualize the motion of chaotic seeds.
We identified chaotic seeds as regions in the lattice which are highly sensitive to initial perturbations.
Because the region where the NMs overlap and exchange the energy in a deterministic chaotic manner is finite, we expected to see chaotic spots (seeds) which are localized in time.
In addition, these chaotic seeds randomly meander within the wave packet interior due to the presence of a stochastic force $f(t)$ in Eq.~\eqref{eq:difussion_rate_flach} acting on the modes when the system is within a thermalized state.
This process is often related to the homogenization of chaos and promote the wave packet thermalization of more lattice sites.

In order to test this hypothesis, we computed the DVD's $\xi_l^D$~\eqref{eq:DVD_definition_num} second moment $m_2^D$~\eqref{eq:second_moment_distr_num}, and participation number $P^D$~\eqref{eq:participation_ratio_num} for both the weak and strong chaos regimes.
Our numerical results suggest that the DVD remains localized within the wave packet with size of about a mode localization volume.
This saturation is reached faster in the strong chaos case compared to the weak chaos where increase of the DVD extent, following $m_2^D \propto t^{0.14}$ and $P^D \propto t^{0.04}$ was still seen up to the largest simulation time.  
In addition, the center of the DVD exhibits random fluctuations with increasing amplitudes inside the wave packet width.

We computed the size  of the region visited by the DVD center and saw that its extent monotonically increases with time, reaching values of $R\approx 10^3$ for the strong chaos case at the final time of integration ($\log_{10}t \approx 10^8$).
This is a clear indication that the DVD is visiting modes (and sites) far outside from the initially excited one.
A power law  $R \propto t^{0.24}$, was noticed for the weak chaos, while for the strong chaos a faster growth $R\propto t^{\alpha_R}$ is reported with $\alpha _R > 0.24$ for several decades, which asymptotically tends to the same $\alpha _R \approx 0.24$ values observed in the weak chaos regime.
This is related to the fact that strong chaos is a transient regime which asymptotically evolves to the weak chaos one.

All these results suggest that a non-zero chaotic strength coupled to random fluctuations of the chaotic seeds are the two ingredients for the subdiffusive spreading of the wave packet in the 1D DDNLS model.
Thus, as long as they are present in the system no crossover of the extended wave packet to regular dynamics is expected to take place.
This fact is general as it has been also observed for the 1D DKG model~\cite{senyange2018characteristics}.

\chapter{\label{chap:chaotic_two_dimensional}Chaotic dynamics of wave packet spreading in the two-dimensional disordered nonlinear Schr\"odinger equation}

\section{\label{sec:introduction_chaos_2d}Introduction}
In this chapter, we will investigate the characteristics of wave packet spreading in the weak and strong chaos regimes conjectured in Secs.~\ref{sec:dynamical_regimes_chirikov} and~\ref{subsec:spreading_mechamism_any_dimension_and_sigma} for the 2D DDNLS model~\eqref{eq:hamilton_complex_ddnls_2d}.
The main obstacle facing this study is related to the long-time numerical integration of the 2D DDNLS model~\eqref{eq:hamilton_complex_ddnls_2d} which is a computationally hard task.
This is seen for instance in the literature where up to date only a few works on the subject are available.
More specifically,~\citep{garcia2009delocalization} studied the wave packet spreading in the 2D DDNLS model~\eqref{eq:hamilton_complex_ddnls_2d} only for a single weak chaos case which requires smaller lattice sizes up to a final time of $t_f = 10^{6}$ time units, averaging over only $10$ disorder realizations.
Further~\citep{sales2018sub} studied the diffusion of localized excitations in a similar model of the 2D DDNLS system, with non-diagonal nonlinear terms, integrating the phase space trajectories up to $t_f = 10^{5}$, and averaging over few disorder realizations for each case.
On the other hand,~\citep{laptyeva2012subdiffusion} performed an analysis of over $400$ disorder realizations of the spreading in the weak and strong chaos regimes for the generalized 2D DKG system $\mathcal{H}_{2K}$~\eqref{eq:hamilton_2d_dkg} whose equations of motion are much easier to simulate numerically.
Thus the numerical investigation of wave packet spreading in the 2D DDNLS lattice~\eqref{eq:hamilton_complex_ddnls_2d} we consider in this thesis remains largely unexplored. 
The novelty of this work is two fold as we not only compute the characteristics of the wave packet spreading, but also we investigate the system's chaotic dynamics in detail, to the best of our knowledge for the first time.

The chapter is structured as follows: In Sec.~\ref{sec:other_numerical_2d} we present additional numerical considerations for the integration of the variational equations of the 2D lattice.
Section~\ref{sec:mle_2d_ddnls_spreading} is devoted to presenting the results of our computations of the MLE for several cases of localized excitations in the weak and strong chaos regimes.
In Sec.~\ref{sec:dvd_2d_ddnls_spreading} we investigate the local nature of chaos, analyzing the dynamical behavior of the DVD in every spreading regime.
The discussion of our findings and the summary of our work are presented in Sec.~\ref{sec:summary_2d_ddnls}.
The outputs of this chapter are heavily based on the research paper~\citep{manda2020chaotic}. 
\section{\label{sec:other_numerical_2d}Some aspects related to our numerical simulations}

\subsection{\label{subsec:dvd_2d_testing}Choice of the initial deviation vector}
\begin{figure}
    \centering
    \includegraphics[scale=0.5]{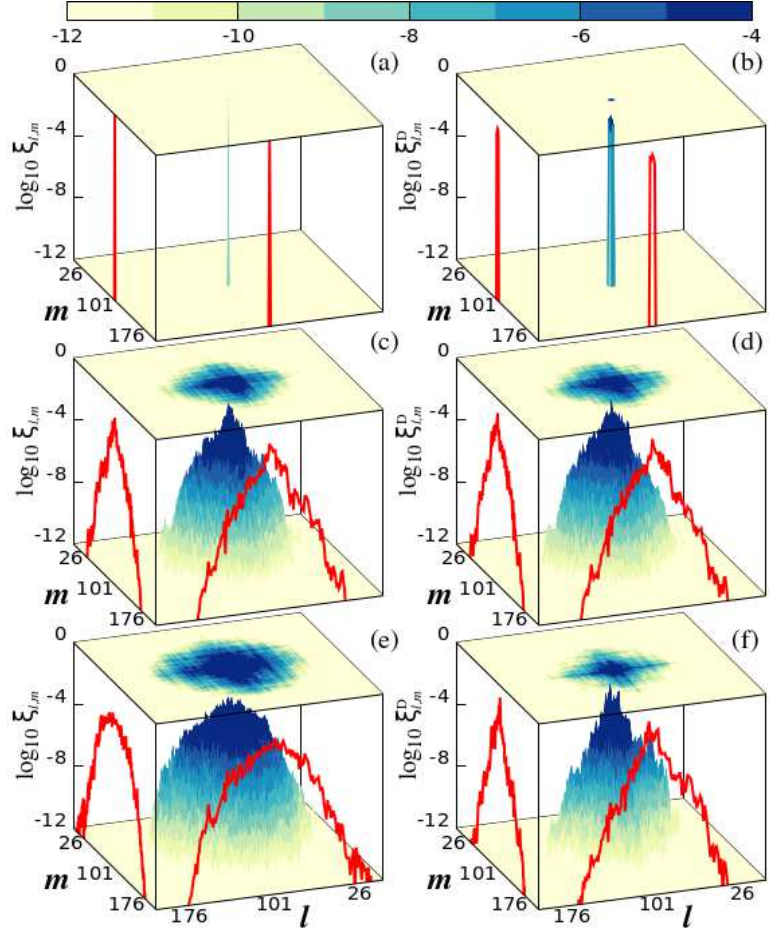}
    \caption{3D spatial distributions and the associated 2D color maps (upper sides) snapshots of the norm density $\xi_{l,m}$~\eqref{eq:norm_density_distribution_2d_chaotic} [panels (a), (c) and (e)] and the DVD $\xi_{l,m}^D$~\eqref{eq:dvd_2d_chaotic} [panels (b), (d) and (f)] profiles for a representative case of localized initial excitation in the 2D DDNLS lattice~\eqref{eq:hamilton_complex_ddnls_2d}. 
        We excite the central site of a $201\times 201$ lattice with a norm~\eqref{eq:norm_complex_dnls_2d} $\mathcal{S}_{2D} = 0.375$ at $W = 10$ and $\beta = 10$.
        The system's total energy is $\mathcal{H}_{2D} \approx 0.7$~\eqref{eq:hamilton_complex_ddnls_2d}.
        The initial condition for the variational equations is DV$_1$ (see text for details).
        Panels (a) and (b) correspond to the configurations of the $\xi_{l,m}$ and $\xi_{l,m}^D$ at time $\log_{10}t = 0$, while panels (c) and (d) at $\log_{10}t \approx 2.45$ and panels (e) and (f) at $\log_{10}t \approx 5.95$.
        The red curves on the sides are the distribution's projections along the $l$ and $m$ directions.  
        The coloring of the distributions is done with respect to the $\log_{10}\xi_{l,m}$ and $\log_{10}\xi_{l,m}^D$ values according to the horizontal color scales presented at the top of the figure. 
    }
    \label{fig:choice_of_dvd_snapshots_2d_a}
\end{figure}

\begin{figure}
    \centering
    \includegraphics[scale=0.5]{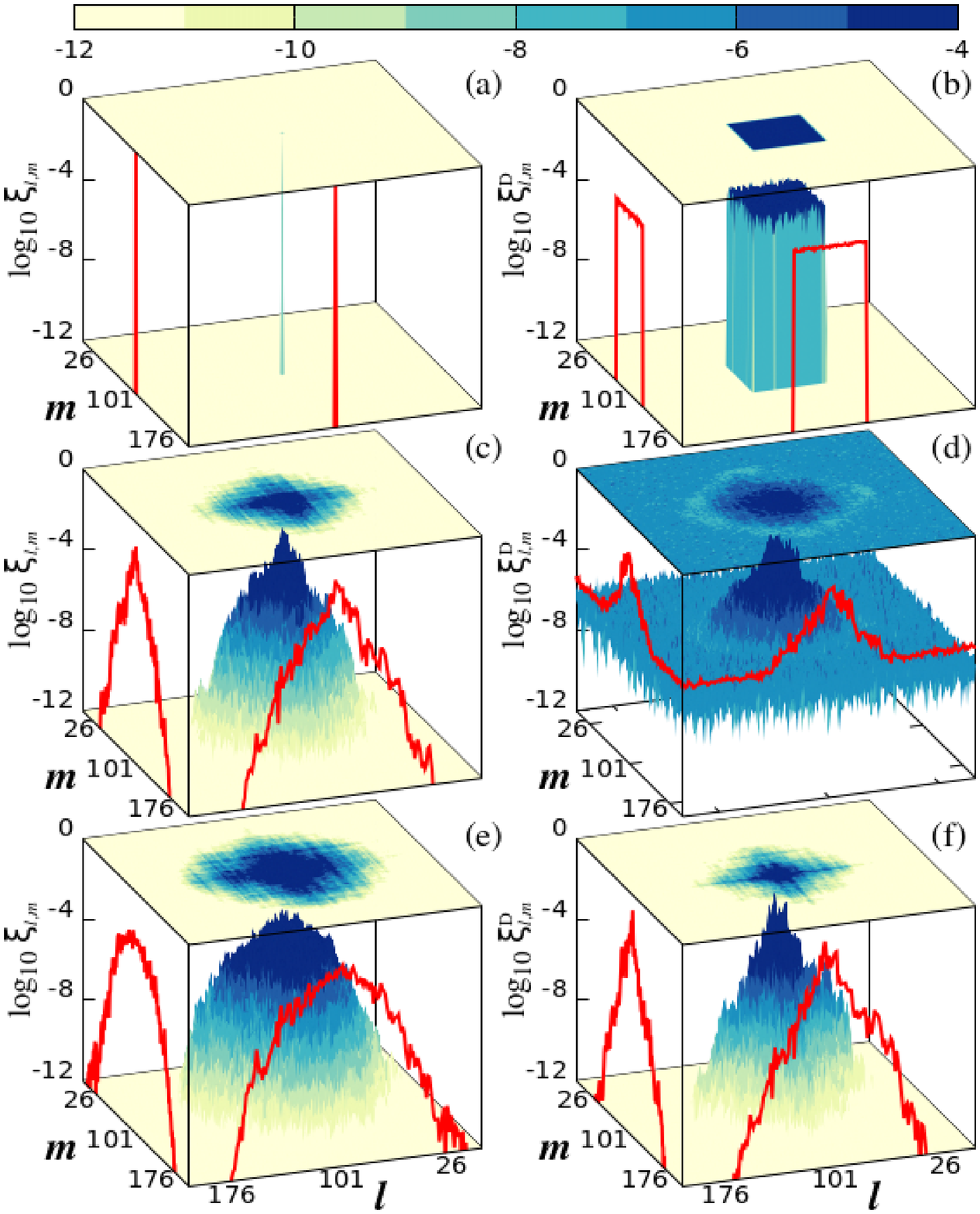}
    \caption{Similar to Fig.~\ref{fig:choice_of_dvd_snapshots_2d_a}, but for the initial condition DV$_2$ (see text for details) for the variational equations of the Hamiltonian function of the 2D DDNLS model $\mathcal{H}_{2D}$~\eqref{eq:hamilton_complex_ddnls_2d}.  
    }
    \label{fig:choice_of_dvd_snapshots_2d_b}
\end{figure}

\begin{figure}
    \centering
    \includegraphics[scale=0.5]{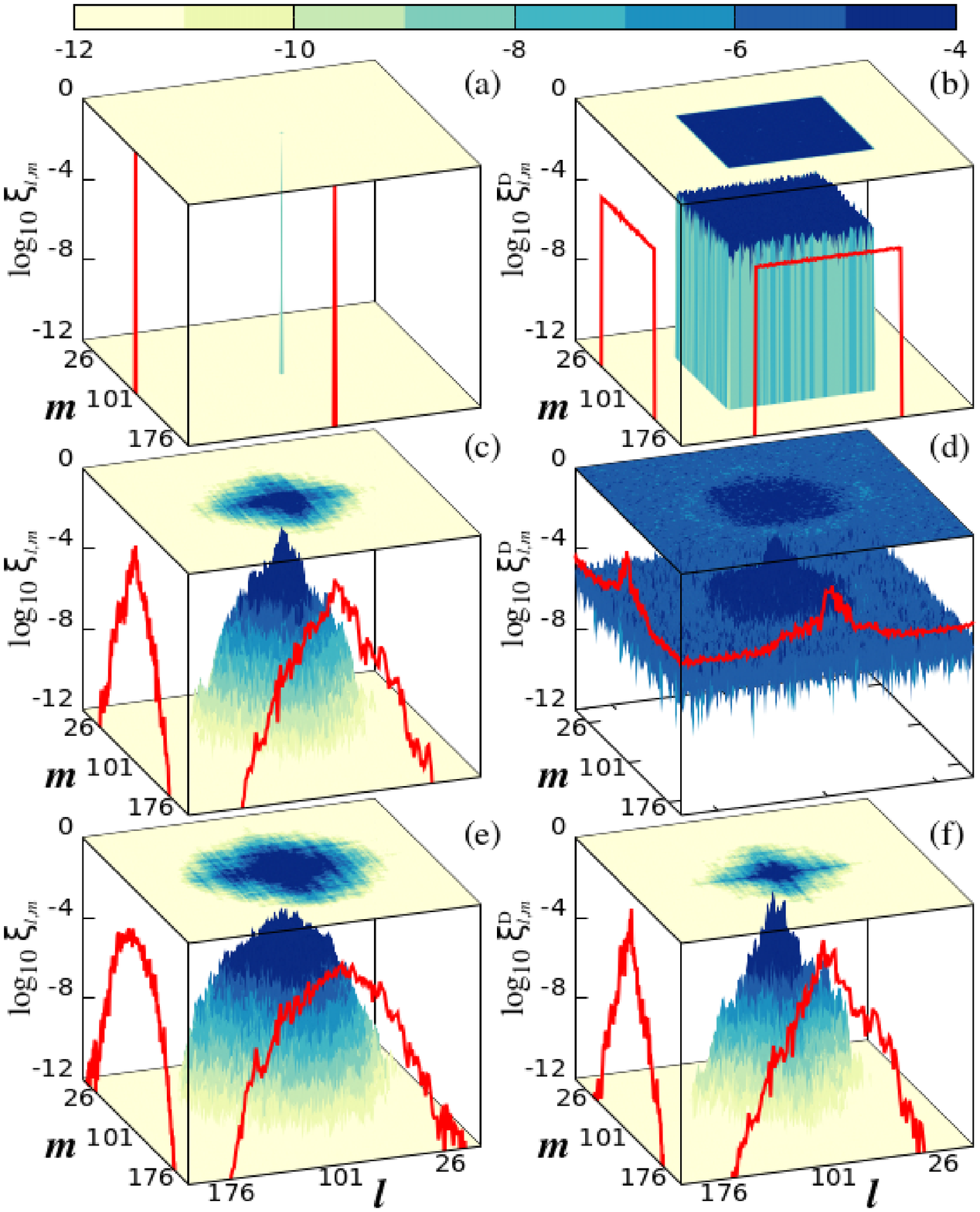}
    \caption{Similar to Fig.~\ref{fig:choice_of_dvd_snapshots_2d_a}, but for the initial condition DV$_3$ (see text for details) for the variational equations  of the Hamiltonian function of the 2D DDNLS model $\mathcal{H}_{2D}$~\eqref{eq:hamilton_complex_ddnls_2d}.}
    \label{fig:choice_of_dvd_snapshots_2d_c}
\end{figure}

\begin{figure}
    \centering
    \includegraphics[scale=0.5]{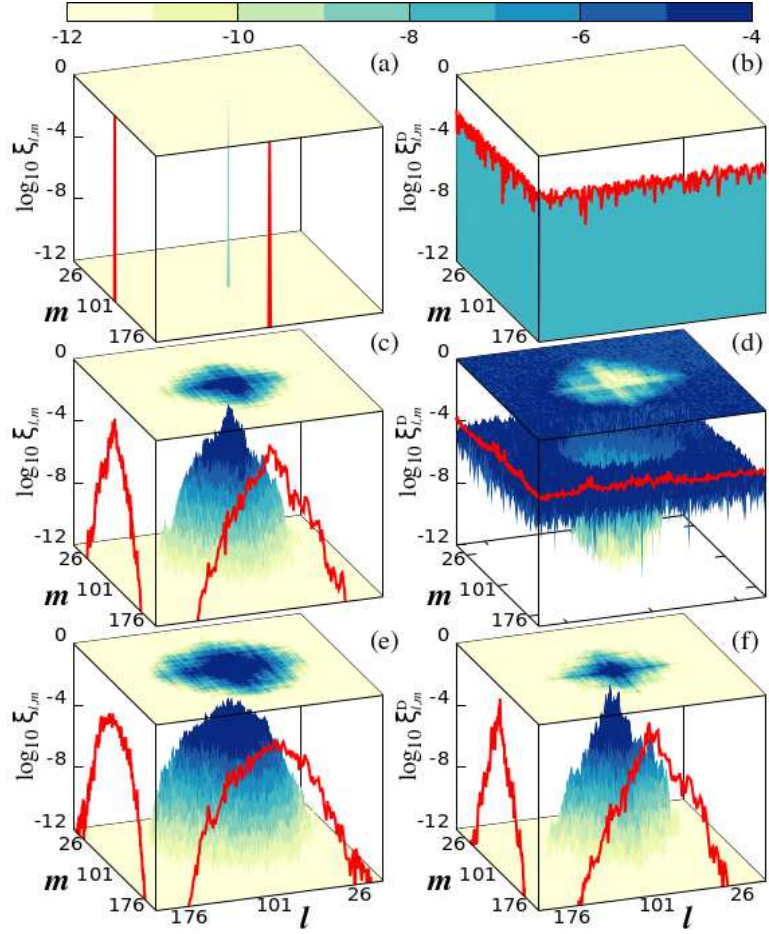}
    \caption{Similar to Fig.~\ref{fig:choice_of_dvd_snapshots_2d_a}, but for the initial condition DV$_4$ (see text for details) for the variational equations  of the Hamiltonian function of the 2D DDNLS model $\mathcal{H}_{2D}$~\eqref{eq:hamilton_complex_ddnls_2d}.
        Note that as only the last sites of the lattice's edges initially receive the $\xi_{l,m}^D$~\eqref{eq:dvd_2d_chaotic} values, it is visible only on the 3D heatmap of panel (b).
        The size of the DVD's excited region is too small compared to the lattice's dimension to be visible in the upper color map of panel (b).
    }
    \label{fig:choice_of_dvd_snapshots_2d_d}
\end{figure}

We start our investigation by first analyzing the dynamical behavior of the tangent dynamics of the 2D DDNLS model~\eqref{eq:hamilton_complex_ddnls_2d} and more specifically its dependence on the type of the used initial DV when a central localized excitation is set as initial state of the lattice.
To that end, we consider a typical initial condition consisting of exciting only the central site of a $201\times 201$ lattice with a total norm~\eqref{eq:norm_complex_dnls_2d} $\mathcal{S}_{2D} = 0.375$.
The total energy~\eqref{eq:hamilton_complex_ddnls_2d} is $\mathcal{H}_{2D} \approx 0.7$ for the particular realization of disorder we choose, with the other parameters set to $W = 10$ and $\beta = 10$.
We integrate the equations of motion and the variational equations of the system up to a final time $t_f = 10^6$, using the $s11\mathcal{ABC}6$ SI~\eqref{eq:s11odr6_general}. 
The integration time step $\tau = 0.24$ used, leads to relative energy~\eqref{eq:rel_energy_error_1dDDNLS} $E_r$ and norm~\eqref{eq:rel_norm_error_1dDDNLS} $S_r$ errors always remaining below $10^{-3}$.
For the initial condition of the variational equations, we implement the following choices of DVs: 
\begin{itemize}
    \item[] {\bf Case DV$_1$.} We give random values to the $\delta q_{l, m}$ and $\delta p_{l, m}$ coordinates of a square with size equal to about the average localization volume of a NM (see Table~\ref{tab:} below) i.e.~$L\times L = 5\times 5$ at the center of the lattice.
        The related spatial distribution of the DVD $\xi_{l,m}^D$~\eqref{eq:dvd_2d_chaotic} is shown in Fig.~\ref{fig:choice_of_dvd_snapshots_2d_a}(b). 
    \item[] {\bf Case DV$_2$.} Similar to the case of DV$_1$ but the size of the initially excited square is approximately $100$ times larger than the average localization volume of a NM i.e. $L \times L = 50\times 50$, leading to the spatial arrangement of the DVD $\xi_{l, m}^D$~\eqref{eq:dvd_2d_chaotic} seen in Fig.~\ref{fig:choice_of_dvd_snapshots_2d_b}(b). 
    \item[] {\bf Case DV$_3$.} Similar to case DV$_2$ but the size of the initially excited square is $4$ times larger, corresponding to a spatial distribution of the DVD $\xi_{l, m}^D$~\eqref{eq:dvd_2d_chaotic} in Fig.~\ref{fig:choice_of_dvd_snapshots_2d_c}(b).
    \item[] {\bf Case DV$_4$.} We randomly excite the $\delta q_{l, m}$ and $\delta p_{l, m}$ coordinates of only the edge layer of the 2D lattice, leading to the spatial arrangement of the DVD $\xi_{l,m}^D$~\eqref{eq:dvd_2d_chaotic} in Fig.~\ref{fig:choice_of_dvd_snapshots_2d_d}(b).      
\end{itemize}
The results of our computations for each considered DV are shown as follows: for DV$_1$ in Fig.~\ref{fig:choice_of_dvd_snapshots_2d_a}, DV$_2$ in Fig.~\ref{fig:choice_of_dvd_snapshots_2d_b}, DV$_3$ in Fig.~\ref{fig:choice_of_dvd_snapshots_2d_c} and DV$_4$ in Fig.~\ref{fig:choice_of_dvd_snapshots_2d_d}.
In each of these figures, panels (a), (c) and (e) present snapshots of the temporal evolution of the spatial extent of the norm density distribution~\eqref{eq:norm_density_distribution_num} 
\begin{equation}
    \xi_{l,m}(t) = \frac{s_{l, m}(t)}{\mathcal{S}_{2D}}, \quad s_{l, m}(t) = \frac{1}{2}\left(q_{l,m}^2(t) + p_{l,m}^2(t)\right),
    \label{eq:norm_density_distribution_2d_chaotic}
\end{equation} 
while panels (b), (d) and (f) show the related DVD~\eqref{eq:DVD_definition_num}
\begin{equation}
    \xi_{l,m}^D(t) = \frac{\delta q_{l,m}^2(t) + \delta p_{l,m}^2(t)}{\sum _{l,m}\delta q_{l,m}^2(t) + \delta p_{l,m}^2(t)},
    \label{eq:dvd_2d_chaotic}
\end{equation}
at times $\log_{10} t = 0$ [panels (a) and (b)], $\log_{10} t \approx 2.45$ [panels (c) and (d)] and $\log_{10} t \approx 5.95$ [panels (e) and (f)].
It is worth noticing that the $s_{l,m}$ in Eq.~\eqref{eq:norm_density_distribution_2d_chaotic} is the local norm at site $(l,m)$ of the 2D DDNLS lattice (as in Chap.~\ref{chap:chaotic_one_dimensional}, we will often use $s$ instead of $s_{l,m}$ when referring to its average value over the entire excited part of the lattice).
Evidently, the time evolution of the $\xi_{l,m}$~\eqref{eq:norm_density_distribution_2d_chaotic} is the same in all simulations [panels (a), (c) and (e)], indicating a chaotic wave packet whose extent increases in time, as this state belongs to the thermal region of the energy-norm parameter space of Fig.~\ref{fig:x_y_thermal_selftrapping_phases}.
On the other hand, as the initial DVs are different in each computation, we expect the tangent dynamics to depict qualitatively different outputs at least in the initial parts of the evolution of the variational equations.
\begin{figure}[!htbp]
    \centering 
    \includegraphics[width=0.5\textwidth, height=0.5\linewidth]{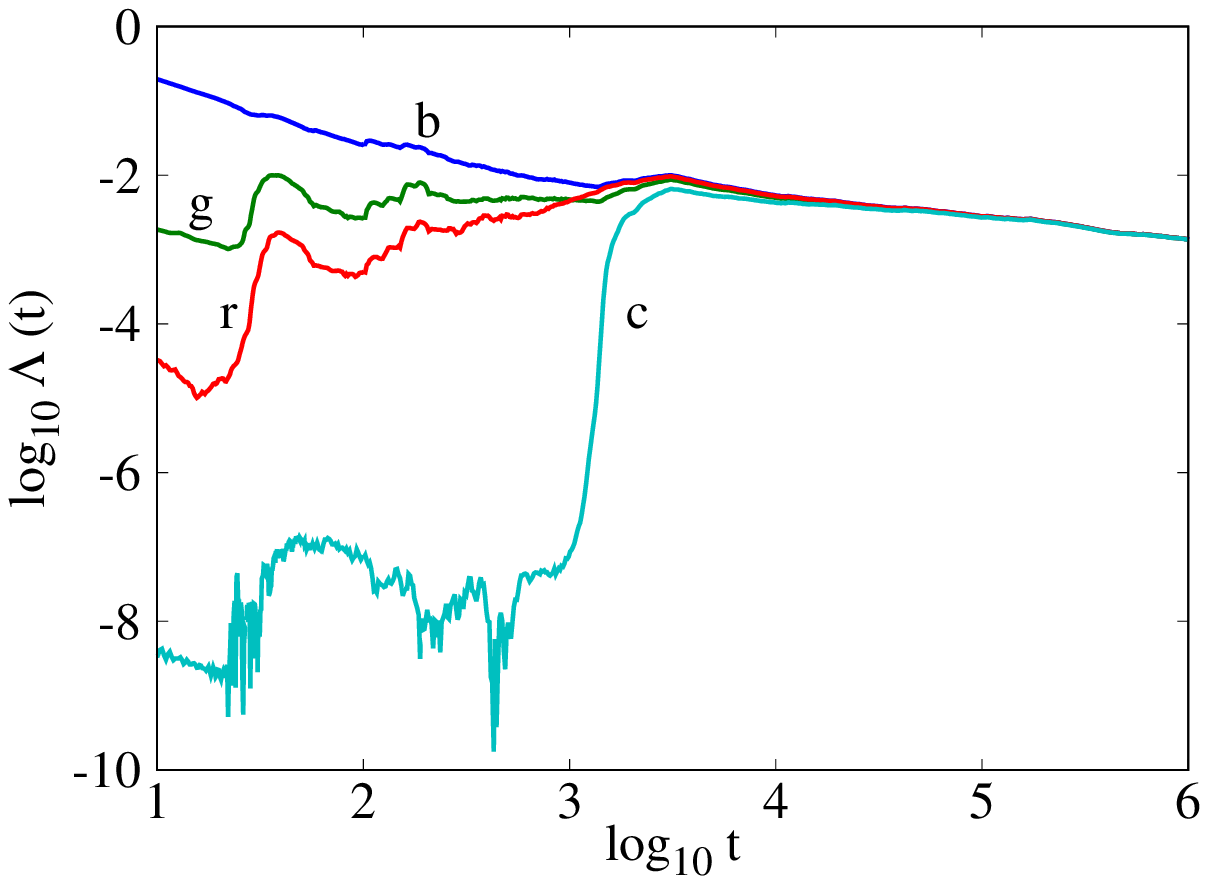}
    \caption{Time evolution of the ftMLE $\Lambda(t)$~\eqref{eq:finite_mle} for the representative initial condition of the 2D DDNLS lattice~\eqref{eq:hamilton_complex_ddnls_2d} used in Figs.~\ref{fig:choice_of_dvd_snapshots_2d_a},~\ref{fig:choice_of_dvd_snapshots_2d_b},~\ref{fig:choice_of_dvd_snapshots_2d_c} and~\ref{fig:choice_of_dvd_snapshots_2d_d}.
        The curves correspond to results obtained for the DV$_1$ [(b)lue], DV$_2$ [(g)reen], DV$_3$ [(r)ed] and DV$_4$ [(c)yan] (see text for details). 
    }
    \label{fig:choice_of_dvd_mle_chaos_2d_chaotic}
\end{figure}

The results of the computation of the DVD $\xi_{l,m}^D$~\eqref{eq:dvd_2d_chaotic} up to times around $\log_{10}t \approx 2.45$ in panels (d) of Figs.~\ref{fig:choice_of_dvd_snapshots_2d_a},~\ref{fig:choice_of_dvd_snapshots_2d_b},~\ref{fig:choice_of_dvd_snapshots_2d_c} and~\ref{fig:choice_of_dvd_snapshots_2d_d}, respectively for the cases DV$_1$, DV$_2$, DV$_3$ and DV$_4$ confirm our predictions, as we see a completely different behavior of the DVs in each case.
For the DV$_1$, a clear expansion of the DVD takes place with the distribution remaining well inside the wave packet extent [see panels (c) and (d) of Fig.~\ref{fig:choice_of_dvd_snapshots_2d_a}], where the deterministic chaotic dynamics is concentrated in the 2D lattice.
For the DV$_2$, the DVD $\xi_{l, m}^D$ spreads over the whole lattice with the unexcited background [from the point of view of $\xi_{l,m}$ in Fig.~\ref{fig:choice_of_dvd_snapshots_2d_b}(c)] having smaller values compared to the chaotic part of the lattice [panel (d) of Fig.~\ref{fig:choice_of_dvd_snapshots_2d_b}].
In the case of DV$_3$, a similar behavior to the one of DV$_2$ is observed.
However, a much stronger contribution of the unexcited background is seen, as the values of the $\xi_{l,m}^D$ there are much larger when compared to the DV$_2$ case.
The DV$_4$ choice depicts a completely different behavior to the three previous cases as only the unexcited part of the lattice possesses $\xi_{l, m}^D$ values during the transient phase of the dynamics.
As we did for the 1D DDNLS model~\eqref{eq:Hamilton_complex_dnls_1d} in Sec.~\ref{subsec:choice_of_init_dvd}, we correlate the dynamics of the DVD to the behavior of the evolution of the ftMLEs~\eqref{eq:finite_mle} plotted in Fig.~\ref{fig:choice_of_dvd_mle_chaos_2d_chaotic}.
Indeed, at short times, the DV$_1$ leads to the largest ftMLE [see (b)lue curve in Fig.~\ref{fig:choice_of_dvd_mle_chaos_2d_chaotic}], as its evolution is contained within the chaotic domain of the lattice from the earlier stages of the evolution.
On the other hand, the contribution of the wide unexcited backgrounds whose dynamics are regular in the cases DV$_2$ and DV$_3$, lowers the value of their ftMLEs compared to the DV$_1$ case, with the DV$_2$ displaying a larger ftMLE than DV$_3$ [respectively the (g)reen and (r)ed curves in Fig.~\ref{fig:choice_of_dvd_mle_chaos_2d_chaotic}]. 
With the DV$_4$ evolving solely on the initially unexcited part of the lattice, we expect its ftMLE values to be very small, at least compared to the other ftMLEs at the first stages of the dynamics, something which is clearly seen on the evolution of the (c)yan curve in Fig.~\ref{fig:choice_of_dvd_mle_chaos_2d_chaotic}.  

During the last two decades of the evolution of the ftMLEs in Fig.~\ref{fig:choice_of_dvd_mle_chaos_2d_chaotic}, all the DV cases present similar values for their ftMLEs, as a result of the fact that all DVs align to the same most chaotic direction related to the system's reference orbit, whose exponential rate of changes measures its MLE.  
Consequently, all the spatial distributions of the DVDs tend to be practically the same as shown in the panels (f) of Figs.~\ref{fig:choice_of_dvd_snapshots_2d_a},~\ref{fig:choice_of_dvd_snapshots_2d_b},~\ref{fig:choice_of_dvd_snapshots_2d_c} and~\ref{fig:choice_of_dvd_snapshots_2d_d} whose snapshots were taken at time $\log_{10}t \approx 5.95$.
Note that the most chaotic direction possesses coordinates $\delta q_{l, m}$ and $\delta p_{l, m}$ only at the lattice's excited sites where the chaotic dynamics of the system takes place.

Based on these numerical experiments, we conclude that initial DVs with random values of the $\delta q_{l, m}$ and $\delta p_{l, m}$ coordinates on a square with area smaller or of the order of the NM average localization volume in the region of the lattice's initially excited sites are good choices. 
Otherwise the DV dynamics has a tendency to initially spread over the unexcited part of the lattice and consequently converging much more slowly to the most chaotic direction of the system's computed trajectory.

\subsection{\label{subsec:dynamic_lattice_2d}The dynamic lattice expansion}
Let us now discuss how we implement the dynamic lattice technique of Sec.~\ref{subsec:dynamic_lattice_chaotic_dynamics_1d} in the 2D system, as a way of saving CPU time.
This is in contrast to using a {\it fixed} large $N\times M$ lattice from the beginning of the dynamical evolution of orbits, so that the lattice edges are practically not reached by the wave packet at the end of the integration.
In the dynamic lattice approach we start with a much smaller lattice around the initially excited sites and expand it whenever it is required by $\mathcal{N}$ sites at the edges of all lattice's sides.  
As for the 1D DDNLS model~\eqref{eq:Hamilton_complex_dnls_1d} this method requires a criterion which checks if the wave packet's front has hit the boundaries. 
For the 2D DDNLS system~\eqref{eq:hamilton_complex_ddnls_2d}, we use the criterion
\begin{equation}
    \max _{l \in \mathcal{D}} s_{l, m} > \delta _{\mathcal{D}},
    \label{eq:condition_expand_2d_chaotic}
\end{equation} 
where $\mathcal{D}$ represents a ribbon at the outermost layer of the lattice and $\delta_{\mathcal{D}}$ is a threshold number.
Notice that this condition is different to the one used in the 1D DDNLS system~\eqref{eq:condition_dynamic_latt_chaos_1d} 
\begin{equation}
    \frac{1}{\mathcal{D}}\sum_{l, m} s_{l,m} > \delta_\mathcal{D}.
    \label{eq:condition_expand_2d_chaotic_02}
\end{equation}
Nevertheless, we numerically checked that results obtained from the two conditions~\eqref{eq:condition_expand_2d_chaotic_02} and~\eqref{eq:condition_expand_2d_chaotic} are similar. 
In our study we prefer using the condition~\eqref{eq:condition_expand_2d_chaotic} as it can be easily applied in the case of 2D lattice. 
\begin{figure}[!htbp]
    \centering 
    \includegraphics[width=0.7\textwidth]{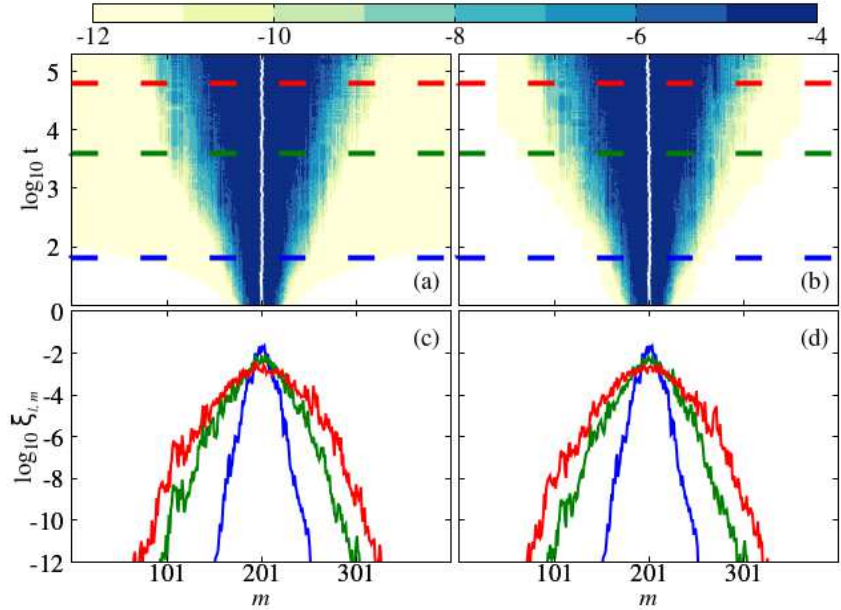}
    \caption{Spatiotemporal dependence of the projection of the norm density distribution $\xi_{l, m}$~\eqref{eq:norm_density_distribution_2d_chaotic} and its related DVD $\xi_{l, m}^D$~\eqref{eq:dvd_2d_chaotic} for $7\times 7$ central square excitation in a $401\times 401$ 2D DDNLS lattice~\eqref{eq:hamilton_complex_ddnls_2d}.
        Each excited site is given a norm $s = 0.72$ such that the total norm $\mathcal{S}_{2D} = 35.28$~\eqref{eq:norm_complex_dnls_2d} with $W = 10$ and $\beta = 1$.
        For the particular disorder realization, the total energy~\eqref{eq:hamilton_complex_ddnls_2d} is $\mathcal{H}_{2D} \approx - 4.714$.
        The color scales at the top of the figure are used for coloring the lattice sites according to their $\log _{10} \xi_{l, m}$ values.     
        In both panels, the white curves represent the motion of the distribution projections' centers.
        Panels (c) and (d) contain snapshots of the spatial dependence of the distribution's projections respectively at times $\log_{10} t \approx 1.8$ [(b)lue curves], $\log_{10} t \approx 3.6$ [(g)reen curves] and $\log_{10} t \approx 4.8$ [(r)ed curves]. 
        These times are also indicated using the same colors for the dashed horizontal lines in panels (a) and (b).
    }
    \label{fig:fix_dyn_2d_chaotic_norm_evol_01}
\end{figure}

The results of a numerical comparison between the fixed and dynamic lattice simulations are shown in Fig.~\ref{fig:fix_dyn_2d_chaotic_norm_evol_01}.
The initial excitation considered there corresponds to all sites in a square of size $L\times L = 7\times 7$ at the center of a $401\times401$ lattice (the center of the square coincides with the lattice center) having the same norm per site $s_{l, m} = 0.72$.
This means that the total norm of the system~\eqref{eq:norm_complex_dnls_2d} is $\mathcal{S}_{2D} = 35.28$, while the value of the Hamiltonian~\eqref{eq:hamilton_complex_ddnls_2d} is $\mathcal{H}_{2D} \approx -4.174$ for the disorder realization we choose, corresponding to a thermal state in the energy-norm parameter space in Fig.~\ref{fig:x_y_thermal_selftrapping_phases} of the system.
In addition, we set $W = 10$ and $\beta = 1$.
We integrate the system using the $s11\mathcal{ABC}6$ SI~\eqref{eq:s11odr6_general} of order $6$ with an integration time step $\tau = 0.135$ which kept the relative energy $E_r$~\eqref{eq:rel_energy_error_1dDDNLS} and norm $S_r$~\eqref{eq:rel_norm_error_1dDDNLS} errors below $10^{-4}$ up to the final time of integration $t_f\approx 10^{5}$.
\begin{figure}[!htbp]
    \centering 
    \includegraphics[width=0.49\textwidth]{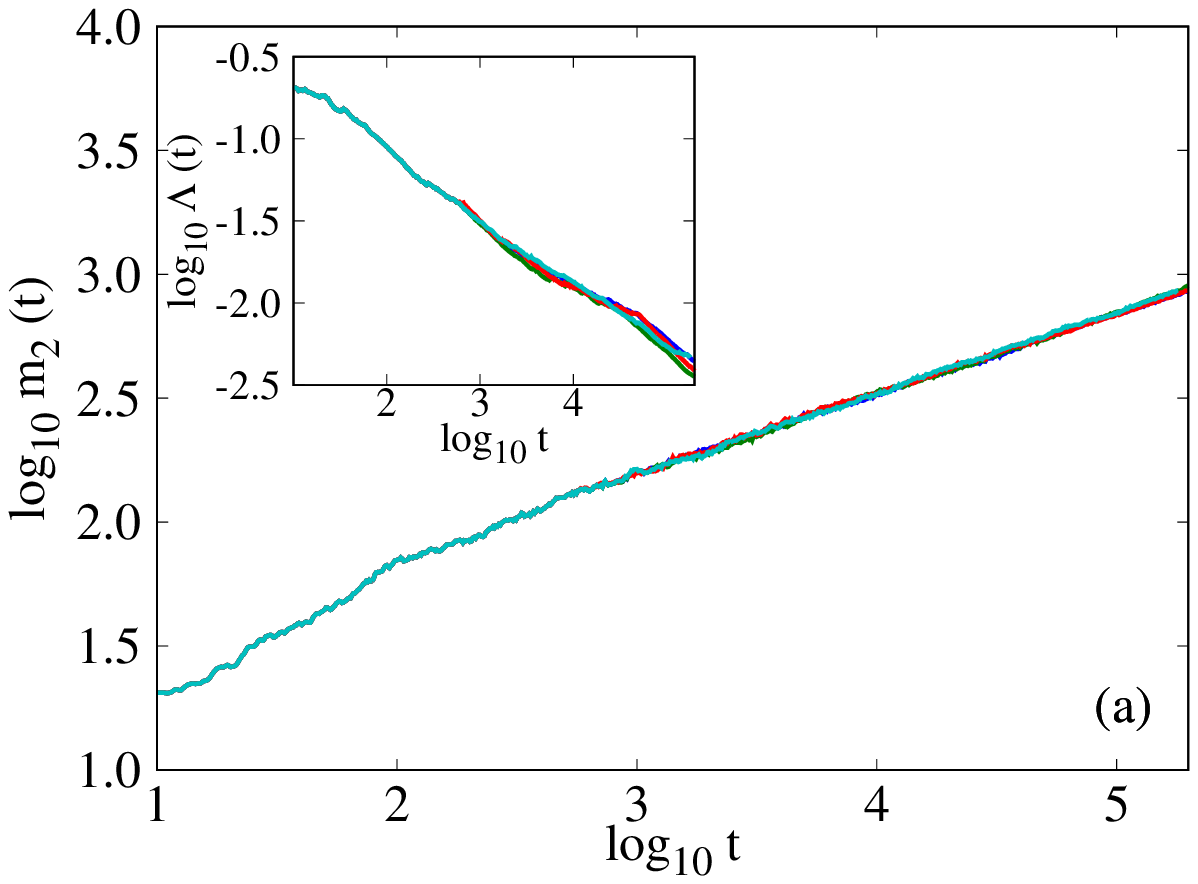}
    \includegraphics[width=0.49\textwidth]{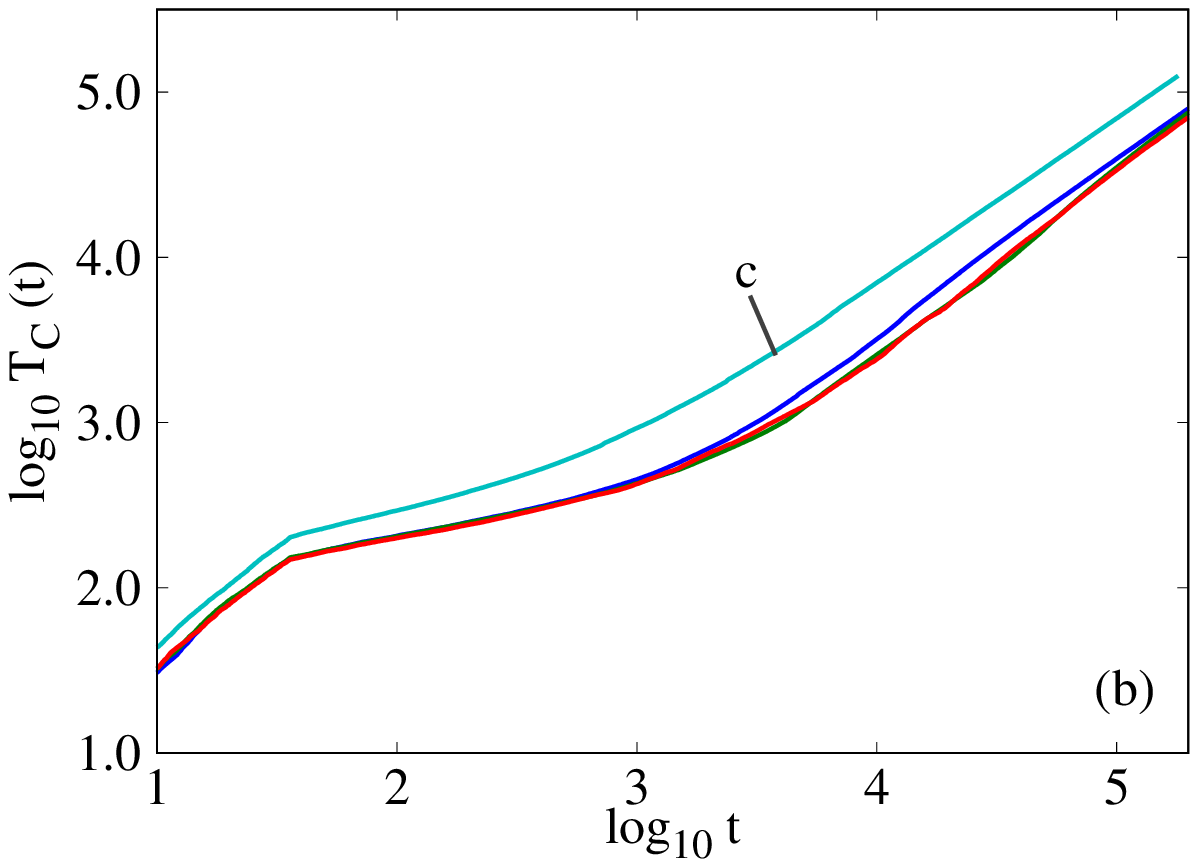}
    \caption{Time evolution of the second moment $m_2(t)$~\eqref{eq:second_moment_2d_chaotic} [panel (a)], the ftMLE $\Lambda(t)$~\eqref{eq:finite_mle} [inset of panel (a)] and the CPU time $T_C$ [panel (b)] for a particular setup of the initial excitation of the 2D DDNLS system~\eqref{eq:hamilton_complex_ddnls_2d} when the dynamic lattice approach is used in comparison to the fixed lattice computation.
        The curve's coloring corresponds to $\delta_\mathcal{D} = 10^{-12}$ [(b)lue curves], $\delta_\mathcal{D} = 10^{-10}$ [(g)reen curves], $\delta_\mathcal{D} = 10^{-8}$ [(r)ed curves] and the fixed lattice [(c)yan curves] computations.       
    }
    \label{fig:fix_dynamics_comparison_2d}
\end{figure}
The initial size of the active dynamic lattice was set to $101\times 101$ which is much larger than the average localization volume of the NMs ($\overline{V} \approx 35$ for $W = 10$ as seen in Table~\ref{tab:} below).
In addition, we use $\delta _\mathcal{D} = 10^{-12}$ and $\mathcal{N} = 10$. 

The spatiotemporal evolution of the projection along the $m$ axis of the system of the norm density distribution $\xi_{l,m}$~\eqref{eq:norm_density_distribution_2d_chaotic} in the case of the fixed lattice computation is presented in Fig.~\ref{fig:fix_dyn_2d_chaotic_norm_evol_01}(a) along with snapshots in Fig.~\ref{fig:fix_dyn_2d_chaotic_norm_evol_01}(c), at times indicated by the dashed lines in Fig.~\ref{fig:fix_dyn_2d_chaotic_norm_evol_01}(a).
A residual norm $s \lesssim 10^{-12}$ spreads over the entire lattice due to numerical fluctuations [see lighter areas in Fig.~\ref{fig:fix_dyn_2d_chaotic_norm_evol_01}(a)].
The same results obtained using the dynamic lattice simulation after $11$ expansions of the initial lattice is shown in Fig.~\ref{fig:fix_dyn_2d_chaotic_norm_evol_01}(b) with its related snapshots taken at the same times [in Fig.~\ref{fig:fix_dyn_2d_chaotic_norm_evol_01}(d)] as for the fixed lattice run.
Note that the white areas in Fig.~\ref{fig:fix_dyn_2d_chaotic_norm_evol_01}(b) correspond to unexcited regions i.e.~these sites possess zero norm ($s_{l,m} = 0$).
Clearly, the two simulations give practically the same wave packet evolution.

We repeat the same simulations, using different thresholds $\delta _{\mathcal{D}}$ [see Eq.~\eqref{eq:condition_expand_2d_chaotic}] for the dynamic lattice simulations.
In Fig.~\ref{fig:fix_dynamics_comparison_2d} we show the result of such numerical computations where the values $\delta _\mathcal{D} = 10^{-12}$ [(b)lue curves], $\delta _\mathcal{D} = 10^{-10}$ [(g)reen curves], $\delta _\mathcal{D} = 10^{-8}$ [(r)ed curves] and the fixed lattice [(c)yan curves] computations are used.
We plot the evolution of the second moment of the norm density distribution $\xi_{l, m}$~\eqref{eq:norm_density_distribution_2d_chaotic}
\begin{equation}
    m_2 = \sum_{l, m} \left[(l - \overline{l})^2 + (m - \overline{m})^2\right] \xi_{l,m}, \quad \overline{l} = \sum_{l, m} l \xi_{l,m}, \quad \overline{m} = \sum_{l,m}m\xi_{l,m},      
    \label{eq:second_moment_2d_chaotic}
\end{equation}
in Fig.~\ref{fig:fix_dynamics_comparison_2d}(a), and the evolution of the associated ftMLEs $\Lambda$~\eqref{eq:finite_mle} for all cases in the inset of Fig.~\ref{fig:fix_dynamics_comparison_2d}(a).
A good correspondence is visible between all the cases.
Similarly to what we did for the 1D DDNLS model~\eqref{eq:Hamilton_complex_dnls_1d} in Sec.~\ref{subsec:choice_of_init_dvd}, our choice of the $\delta_\mathcal{D}$ depends on the final values of the norm per site attained in our widest wave packet. 
In the 2D DDNLS model~\eqref{eq:hamilton_complex_ddnls_2d}, we are reaching wave packet width $ m_{2, f} \propto 10^3$ sites.
Since the constant total norm~\eqref{eq:norm_complex_dnls_2d} $\mathcal{S}_{2D}\approx 10^{0}-10^2$ is shared among all the excited sites, the final average norm per site 
\begin{equation}
    s_f \propto \frac{\mathcal{S}_{2D}}{m_{2, f}} \sim 10^{-3} - 10^{-1},
    \label{eq:final_norm_per_site_longest_2d_chaotic}
\end{equation}
such that using 
\begin{equation}
    \delta_{\mathcal{D}} = 10^{-8},
    \label{eq:used_threshold_2d_chaotic}
\end{equation}
we capture the wave packet spreading with a rather good precision as shown in Fig.~\ref{fig:fix_dynamics_comparison_2d}(a) up to the final time of integration $t_f \approx 2\times 10^{6}$.

In Fig.~\ref{fig:fix_dynamics_comparison_2d}(b) we plot the time evolution of the cumulative CPU time $T_C(t)$ necessary for the simulations.
As in the 1D case  [Figs.~\ref{fig:time_evol_m2_P_mD_PD_MLE_CPU_1d_spreading}(d) and~\ref{fig:time_evol_m2_MLE_CPU_various_threshold_1d_spreading}(b)], the fixed lattice simulation takes longer to complete than the dynamic lattice ones.
The dynamic lattice approach with a threshold~\eqref{eq:condition_expand_2d_chaotic} $\delta _\mathcal{D} = 10^{-8}$ gives the smallest $T_C$ value as expected.
Overall, a gain~\eqref{eq:speedup_factor} $g \approx 36\%$, $40\%$ and $43\%$ is seen for the $\delta _{\mathcal{D}} = 10^{-12}$, $\delta _{\mathcal{D}} = 10^{-10}$ and $\delta _{\mathcal{D}} = 10^{-8}$ threshold values respectively. 
In the remaining part of this chapter, we will use $\delta_\mathcal{D} = 10^{-8}$ for our simulations.

\section{\label{sec:mle_2d_ddnls_spreading}The maximum Lyapunov exponent}
In order to avoid repetition of similar information in the rest of this chapter, let us note that we perform simulations of lattice up to sizes $451\times 451$ in order to avoid finite size effects, integrating the equations of motion and the variational equations up to final time $t_f \approx 2\times 10^{6}$ by the $s11\mathcal{ABC}6$ SI~\eqref{eq:s11odr6_general}.
The choice of the integration time step, $\tau = 0.1-0.25$, leads to relative energy $E_r$~\eqref{eq:rel_energy_error_1dDDNLS} and norm $S_r$~\eqref{eq:rel_norm_error_1dDDNLS} errors always remaining below $10^{-3}$ and $10^{-2}$ respectively. 
We excite all sites in a square of side $L$ at the center of the lattice with the same norm per site $s$ (or the same rescaled norm $x = \beta s$), averaging our observables over $50$ disorder realizations. 
Notice that similar to what was done in the 1D DDNLS system, when $\beta>0$ in $\mathcal{H}_{2D}$~\eqref{eq:hamilton_complex_ddnls_2d}, we set the disorder parameter $\epsilon_{l, m}$ of lattice's central site to zero.
In that way, single site excitation of all disorder realizations with the same total norm $\mathcal{S}_{2D}$~\eqref{eq:norm_complex_dnls_2d} has the same total energy value~\eqref{eq:hamilton_complex_ddnls_2d} $\mathcal{H}_{2D} = \beta \mathcal{S}_{2D}/2$, which always belongs to the thermal phase of the energy-norm space in Fig.~\ref{fig:x_y_thermal_selftrapping_phases}.
On the other hand, for multi-site wave packet excitations, all disorder realizations have the same value of the total norm $\mathcal{S}_{2D}$~\eqref{eq:norm_complex_dnls_2d} and we ensure that the values of the total energy $\mathcal{H}_{2D}$~\eqref{eq:hamilton_complex_ddnls_2d}  remain bounded within the thermal phase of the energy-norm space of the 2D DDNLS system.
As an additional remark, let us note that we are using the energy-norm space of Fig.~\ref{fig:x_y_thermal_selftrapping_phases} of the 1D DDNLS model~\eqref{eq:Hamilton_complex_dnls_1d} for the 2D case~\eqref{eq:hamilton_complex_ddnls_2d}, as they are very similar to each other.
In fact, the only difference is that the extra dimension of the 2D system renormalizes the ground states of Fig.~\ref{fig:x_y_thermal_selftrapping_phases} in the thermodynamic limit, which is now located well below that of the 1D system.
Nevertheless, as we are not interested to the dynamics of the system close to the ground states, it makes Fig.~\ref{fig:x_y_thermal_selftrapping_phases} sufficient for exploring the thermal states of the 2D DDNLS Hamiltonian model~\eqref{eq:hamilton_complex_ddnls_2d}.

\subsection{\label{subsec:weak_chaos_2d_ddnls_spreading}The weak chaos regime}
Let us now investigate the chaotic dynamics of wave packet spreading in the weak chaos regime of the 2D DDNLS system~\eqref{eq:hamilton_complex_ddnls_2d}. 
We consider the following cases:
\begin{enumerate}
    \item[$\bullet$] {\bf Case W$1_{2D}$.} $W = 10$, $\beta = 0.15$, $L = 2$, $x = 0.15$.
    \item[$\bullet$] {\bf Case W$2_{2D}$.} $W = 10$, $\beta = 0.92$, $L =1$, $x = 0.92$.
    \item[$\bullet$] {\bf Case W$3_{2D}$.} $W = 10$, $\beta = 4$, $L  = 1$, $x = 1$.
    \item[$\bullet$] {\bf Case W$4_{2D}$.} $W = 12$, $\beta = 1.75$, $L = 1$, $x = 1.75$.
\end{enumerate}
Let us explain why these cases theoretically belong to the weak chaos regime of the 2D DDNLS model of Figs.~\ref{fig:parameter_space_dnls_2d_delta_W} and~\ref{fig:x_y_thermal_selftrapping_phases}.
In the case of W$1_{2D}$, which is a multi-site initial excitation, we apply the first condition of Eq.~\eqref{eq:spreading_regime_1d_ddnls_1}, 
\begin{equation}
    x \frac{L^2}{\overline{V}} < d,
    \label{eq:condition_weak_chaos_2d_chaotic}
\end{equation}
where $x = \beta s$ is the initial frequency shift, $\overline{V}$ is the average localization volume, $d$ is the average spacing between NMs in a NM volume and $L^2$ is the area of the initial excitation (note that for the 1D DDNLS model, we have $L$ instead of $L^2$).
The estimations of $\overline{V}$~\eqref{eq:loc_vol_loc_length_2d_ddnls_estimation}, and $d$~\eqref{eq:average_spacing_between_freq_anderson_model} are listed in Table~\ref{tab:} for the values of the disorder strength $W$ we use. 
It is worth mentioning that the $\overline{V}$ is not obtained through the analysis of the NM themselves, but rather by evolving a single site excitation within the 2D DDLS model (i.e. $\mathcal{H}_{2D}$~\eqref{eq:hamilton_complex_ddnls_2d} with $\beta = 0$), up to a time when a halt in the evolution is observed.
Then, the average participation number $P$~\eqref{eq:participation_ratio_num} over several disorder realizations of the norm density distribution $\xi_{l, m}$~\eqref{eq:norm_density_distribution_2d_chaotic} is a rough measure of the NM localization volume. 
Thus, using the values $x = 0.15$, $L = 2$, $\overline{V} =35$ and $d = 0.51$ we obtain the inequality $0.017<0.51$ from Eq.~\eqref{eq:condition_weak_chaos_2d_chaotic} which is true.
\begin{table}[!htbp]
    \centering 
    \begin{tabular}{c|rrrrr}
         2D NM properties & $W = 10$ & $W = 10.5$ & $W = 11$ & $W = 12$ & $W = 14$ \\
        \toprule
        $\overline{V}$ & $35$ & $28$ & $23$ & $16$ & $10$\\
        \midrule 
        $\Delta$ & $18$ & $18.5$ & $19$ & $20$ & $22$ \\
        \midrule
        $d = \Delta/\overline{V}$ & $0.51$ & $0.66$ & $0.83$ & $1.25$ & $2.2$ \\
        \bottomrule
    \end{tabular}
    \caption{Average NM localization volume $\overline{V}$, frequency bandwidth $\Delta = W + 8$ and average spacing between NM $d = \Delta/\overline{V}$~\eqref{eq:average_spacing_between_freq_anderson_model} in the case of a central multi-site excitation of width $L^2 \approx \overline{V}$ sites at the center of the 2D DDNLS lattice system~\eqref{eq:hamilton_complex_ddnls_2d}.
        Note that the values of $\overline{V}$ are obtained by averaging the participation number $P$~\eqref{eq:participation_ratio_num} of the excited part at the end of the integration of the 2D DDLS model (i.e.~$\beta = 0$ in $\mathcal{H}_{2D}$~\eqref{eq:hamilton_complex_ddnls_2d}) up to final times $t_f \approx 10^{5}-10^{6}$ over $10$ disorder realizations.
    }
    \label{tab:}
\end{table}
The remaining initial conditions of cases W$2_{2D}$, W$3_{2D}$ and W$4_{2D}$ correspond to single site excitations.
Consequently, we apply the first condition of Eq.~\eqref{eq:spreading_regime_1d_ddnls_3}  
\begin{equation}
    x < \Delta,
    \label{eq:condition_weak_chaos_2d_chaotic_2}
\end{equation}
such that the $x = 0.92$, $x = 1$ and $x = 1.75$ of cases W$2_{2D}$, W$3_{2D}$ and W$4_{2D}$ are respectively always smaller than the frequency bandwidths $\Delta = 18$, $18$ and $20$ of $W = 10$, $10$ and $12$ seen in Table~\ref{tab:}.

\begin{figure}[!htb]
    \centering 
    \includegraphics[width=0.4\textwidth, height=0.4\linewidth]{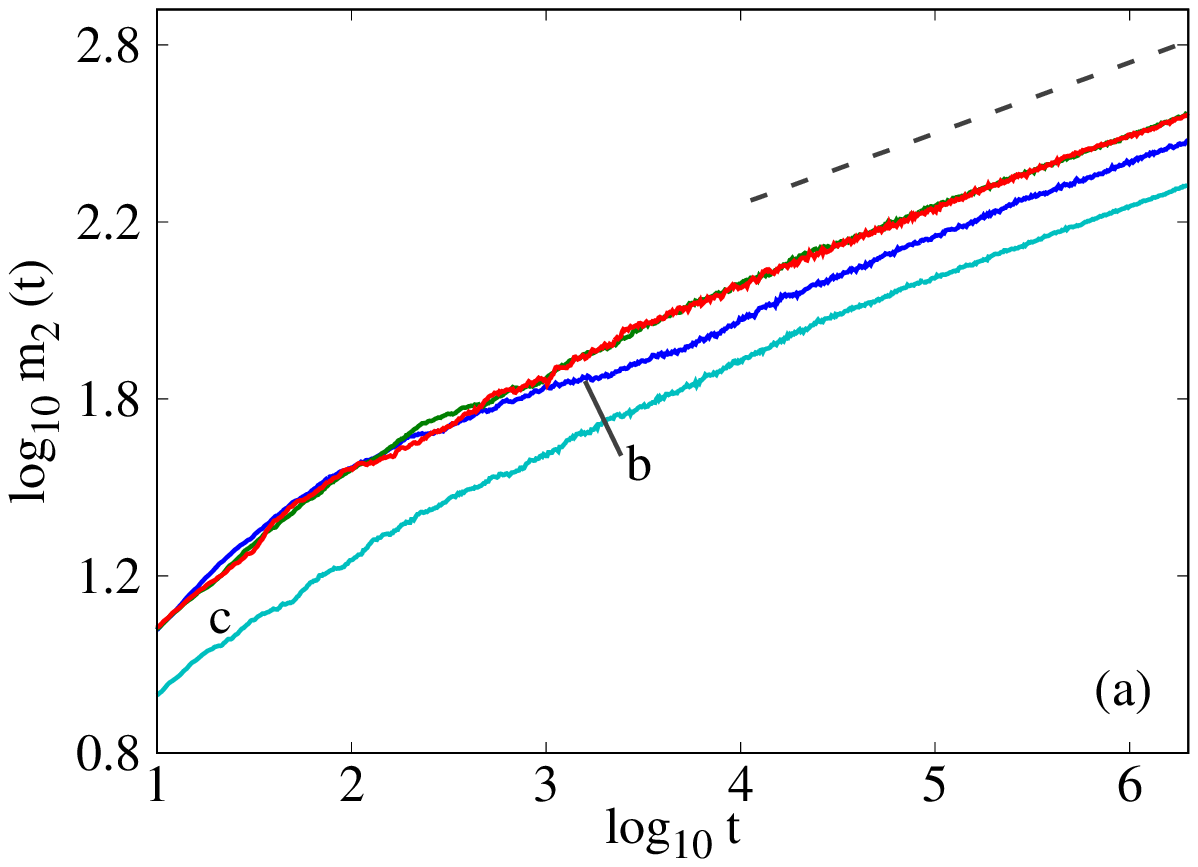}
    \includegraphics[width=0.4\textwidth, height=0.4\linewidth]{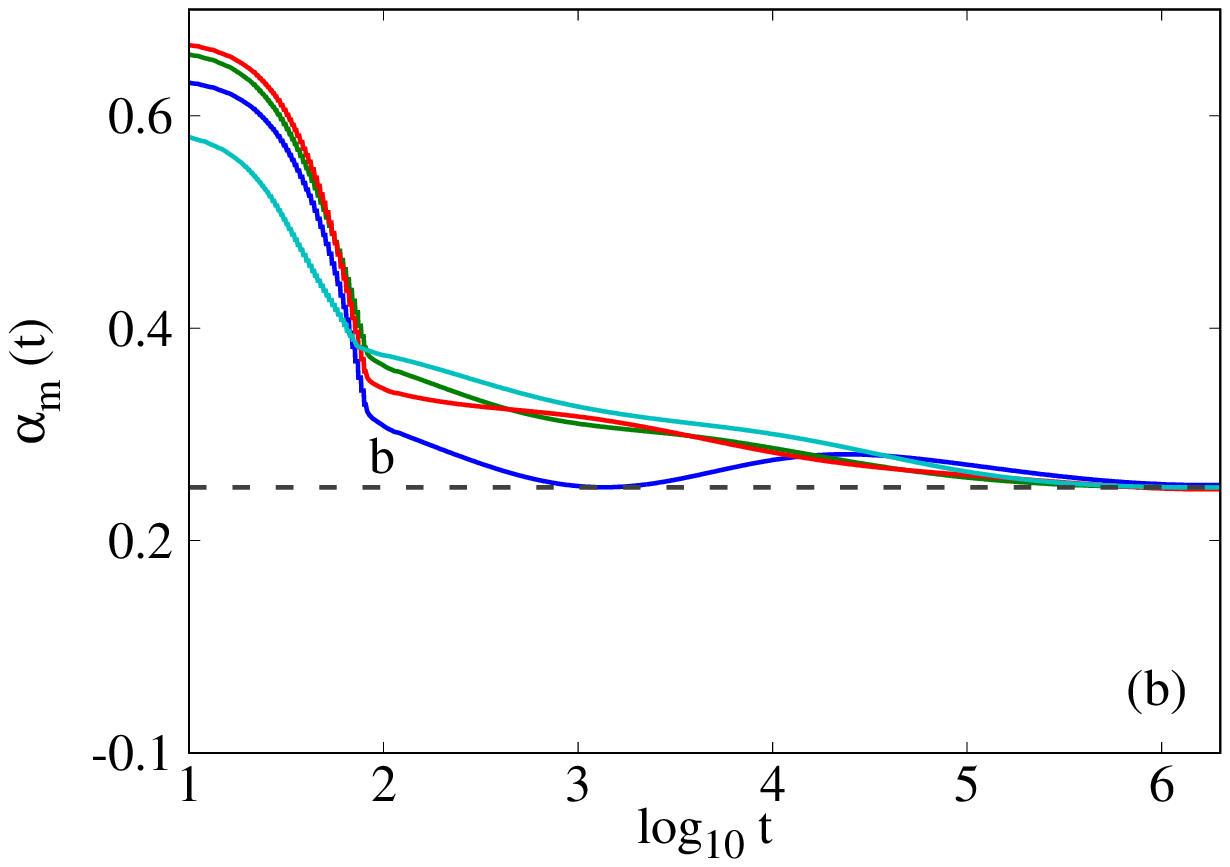}
    \includegraphics[width=0.4\textwidth, height=0.4\linewidth]{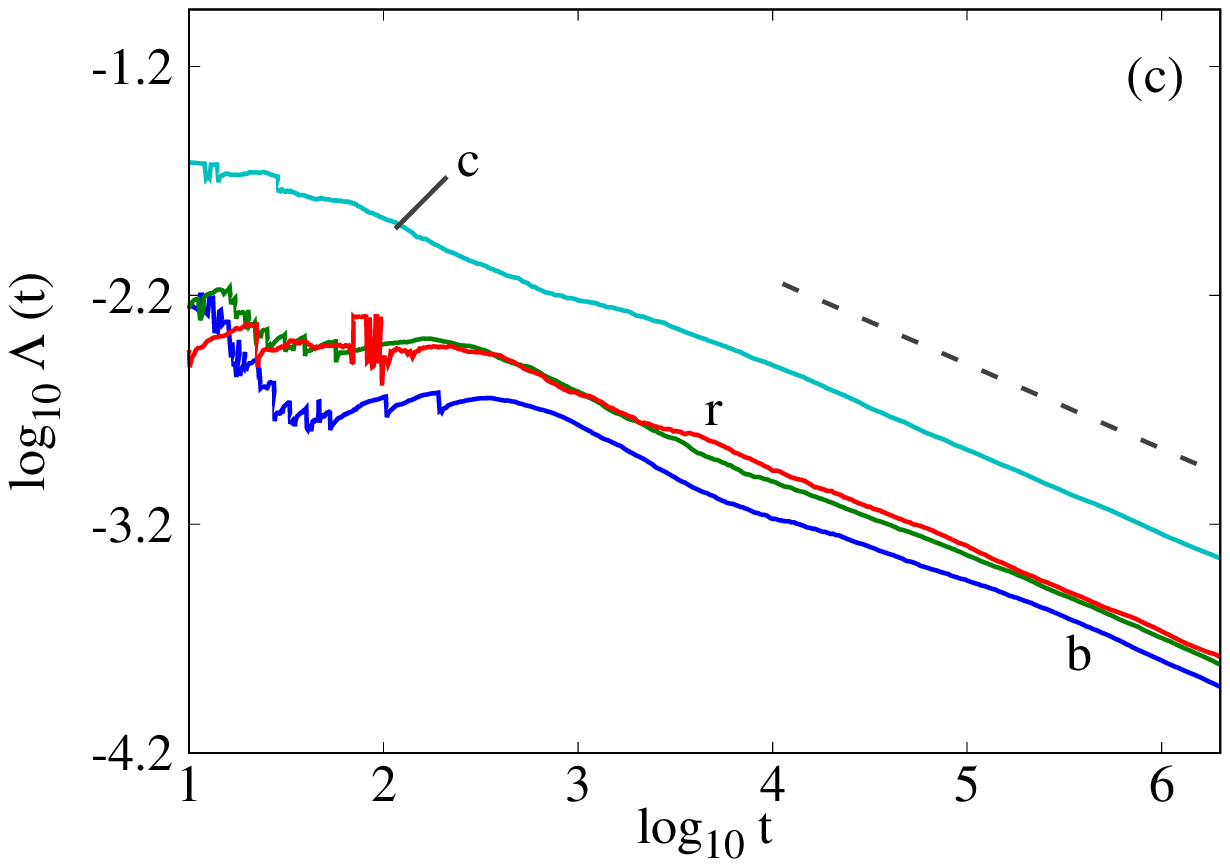}
    \includegraphics[width=0.4\textwidth, height=0.4\linewidth]{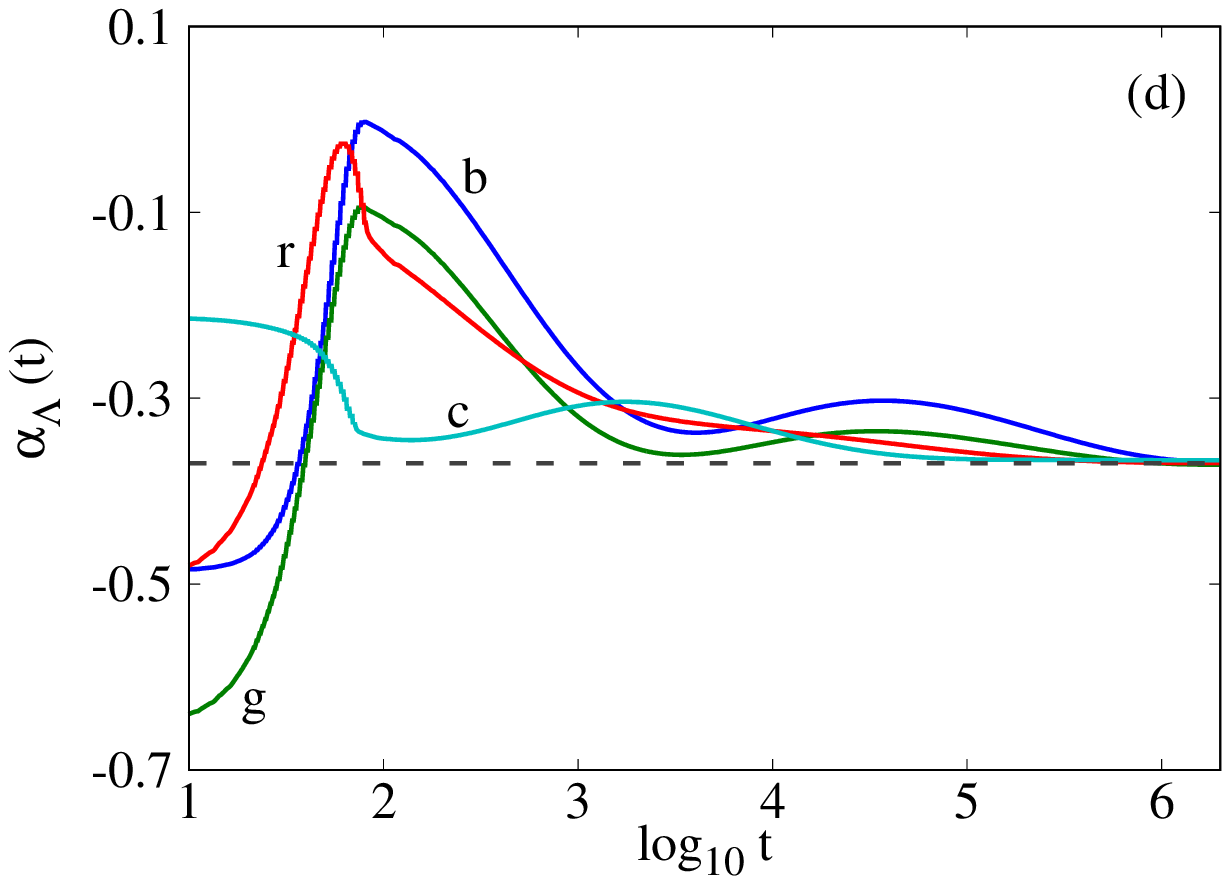}
    \caption{Results of the time evolution of the second moment $m_2(t)$~\eqref{eq:second_moment_2d_chaotic} [panel (a)] of the norm density distribution $\xi_{l,m}$~\eqref{eq:norm_density_distribution_2d_chaotic} and of the related ftMLEs $\Lambda(t)$~\eqref{eq:finite_mle} [panel (c)] for the weak chaos cases W$1_{2D}$ [(b)lue curves], W$2_{2D}$ [(g)reen curves], W$3_{2D}$ [(r)ed curves] and W$4_{2D}$ [(c)yan curves] (see text for details).
        In panels (b) and (d) we see the associated power law exponents when the data are fitted with the relations $m_2(t) \propto t^{\alpha _m}$ and $\Lambda(t) \sim t^{\alpha_\Lambda}$ respectively.
        The black dashed lines guide the eye for slopes $0.2$ [panel (a)] and $-0.37$ [panel (c)].
        The horizontal dashed lines indicate the values $0.2$ in panel (b) and $-0.37$ in panel (d).
        Each case has been averaged over $50$ disorder realizations.
    }
    \label{fig:weak_chaos_m2_2d_ddnls}
\end{figure}

In Fig.~\ref{fig:weak_chaos_m2_2d_ddnls}(a), we plot the time evolution of the second moment $m_2(t)$~\eqref{eq:second_moment_2d_chaotic} of the norm distribution $\xi _{l,m}$~\eqref{eq:norm_density_distribution_2d_chaotic}.
For all cases, a diffusive behavior of the wave packet is seen, which is well fitted by the monotonic power law $m_2 \propto t^{\alpha_m}$ with $0<\alpha_m < 1$ in the last decades of the evolution.
The time dependence of the power law exponents $\alpha _m(t)$ is shown in Fig.~\ref{fig:weak_chaos_m2_2d_ddnls}(b) for each studied case.
A clear saturation of the $\alpha_m(t)$ values to  $\alpha_m \approx 0.2$ is visible, in agreement with the predictions of~\citep{flach2010spreading,laptyeva2012subdiffusion} (see also Sec.~\ref{sec:spreading_mechanism}).
This power exponent value is numerically retrieved from our data, averaging all the $\alpha_m(t)$ values in Fig.~\ref{fig:weak_chaos_m2_2d_ddnls}(b) after the $6^\text{th}$ decade where a clear convergence is seen.  
This power law exponent was also numerically obtained by~\citep{laptyeva2012subdiffusion} for the 2D DKG system~\eqref{eq:hamilton_2d_dkg}. 
For all these weak chaos cases, we evaluate the wave packet's chaotic strength, through the computation of their ftMLEs $\Lambda(t)$~\eqref{eq:finite_mle} depicted in Fig.~\ref{fig:weak_chaos_m2_2d_ddnls} (c).
A clear decay of the $\Lambda(t)$ values is visible for all the cases following a power law $\Lambda \propto t^{\alpha_\Lambda}$ ($\alpha_\Lambda < 0$) similarly to what was observed in the 1D case [Sec.~\ref{subsec:weak_chaos_1d}].
The values of the time evolution of $\alpha _\Lambda(t)$ are presented in Fig.~\ref{fig:weak_chaos_m2_2d_ddnls}(d), showing a tendency to saturate around $\alpha_\Lambda \approx - 0.37$.
This result of the wave packet chaoticity is not particular to the 2D DDNLS system~\eqref{eq:hamilton_complex_ddnls_2d}, as it was also found for the 2D DKG model~\eqref{eq:hamilton_2d_dkg}~\citep{manda2020chaotic}.

\subsection{\label{subsec:strong_chaos_2d_ddnls_spreading}The strong chaos regime}
We also study the strong chaos regime of the 2D DDNLS system~\eqref{eq:hamilton_complex_ddnls_2d} of Figs.~\ref{fig:parameter_space_dnls_2d_delta_W} and~\ref{fig:x_y_thermal_selftrapping_phases}, by considering the following cases:
\begin{enumerate}
    \item[$\bullet$] {\bf Case S$1_{2D}$.}  $W = 10.5$, $\beta = 0.145$, $L = 21$, $x = 0.145$.
    \item[$\bullet$] {\bf Case S$2_{2D}$.}  $W =11$, $\beta = 0.68$, $L =10$, $x = 0.68$.
    \item[$\bullet$] {\bf Case S$3_{2D}$.} $W = 12$, $\beta = 0.35$, $L = 21$, $x = 0.35$.
    \item[$\bullet$] {\bf Case S$4_{2D}$.} $W = 14$, $\beta = 6$, $L = 15$, $x = 0.72$.
\end{enumerate}
\begin{figure}[!htb]
    \centering 
    \includegraphics[width=0.4\textwidth, height=0.4\linewidth]{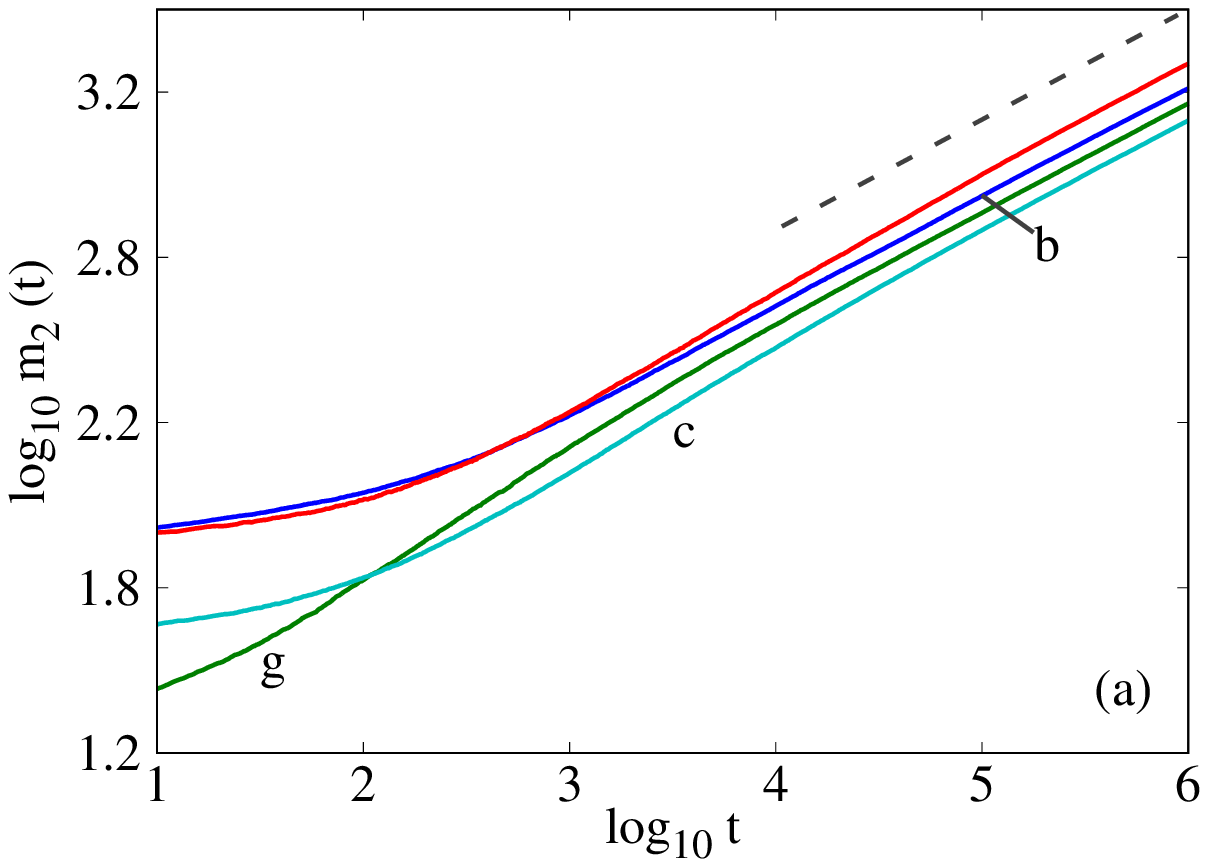}
    \includegraphics[width=0.4\textwidth, height=0.4\linewidth]{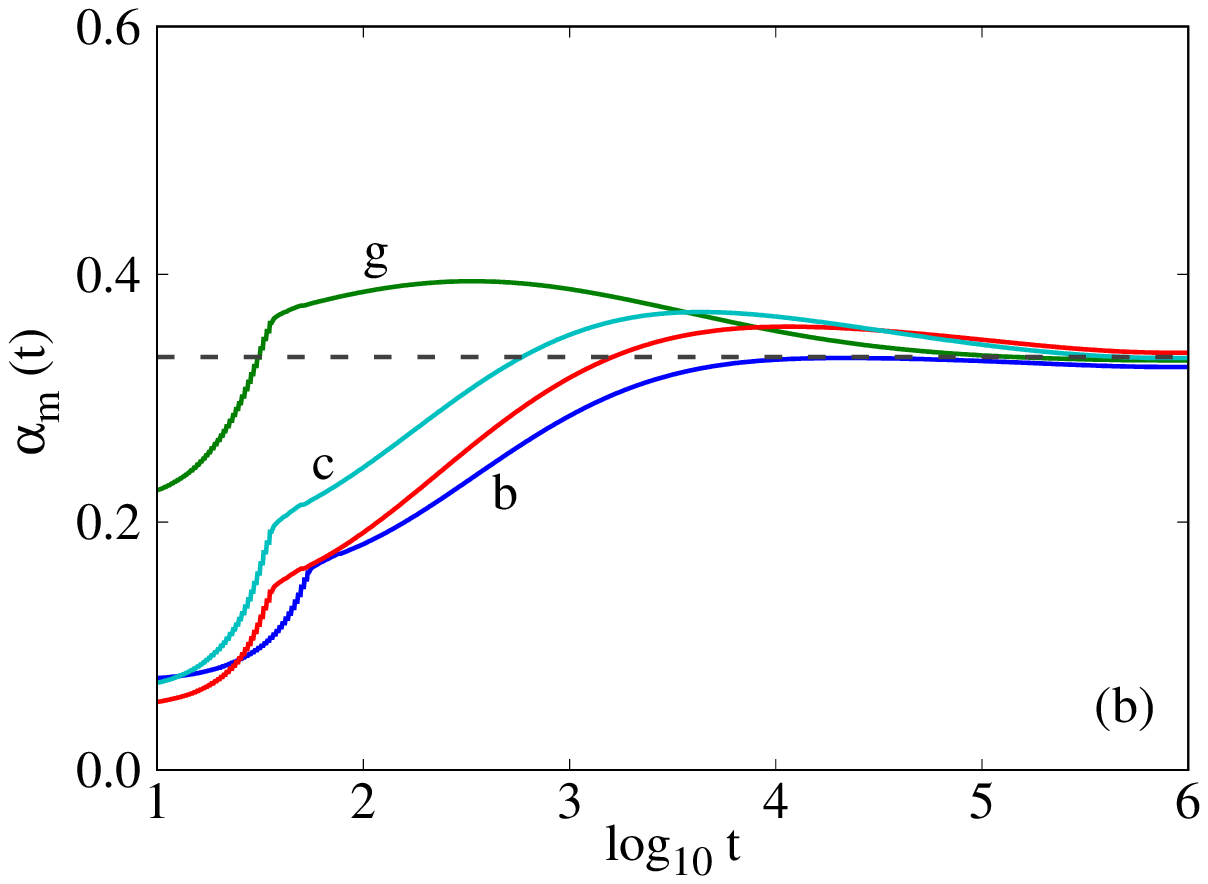}  
    \includegraphics[width=0.4\textwidth, height=0.4\linewidth]{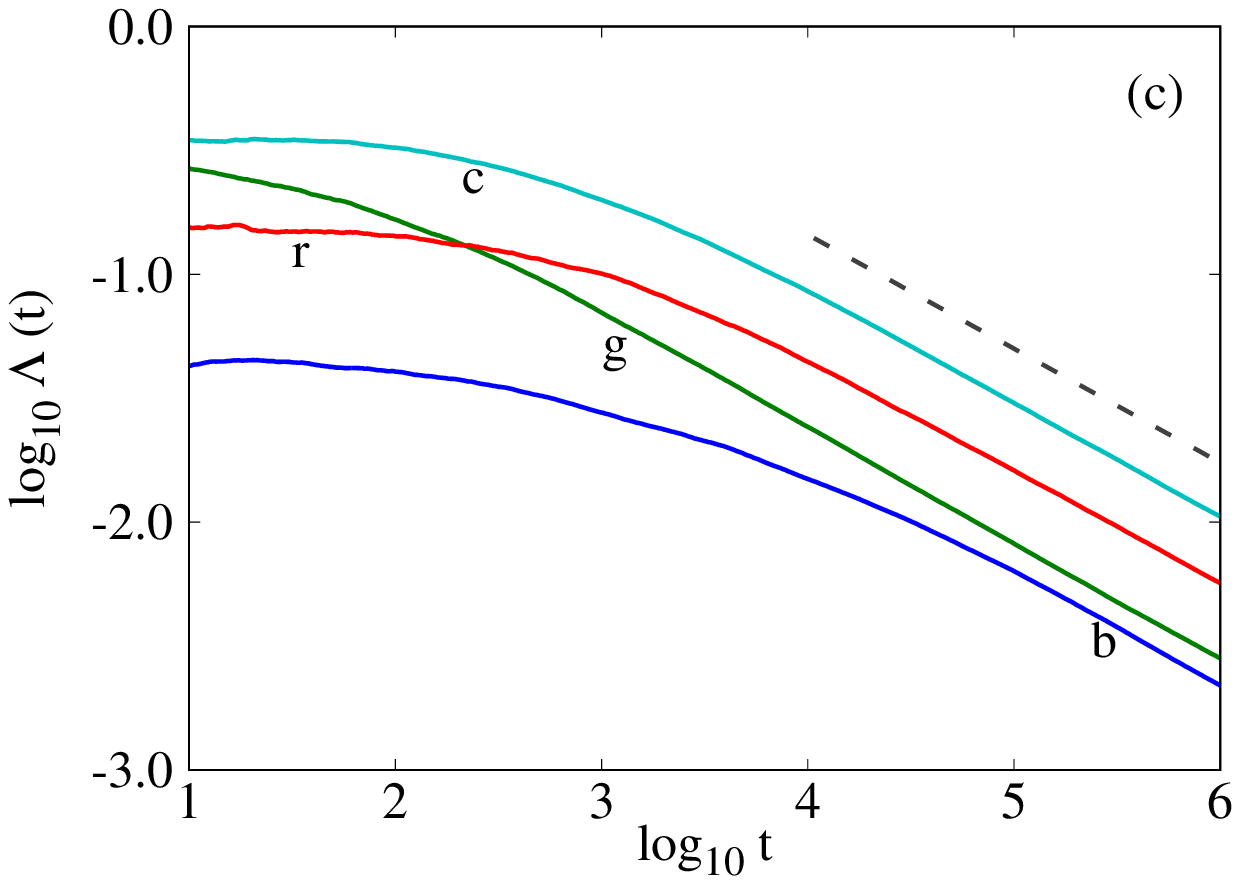}
    \includegraphics[width=0.4\textwidth, height=0.4\linewidth]{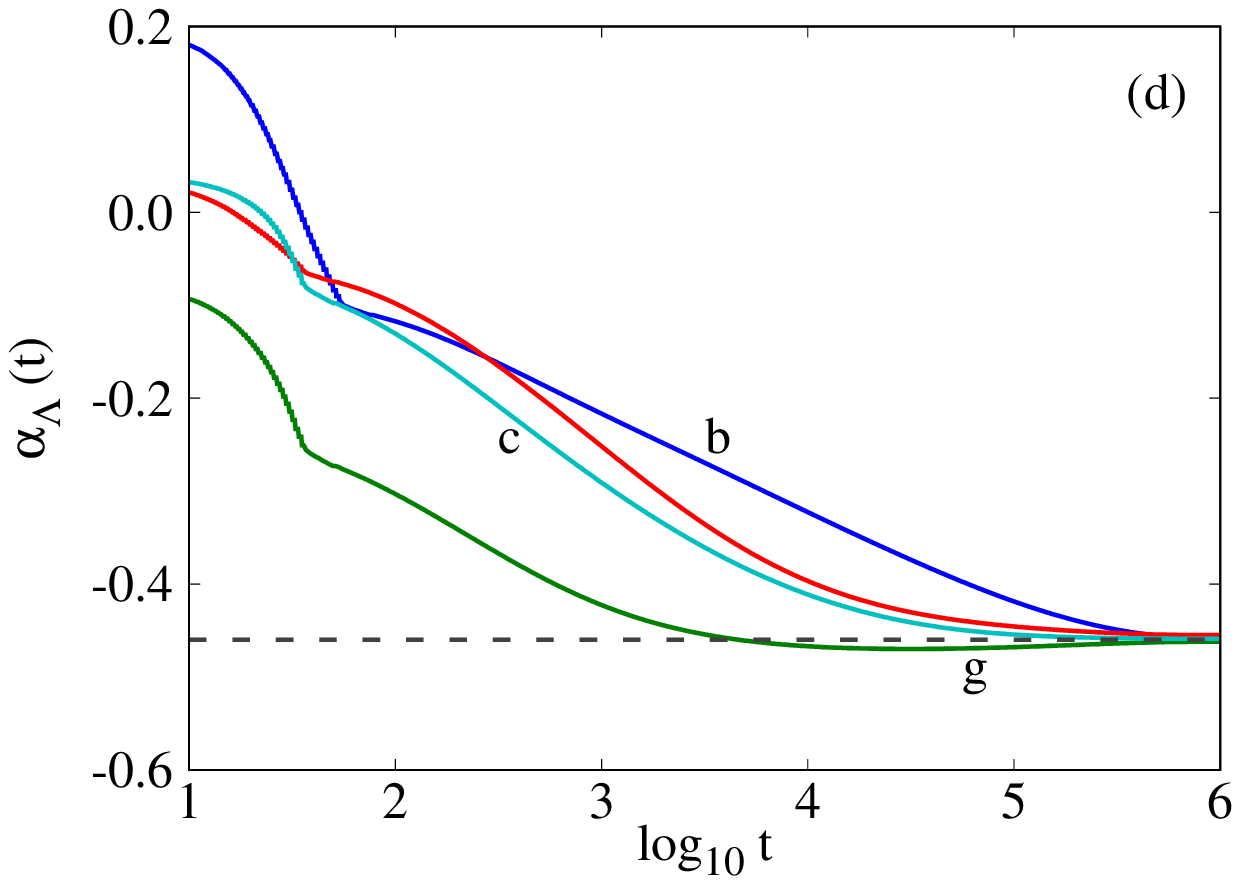}   
    \caption{Similar to Fig.~\ref{fig:weak_chaos_m2_2d_ddnls}, but for the strong chaos cases S$1_{2D}$ [(b)lue curves], S$2_{2D}$ [(g)reen curves], S$3_{2D}$ [(r)ed curves] and S$4_{2D}$ [(c)yan curves] (see text for details).
        The dashed lines in panels (a) and (c) guide the eye for slopes $0.33$ and $-0.46$ respectively.
        The horizontal dashed lines indicate the value $0.33$ in panel (b) and $-0.46$ in panel (d).
        The presented results are averaged over $50$ disorder realizations.  
    }
    \label{fig:strong_chaos_m2_2d_ddnls}
\end{figure}

Note that here, we use extended localized initial excitations as the value of $L^2$ is much greater than our estimation of the NM average localization volume shown in Table~\ref{tab:}.
In Sec.~\ref{sec:incommensureate_wave_packet}, we discussed a particular condition~\eqref{eq:size_incommensurate_wave_2d} for setting up the initial wave packet within the strong chaos region of Figs.~\ref{fig:parameter_space_dnls_2d_delta_W} and~\ref{fig:x_y_thermal_selftrapping_phases}, which for the 2D DDNLS system becomes
\begin{equation}
    L^2 \gtrsim \frac{W}{x}.
    \label{eq:condition_strong_chaos_2d_02}
\end{equation}
The advantage of this approach is that it does not depend on the characteristics of the NMs of the 2D DDLS system (i.e.~$\beta =0$ in $\mathcal{H}_{2D}$~\eqref{eq:hamilton_complex_ddnls_2d}), which have not yet been extensively studied.
Consequently, using the parameter values of the cases S$1_{2D}$, S$2_{2D}$, S$3_{2D}$ and S$4_{2D}$ we compute $L^2$ to respectively be $441,~ 100, 441$ and $225$.
These values are always much greater than $W/x \approx 72$, $16$, $34$ and $19$ respectively so that condition~\eqref{eq:condition_strong_chaos_2d_02} is satisfied.  

Let us note that, both the conditions of strong chaos in Eq.~\eqref{eq:spreading_regime_1d_ddnls_1}
\begin{equation}
    x \frac{L^2}{\overline{V}}> d,
    \label{eq:condition_strong_chaos_2d_01}
\end{equation}
which depends on the NM characteristics ($\overline{V}$~\eqref{eq:loc_vol_loc_length_2d_ddnls_estimation} and $d$~\eqref{eq:average_spacing_between_freq_anderson_model}), and the one of Eq.~\eqref{eq:condition_strong_chaos_2d_02} are equivalent.
Indeed, using the following simple algebraic manipulations
\begin{equation}
    x \frac{L^2}{\overline{V}}> d \Rightarrow x \frac{L^2}{\overline{V}} > \frac{\Delta}{\overline{V}} \Rightarrow x L^2 > \Delta \Rightarrow L^2 \gtrsim \frac{W}{x},
\end{equation}  
[where we assumed $\Delta \sim W$ for the 2D system~\eqref{eq:hamilton_complex_ddnls_2d}] we can relate the two inequalities.
In order to be completely certain about our argumentation, we have also checked using Eq.~\eqref{eq:condition_strong_chaos_2d_01} that our cases qualitatively belong to the strong chaos region of the 2D DDNLS model~\eqref{eq:hamilton_complex_ddnls_2d}. 
Indeed, using the parameter values for the cases S$1_{2D}$, S$2_{2D}$, S$3_{2D}$ and S$4_{2D}$ and the NM average localization volume $\overline{V}$ in Table~\ref{tab:}, we compute $xL^2/\overline{V}$ to respectively get $0.145\cdot 21^2/28\approx 2.28$, $0.68\cdot 10^2/23 \approx 2.96$, $0.35\cdot 21^2/16 \approx 9.65$ and $0.72\cdot 15^2/10 \approx 16.2$.
These numbers are always much higher than the $d$ values $0.66$, $0.83$, $1.25$ and $2.2$ for the disorder strengths $W = 10.5$, $11$, $12$ and $14$. 

The result of the computation of the second moment $m_2(t)$~\eqref{eq:second_moment_2d_chaotic} of the norm distribution $\xi_{l, m}$~\eqref{eq:norm_density_distribution_2d_chaotic} is shown in Fig.~\ref{fig:strong_chaos_m2_2d_ddnls} (a) for all the strong chaos cases.
Again, a spreading of wave packets appears for all the cases as the $m_2(t)$ values monotonically increase for all times.
These increases are well fitted by a power law $m_2(t) \propto t^{\alpha _m}$, with $0< \alpha_m < 1$.
The time evolution of the power law exponents $\alpha_m(t)$ are shown in Fig.~\ref{fig:strong_chaos_m2_2d_ddnls}(d) for all these cases.
Clearly, the values of $\alpha _m(t)$ reach $\alpha_m \approx 0.33$ at some point of their evolution, in agreement with the theory~\citep{flach2010spreading,laptyeva2012subdiffusion} (see also Sec.~\ref{sec:spreading_mechanism}). 
Results about of the time dependence of the ftMLEs $\Lambda(t)$~\eqref{eq:finite_mle} for all the strong chaos cases are shown in Fig.~\ref{fig:strong_chaos_m2_2d_ddnls}(c).
Again, a monotonic decay is evident for the ftMLEs, well fitted by $\Lambda(t) \propto t^{\alpha_\Lambda}$ with $\alpha_\Lambda < 0$.
The values of the power law exponents converge toward $\alpha _\Lambda \approx -0.46$ [Fig.~\ref{fig:strong_chaos_m2_2d_ddnls}(d)].
Both estimations $\alpha_m = 0.33$ and $\alpha_\Lambda = -0.46$, are obtained from averaging results from all the cases studied in the interval of time where the saturation is seen. 
These behaviors of the wave packet's chaoticity in the strong chaos regime were also found for the 2D DKG model~\eqref{eq:hamilton_2d_dkg} in~\citep{manda2020chaotic}.


\section{\label{sec:dvd_2d_ddnls_spreading}The deviation vector distribution}
The computation of the spatiotemporal evolution of the DVD obtained during the computation of the ftMLEs was used in the case of 1D DDNLS model~\eqref{eq:Hamilton_complex_dnls_1d} to visualize the motion of chaotic seeds in the lattice's excited part.
We perform a similar dynamical investigation for the DVD~\eqref{eq:dvd_2d_chaotic} in the 2D DDNLS model~\eqref{eq:hamilton_complex_ddnls_2d}.

\subsection{\label{subsec:weak_chaos_2d_ddnls_dvd}The weak chaos regime}
\begin{figure}[!htb]
    \centering
    \includegraphics[scale=0.5]{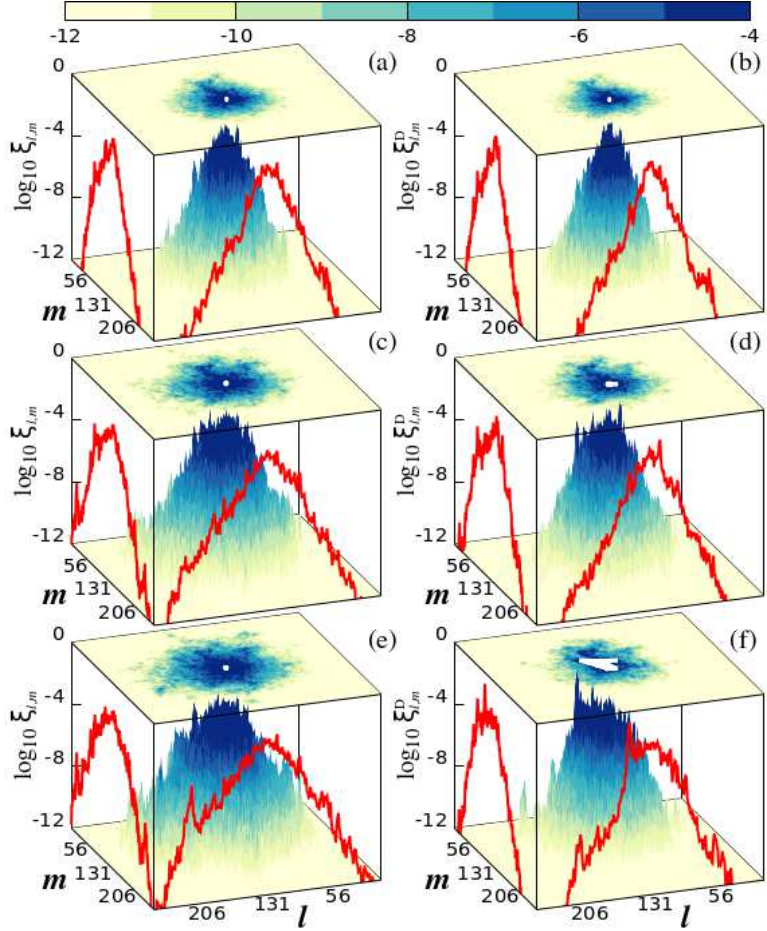}
    \caption{3D spatial distribution and the associated 2D color map (upper sides) snapshots of the norm density $\xi_{l,m}$~\eqref{eq:norm_density_distribution_2d_chaotic} [panels (a), (c) and (e)] and the DVD $\xi_{l,m}^D$~\eqref{eq:dvd_2d_chaotic} for a representative realization of the case W$1_{2D}$ (see text for details). 
        Panels (a) and (b) correspond to the configurations of the $\xi_{l,m}$ and $\xi_{l,m}^D$ at time $\log_{10}t \approx 3.15$, panels (c) and (d) at $\log_{10}t \approx 5.04$ and panels (e) and (f) at $\log_{10}t \approx 6.09$.
        The red curves on the sides are the distributions' projections along the $l$ and $m$ directions.  
        The white areas in the upper 2D color map represent the motion of the distribution's center. 
        The coloring of the distributions is done with respect to the $\log_{10}\xi_{l,m}$ and $\log_{10}\xi_{l,m}^D$ values according to the horizontal color scales presented at the top of the figure. 
    }
    \label{fig:heatmap_3d_plot_dvd_norm_weak_2d}
\end{figure}

Let us first study the behavior of the DVD $\xi_{l,m}^D$~\eqref{eq:dvd_2d_chaotic} in the weak chaos regime.
In Fig.~\ref{fig:heatmap_3d_plot_dvd_norm_weak_2d} we show the time evolution of the norm density distribution $\xi_{l,m}$~\eqref{eq:norm_density_distribution_2d_chaotic} [panels (a), (c) and (e)] and the DVD $\xi_{l, m}^D$~\eqref{eq:dvd_2d_chaotic} [panels (b), (d) and (f)] for a representative disorder realization of the $W1_{2D}$ case.
The snapshots of the distributions are taken at times $\log_{10}t \approx 3.15$ [panels (a) and (b)], $\log_{10}t \approx 5.04$ [panels (c) and (d)] and $\log_{10}t \approx 6.09$ [panels (e) and (f)].
The results of Figs.~\ref{fig:heatmap_3d_plot_dvd_norm_weak_2d}(a), (c) and (e) clearly show that the norm distribution is spreading as its extent (represented for instance by the red curves of the side projections of $\xi_{l,m}$) grows rather smoothly and symmetrically about the wave packet center which remains practically fixed [see white region at the center of the 2D color map on the upper sides of Figs.~\ref{fig:heatmap_3d_plot_dvd_norm_weak_2d}(a), (c) and (e)].
On the other hand, the DVD $\xi_{l,m}^D$~\eqref{eq:dvd_2d_chaotic} keeps a rather localized pointy shape (although its extent slightly increases in time), which remains well inside the wave packet width [see for example the  red curves of the side projections in Figs.~\ref{fig:heatmap_3d_plot_dvd_norm_weak_2d}(b), (d) and (f)].
Its center is meandering inside the lattice's excited part with increasing amplitudes [see white areas at the upper sides of the 2D color maps in Figs.~\ref{fig:norm_and_dvd_distributoin_weak_1d}(b), (d) and (f)].
\begin{figure}[!htb]
    \centering 
    \includegraphics[width=0.49\textwidth, height=0.5\linewidth]{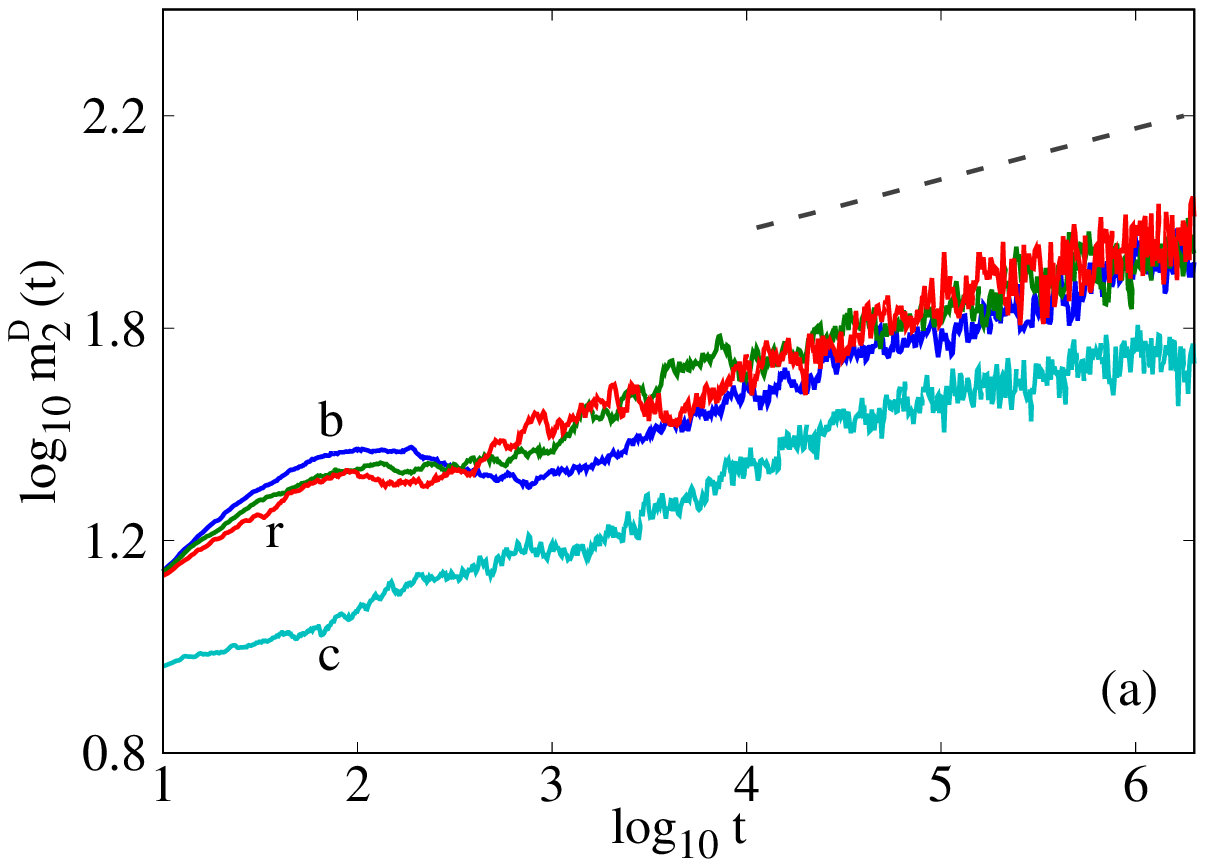}
    \includegraphics[width=0.49\textwidth, height=0.5\linewidth]{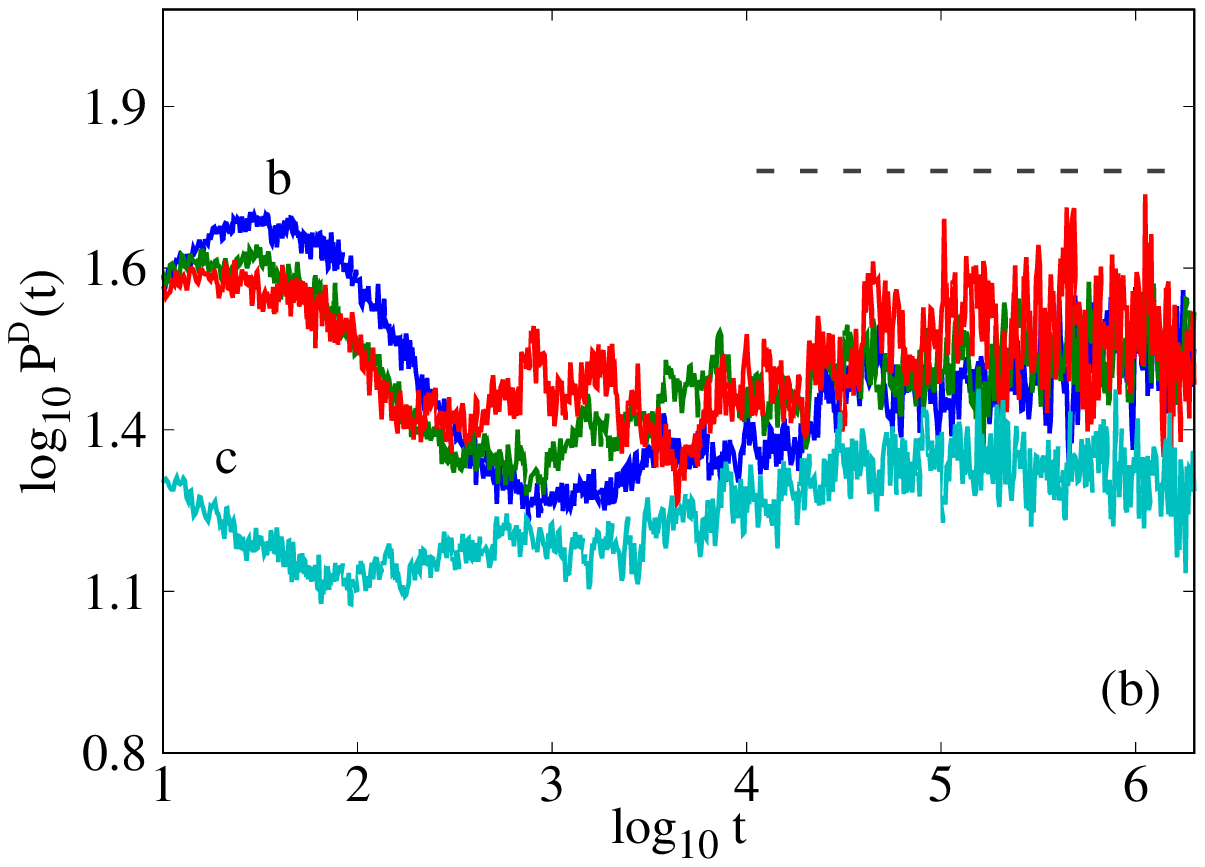}
    \caption{Time evolution of observables of the tangent dynamics in the weak chaos regime.
        (a) The second moment $m_2^D$~\eqref{eq:second_moment_2d_chaotic} and (b) the participation number $P^D$~\eqref{eq:participation_ratio_num} of the DVD $\xi_{l,m}^D$~\eqref{eq:dvd_2d_chaotic}.
        The dashed lines in panels (a) and (b) guide the eye for slopes $0.12$ and $0.0$ respectively.
        The curve's colors correspond to the ones used in Fig.~\ref{fig:weak_chaos_m2_2d_ddnls}.
    }
    \label{fig:weak_chaos_mdpdarr_and_power_exponent_2d_ddnls}
\end{figure}
\begin{figure}[!htb]
    \centering 
    \includegraphics[width=0.49\textwidth, height=0.5\linewidth]{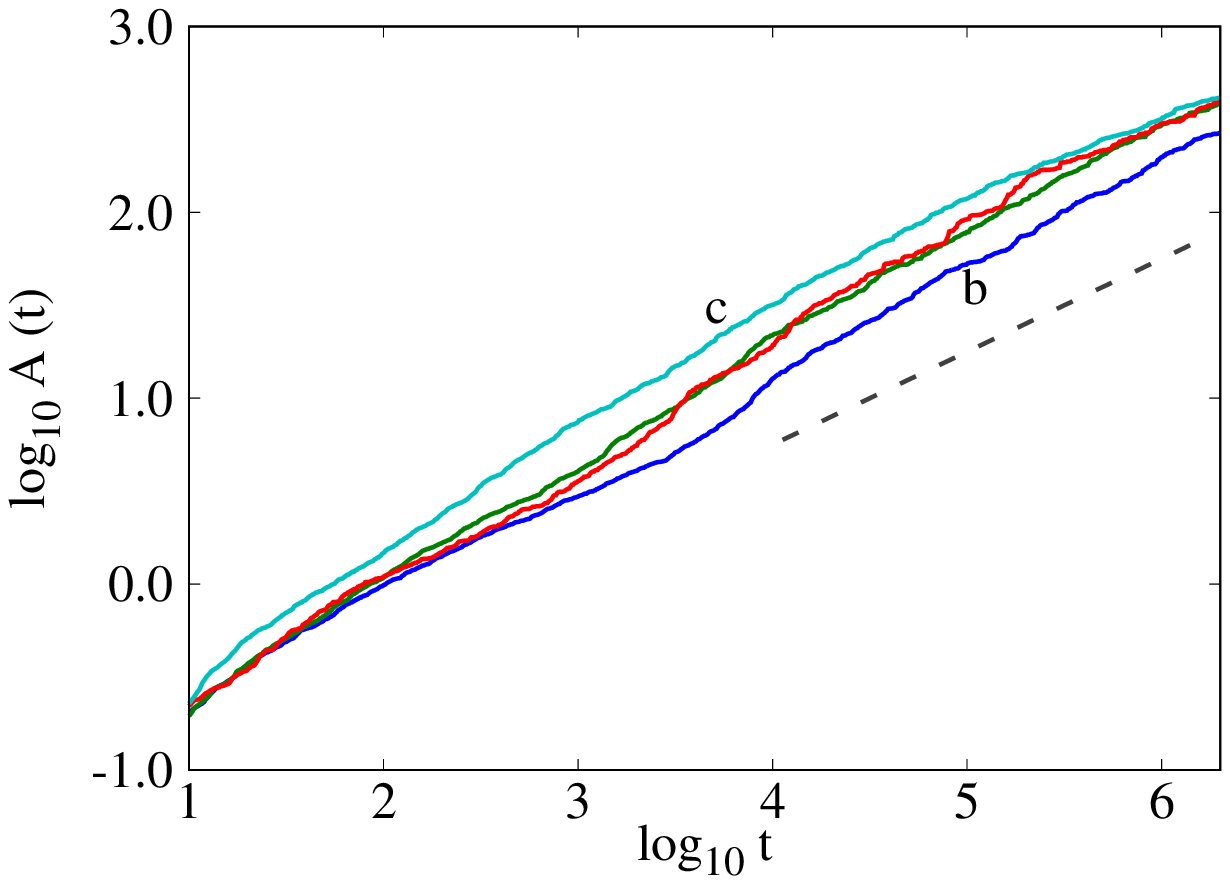}
    \caption{Time dependence of the area $A$~\eqref{eq:range_covered_by_DVD_2d} covered by the DVD's center $\bm{r}^D = \left(\overline{l}^D, \overline{m}^D\right)$ in cases belonging to the weak chaos regime of the 2D DDNLS system~\eqref{eq:hamilton_complex_ddnls_2d}.
        The dashed line indicates slope $0.5$.
        The curve's colors correspond to the ones used in Fig.~\ref{fig:weak_chaos_m2_2d_ddnls}.
    }
    \label{fig:weak_chaos_mdpdarr_and_power_exponent_2d_ddnls_02}
\end{figure}

In order to quantify the observations made about the DVD's behavior in Fig.~\ref{fig:heatmap_3d_plot_dvd_norm_weak_2d}, we compute the second moment $m_2^D(t)$~\eqref{eq:second_moment_2d_chaotic} and participation number $P^D(t)$~\eqref{eq:participation_ratio_num} of $\xi_{l,m}^D$~\eqref{eq:dvd_2d_chaotic}.
The obtained results are shown in Figs.~\ref{fig:weak_chaos_mdpdarr_and_power_exponent_2d_ddnls}(a) and (b) respectively.
A steady increase of the $m_2^D(t)$ values [Fig.~\ref{fig:weak_chaos_mdpdarr_and_power_exponent_2d_ddnls}(a)] is observed for all the weak chaos cases, which is well fitted by $m_2^D(t) \propto t^{0.12}$, exhibiting a much slower growth than the $m_2(t) \propto t^{0.2}$ behavior of the wave packet expansion observed in Figs.~\ref{fig:weak_chaos_m2_2d_ddnls}(a) and~(b).
In addition, a clear saturation of the $P^D(t)$ values is visible at the later stage of the evolution, which tend to fluctuate around numbers of the order of the NM average localization volume.
For instance, for cases W$1_{2D}$ [(b)lue curve], W$2_{2D}$ [(g)reen curve] and W$3_{2D}$ [(r)ed curve] in Fig.~\ref{fig:weak_chaos_m2_2d_ddnls}(b), the DVD's participation number attains values $P^D \approx 10^{1.5} \approx 31$ similar to the $\overline{V} \approx 35$ observed for $W = 10$ in Table~\ref{tab:}.
This behavior is due to the localized pointy shapes of the $\xi_{l,m}^D$~\eqref{eq:dvd_2d_chaotic} seen in Figs.~\ref{fig:heatmap_3d_plot_dvd_norm_weak_2d}(b), (d) and (f) whose extents approximately span the average volume of a NM.
Note that larger fluctuations of the $P^D(t)$ values in Figs.~\ref{fig:weak_chaos_m2_2d_ddnls}(b) compared to the 1D case in Fig.~\ref{fig:dvd_m2_P_R_slopes_R_weak_chaos_1d}(b) can be attributed to the smaller number of disorder realizations used in the 2D case and to the fact that the geometrical characteristics of the NMs in 2D lattices is more complex compared to the 1D case, leading for instance to larger discrepancies in the measurement of the NM localization length.  

We also investigate the motion of the DVD's center $\overline{\bm{r}}^D = \left(\overline{l}^D, \overline{m}^D\right)$, by computing the quantity  
\begin{equation}
    A(t) = R_x (t)\cdot R_y(t),
    \label{eq:range_covered_by_DVD_2d}
\end{equation}
at each time $t>0$ of the evolution.
Here, 
\begin{equation}
    R_x(t) = \max _{[0, t]} \left\{\overline{l}^D(t)\right\} - \min _{[0, t]} \left\{\overline{l}^D(t)\right\}, \qquad     R_y(t) = \max _{[0, t]} \left\{\overline{m}^D(t)\right\} - \min _{[0, t]} \left\{\overline{m}^D(t)\right\}.
    \label{eq:range_covered_by_DVD_2d_02}
\end{equation}
similar to the quantity $R$~\eqref{eq:range_covered_by_DVD} used in the 1D case.
The $A$ quantity~\eqref{eq:range_covered_by_DVD_2d} tries to quantify the area visited by the DVD (whose extent possesses fuzzy borders), by measuring the surface of the smallest rectangle containing the white area in panels (b), (d) and (f) of Figs.~\ref{fig:heatmap_3d_plot_dvd_norm_weak_2d} and~\ref{fig:heatmap_3d_plot_dvd_norm_strong_2d}.

The fate of the wave packet can be conjectured based on the behavior of the area $A$~\eqref{eq:range_covered_by_DVD_2d} in a similar way to the one followed in the 1D case for the length $R$~\eqref{eq:range_covered_by_DVD}.
Indeed, if 
\begin{equation}
    A \lesssim \overline{V}, \quad A(t) \sim t^{0},
\end{equation}
the DVD does not visit sites outside the initially excited central ones.
In this case, no spreading of the wave packet will be observed. 
On the other hand, when 
\begin{equation}
    A(t) > \overline{V},
\end{equation}
after a certain amount of time, two scenarios are possible.
If the values of $A$ saturate i.e~$A(t) \sim t^{0}$ after some time, then the DVD's center has only reached sites in close proximity of the central ones.
Thus, a subdiffusion of the wave packet will be visible, followed by a slowing down of the wave packet spreading which will eventually halt forever.   
This corresponds to the situation discussed in~\citep{johansson2010kam,aubry2011kam,pikovsky2011scaling}.
On the other hand, if the $A(t)$ values show a monotonic increase such that at large times $A(t)\gg V$, then the DVD center is visiting sites far away from the initially excited ones, leading to the case of wave packet which is spreading indefinitely as described in~\citep{pikovsky2008destruction,garcia2009delocalization,flach2010spreading,laptyeva2012subdiffusion}. 

The time evolution of the $A(t)$~\eqref{eq:range_covered_by_DVD_2d_02} values is depicted in Fig.~\ref{fig:weak_chaos_mdpdarr_and_power_exponent_2d_ddnls_02} for all the weak chaos cases of Fig.~\ref{fig:weak_chaos_m2_2d_ddnls}.
A clear monotonic increase of $A(t)$ is visible.
The values of $A(t)$ are well fitted by a power law $A(t) \propto t^{\alpha_A}$ with $\alpha_A \approx 0.5$ [black dashed line in Fig.~\ref{fig:weak_chaos_mdpdarr_and_power_exponent_2d_ddnls_02}].
The slight slowing down observed for cases W$3_{2D}$ and W$4_{2D}$ [(r)ed and (c)yan curves in Fig.~\ref{fig:weak_chaos_mdpdarr_and_power_exponent_2d_ddnls_02}] at the end of the integration is likely due to the fact that the integration time is not enough to display a firm asymptotic behavior of the chaotic seeds.
This is supported by the fact that for all cases, the second moment $m_2(t)$~\eqref{eq:second_moment_2d_chaotic} in Figs.~\ref{fig:weak_chaos_m2_2d_ddnls}(a) and (b) showed that the system has truly reached the weak chaos regime around the last decade of the evolution.    
Nevertheless, the fact that the $A(t)$ values monotonically increase up to $A(t) \propto 10^{2.5} \approx 300$ sites at the final stages of the simulations, which correspond to areas much larger than the estimates of the NM localization volume is a strong evidence that the chaotic seeds are visiting sites far away from the lattice's center.

\subsection{\label{subsec:sttrong_chaos_2d_ddnls_dvd}The strong chaos regime}
We also investigate the behavior of the DVD~\eqref{eq:dvd_2d_chaotic} in the strong chaos regime.
\begin{figure}[!htb]
    \centering
    \includegraphics[scale=0.5]{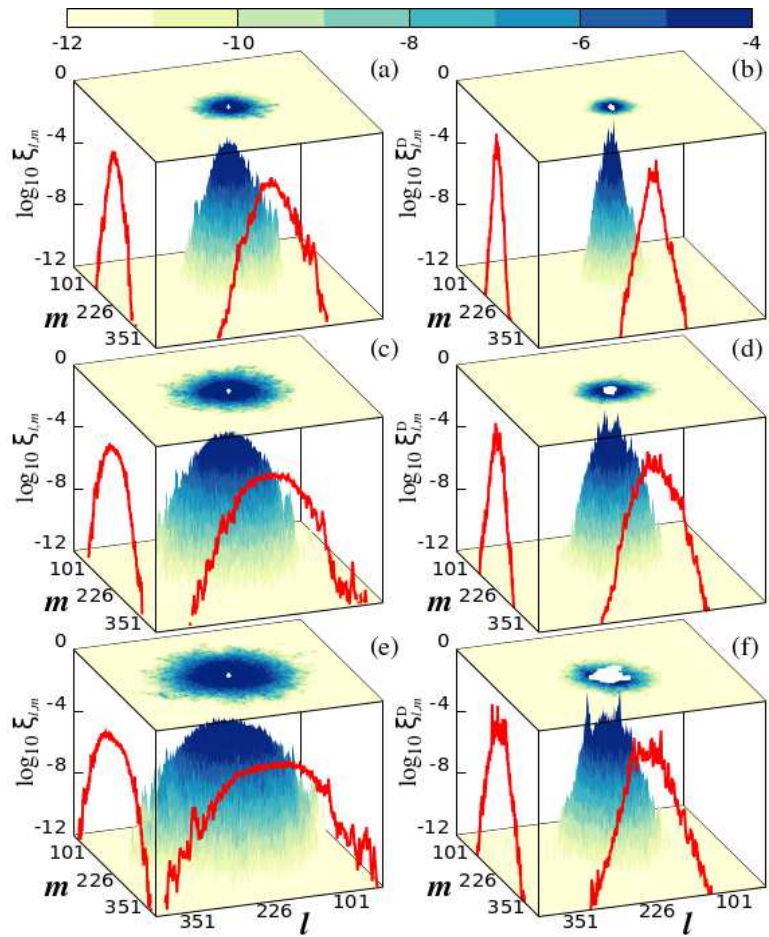}
    \caption{Similar to Fig.~\ref{fig:heatmap_3d_plot_dvd_norm_weak_2d} but for a representative realization  of the S$1_{2D}$ strong chaos case.
        Panels (a) and (b) correspond to the configurations of the $\xi_{l,m}$~\eqref{eq:norm_density_distribution_2d_chaotic} and $\xi_{l,m}^D$~\eqref{eq:dvd_2d_chaotic} at $\log_{10}t \approx 3$, panels (c) and (d) at $\log_{10}t \approx 4.8$ and panels (e) and (f) at $\log_{10}t \approx 5.90$.
    }
    \label{fig:heatmap_3d_plot_dvd_norm_strong_2d}
\end{figure}
In Fig.~\ref{fig:heatmap_3d_plot_dvd_norm_strong_2d}, we show the evolution of the norm density $\xi_{l, m}$~\eqref{eq:norm_density_distribution_2d_chaotic} and its related DVD $\xi_{l,m}^D$~\eqref{eq:dvd_2d_chaotic} at three different times $\log_{10} t \approx 3.0$ [panels (a) and (b)], $\log_{10} t \approx 4.8$ [panels (c) and (d)] and $\log_{10} t \approx 5.9$ [panels (e) and (f)]  for a representative disorder realization of the $S1_{2D}$ case.
Similarly to the weak chaos case [Figs.~\ref{fig:heatmap_3d_plot_dvd_norm_weak_2d}(a),~(c) and (e)], in Figs.~\ref{fig:heatmap_3d_plot_dvd_norm_strong_2d}(a),~(c) and (e) we see that the norm density $\xi_{l, m}$~\eqref{eq:norm_density_distribution_2d_chaotic} spreads more or less evenly around its center [white region in the color map of the upper sides of Figs.~\ref{fig:heatmap_3d_plot_dvd_norm_strong_2d}(a), (c) and (e)] which remains practically fixed.
On the other hand the DVD $\xi _{l, m}^D$~\eqref{eq:dvd_2d_chaotic} maintains a rather localized pointy shape.
In addition, its center meanders inside the wave packet interior with increasing amplitude, as indicated by the always growing white regions at the center of the upper sides of Fig.~\ref{fig:heatmap_3d_plot_dvd_norm_strong_2d} (b), (d) and (f).
These oscillations are much larger compared to what was observed in the weak chaos case [see Figs.~\ref{fig:heatmap_3d_plot_dvd_norm_weak_2d}(b), (d) and (f)].

\begin{figure}[!htb]
    \centering 
    \includegraphics[width=0.49\textwidth, height=0.5\linewidth]{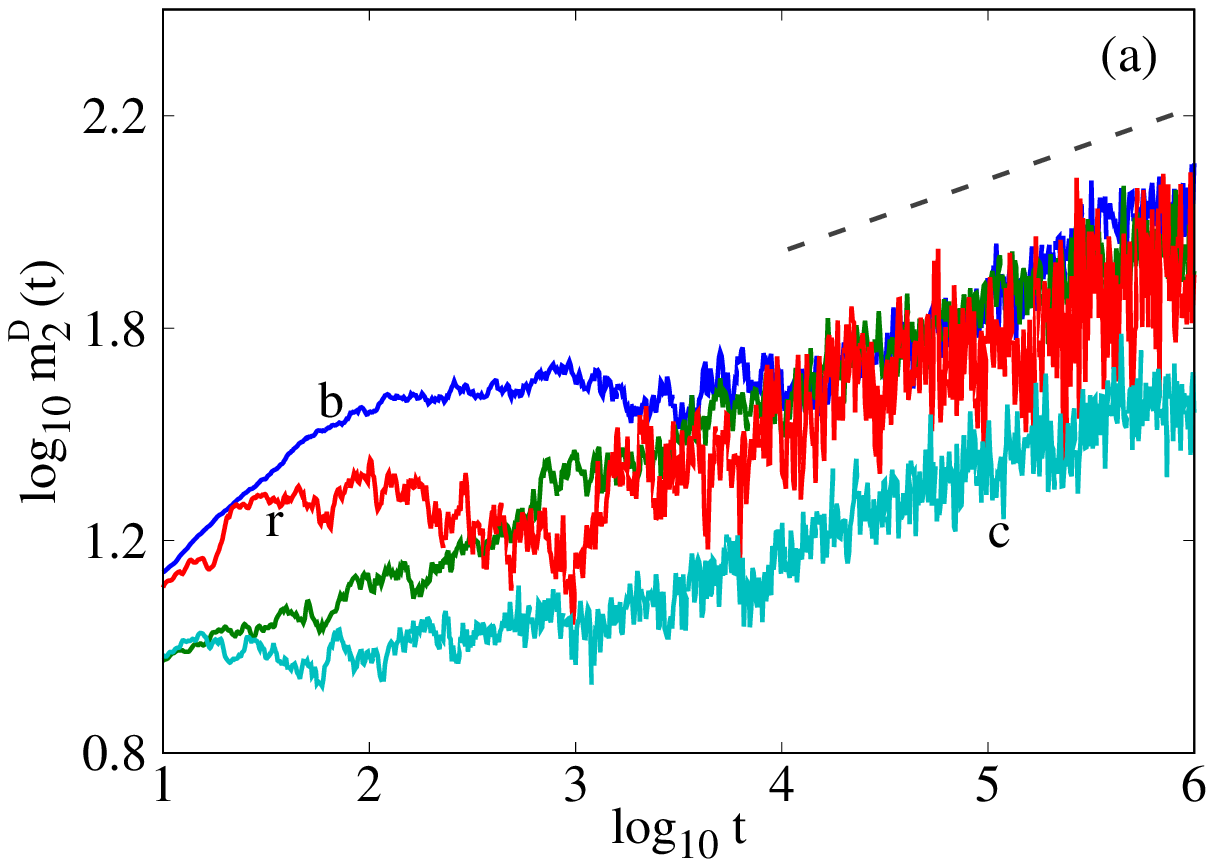}
    \includegraphics[width=0.49\textwidth, height=0.5\linewidth]{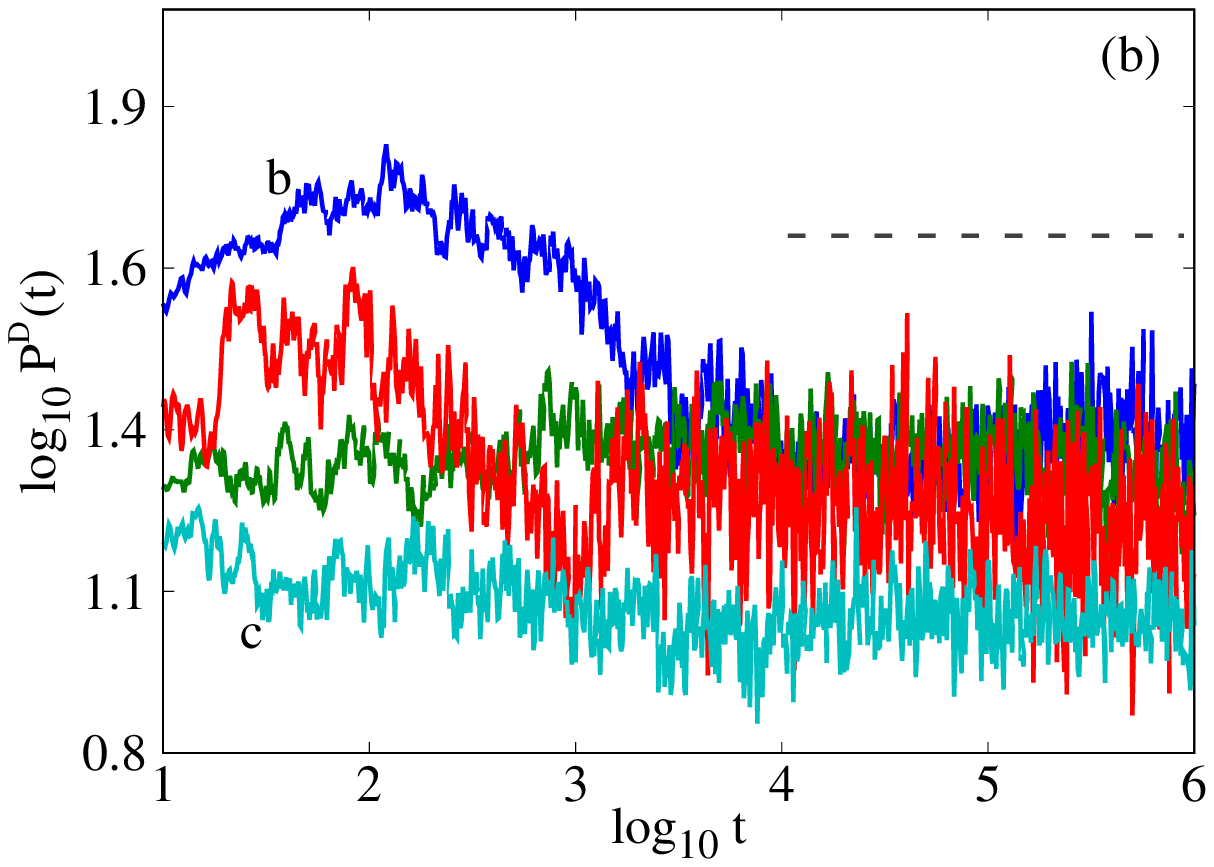}
    \caption{Similar to Fig.~\ref{fig:weak_chaos_mdpdarr_and_power_exponent_2d_ddnls}, but for the strong chaos regime of the 2D DDNLS system~\eqref{eq:hamilton_complex_ddnls_2d}.
        The dashed lines in panels (a) and (b) guide the eye for slopes $0.17$ and $0.0$ respectively.
        The curve's colors correspond to the ones used in Fig.~\ref{fig:strong_chaos_m2_2d_ddnls}.
    }
    \label{fig:strong_chaos_mdpdarr_and_power_exponent_2d_ddnls}
\end{figure}

In order to substantiate these observations, in Figs.~\ref{fig:dvd_m2_P_R_slopes_R_strong_chaos_1d}(a) and (b) we respectively show the time evolution of the DVD's second moment $m_2^D(t)$~\eqref{eq:second_moment_2d_chaotic}, and participation number $P^D(t)$~\eqref{eq:participation_ratio_num}.
A slow increase of the values of the $m_2^D(t)$, which are well fitted by the power law $m_2^D(t) \propto t^{0.17}$ is observed, much smaller than the wave packet's second moment evolution $m_2(t)\propto t^{0.33}$ in Figs.~\ref{fig:strong_chaos_m2_2d_ddnls}(a) and (b).
The saturation of the $P^D$ values is clearly seen in Fig.~\ref{fig:strong_chaos_mdpdarr_and_power_exponent_2d_ddnls}(b), where fluctuations in the interval $P^D \approx 10$ for S$4_{2D}$ [(c)yan curve in Fig.~\ref{fig:strong_chaos_mdpdarr_and_power_exponent_2d_ddnls}(b)] up to $P^D \approx 10^{1.4} \approx 25$ in the case of S$1_{2D}$ [(b)lue curve in Fig.~\ref{fig:strong_chaos_mdpdarr_and_power_exponent_2d_ddnls}(b)] are observed.
In both cases the $P^D$ values are of the order of the NM localization volume $\overline{V} \approx 10$ and $28$, respectively at $W = 14$ and $10.5$ [Table~\ref{tab:}]. 
Consequently, the DVD $\xi_{l,m}^D$~\eqref{eq:dvd_2d_chaotic} remains localized during the system's evolution, with its extent reaching approximately the NM's average localization volume.
Similarly to the weak chaos case, we understand the larger fluctuations in the calculations of the $m_2^D(t)$~\eqref{eq:second_moment_2d_chaotic} and $P^D(t)$~\eqref{eq:participation_ratio_num} based on the rather small number of considered disorder realizations and the complex geometry of 2D NMs. 

\begin{figure}[!htb]
    \centering 
    \includegraphics[width=0.49\textwidth, height=0.5\linewidth]{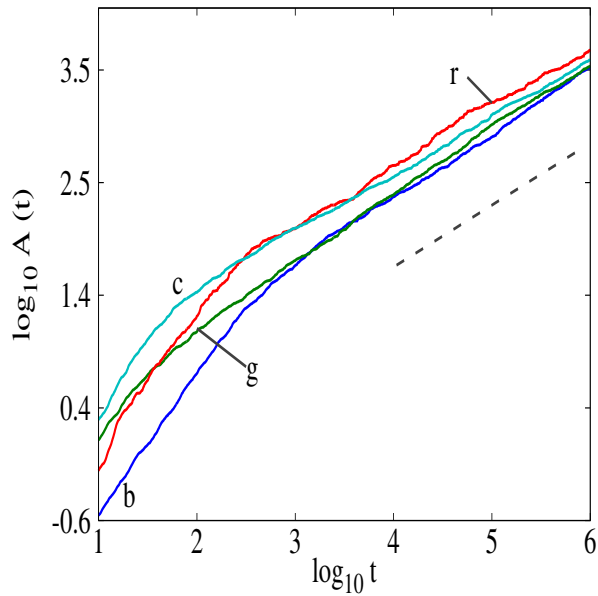}
    \caption{Similar to Fig.~\ref{fig:weak_chaos_mdpdarr_and_power_exponent_2d_ddnls_02}, but for the strong chaos regime of the 2D DDNLS system~\eqref{eq:hamilton_complex_ddnls_2d}.
    The dashed line indicates slope $0.55$.
    The curve's colors correspond to the ones used in Fig.~\ref{fig:strong_chaos_m2_2d_ddnls}.
}
    \label{fig:strong_chaos_mdpdarr_and_power_exponent_2d_ddnls_02}
\end{figure}

The time dependence of the area $A(t)$~\eqref{eq:range_covered_by_DVD_2d_02} visited by the DVD center is shown in Fig.~\ref{fig:strong_chaos_mdpdarr_and_power_exponent_2d_ddnls_02} for all the strong chaos cases of Fig.~\ref{fig:strong_chaos_mdpdarr_and_power_exponent_2d_ddnls}.
As in the case of weak chaos, a monotonic increase of the $A$ values is observed for all cases.
At the final integration time ($\log_{10} t = 10^6$), this quantity has reached values $A \approx 10^{3.5} \approx 3162$ for all cases, which are much larger than any NM average localization volume [see Table~\ref{tab:}].
In addition, the values of $A(t)$ are well fitted by a power law $A \propto t^{\alpha _A}$ with $\alpha _A \approx 0.55$.
Consequently, this provides a strong indication that the chaotic seeds are consistently visiting sites far from the central part of the lattice.
It is worth mentioning that similar results was also obtained for the 2D DKG system~\eqref{eq:hamilton_2d_dkg} in~\citep{manda2020chaotic}.

\section{\label{sec:summary_2d_ddnls}Discussion and summary}
Let us now summarize the main features of the chaotic dynamics of the wave packet spreading in the 2D DDNLS lattice~\eqref{eq:hamilton_complex_ddnls_2d}.  
The chaos strength of the wave packet spreading was performed through the computation of the system's ftMLE $\Lambda(t)$~\eqref{eq:finite_mle}.
More specifically, we found that the $\Lambda(t) \propto t^{-0.37}$ in the weak and $\Lambda(t)\propto t^{-0.46}$ for the strong chaos regimes.
These behavior are clearly different from what is observed in the case of regular motion ($\Lambda \propto t^{-1}$), confirming that the wave packet remains chaotic although its chaos strength decreases in time.
As to the question of whether wave packet chaotization is fast enough to support the wave packet subdiffusion, we can compare the wave packet's spreading timescale $T_M = 1/D$, with $D = dm_2/dt$ to its Lyapunov time $T_L = 1/\Lambda$~\eqref{eq:lyapunov_time_1d}.
By doing that, we find that since $m_2\sim t^{\alpha_m}$ and $\Lambda \sim t^{\alpha_\Lambda}$ 
\begin{equation}
    \frac{T_M}{T_L} \sim t^{1 - \alpha _\Lambda - \alpha _m},
\end{equation}
which leads to 
\begin{equation}
    \frac{T_M}{T_L} \sim t^{0.43},
    \label{eq:2d_weak_chaos_ratio}
\end{equation}
for the weak chaos regime where $\alpha _m = 1/5$ and $\alpha _\Lambda = -0.37$, and to 
\begin{equation}
    \frac{T_M}{T_L} \sim t^{0.21},
    \label{eq:2d_strong_chaos_ratio}
\end{equation}
for the strong chaos regime where $\alpha _m = 1/3$ and $\alpha _\Lambda = -0.46$.
Thus, $T_L$ is always smaller than $T_M$, such that the wave packet chaotization happens faster than its spreading to more sites.

By examining the ratios $T_M/T_L$ obtained in all spreading dynamical regimes for both the 1D and 2D DDNLS models i.e.
\begin{equation}
    \frac{T_M}{T_L} \sim t^{0.43} \quad  \mbox{(weak chaos), } \quad \frac{T_M}{T_L} \sim t^{0.21} \quad \mbox{(strong chaos)}
    \label{eq:tm_tl_1d}
\end{equation}
obtained in Eq.~\eqref{eq:2d_weak_chaos_ratio} and~\eqref{eq:2d_strong_chaos_ratio} for the 2D DDNLS model~\eqref{eq:hamilton_complex_ddnls_2d} and 
\begin{equation}
    \frac{T_M}{T_L} \sim t^{0.42} \quad  \mbox{(weak chaos),  } \quad \frac{T_M}{T_L} \sim t^{0.20} \quad \mbox{(strong chaos)},
    \label{eq:tm_tl_2d}
\end{equation}
observed for the 1D DDNLS system~\eqref{eq:Hamilton_complex_dnls_1d} in Sec.~\ref{sec:discussion_1d_spreading}, we see a remarkable closeness of exponents in Eqs.~\eqref{eq:tm_tl_1d} and~\eqref{eq:tm_tl_2d}.
This is a strong indication that nonlinear interactions of the same nature are responsible for the wave packet spreading in the cases of both spatial dimensions within the thermal region of the energy-norm parameter space. 
This is in agreement with the theoretical calculations performed in Sec.~\ref{sec:spreading_mechanism}.
Indeed, for both the 1D and 2D DDNLS systems, we ended up with the same reduced equation of the asymptotic dynamics 
\begin{equation}
    i\frac{\partial \phi_\nu }{\partial t} \sim E_\nu \phi_\nu + \beta \mathcal{P} \left(\beta \phi_{\nu_1}^2\right) \phi_{\nu_1}^3 f(t).
    \label{eq:asymp_dyn_eq_motion_2d_chaotic}
\end{equation} 
Here, the nonlinear term $\beta \mathcal{P} \left(\beta \phi_{\nu_1}^2\right) \phi_{\nu_1}^3 f(t)$ is responsible of the thermalization of more sites in both lattices.
Note that in Eq.~\eqref{eq:asymp_dyn_eq_motion_2d_chaotic} $\nu$ and $\nu _1$ represent respectively the non-thermalized and thermalized modes.
Thus, only the geometry of the problem affects the spreading and the chaotic dynamics of the wave packet.

Notice that equating the $T_M/T_L$ ratios of Eqs.~\eqref{eq:tm_tl_1d} and~\eqref{eq:tm_tl_2d} for each dynamical regime, we can write
\begin{equation}
    \frac{\Lambda^1}{m_2^1} = \frac{\Lambda}{m_2},
    \label{eq:scaling_chaotic_spreading_1d_&_2d}
\end{equation}
where the subscript `1' refers to the 1D lattice.
Equation~\eqref{eq:scaling_chaotic_spreading_1d_&_2d} allows the prediction of the characteristics of chaos of the wave packet dynamics.
For example, using the relations describing the time evolution of the wave packet second moment $m_2$ in Chap.~\ref{chap:spreading} and the results obtained for the computation of the ftMLEs of the 1D DDNLS system in Chap.~\ref{chap:chaotic_one_dimensional}, we get from Eq.~\eqref{eq:scaling_chaotic_spreading_1d_&_2d}
\begin{equation}
    \Lambda \sim t^{\alpha _m - \alpha _m^1 + \alpha_\Lambda^1}.
\end{equation}
Then, we obtain for the behavior of the ftMLE in the case of the 2D lattice
\begin{equation}
    \Lambda \sim t^{-0.38},
    \label{eq:law_weak_chaos_scale}
\end{equation}
for the weak chaos regime where $\alpha _m^1 = 1/3$, $\alpha _m = 1/5$ and $\alpha _\Lambda = -0.25$, and 
\begin{equation}
    \Lambda \sim t^{-0.47},
    \label{eq:law_strong_chaos_scale}
\end{equation}
in the strong chaos case where $\alpha _m^1 = 1/2$, $\alpha _m = 1/3$ and $\alpha _\Lambda = -0.3$.
Both laws~\eqref{eq:law_weak_chaos_scale} and~\eqref{eq:law_strong_chaos_scale} are extremely similar to the $\Lambda \sim t^{-0.37}$ and $\Lambda \sim t^{-0.46}$ found numerically in Sec.~\ref{sec:mle_2d_ddnls_spreading}.

Similarly to what was seen in the 1D case, the evolution of the DVD revealed the mechanism of wave packet chaotization: Localized chaotic seeds randomly oscillate in the wave packet's interior, homogenizing chaos for the lattice's excited region and thermalizing neighboring initially unexcited sites.
Indeed, our numerical computations showed a DVD's second moment $m_2^D(t)$~\eqref{eq:second_moment_2d_chaotic} which slowly increases in time following a $m_2(t) \propto t^{0.12} - t^{0.17}$ law.
In addition, the DVD's participation number $P^D$ saturated at values of the order of the NM average localization volume.
Thus, the DVD always retains a rather localized and pointy shape, which randomly meanders inside the wave packet's interior in both the weak and strong chaos regimes.

The numerical computations of the area $A$~\eqref{eq:range_covered_by_DVD_2d} visited by the the DVD's center showed a monotonically increasing relation with time, reaching values $A \approx 300$ sites in the weak and $A \approx 3000$ sites in the strong chaos regimes for final integration time $t_f \approx 2\times 10^{6}$.
Fitting the variations of $A$ with the power law $t^{\alpha_R}$, we found $\alpha _A \approx 0.5$ (weak chaos) and $\alpha _A \approx 0.55$ (strong chaos).
The fact that we observed larger values of $A$ and larger increase rates in the strong chaos case compared to the weak chaos one is in agreement with the results of Chap.~\ref{chap:spreading}~\citep[see~e.g.][]{flach2010spreading}.
Indeed, as in the 1D DDNLS lattice, the faster spreading in the strong chaos case implies that chaotic seeds visit larger regions of the lattice as faster homogenization of chaos and thermalization of more lattice sites is required.

The fact that all these behaviors have been found in both the 1D and 2D lattices of the DDNLS and DKG models is a clear indication of the generality of our findings.

\chapter{\label{chapter:conclusion and future outlook}Summary and outlook}

This work presented results regarding the contribution of chaotic dynamics to wave packet spreading in disordered multidimensional Hamiltonian models.
In particular, we focussed on the well known one (1D) and two dimensional (2D) disordered discrete nonlinear Schr\"odinger (DDNLS) models, which are generic in modelling physical systems such as the conductivity of metals, the propagation of light in optical media or the expansion of trapped Bose-Einstein condensates.
The aim of this work was to give more insights into the long standing question of whether a wave packet spreading in a disordered medium will expand forever or halt after a certain amount of time, based on the computation of observables of the system's tangent dynamics such as the maximum Lyapunov exponent (MLE) and its related deviation vector distribution (DVD).
This question is of utmost importance as it links up to several physical aspects like the ergodic, chaotic and statistical properties of multidimensional nonlinear systems which are still widely debated.
In the next paragraphs, we recapitulate our work and give some of the possible open avenues along which this study could be taken forward.

As a preliminary step to reaching our goal, in Chap.~\ref{chap:spreading} we presented the theory behind the wave packet spreading in the 1D and 2D DDNLS systems.
Firstly, we outlined the so called Anderson localization (AL), whereby in a disordered linear model, localized excitations remain in that state forever, not diffusing through the lattice. 
Then, we discussed what happens to the AL when nonlinearity is introduced into the system.
Upon examining the dynamical behavior of normal modes (NMs), we found that when AL prevails, the NMs possess a discrete frequency spectrum coupled to a spatially localized nature which gets more pronounced when the disorder becomes stronger in the system.
On the other hand, when nonlinearity is switched on, the NMs overlap and interact, allowing the energy to move from mode to mode in a deterministic chaotic manner.
Consequently, the AL does not survive, leading to an avenue for a subdiffusive spreading of the initially localized excitation. 
Depending on the strength of the overlapping part, which is expressed as a function of the norm $\mathcal{S}$, the total energy $\mathcal{H}$, the magnitude of the nonlinear coefficient $\beta$ and the disorder strength $W$, we discussed how two different subdiffusive behaviors of the localized excitation can take place.
In particular, the so-called weak chaos regime where the wave packet extent, measured through the second moment $m_2$~\eqref{eq:second_moment_distr_num} [participation number $P$~\eqref{eq:participation_ratio_num}], increases according to $m_2 \sim t^{\alpha_m}$ [$P\sim t^{\alpha_P}$] with $\alpha_m = 1/3$ [$\alpha_P = 1/6$] for the 1D DDNLS model and $\alpha_m = 1/5$ [$\alpha _P= 1/5$] in the case of the 2D system. 
In addition, the so-called strong chaos regime, with a faster wave packet expansion where the power law exponents for the $m_2$ [$P$] are $\alpha_m = 1/2$ [$\alpha_P = 1/4$] in the case of the 1D and $\alpha_m = 1/3$ [$\alpha_P = 1/3$] for the 2D DDNLS system can also appear.

In order to investigate the chaotic dynamics of the spreading regimes mentioned above, we mainly relied on numerical experiments.
This is motivated by the difficulty involved with an analytical treatment of the stochasticity appearing both in the system's equations of motion and tangent dynamics.
Due to the complexity of integrating the equations of motion of the DDNLS models compared to other multidimensional lattice models like the disordered Klein-Gordon (DKG) and Fermi-Pasta-Ulam-Tsingou (FPUT) chains of nonlinear oscillators, no particular effort in designing efficient numerical methods for integrating the variational equations of the former were undertaken to date.
Thus, a large part of Chap.~\ref{chapter:num_integration} is dedicated to designing efficient, accurate and fast numerical integration techniques for the long time integration of the system's equations of motion and the variational equations required in this study.
We exploited symplectic integrators (SIs), which are numerical schemes specifically built for integrating the equations of motion of Hamiltonian models, along with the tangent map method~\citep[see~e.~g.][]{skokos2010numerical} in order to evolve the variational equations.
Note that SIs keep the computed energy of the system constant up to a certain level of accuracy which can be associated with the precision of the integrator.
We presented methods based on the two part (only for the 1D DDNLS model) and three part split approaches of the Hamiltonian functions of the 1D and 2D DDNLS systems and compared their performances based on standard schemes like the $DOP853$~\citep{hairer1993solving}, a Runge-Kutta (RK) integrator of order $8$ based on the Dormand and Prince method and the \textit{TIDES} algorithm based on Taylor/Lie series expansions of the system's field equations~\citep{abad2012algorithm,abad2014tides}. 

Then, we studied the performance of the designed SIs in Sec.~\ref{sec:num_comparison_num} of Chap.~\ref{chapter:num_integration} based on their ability to accurately reproduce the system's dynamical behavior.
For instance, this was quantified by checking the constancy of the total energy $\mathcal{H}$ and norm $\mathcal{S}$ through the computation of the relative energy and norm errors, respectively denoted as $E_r$ and $S_r$.
We also calculated the characteristics of the wave packet spreading, i.e.~the time evolution of the second moment $m_2$ and the participation number $P$, and evaluated the precision of the evolution of the variational equations based on the computed finite time maximum Lyapunov exponent (ftMLE) $\Lambda$ for several initial conditions.  
We focussed on two energy accuracy levels: moderate accuracy with $E_r \lesssim 10^{-5}$, which is often considered to be a good precision in nonlinear lattice studies~\citep[see~e.g.][]{laptyeva2010crossover,skokos2013nonequilibrium}, as well as high accuracy where the threshold values $E_r \lesssim 10^{-9}$ were used~\citep[see~e.g.][]{sales2018sub}.
All the tested numerical integrators produced similar results for both the evolution of system's phase space and tangent dynamics.
Although the $DOP853$ and the \textit{TIDES} are the most accurate of all the integrators we tested, the fact that they do not preserve the values of the computed total energy and norm ($E_r$ and $S_r$ are not bounded from above) makes them a not ideal choice for the long time evolution of the system's dynamical equations.
On the other hand, we found that SIs in general take less CPU time than the $DOP853$ and \textit{TIDES} in order to reproduce the same system dynamics with the three part split SIs performing better than the two part split SIs for large lattice sizes, i.e.~when the number of degrees of freedom $N\gtrsim 70$.
In particular the $s11\mathcal{ABC}6$ and $s9\mathcal{ABC}6$ SIs of order $6$ along with the $s17\mathcal{ABC}8$ and $s19\mathcal{ABC}8$ SIs of order $8$ proved to be the best integrators respectively for moderate ($E_r \lesssim 10^{-5}$) and high ($E_r \lesssim 10^{-9}$) accuracies.
The results of these investigations were presented in a recent research paper~\citep{danieli2019computational}.

In Chap.~\ref{chap:chaotic_one_dimensional}, we computed the time evolution of the finite time MLE (ftMLE) and the spatiotemporal evolution of its related DVD for the 1D DDNLS model in the weak and strong chaos spreading regimes.
We considered a total of $8$ cases, $4$ cases in the weak chaos and $4$ in the strong chaos regimes.   
For each case, we averaged the observables of the system's phase space and tangent dynamics over $100$ different realizations i.e.~different values of the lattice's disorder parameters $\epsilon_l$, for norm and energy values located in the same thermal regime. 
In all cases, a decay of the ftMLE $\Lambda$ following the power law $\Lambda \sim t^{\alpha \Lambda}$ was seen. 
For the weak chaos case the power law exponent was fitted to $\alpha _\Lambda \approx -0.25$ while in the strong chaos regime, a smaller exponent value $\alpha _\Lambda \approx -0.3$ was found.
As both exponents are different from $\alpha_\Lambda = -1$ observed for regular motion, we concluded that these wave packet dynamics remain always chaotic up to the largest times of our simulations ($t_f \approx 10^8$).
We correlated the different values of $\alpha_\Lambda$ of the weak and strong chaos regimes to the wave packet spreading dynamics.
Indeed, as the wave packet spreading is faster in the strong chaos, the constant total norm is shared to a larger number of thermalized lattice sites resulting in a faster decay of the norm per site $s$ and consequently of the effective nonlinearity, which is approximately $\beta s^2$.
As a result, the system is driven to the integrable limit quicker in the strong chaos regime, and the $\alpha _\Lambda = -0.3$ value is closer to the regular case ($\alpha_\Lambda = -1$) compared to $\alpha_\Lambda = -0.25$ seen in the weak chaos case.

The motion of the DVD related to the computation of the MLE showed random fluctuations of chaotic spots (called chaotic seeds) which are known to homogenize chaos inside the excited part of the lattice and contribute to the wave packet thermalization.
The computation of the second moment $m_2^D$ and the participation number $P^D$ of the DVD in the weak chaos regime showed that the DVD extent increases slowly in time as $m_2^D \sim t^{0.14}$ and $P^D \sim t^{0.04}$, with $P^D$ showing a tendency to saturate to values of the order of the NM's localization volume.
This process is faster in the strong chaos case where a clear constancy of the $m_2^D$ and $P^D$ was visible from time $t \gtrsim 10^{5}$.
In addition, we computed the length $R$ of the region visited by the DVD center and found that this quantity monotonically increases in time for all cases, to values $R\approx 10^2$ and $R\approx 10^3$ at the largest times of our simulations in the weak and strong chaos regimes respectively.
These values of $R$ are much larger than the NM localization volume and therefore constitute clear numerical evidence that modes other than the initially excited ones also contribute to the wave packet thermalization.
The obtainment of larger $R$ values in the strong chaos case is likely related to the fact that a faster spreading of the wave packet is seen in that regime, so that chaotic seeds have to travel to larger excited regions of the lattice.
In addition, we found that the time dependence of $R$ is well fitted by $R \sim t^{\alpha _R}$, where $\alpha \approx 0.24$ for the weak chaos case, while the strong chaos case, exhibits higher values of $\alpha _R$ which decrease in time toward $\alpha_R \approx 0.24$.
This result can be explained by taking into account the fact that the strong chaos regime is a transient one, meaning that an asymptotic crossover of the chaotic dynamics of the system to the weak chaos behavior has to be seen after a finite amount of time.
A large part of the results of Chap.~\ref{chap:chaotic_one_dimensional} have been reported in~\citep{senyange2018characteristics}.

Chapter~\ref{chap:chaotic_two_dimensional}, focussed to the same questions as in Chap.~\ref{chap:chaotic_one_dimensional} but this time for the 2D DDNLS model.
Notice that, to the best of our knowledge, this is the first time that such a detailed investigation has been performed for this model mainly due to the fact that the integration of the equations of motion and variational equations of this system is a computationally very difficult task.
In our study, we considered $4$ initial conditions in the weak chaos and $4$ in the strong chaos regions.
For each case, we averaged the observables of the system's phase space and tangent dynamics over $50$ disorder realizations.
Our numerical computations confirmed the theoretical expected values of the wave packet spreading characteristics i.e.~$m_2 \sim t^{\alpha_m}$ with $\alpha _m = 1/5$ in the weak  and $\alpha _m  = 1/3$ for the strong chaos regimes respectively.
Furthermore, the computation of $\Lambda$ showed a decaying chaos strength, whose time dependence was well fitted by $\Lambda \sim t^{\alpha_\Lambda}$ with $\alpha_\Lambda < 0$ being  similar to the 1D system.
Actually, the values $\alpha_\Lambda \approx -0.37$ and $\alpha _\Lambda \approx -0.46$ were found respectively for the weak and strong chaos cases.
It is worth noticing again that these values are different from $\alpha_\Lambda = -1$ which is seen in the case of regular motion.
Thus the wave packet always remains chaotic in the 2D DDNLS model as well.   
Furthermore we found a dimension independent scaling between the wave packet dynamics (measured through the $m_2$) and its degree of chaoticity (measured through the $\Lambda$) between the 1D and 2D models.
It will be interesting to check if this relation also holds for three dimensional models as well.

Chaos is not the only ingredient needed for wave packet spreading, as the other necessary component is the random oscillations of chaotic seeds within the lattice's excited part.
The evolution of the DVD obtained during the computation of the ftMLEs in the 2D DDNLS system also revealed the existence of localized chaotic spots which randomly wander inside the excited part of the lattice.
Indeed, in the weak chaos case a timid spatial expansion of the DVD was observed over time as its second moment grew as $m_2^D \sim t^{0.12}$ and the participation number $P^D$ showed a tendency to saturate to values similar to the 2D NM's localization volume at later stages of the evolution.
The saturation of these observables was again much faster for the strong chaos cases as a relation of the form $P^D \sim t^{0}$ could be fitted to the data over a larger period of time compared to the weak chaos case.
In addition, we computed the area $A$ ($A\sim R^2$) visited by the DVD center and found a monotonic increase, leading to values much larger than the NM's localization volume.
Specifically, the values $A\approx 10^2$ and $A\approx 10^4$ are found for the weak and strong chaos regimes at the final time of integration, suggesting that the DVD visited sites far from the lattice's center.
Further, the temporal dependence of $A$ was well fitted by $A \sim t^{\alpha_A}$, where the values $\alpha_A \approx 0.5$ and $\alpha _A \approx 0.55$ were obtained for the weak and strong chaos regimes respectively.
Similarly to the 1D DDNLS model, the higher values of $A$ and stronger variations of $\alpha_A$ observed in the strong chaos regime can be related to the faster spreading of the wave packet in that regime, meaning that the DVD has to cover larger areas of the excited part of the lattice than in the weak chaos case.
Consequently, the mechanisms of chaos responsible of the wave packet spreading in the 1D and 2D DDNLS systems are similar.
Most of findings of this chapter have been reported in a recent research paper~\citep{manda2020chaotic}.

It is worth noting that similar results to the ones presented in Chaps.~\ref{chap:chaotic_one_dimensional} and~\ref{chap:chaotic_two_dimensional} of this thesis have been obtained for the 1D and 2D DKG models in~\citep{skokos2013nonequilibrium,senyange2018characteristics,manda2020chaotic}, emphasizing the generality of our findings.

Although our study contributed in comprehending several aspects of the dynamical behavior of localized initial excitations in disordered systems, several issues still need to be investigated.
For instance the energy-norm density spaces of the 1D and 2D DDNLS systems discussed in Sec.~\ref{sec:statistical_physics_spreading} revealed a more complex partition between the wave packet spreading regions and the thermal and non-thermal phases, also underlined in a recent work~\citep{kati2020density}.
Thus, the techniques used in this work to study the weak and strong chaos spreading regimes could be used in further investigating the connection between the wave packet spreading and the ergodic properties of these two models.
Another important aspect is the theoretical determination of the observed power law exponents of the ftMLE seen in this work for both the 1D and 2D DDNLS systems.
Achieving this goal could help, for instance in predicting the system's dynamical characteristics for nonlinear strengths and model's dimensionalities far from the reach of current computer simulations. 
In addition, we pinned the motion of the DVD as random and therefore the evolution of the DVD center could be approximated by the spatial evolution of a random walker evolving in a field of one or two spatial dimensions.
This means that, the total length/area covered represents only a single possible measurement among many others of the characteristics of the DVD as a wide theory exists around random walks.
Furthermore, as the DVD generated during the computation of the MLE shows the location of only the chaotic spot with the largest instantaneous chaoticity, another important challenge is related to the construction of a method to investigate most, if not all, of the chaotic spots, as they are central in understanding the NM interactions.
Pursuing methods to extend all these ideas could potentially reveal more subtle changes in the wave packet spreading dynamical behavior, allowing a better in depth understanding of chaos properties in multidimensional systems.

\chapter*{\label{chap:list_of_abbreviations}List of abbreviations}
\addcontentsline{toc}{chapter}{Abbreviations}
\pagestyle{fancy}
\fancyhf{}
\fancyhead[OC]{\leftmark}
\fancyhead[EC]{\rightmark}
\cfoot{\thepage}
The following abbreviations are used throughout the content of this thesis.
\begin{itemize}
    \item 1D -- One-Dimensional/One-Dimension.
    \item 2D -- Two-Dimensional/Two-Dimension.
    \item AL -- Anderson Localization.
    \item BCH -- Baker-Campbell-Hausdorff.
    \item CPU -- Central Processing Unit.
    \item DDLS -- Disordered Discrete Linear Schr\"odinger.
    \item DDLSE -- Disordered Discrete Linear Schr\"odinger Equation.
    \item DDNLS -- Disordered Discrete NonLinear Schr\"odinger.
    \item DKG -- Disordered Klein-Gordon.
    \item DOP -- Dormand and Prince.
    \item DV -- Deviation Vector.
    \item DVD -- Deviation Vector Distribution.
    \item FPUT -- Fermi-Pasta-Ulam-Tsingou.
    \item ftMLE -- finite-time Maximum Lyapunov Exponent.
    \item KAM -- Kolmogorov-Arnol'd-Moser.
    \item MLE -- Maximum Lyapunov Exponent.
    \item NM -- Normal Mode.
    \item ODE -- Ordinary Differential Equation.
    \item RK -- Runge-Kutta.
    \item SI -- Symplectic Integration/Symplectic Integrator.
\end{itemize}



\begin{appendix}
\chapter{\label{app:conv_linear_ddnls_dkg}Frequency spectrum of the 1D DDLS model and the 1D DKG system of harmonic oscillators}

\section{\label{sec:linear_schroedinger_app}The 1D DDLS model}
Here we analytically determine the characteristics of the bandwidth of the frequency spectrum of the DDLS model.
Its Hamiltonian takes the form [see Eq.~\eqref{eq:hamilton_anderson_model}]
\begin{equation}
    \mathcal{H}_L = \sum _l \epsilon _l \lvert \psi _l\rvert ^2 - \left(\psi _l ^\star\psi _{l + 1} + \psi _l \psi _{l  + 1}^\star\right),
    \label{eq:hamilton_Anderson_app}
\end{equation}
where $\psi _l\in \mathbb{C}$ is the wave function and $\epsilon _l$ the disorder parameter at site $l$.
The values of ${\epsilon _l}$ in~\eqref{eq:hamilton_Anderson_app} are chosen from a uniform probability distribution $\mathcal{P}_W = 1/W$ on the interval $[-W/2, W/2]$, with $W$ denoting the system's disorder strength.
As has already been mentioned many times in this thesis, plane waves describe the solution of Hamiltonian~\eqref{eq:hamilton_Anderson_app}.
Consequently we write the wave function at site $l$ with norm $s_l$ and phase $\theta _l$ (which modulates the value of the energy $h_l$), as 
\begin{equation}
    \psi _l = \sqrt{s_l} \exp \left(i \theta _l\right).
    \label{eq:sol_ddls_plane_wave_general}
\end{equation}
In this thesis we have been working (without loss of generality) with homogeneous initial wave packets so that $s = s_l$, for all the initially excited sites.
In this situation, the Hamiltonian~\eqref{eq:hamilton_Anderson_app} becomes
\begin{equation}
    \mathcal{H}_L = \sum _l \epsilon _l s - 2s \cos \left(\theta _{l + 1} - \theta _l\right).
    \label{eq:hamilton_anderson_app_2}
\end{equation}
Since the cosine function is bounded, the value of $\mathcal{H}_L$ can also be bounded as 
\begin{equation}
   \sum _l \left(\epsilon _l -2\right)s \leq \sum _l \epsilon _l s - 2s \cos \left(\theta _{l + 1} - \theta _l\right)\leq \sum _l \left(\epsilon _l +2\right)s.
    \label{eq:hamilton_bounded_app}
\end{equation}
In particular, the extreme case for the lower bound of~\eqref{eq:hamilton_bounded_app} leads to a minimum value when all sites have $\epsilon _l ^{-} = -W/2$.
In the same way, the scenario corresponding to the maximum possible value of the upper bound in~\eqref{eq:hamilton_bounded_app} is all sites having $\epsilon _l ^{+} = W/2$.
Thus,
\begin{equation}
    \sum _l \left(-\frac{W}{2} -2\right)s \leq \sum _l \epsilon _l s - 2s \cos \left(\theta _{l + 1} - \theta _l\right)\leq \sum _l \left(\frac{W}{2} +2\right)s.
    \label{ineq:bounded_linear_anderson_model_app}
\end{equation}
Summing up each term of~\eqref{ineq:bounded_linear_anderson_model_app}, we get 
\begin{equation}
    \mathcal{H}_{L}^{-} = E ^{-} \mathcal{S}_{1D} \leq \mathcal{H}_L \leq E ^{+} \mathcal{S}_{1D} = \mathcal{H}_{L}^{+},
    \label{eq:bounded_hamilton_anderson_full_app}
\end{equation}
with 
\begin{equation}
    E ^{-}  = -2 - \frac{W}{2}, \quad E ^{+} = 2 + \frac{W}{2},
    \label{eq:boundaries_freq_spectrum_ddls_app}
\end{equation}
and $\mathcal{S}_{1D} = \sum _l s$ being the total norm of the system which is a conserved quantity. 
It is worth noticing that the Hamiltonian functions $\mathcal{H}^{\pm}_L = E ^{\pm}S_{1D}$ model independent harmonic oscillators, each one of which has the same proper frequency $E^{\pm}$.
The inequality in~\eqref{eq:bounded_hamilton_anderson_full_app} involves solely macroscopic quantities and therefore is valid for all kinds of initial states of the system.

Applying the linear differentiation operator $\frac{\partial }{\partial (i \psi_l ^\star)}$ in~\eqref{eq:bounded_hamilton_anderson_full_app}, we obtain 
\begin{equation}
    \begin{split}
        \begin{pmatrix}
            E^{-} & 0 &  \ldots & 0 & 0 \\
            0 & E^{-} & \ldots & 0 & 0 \\
            \vdots & \vdots & \ddots & \vdots & \vdots \\
            0 & 0 & \ldots & E^{-} & 0 \\
            0 & 0& \ldots & 0 & E^{-}
        \end{pmatrix}
        \begin{pmatrix}
            \psi _1 \\
            \psi _2 \\
            \vdots \\
            \psi _{N - 1} \\
            \psi _N
        \end{pmatrix}
        &\leq 
        \begin{pmatrix}
            \epsilon _ 1 & -1 & 0 & \ldots & 0 & 0 \\
            -1 & \epsilon _2 & -1 & \ldots & 0 & 0 \\
            \vdots & \vdots & \vdots & \ddots & \vdots & \vdots \\
            0 & 0 & 0 & \ldots & \epsilon _{N - 1} & 0  \\
            0 & 0 & 0 & \ldots & -1 & \epsilon _{N} \\
        \end{pmatrix}
        \begin{pmatrix}
            \psi _1 \\
            \psi _2 \\
            \vdots \\
            \psi _{N - 1} \\
            \psi _N
        \end{pmatrix}
        \\
        &\leq 
        \begin{pmatrix}
            E^{+} & 0 &  \ldots & 0 & 0 \\
            0 & E^{+} & \ldots & 0 & 0 \\
            \vdots & \vdots & \ddots & \vdots & \vdots \\
            0 & 0 & \ldots & E^{+} & 0 \\
            0 & 0& \ldots & 0 & E^{+}
        \end{pmatrix}
        \begin{pmatrix}
            \psi _1 \\
            \psi _2 \\
            \vdots \\
            \psi _{N - 1} \\
            \psi _N
        \end{pmatrix},
    \end{split}
    \label{ineq:bounded_eq_motion_anderson_model_1}
\end{equation}
where the `$\leq$' compares the system's field vectors element by element.
Equation~\eqref{ineq:bounded_eq_motion_anderson_model_1}, can be reduced to a matrix form 
\begin{equation}
    \bm{D}^{-} \bm{\Psi} \leq \tilde{\bm{A}} \bm{\Psi} \leq \bm{D}^{+} \bm{\Psi},
    \label{ineq:bounded_eq_motion_anderson_model_2}
\end{equation}
with 
\begin{equation}
    \bm{D}^{\pm} =     
    \begin{pmatrix}
        E^{\pm} & 0 &  \ldots & 0 & 0 \\
        0 & E^{\pm} & \ldots & 0 & 0 \\
        \vdots & \vdots & \ddots & \vdots & \vdots \\
        0 & 0 & \ldots & E^{\pm} & 0 \\
        0 & 0& \ldots & 0 & E^{\pm}
    \end{pmatrix}, 
    \quad \text{and} \quad 
    \bm{\Psi}  = 
    \begin{pmatrix}
        \psi _1 \\
        \psi _2 \\
        \vdots \\
        \psi _{N - 1} \\
        \psi _N
    \end{pmatrix}.
\end{equation}
In addition in~\eqref{ineq:bounded_eq_motion_anderson_model_2} $\bm{\tilde{A}}$ is a tridiagonal matrix with entries
\begin{equation}
    \tilde{\bm{A}} = 
    \begin{pmatrix}
        \epsilon _ 1 & -1 & 0 & \ldots & 0 & 0 \\
        -1 & \epsilon _2 & -1 & \ldots & 0 & 0 \\
        \vdots & \vdots & \vdots & \ddots & \vdots & \vdots \\
        0 & 0 & 0 & \ldots & \epsilon _{N - 1} & -1  \\
        0 & 0 & 0 & \ldots & -1 & \epsilon _{N} \\
    \end{pmatrix}.
\end{equation}
The matrix $\tilde{\bm{A}}$ is diagonalizable~\citep{gershgorin1931uber,wolkowicz1980bounds,smith1985numerical,losonczi1992eigenvalues,yueh2005eigenvalues,da2019eigenpairs}.
Its corresponding diagonal matrix $\bm{D}$ has as diagonal elements the eigenvalues $E_\nu$, $\nu =1, 2, \ldots, N$ of $\tilde{\bm{A}}$.
Defining $\bm{P}$ a matrix having as columns the normalized eigenvectors related to the eigenvalues of $\bm{D}$, we get
\begin{equation}
    \bm{D}^{-} \bm{\Psi} \leq \bm{P} \bm{D} \bm{P}^{-1} \bm{\Psi}  \leq \bm{D}^{+} \bm{\Psi}.
    \label{ineq:bounded_eq_motion_anderson_model_3}
\end{equation}
Multiplying from the left side~\eqref{ineq:bounded_eq_motion_anderson_model_3} by $\bm{P}^{-1}$ leads to 
\begin{equation}
    \bm{D}^{-} \bm{P}^{-1}\bm{\Psi} \leq \bm{D} \bm{P}^{-1} \bm{\Psi} \leq \bm{D}^{+} \bm{P}{-1}\bm{\Psi},
    \label{ineq:bounded_eq_motion_anderson_model_4}
\end{equation}
where we used $\bm{P}^{-1}\bm{P} = \bm{P}\bm{P}^{-1}= \bm{I}_N$ with $\bm{I}_N$ being the $N\times N$ identity matrix.
Setting the state vector $\bm{U} = \bm{P}^{-1}\bm{\Psi}$, we therefore end up with a set of uncoupled oscillators 
\begin{equation}
    \bm{D}^{-} \bm{U} \leq \bm{D} \bm{U}  \leq \bm{D}^{+} \bm{U},
    \label{ineq:bounded_eq_motion_anderson_model_5}
\end{equation}
whose frequency spectra can be compared wave number by wave number in the matrices of~\eqref{ineq:bounded_eq_motion_anderson_model_5}. 
What comes out from this comparison is the inequality
\begin{equation}
    E ^{-} \leq E_\nu \leq E ^{+},
    \label{ineq:bounded_freq_anderson_model_app}
\end{equation}
with $E_\nu$, $\nu = 1, 2, \ldots, N$ representing the eigenvalues of matrix $\tilde{\bm{A}}$.
Therefore, the width of the frequency spectrum is 
\begin{equation}
    \Delta = E ^{+} - E ^{-} = W + 4,
\end{equation}
since
\begin{equation}
    E_\nu \in \left[-2 - \frac{W}{2}, 2 + \frac{W}{2}\right].
\end{equation}

\section{\label{sec:linear_kg_model_app}The 1D DKG system of harmonic oscillators}
We can express the linear characteristics of the DKG model in terms of the DDLS model.
The Hamiltonian function of the linear DKG model of harmonic oscillators reads~\citep[see~e.g.][]{skokos2009delocalization}
\begin{equation}
    \mathcal{H}_{LK} =\sum _{l } \frac{p_l^2}{2} + \frac{\tilde{\epsilon} _l }{2} q_l ^2 + \frac{1}{2W} \left(q_{l+1} - q_{l} \right)^2,
    \label{eq:hamilton_1d_linear_dkg_app}
\end{equation}
and its related equations of motion can be written as
\begin{equation}
    \ddot{q}_l = \tilde{\epsilon}_l q_l + \frac{1}{W} \left(2q_l  - q_{l - 1} - q_{l + 1}\right), 
    \label{eq:eqtmot_1d_linear_dkg_app}
\end{equation}
where $q_l$ and $p_l$ are respectively the generalized position and conjugate momentum and $\tilde{\epsilon}_l$ denotes disorder parameter at site $l$.
The values of $\epsilon_l$ are chosen on the interval $[1/2, 3/2]$ following a random uniform distribution.
Here $W$ denotes the disorder strength because it controls the size of the normal mode's localization volume $\overline{V}$~\citep{skokos2009delocalization,senyange2020properties}.

The solutions of~\eqref{eq:eqtmot_1d_linear_dkg_app} are of the form 
\begin{equation}
    q_l = A_l \exp \left(i \omega t\right),
    \label{eq:plane_wave_dkg_1d_app}
\end{equation}
with $A_l$ being the amplitude at site $l$ and $\omega$ the frequency of oscillation.
By substituting~\eqref{eq:plane_wave_dkg_1d_app} to~\eqref{eq:eqtmot_1d_linear_dkg_app} we obtain the following eigenvalue problem
\begin{equation}
    \left(\omega ^2 W - W - 2\right)A_l = W\left(\tilde{\epsilon}_l - 1\right)A_l - A_{l - 1} - A_{l + 1}.
    \label{eq:eqtmot_1d_linear_dkg_app_2}
\end{equation}
The normal modes of~\eqref{eq:eqtmot_1d_linear_dkg_app_2} $\bm{\Psi}_\nu = (A_{\nu, 1}, A_{\nu, 2}, \ldots, A_{\nu, N})$, match the ones of the DDLS system~\eqref{eq:Hamilton_complex_dnls_1d} given in~\eqref{eq:eigenvalue-vectors_eq} by setting 
\begin{align}
    \label{eq:condition_eqv_ddnls_dkg_app_1}
    E &= \omega ^2 W - W - 2, \\
    \label{eq:condition_eqv_ddnls_dkg_app_2}
    \epsilon _l &= W\left(\tilde{\epsilon}_l - 1\right).
\end{align}
Consequently Eq.~\eqref{eq:condition_eqv_ddnls_dkg_app_1} allows us to determine the limits of the frequency spectrum for the DKG system of linear oscillators~\eqref{eq:hamilton_1d_linear_dkg_app} based on the results obtained for the DDLS model in Sec.~\ref{sec:linear_schroedinger_app}.
In particular, using~\eqref{eq:boundaries_freq_spectrum_ddls_app} the lower squared frequency $\omega_{-}^2$ of the bandwidth is 
\begin{equation}
    E^{-} = \omega_{-}^2 W - W - 2 \quad \Rightarrow \quad -2 - \frac{W}{2} = \omega_{-}^2W - W - 2 \quad \Rightarrow \quad \omega_{-}^2 = \frac{1}{2},
\end{equation}
and the upper bound $\omega _{+}^2$ is given by 
\begin{equation}
    E^+ = \omega_{+}^2 W - W - 2 \quad \Rightarrow \quad 2 + \frac{W}{2} = \omega_{+}^2 W - W - 2 \quad \Rightarrow \quad \omega_{+}^2 = \frac{3}{2} + \frac{4}{W}.
\end{equation}
Thus the size of the squared frequency spectrum of the DKG system of linear oscillators is 
\begin{equation}
    \Delta _{K} = \frac{3}{2} + \frac{4}{W} - \frac{1}{2} \quad \Rightarrow \quad  \Delta _{K} = 1 + \frac{4}{W}.
\end{equation}
A more general derivation of this relation can be found in~\citep{senyange2020properties}.

\chapter{\label{app:eq_mot_normal_mode_space}Equations of motion of the 1D DDNLS system in the NM space}

We recall the equations of motion of the 1D DDNLS system~\eqref{eq:eq_mot_1d_ddnls} are 
\begin{equation}
    i\dot{\psi} _l = \epsilon _l \psi _l + \beta \lvert\psi _{l}\rvert^2 \psi _l - \left(\psi_{l - 1} + \psi _{l+1} \right).
    \label{eq:eq_mot_1d_ddnls_app}
\end{equation}
Writing the solution of Eq.~\eqref{eq:eq_mot_1d_ddnls_app} in the NM space of the system as a weighted sum over its linear NMs $\bm{\Psi} _{\nu} \propto \overline{\zeta}^{-1/2}\exp\left[-(l - \overline{l}_\nu)/\overline{\zeta}\right]$
\begin{equation}
    \psi _l = \sum _\nu \phi _\nu \bm{\Psi}_{\nu} = \sum _\nu \phi _\nu A_{\nu, l},
    \label{eq:general_solution_psi_linear_app}
\end{equation}
where $\bm{\Psi} _{\nu} = \left(A_{\nu, 1}, A_{\nu, 2}, \ldots,A_{\nu, N}\right)$ is the NM and $\phi _\nu$ the time dependent amplitude at mode $\nu$ with $\nu = 1, 2, \ldots, N$ ($N$ being the lattice size), and substituting this expression in Eq.~\eqref{eq:eq_mot_1d_ddnls_app}, we obtain 
\begin{equation}
    i \sum _\nu A_{\nu, l}\dot{\phi}_\nu = \sum _\nu \epsilon _l  A_{\nu, l}\phi _\nu - \sum _\nu \phi _\nu A_{\nu, l - 1} -  \sum _\nu \phi _\nu A_{\nu, l + 1} + \beta \sum _{\nu _1 \nu _2 \nu _3} A_{\nu _1, l}^\star A_{\nu _2, l}A_{\nu _3, l} \phi _{\nu_1}^\star \phi _{\nu_2} \phi _{\nu_3}.
    \label{eq:eqt_motion_transit_nm_space_app}
\end{equation}
Taking into account Eq.~\eqref{eq:eigenvalue-vectors_eq} i.e.~$EA_l = \epsilon_l A_l - A_{l-1} - A_{l+1}$, we rewrite~\eqref{eq:eqt_motion_transit_nm_space_app} as 
\begin{equation}
    i \sum _\nu A_{\nu, l}\dot{\phi}_\nu = \sum _\nu E_\nu A_{\nu, l}\phi _\nu + \beta \sum _{\nu _1 \nu _2 \nu _3} A_{\nu _1, l}^\star A_{\nu _2, l}A_{\nu _3, l} \phi _{\nu_1}^\star \phi _{\nu_2} \phi _{\nu_3},
    \label{eq:eqt_motion_transit_nm_space_app_2}
\end{equation}
where $E_\nu$ is the frequency of the normal mode $\nu$.
We then multiply Eq.~\eqref{eq:eqt_motion_transit_nm_space_app_2} by $\bm{\Psi}_{\nu^\prime}^\star$, and get 
\begin{equation}
    i \sum _\nu \sum_l A_{\nu^\prime, l}^{\star}A_{\nu, l}\dot{\phi}_\nu = \sum _\nu \sum_l A_{\nu^\prime, l}^{\star} E_\nu A_{\nu, l}\phi _\nu + \beta \sum _{\nu _1 \nu _2 \nu _3} \sum_l A_{\nu^\prime, l}^{\star} A_{\nu _1, l}^\star A_{\nu _2, l}A_{\nu _3, l} \phi _{\nu_1}^\star \phi _{\nu_2} \phi _{\nu_3}.
    \label{eq:eqt_motion_transit_nm_space_app_3}
\end{equation}
At this point it is useful to emphasize few things:
\begin{enumerate}
    \item[](a) The orthogonality of the NMs leads to
        \begin{equation}
            \bm{\Psi}_\nu ^\star \bm{\Psi}_\nu = \sum _l A_{\nu^\prime, l}^{\star} A_{\nu, l} = \delta _{\nu^\prime, \nu}.
            \label{eq:orthogonalization_of_freq_modes_app}
        \end{equation}
    \item[](b) In the considered set up, initial conditions are chosen so that the total norm of any initial wave packet is set to unity.
        This leads to the condition
        \begin{equation}
            \sum _l \lvert \psi _l\rvert ^2 = \sum _l \sum _{\nu^\prime}\sum _{\nu} \phi _{\nu^\prime} ^\star \phi _\nu \bm{\Psi}_{\nu^\prime}^\star \bm{\Psi}_\nu = \sum _{\nu} \sum _{l} A_{\nu, l}^\star A_{\nu, l} \phi _\nu^\star \phi _{\nu} = \sum _{\nu} \lvert \phi _\nu \rvert ^2 = 1.
            \label{eq:normalization_of_freq_amplitudes_app}
        \end{equation}
\end{enumerate}
Substituting Eq.~\eqref{eq:orthogonalization_of_freq_modes_app} into Eq.~\eqref{eq:eqt_motion_transit_nm_space_app_3} leads to
\begin{equation}
    i \dot{\phi}_{\nu^\prime} =  E_{\nu^\prime} \phi _{\nu^\prime} + \beta \sum _{\nu _1 \nu _2 \nu _3} I_{\nu^\prime, \nu _1, \nu_2, \nu_3} \phi _{\nu_1}^\star \phi _{\nu_2} \phi _{\nu_3},
    \label{eq:eqt_motion_transit_nm_space_app_4}
\end{equation}
where 
\begin{equation}
    I_{\nu^\prime, \nu _1, \nu_2, \nu_3}  = \sum_l A_{\nu^\prime, l}^{\star} A_{\nu _1, l}^\star A_{\nu _2, l}A_{\nu _3, l},
\end{equation}
is the so-called overlap-integral. 
Note that similar calculation can be performed for the DKG system~\citep{laptyeva2013nonlinear}.

\chapter{\label{app:statistical_mech}Statistical physics analytical computations}
\label{app_ex2}

\section{\label{sec:app_yb}Derivation of Eq.~\eqref{eq:calculated_y_stats}}
We give here the detailed calculation of the integral 
\begin{equation}
    \tilde{y}_{\beta} \left(\tilde{\nu}, \mu _l\right) = \int _{0}^{\infty} dx e^{-\tilde{\nu} \mu _l x} e^{-\tilde{\nu} \frac{\beta}{2} x^2},
    \label{eq:yt_full}
\end{equation}
appearing in Eq.~\eqref{eq:integral_over_a_norm_coord}.
In~\eqref{eq:yt_full}, $\tilde{\nu} = 1/T$ with $T$ being the temperature of the system, $\beta$ is the nonlinear coefficient of the 1D DDNLS model~\eqref{eq:Hamilton_complex_dnls_1d} and $\mu_l = \mu + \epsilon _l$ represents a kind of local chemical potential at site $l$ of the lattice with $\epsilon_l$ being the disorder parameter of $\mathcal{H}_{1D}$~\eqref{eq:Hamilton_complex_dnls_1d}.
From the Taylor expansion of the exponential function $e^x = \sum _{n = 0}^{\infty} x^n/n!$ we get
\begin{equation}
    e^{-\tilde{\beta}\frac{x^2}{2}} = 1 - \frac{\tilde{\beta}x^2}{2} + \frac{\tilde{\beta}^2 x^4}{8} + \ldots,
    \label{eq:exp_expansion_1}
\end{equation}
with $\tilde{\beta} = \beta \tilde{\nu}$.
Substituting~\eqref{eq:exp_expansion_1} into~\eqref{eq:yt_full} leads to 
\begin{equation}
    \tilde{y}_{\beta} \left(\tilde{\nu}, \mu _l\right) = \int _{0}^{\infty} dx e^{-\tilde{\nu} \mu _l x} \left[1 - \frac{\tilde{\beta}x^2}{2} + \frac{\tilde{\beta}^2 x^4}{8} + \ldots\right].
    \label{eq:yt_full_1}
\end{equation}
Consequently, we can write
\begin{equation}
    \tilde{y}_{\beta} \left(\tilde{\nu}, \mu _l\right) \approx \int _{0}^{\infty} dx e^{-\tilde{\nu} \mu _l x} - \int _{0}^{\infty} dx \frac{\tilde{\beta}x^2}{2} e^{-\tilde{\nu} \mu _l x} + \ldots.
    \label{eq:yt_full_3}
\end{equation}
The first integral in~\eqref{eq:yt_full_3} can be computed straightforwardly.
For the following terms, we use the fact that~\citep{abramowitz1964handbook,johansson2004statistical}
\begin{equation}
    \frac{\Gamma (n + 1)}{a^{n + 1}} = \int _{0}^{\infty} dx x^{n} e^{-ax},
    \label{eq:integral_exp_xn_1}
\end{equation}
where $a \in \mathbb{R}$ is a constant and $\Gamma$ is the usual $\Gamma$ function defined such that $\Gamma (n + 1) = n\Gamma (n)$ if $n \in \mathbb{N}$.
Therefore 
\begin{equation}
    \tilde{y}_{\beta}  \left(\tilde{\nu}, \mu _l\right) = \frac{1}{\tilde{\nu} \mu_l} - \frac{\tilde{\beta}}{2}\frac{\Gamma(3)}{(\tilde{\nu} \mu)^3} + \frac{\tilde{\beta}^2}{8}\frac{\Gamma (5)}{(\tilde{\nu} \mu_l)^5} + \ldots.
    \label{eq:yt_full_4}
\end{equation}
In the limit of $\tilde{\nu} \rightarrow 0$, because $\beta$ is finite $\tilde{\beta}\nu \rightarrow 0$ we can neglect higher order terms in $\tilde{\beta}$ and obtain 
\begin{equation}
    \tilde{y}_{\beta}  \left(\tilde{\nu}, \mu _l\right) \approx \frac{1}{\tilde{\nu} \mu _l} \left(1 - \frac{\beta \tilde{\nu}}{(\tilde{\nu} \mu _l)^2}\right).
    \label{eq:yt_full_5}
\end{equation}

\section{\label{secapp:ensemble_average}Derivation of Eq.~\eqref{eq:ham_norm_inf_temp}}
The total energy $\mathcal{H}_{1D}$ and the total norm $\mathcal{S}_{1D}$ of the 1D DDNLS system~\eqref{eq:Hamilton_complex_dnls_1d} can be calculated from the grand partition function $\mathcal{Z}$~\eqref{eq:partition_function_z_ln}~\citep{landau1958lifshitz,tong2012statistical}.
Here we evaluate both the average norm $s = \mathcal{S}_{1D}/N$ and energy $h=\mathcal{H}_{1D}/N$ per site in terms of the system's parameters at equilibrium.
We recall the partition function $\ln \mathcal{Z}$~\eqref{eq:partition_function_z_ln} containing all the information on the macroscopic states of the system was presented to be
\begin{equation}
    \ln \mathcal{Z} \approx N \ln 2\pi - \sum _{l =1}^{N} \left[\ln\left(\tilde{\nu} \mu _l\right) + \frac{\tilde{\nu} \beta}{\left(\tilde{\nu} \mu _l\right)^2} \right],
    \label{eq:part_function_ln_app}
\end{equation}
where $\tilde{\nu}=1/T$, with $T$ being the system temperature, $\beta$ the nonlinear coefficient in the expression of $\mathcal{H}_{1D}$~\eqref{eq:Hamilton_complex_dnls_1d} and $\mu_l = \mu + \epsilon _l$, where $\mu$ is a chemical potential and $\epsilon_l$ the disorder parameter in~\eqref{eq:Hamilton_complex_dnls_1d}. 
The total norm $\mathcal{S}_{1D}$ is given by
\begin{equation}
    \begin{split}
        \mathcal{S}_{1D} &=- \frac{1}{\tilde{\nu}} \frac{\partial}{\partial \mu _l} \ln \mathcal{Z}, \\
            &= - \frac{1}{\tilde{\nu}} \left\{-\sum _{l} \left[ \frac{\tilde{\nu}}{\tilde{\nu} \mu _l} - \frac{2 \tilde{\nu} ^2 \beta}{(\tilde{\nu} \mu _l)^3} \right] \right\}, \\
            &= \sum _{l} \frac{1}{\tilde{\nu} \mu _l} - \sum _{l} \frac{2\beta}{(\tilde{\nu} \mu _l)^2\mu _l}.  
    \end{split}
    \label{eq:norm_cal_4} 
\end{equation}
We note that in the infinite temperature limit $\tilde{\nu} \rightarrow 0$, while the product $\tilde{\nu} \mu$ remains constant by definition~\citep{landau1958lifshitz,rasmussen2000statistical,johansson2004statistical,tong2012statistical,flach2016spreading}.
Therefore the chemical potential $\mu \sim 1/\tilde{\nu}$.
In addition since $\epsilon _l$ is finite, and $\tilde{\nu}$ is large, we also assume that $\mu _l \sim 1/\tilde{\nu}$.
Consequently the second term in~\eqref{eq:norm_cal_4} becomes
\begin{equation}
    \sum _{l} \frac{2\beta}{(\tilde{\nu} \mu _l)^2\mu _l} \sim \tilde{\nu} \sum _{l} \frac{2\beta }{(\tilde{\nu} \mu _l)^2} \rightarrow 0,
    \label{eq:norm_second_term_approx_cal_4}
\end{equation}
for $\tilde{\nu} \rightarrow 0$.
Thus,~\eqref{eq:norm_second_term_approx_cal_4} can be used to provide the approximation
\begin{equation}
    \mathcal{S}_{1D} \approx \sum _l \frac{1}{\tilde{\nu} \mu _l} = \sum _l \frac{1}{\tilde{\nu} (\mu + \epsilon _l)} = \sum _l \frac{1}{\tilde{\nu} \mu + \tilde{\nu}\epsilon_l} \sim \frac{N}{\tilde{\nu} \mu},
    \label{eq:norm_av_stats}
\end{equation}
as $\tilde{\nu} \epsilon _l \rightarrow 0$ when $\tilde{\nu} \rightarrow 0$.
From~\eqref{eq:norm_av_stats}, we conclude that the expression of the average norm per site is 
\begin{equation}
    s = \frac{\mathcal{S}_{1D}}{N} \approx \frac{1}{\tilde{\nu}\mu}.
\end{equation}
Let us now estimate the average energy per site.
The total energy $\mathcal{H}_{1D}$ is given by
\begin{equation}
    \mathcal{H}_{1D} = \left(\frac{\mu}{\tilde{\nu}} \frac{\partial }{\partial \mu _l} - \frac{\partial}{\partial \tilde{\nu}} \right)  \ln \mathcal{Z}.
    \label{eq:ens_av_energy_main_formula_app} 
\end{equation}
Then~\eqref{eq:ens_av_energy_main_formula_app}, using the expression of $\mathcal{Z}$ from~\eqref{eq:part_function_ln_app} gives   
\begin{equation}
    \begin{split}
        \mathcal{H}_{1D} &= \frac{\mu}{\tilde{\nu}} \left( -\sum _l \left[\frac{\tilde{\nu}}{\tilde{\nu} \mu _l} - \frac{2\tilde{\nu} ^2 \beta}{(\tilde{\nu} \mu _l)^3} \right]\right) + \sum _l \left[\frac{\mu _l}{\tilde{\nu} \mu _l}  - \frac{\beta}{(\tilde{\nu} \mu _l)^2} \right], \\
            &= -\sum _l \frac{\mu}{\tilde{\nu} \mu _l} + \sum _{l} \frac{2\beta \mu}{(\tilde{\nu}\mu_l)^2 \mu _l} + \sum_{l} \frac{\mu _l}{\tilde{\nu} \mu _l} - \sum _{l}\frac{\beta}{(\tilde{\nu} \mu _l)^2}, \\
            &= \sum _l \frac{\epsilon _l}{\tilde{\nu} \mu _l} - \sum _l \frac{\beta \mu _l}{(\tilde{\nu} \mu _l)^2 \mu_l} + \sum _{l} \frac{2\beta \mu}{(\tilde{\nu}\mu_l)^2 \mu _l}, \\
            &= \sum _l \frac{\epsilon _l}{\tilde{\nu} \mu _l} + \sum _{l} \frac{\beta}{(\tilde{\nu} \mu _l)^2} - \sum _{l} \frac{\beta \epsilon _l}{(\tilde{\nu} \mu _l)^2 \mu _l}. 
    \end{split}
    \label{eq:energy_cal_4}
\end{equation}
The last term in the final expression of $\mathcal{H}_{1D}$ in~\eqref{eq:energy_cal_4} is similar to the one in the final expression of $\mathcal{S}_{1D}$ in~\eqref{eq:norm_cal_4}, which converges to zero in the limit of $\tilde{\nu} \rightarrow 0$.
Thus
\begin{equation}
    \mathcal{H}_{1D} \approx \sum _l \frac{\epsilon _l}{\tilde{\nu} \mu _l} + \sum _{l} \frac{\beta}{(\tilde{\nu} \mu)^2}.
    \label{eq:energy_final_v1}
\end{equation}
Let us look a bit closer each term of $\mathcal{H}_{1D}$~\eqref{eq:energy_final_v1}.
The first one can be written as
\begin{equation}
    \begin{split}
        \sum _l \frac{\epsilon _l}{\tilde{\nu} \mu _l} &= \sum _l \frac{\epsilon _l}{\tilde{\nu} (\mu + \epsilon _l)} = \frac{1}{\tilde{\nu} \mu} \sum _l \left(\frac{\epsilon _l}{1 + \frac{\epsilon_l}{\mu}}\right). 
    \end{split}
    \label{eq:first_ens_av_term_2}
\end{equation}
Then, applying the Taylor expansion 
\begin{equation}
    \frac{\epsilon _l}{1 + \frac{\epsilon_l}{\mu}} = \epsilon _l \left(1 - \frac{\epsilon _l}{\mu} + \frac{\epsilon _l^2}{\mu ^2} + \ldots\right),
    \label{eq:taylor_energy_stat_first_1}
\end{equation}
and assuming that $\epsilon_l/\mu\ll 1$  as $\mu \sim 1/\tilde{\nu}\gg 1$, for $\tilde{\nu}\rightarrow 0$ in~\eqref{eq:first_ens_av_term_2}, we obtain 
\begin{equation}
    \sum _l \frac{\epsilon _l}{\tilde{\nu} (\mu + \epsilon _l)} \approx \frac{1}{\tilde{\nu} \mu} \left(\sum_l \epsilon _l - \frac{1}{\mu} \sum _l \epsilon _l ^2 \right),
    \label{eq:rhs_1_approx}
\end{equation}
where we considered only the first two leading terms in~\eqref{eq:taylor_energy_stat_first_1} [see also~\citep{flach2016spreading}].
In addition, in order for the value of $\sigma_\epsilon$ (see~\eqref{eq:var_random_on_site_disorder_1}) to be finite i.e.~$\sup \sigma _\epsilon < \infty$, we must have 
\begin{equation}
    \sum _{l} \epsilon _{l} ^{2} \sim N \quad \Rightarrow \quad \sum _l \epsilon_l \sim \sqrt{N}.
    \label{eq:var_sum_term_app}
\end{equation}
With these two approximations of~\eqref{eq:var_sum_term_app},~\eqref{eq:rhs_1_approx} gives
\begin{equation}
    \sum _l \frac{\epsilon _l}{\tilde{\nu} (\mu + \epsilon _l)} \sim \frac{\sqrt{N}}{\tilde{\nu} \mu} - \frac{N\tilde{\nu}}{\left(\tilde{\nu}\mu\right)^2}.
    \label{eq:tayplor_final_approx_rhs}
\end{equation}
The second term in the right hand side of~\eqref{eq:tayplor_final_approx_rhs} can be neglected when $\tilde{\nu} \rightarrow 0$.
In addition the variation of $\sqrt{N}/(\tilde{\nu} \mu)$ is too slow to contribute to the final energy-norm relation at infinite temperature~\citep{flach2016spreading}.
Finally, only the second term in the right hand side of~\eqref{eq:energy_final_v1} remains, leading to 
\begin{equation}
    h =  \frac{\mathcal{H}_{1D}}{N} \approx \frac{\beta}{(\tilde{\nu} \mu)^2} \quad \Rightarrow \quad h \approx \beta \left(\frac{1}{\tilde{\nu}\mu}\right)^2 \quad \Rightarrow \quad h \approx \beta s^2.
    \label{eq:}
\end{equation}

\chapter{\label{app:diff_op_ddnls}Symplectic integration of the equations of motion and variational equations of the 1D and 2D DDNLS models}

We present here the expression of the various operators needed for the implementation of SIs for the integration of the equations of motion and the variational equations for the 1D and 2D DDNLS systems.
We note that the Hamiltonian of the 1D DDNLS model~\eqref{eq:Hamilton_complex_dnls_1d}
\begin{equation}
  \mathcal{H}_{1D} = \sum _{l = 1}^{N} \epsilon _l \lvert \psi _l \rvert ^2 + \frac{\beta }{2}\lvert \psi _l \rvert ^{4} - \left(\psi_{l+1} \psi _l ^\star + \psi_{l+1}^\star \psi _l\right),
  \label{eq:Hamilton_complex_dnls_1d_app}
\end{equation}
becomes
\begin{equation}
  \mathcal{H}_{1D} = \sum _{l} \left\{\frac{\epsilon _l }{2}\left(q_l ^2 + p_l ^2 \right) + \frac{\beta}{8}\left(q_l ^2 + p_l ^2 \right)^4 - p_{l + 1}p_l - q_{l + 1} q_l \right\},
  \label{eq:ham_real_1d_ddnls_app}
\end{equation}
with $q_l$ and $p_l$ being real variables, through the implementation of the transformation~\eqref{eq:transformation_complex_real_1d_num}
\begin{equation}
  \psi_l = \frac{1}{\sqrt{2}}\left(q_l + ip_l\right).
\end{equation}
In a similar manner, the 2D DDNLS Hamiltonian function~\eqref{eq:hamilton_complex_ddnls_2d}
\begin{equation}
  \mathcal{H}_{2D} = \sum _{l=1, m=1}^{N, M} \epsilon _{l, m} \lvert \psi _{l, m} \rvert ^2 + \frac{\beta}{2}\lvert \psi _{l, m} \rvert ^{4} - \left(\psi_{l+1, m} \psi _{l, m} ^\star + \psi _{l,m+1}\psi _{l, m}^\star + \psi_{l+1, m}^\star \psi _{l, m} + \psi _{l,m+1}^\star\psi _{l, m} \right),
  \label{eq:hamilton_complex_ddnls_2d_app}
\end{equation}
becomes
\begin{equation}
  \mathcal{H}_{2D} = \sum _{l, m} \left\{ \frac{\epsilon _{l, m} }{2} \left(q_{l, m} ^2 + p_{l, m}^2\right) + \frac{\beta}{8} \left(q_{l, m} ^2 + p_{l, m}^2\right) ^2 - q_{l, m + 1} q _{l, m} - q_{l + 1, m} q _{l, m} -  p _{l, m + 1} p _{l, m} - p _{l + 1, m}p _{l, m} \right\},
  \label{eq:ham_real_2d_ddnls_app_oo}
\end{equation} 
after the implementation of the transformation 
\begin{equation}
  \psi_{l,m} = \frac{1}{\sqrt{2}} \left(q_{l,m} + ip_{l,m}\right).
\end{equation}
We note that for both Hamiltonians~\eqref{eq:ham_real_1d_ddnls_app} and~\eqref{eq:ham_real_2d_ddnls_app_oo} we use fixed boundary conditions.

\section{\label{sec:two-part_split_app}Two-part split symplectic integration of the 1D DDNLS model}
As was mentioned in Sec.~\ref{subsec:two_part_split}, the Hamiltonian function of the 1D DDNLS system can be split into two integrable parts as
\begin{equation}
  \mathcal{H}_{1D} = A + B,
\end{equation}
with
\begin{equation}
  A = \sum _{l} \frac{\epsilon _l}{2}(q_l ^2 + p_l ^2) + \frac{\beta}{8}(q_l ^2 + p_l ^2)^2, \quad B = \sum _{l} - p_{l+1}p_l - q_{l+1}q_l,
  \label{eq:ham_2ps_gen}
\end{equation}
whose solutions can be obtained in closed forms.

\subsection{\label{subsec:e_lav_app}Analytic expression for the operator $e^{\tau {\bf L}_{A_V}}$}
Let us first consider the $A$ part of the decomposition~\eqref{eq:ham_2ps_gen}
\begin{equation}
  A = \sum _{l} \frac{\epsilon _l}{2}(q_l ^2 + p_l ^2) + \frac{\beta}{8}(q_l ^2 + p_l ^2)^2.
  \label{eq:a_twp_app_1}
\end{equation}
Its equations of motion are
\begin{align}
  \frac{d\bm{z}}{dt} = \bm{L}_{A} \bm{z} 
  \colon 
  \begin{dcases}
    \dot{q_l} &= p_l \left(\epsilon _l + \beta\frac{q_l ^ 2 + p_l ^2}{2}\right) \\
    \dot{p_l} & = -q_l  \left(\epsilon _l + \beta\frac{q_l ^ 2 + p_l ^2}{2} \right)          
  \end{dcases},
  \label{eq:eq_of_motion_a1}
\end{align}
which conserve the value of $A$~\eqref{eq:a_twp_app_1}.

Let us express~\eqref{eq:a_twp_app_1} and~\eqref{eq:eq_of_motion_a1} in action-angles variables $J$, $\Theta$ through the canonical transformation
~\citep{herbert1980classical,goldstein2002classical,tong2004classical}
\begin{equation}
  \Omega _A: \left(q_l, p_l\right)\mapsto \left(J_l, \Theta_l\right) \colon 
  \left\{
  \begin{array}{ll}
    J _l = \frac{1}{2} (q_l ^2 + p_l ^2 ) \\
    \Theta _l = \arctan \frac{q_l }{p_l} \\
  \end{array}
  \right..
  \label{eq:transf_action_angle_var}
\end{equation}
It is easily seen that the inverse of transformation $\Omega_A$~\eqref{eq:transf_action_angle_var} is given by
\begin{equation}
  \Omega _A^{-1}: \left(J_l, \Theta_l\right) \mapsto \left(q_l, p_l\right) \colon 
  \left\{
  \begin{array}{ll}
    q_l = \sqrt{2J_l} \sin \Theta _l \\
    p_l = \sqrt{2J_l} \cos \Theta _l
  \end{array}
  \right..
  \label{eq:inv_transf_action_angle_var}
\end{equation}
In~\eqref{eq:transf_action_angle_var} and~\eqref{eq:inv_transf_action_angle_var}, the variable $\Theta_l$ plays the role of a generalized position and $J_l$ of the generalized momentum.
The application of~\eqref{eq:inv_transf_action_angle_var} on Hamiltonian $A$~\eqref{eq:a_twp_app_1} brings it to the form
\begin{equation}
  A = \sum _l \epsilon _l J_l + \frac{\beta}{2}J_l ^2.
  \label{eq:a_twp_app_a1_act_angle_var}
\end{equation}
The corresponding equations of motion generated through the Hamilton equations of motion
\begin{equation}
  \dot{\Theta} _l = \frac{\partial A}{\partial J_l}, \quad \dot{J}_l = - \frac{\partial A}{\partial \Theta _l}, 
  \label{eq:hamilton_eq_of_motion_a_2ps_app}
\end{equation}
are
\begin{align}
  \begin{dcases}
    \dot{\Theta} _l &= \epsilon _l + \beta J_l \\
    \dot{J}_l &= 0         
  \end{dcases}.
  \label{eq:eq_mot_a1_action_angl_var}
\end{align}
From the second equation of~\eqref{eq:eq_mot_a1_action_angl_var}, we see that the actions $J_l$, $l=1, 2, \ldots, N$ are integrals of motion and their values are conserved along the system's evolution.
Given an initial condition $(\Theta _l (0), J_l(0))$, the solution of~\eqref{eq:eq_mot_a1_action_angl_var} is given by 
\begin{align}
  \begin{dcases}
    \Theta _l (t) &= (\epsilon _l + \beta J_l (0) )t + \Theta _l (0) \\
    J_l (t) &= J_l (0)
  \end{dcases}.
  \label{eq:sol_a1_action_angl_var}
\end{align}
Using the transformation $\Omega _A^{-1}$~\eqref{eq:inv_transf_action_angle_var}, we rewrite the solution~\eqref{eq:sol_a1_action_angl_var} in terms of the $q_l$ and $p_l$ variables as
\begin{align}
  \begin{dcases}
    q_l (t) &= \sqrt{2J_l (0)} \sin \left[ \left(\epsilon _l + \beta J_l (0) \right)t + \Theta _l (0) \right] \\
    p _l (t)&=\sqrt{2J_l (0)}\cos \left[ \left(\epsilon _l + \beta J_l (0) \right)t + \Theta _l (0) \right]      
  \end{dcases},
  \label{eq:sol_A1_qp_gen}
\end{align}
with
\begin{equation}
  J_l (0) = \frac{1}{2}\left(q_l (0) ^2 + p_l (0)^2\right),
  \quad 
  \Theta _l (0) = \arctan \frac{q_l (0)}{p_l (0) }.
  \label{eq:coefs_sol_A1_qp_gen}
\end{equation}
Let us now discuss some particular cases related to~\eqref{eq:sol_A1_qp_gen}. 
\begin{enumerate}
    \item[] {\bf Case A: $p_l (0) \neq 0$.}
        No problems arise in this case and the substitution of~\eqref{eq:coefs_sol_A1_qp_gen} onto Eq.~\eqref{eq:sol_A1_qp_gen} leads to
        \begin{align}
            \begin{dcases}
                q_l (t) &= \sqrt{q_l (0) ^2 + p_l (0) ^2} \left[ \sin \left[(\epsilon _l + \beta J_l (0) )t\right] \cos \Theta _l (0)  + \cos \left[(\epsilon _l + \beta J_l (0) )t\right] \sin \Theta _l (0)\right] \\
                p_l (t) &= \sqrt{q_l (0) ^2 + p_l (0) ^2} \left[ \cos \left[(\epsilon _l + \beta J_l (0) )t\right] \cos \Theta _l (0) - \sin \left[(\epsilon _l + \beta J_l (0) )t \right] \sin \Theta _l (0)\right]    
            \end{dcases},
            \label{eq:sol_A1_qp_gen_2}
        \end{align}
        which can further be rewritten as
        \begin{align}
            \begin{dcases}
                q_l (t) &= \sqrt{q_l (0) ^2 + p_l (0) ^2} \left[ \sin \left[(\epsilon _l + \beta J_l (0) )t\right] \frac{1}{\sqrt{1 + \frac{q_l(0)^2}{p_l (0)^2}}}  + \cos \left[(\epsilon _l + \beta J_l (0) )t\right]\frac{\frac{q_l(0)}{p_l(0)}}{\sqrt{1 + \frac{q_l(0)^2}{p_l (0)^2}}}\right] \\
                p_l (t) &= \sqrt{q_l (0) ^2 + p_l (0) ^2} \left[ \cos \left[(\epsilon _l + \beta J_l (0) )t\right] \frac{1}{\sqrt{1 + \frac{q_l(0)^2}{p_l (0)^2}}} - \sin \left[(\epsilon _l + \beta J_l (0) )t\right] \frac{\frac{q_l(0)}{p_l(0)}}{\sqrt{1 + \frac{q_l(0)^2}{p_l (0)^2}}}\right]    
            \end{dcases}.
            \label{eq:sol_A1_qp_gen_3}
        \end{align}
        We note that in deriving Eq.~\eqref{eq:sol_A1_qp_gen_3}, we used the following trigonometric identities $\sin (a + b) = \sin a \cos b + \cos a \sin b$, $\cos (a + b) = \cos a \cos b - \sin a \sin b$, $\sin a  = \tan a/\sqrt{1 + \tan ^2 a}$ and $\cos a  = 1/\sqrt{1 + \tan ^2 a}$.
        In this case the solution of the Hamilton equations of motion of $A$~\eqref{eq:a_twp_app_1} is given by~\citep{skokos2014high,gerlach2016symplectic,danieli2019computational}
        \begin{align}
          \begin{dcases}
            q_l (t) &= q_l (0) \cos \alpha _l t + p_l (0) \sin \alpha _l t \\
            p_l (t) &=  p_l (0) \cos \alpha _l t - q_l (0) \sin \alpha _l t
          \end{dcases},
          \label{eq:sol_A1_pq_eq_motion_final}
        \end{align}
        with  
        \[ 
            \alpha _l = \epsilon _l + \beta J_l (0) = \epsilon _l + \beta \frac{q_l (0)^2 + p_l (0)^2}{2}, 
        \]
        being constant coefficients.

    \item[] {\bf Case B: $p_l (0) = 0$.}
        In this case, we have to consider two different options: $q_l (0) \neq 0$ and $q_l (0) = 0$.
        \begin{enumerate}
            \item[] {\bf Case B$1$: $q_l (0) = 0$.}
                The equations of motion~\eqref{eq:eq_of_motion_a1} become
                \begin{align}
                    \begin{dcases}
                        \dot{q} _l = 0 \\
                        \dot{p} _l = 0
                    \end{dcases},
                \end{align}
                and can be trivially solved to get $q_l (t) = 0$ and $p_l (t) = 0$.
            \item[] {\bf Case B$2$: $q_l (0) \ne 0$.}
                In this case, we can have two possible values for $\Theta _l (0)$, namely
                \begin{equation}
                    \left\{
                        \begin{array}{lcr}
                            \Theta _l (0) =& \frac{\pi}{2}  &\mbox{ if $q_l (0) > 0$} \\
                            \Theta _l (0) =& -\frac{\pi}{2}  &\mbox{ if $q_l (0) < 0$}    
                        \end{array}
                    \right..
                    \label{eq:cond_subcases_eq_mot_1}
                \end{equation}
                For the first case in~\eqref{eq:cond_subcases_eq_mot_1}, the solution~\eqref{eq:sol_A1_qp_gen} is written as
                \begin{align}
                    \begin{dcases}
                        q_l (t) &= \sqrt{2J_l (0)} \sin \left[ (\epsilon _l + \beta J_l (0) )t + \frac{\pi}{2} \right] \\
                        p _l (t)&=\sqrt{2J_l (0)}\cos\left[ (\epsilon _l + \beta J_l (0) )t + \frac{\pi}{2} \right]
                    \end{dcases},
                    \label{eq:sol_A1_qp_p=0_q>0_1}
                \end{align}
                or simply as
                \begin{align}
                    \begin{dcases}
                        q_l (t) &=  q_l (0) \cos \left[ (\epsilon _l + \beta J_l (0) )t \right] \\
                        p _l (t)&=- q_l (0)  \sin \left[ (\epsilon _l + \beta J_l (0) )t\right]  
                    \end{dcases}.
                    \label{eq:sol_A1_qp_p=0_q>0_2}
                \end{align}
                We note that going from~\eqref{eq:sol_A1_qp_p=0_q>0_1} to~\eqref{eq:sol_A1_qp_p=0_q>0_2}, we used the trigonometric identities $\sin (a + \pi /2) = \cos a$ and $\cos (a + \pi /2) = - \sin a$, as well as the facts that  $J_l (0) = \frac{1}{2} q_l (0)^2$ and $\lvert q_l (0)\rvert =  q_l (0)~(q_l (0) > 0 )$.
                Following a similar process as in~\eqref{eq:sol_A1_qp_p=0_q>0_1} and~\eqref{eq:sol_A1_qp_p=0_q>0_2} in the second case of~\eqref{eq:cond_subcases_eq_mot_1}, and using the identities $\sin (a - \pi /2) = -\cos a$  and $\cos (a - \pi /2) =  \sin a$, along with $\lvert q_l (0) \rvert = - q_l (0)$ on~\eqref{eq:sol_A1_qp_p=0_q>0_1}, we finally obtain the solution
                \begin{align}
                    \begin{dcases}
                        q_l (t) &= q_l (0) \cos \left[ (\epsilon _l + \beta J_l (0) )t \right] \\
                        p _l (t)&=- q_l (0)  \sin \left[ (\epsilon _l + \beta J_l (0) )t\right]
                    \end{dcases}.
                    \label{eq:sol_A1_qp_p=0_q<0_2}
                \end{align}
                The solutions~\eqref{eq:sol_A1_qp_p=0_q>0_2} and~\eqref{eq:sol_A1_qp_p=0_q<0_2} are identical. 
                Consequently, no distinction between the cases B$1$ and B$2$ needs to be implemented in our computer program.
        
        \end{enumerate}
\end{enumerate} 
       
Let us now discuss the solution of the variational equations of the Hamiltonian $A$~\eqref{eq:a_twp_app_1} which have the form
\begin{equation}
    \frac{d\delta\bm{z}}{dt} = \bm{L}_{A_V} \delta \bm{z} 
    \colon 
    \begin{dcases}
        \dot{\delta q_l} &= \beta q_l p_l \delta q_l + \left[\epsilon _l + \beta\frac{q_l ^ 2 + p_l ^2}{2}  +  \beta p_l ^2\right] \delta p_l\\
        \dot{\delta p_l} &= - \beta q_l p_l \delta p_l - \left[\epsilon _l + \beta\frac{q_l ^ 2 + p_l ^2}{2} + \beta q_l^2\right] \delta q_l 
    \end{dcases}.
    \label{eq:2ps_3ps_A_var_equation_qp-frame}
\end{equation}
In the action-angle variables, ~\eqref{eq:2ps_3ps_A_var_equation_qp-frame} obtains a much simpler form
\begin{equation}
    \begin{dcases}
        \dot{\delta \Theta _l} &= \beta \delta J_l    \\
        \dot{\delta J_l} &= 0
    \end{dcases},
    \label{eq:2ps_3ps_A_diff_var_eq_oj-frame}
\end{equation}
whose analytic solution can be easily found. 
In particular, the solution of~\eqref{eq:2ps_3ps_A_diff_var_eq_oj-frame} for initial conditions $(\delta \Theta _l (0), \delta J_l (0))$, is
\begin{equation}
    \begin{dcases}
        \delta \Theta _l (t) &= \beta \delta J_l (0)  t + \delta \Theta  _l (0) \\
        \delta J_l (t) &= \delta J_l (0)
    \end{dcases}.
    \label{eq:2ps_3ps_A_solution_var_equation_oj-frame}
\end{equation}
The transformation relating~\eqref{eq:2ps_3ps_A_var_equation_qp-frame} to~\eqref{eq:2ps_3ps_A_diff_var_eq_oj-frame}, can be obtained via total differentiation of~\eqref{eq:transf_action_angle_var} and~\eqref{eq:inv_transf_action_angle_var} and has the form
\begin{equation}
    \delta \Omega_A:\left(\delta q_l, \delta p_l\right) \mapsto \left(\delta\Theta_l, \delta J_l\right)
    \colon
    \begin{dcases}
        \delta \Theta _l(t) &= \frac{p_l (t)}{2J_l(t)}\delta q_l(t) - \frac{q_l (t)}{2J_l(t)} \delta p_l(t) \\
        \delta J_l (t)&= q_l(t) \delta q_l(t) + p_l (t)\delta p_l(t)
    \end{dcases},
    \label{eq:2ps_3ps_A_sol_var_eq_oj-frame}
\end{equation}
while its inverse transformation is given by
\begin{equation}
    \delta \Omega^{-1}_A:\left(\delta\Theta_l, \delta J_l\right) \mapsto \left(\delta q_l, \delta p_l\right)
    \colon
    \begin{dcases}
        \delta q_l (t) &= \left[\frac{1}{\sqrt{2J_l (t)}} \sin \Theta _l (t)\right] \delta J_l (t)  +\left[ \sqrt{2J_l (t)} \cos \Theta _l(t) \right] \delta \Theta _l (t) \\
        \delta p_l(t) &= \left[\frac{1}{\sqrt{2J_l(t)}} \cos \Theta _l(t) \right] \delta J_l (t) - \left[ \sqrt{2J_l(t)} \sin \Theta _l (t)\right] \delta \Theta _l (t)
    \end{dcases}. 
    \label{eq:2ps_3ps_A_sol_var_eq_qp-frame}
\end{equation}
The equivalence of solution~\eqref{eq:2ps_3ps_A_solution_var_equation_oj-frame} (in action-angle variables) and~\eqref{eq:2ps_3ps_A_sol_var_eq_oj-frame} (in position and momentum variables) can be easily seen by differentiating~\eqref{eq:2ps_3ps_A_sol_var_eq_oj-frame}.
Let us see that in more detail for the $\delta \Theta _l$ variable.
Applying the time derivative to the first equation of~\eqref{eq:2ps_3ps_A_sol_var_eq_oj-frame}
\begin{equation}
    \label{eq:proof_domega_sol_var_oj}
    \begin{split}
    \frac{d\delta \Theta_l (t)}{dt} &= \frac{1}{2J_l(t)} \left[ \dot{p_l }(t) \delta q_l (t) + p_l (t) \dot{\delta q_l } (t) - \dot{q_l} (t) \delta p_l (t) - q \dot{\delta p_l} (t)  \right], \\
            &=  \frac{1}{2J_l(t)} \left\{ \left[-q_l  \left(\epsilon _l + \beta\frac{q_l ^ 2 + p_l ^2}{2} \right) \right] \delta q_l (t) +  p_l (t) \left[ \beta q_l p_l \delta q_l + \left(\epsilon _l + \beta\frac{q_l ^ 2 + p_l ^2}{2}  +  \beta p_l ^2\right) \delta p_l\right] \right.  \\ 
            &  - \left. \left[p_l \left(\epsilon _l + \beta\frac{q_l ^ 2 + p_l ^2}{2}\right) \right] \delta p_l (t)  -  q_l (t) \left[ - \left(\epsilon _l + \beta\frac{q_l ^ 2 + p_l ^2}{2} + \beta q_l^2\right) \delta q_l - \beta q_l p_l \delta p_l\right] \right\}, \\
            &= \frac{1}{2J_l(t)} \left[2J_l \beta \left(q_l \delta q_l + p_l \delta p_l \right)\right], \\
            &= \beta \delta J_l (t),
    \end{split}
\end{equation}
which is just the first equation of~\eqref{eq:2ps_3ps_A_diff_var_eq_oj-frame}. 
A similar result for $\delta J_l (t)$~\eqref{eq:2ps_3ps_A_sol_var_eq_oj-frame} can also be obtained. 

In practice in order to obtain the solution of the system's variational equations we do the following: given an initial condition $\left(q_l(0), p_l(0), \delta q_l (0), \delta p_l (0) \right)$, we compute the quantities
\[ 
    J_l (0) = \frac{1}{2} (q_l (0)^2 + p_l(0)^2), \quad \alpha _l = \epsilon _l + \beta   J_l (0), \quad l =1, 2, \ldots, N, 
\]
and
\[
    \delta J_l (0) = q_l (0) \delta q_l (0) + p_l (0) \delta p_l (0), \quad \delta \Theta _l(0) = \frac{p_l (0)}{2J_l(0)}\delta q_l(0) - \frac{q_l (0)}{2J_l(0)} \delta p_l(0), \quad l =1, 2, \ldots, N. 
\]
Thus the solution~\eqref{eq:2ps_3ps_A_sol_var_eq_qp-frame} of the variational equations takes the form
\begin{equation}
    \begin{dcases}
        \delta q_l (t) &= \frac{q_l (0) \cos \alpha _l t + p_l (0) \sin \alpha _l t}{ 2 J_l(0)} \delta J_l (0) + \left(p_l (0) \cos \alpha _l t - q_l (0) \sin \alpha _l t\right) \left(\beta \delta J_l (0)  t + \delta \Theta  _l (0) \right) \\
        \delta p_l (t) & = \frac{p_l (0) \cos \alpha _l t - q_l (0) \sin \alpha _l t}{2J_l (0)} \delta J_l (0) - \left( q_l (0) \cos \alpha _l t + p_l (0) \sin \alpha _l t \right) \left(\beta \delta J_l (0)  t + \delta \Theta  _l (0)\right)
    \end{dcases},
    \label{eq:2ps_3ps_sol_var_eq_A1}
\end{equation}
using the equalities 
\[ 
    \sin \Theta_l(t) =  \frac{q_l(t)}{\sqrt{2J_l(t)}} = \frac{q_l (0) \cos \alpha _l t + p_l (0) \sin \alpha _l t}{\sqrt{2J_l (0)}},  
\]
and 
\[ 
    \sqrt{2J_l(t)}\cos \Theta(t) = p_l(t) =   p_l (0) \cos \alpha _l t - q_l (0) \sin \alpha _l t.   
\]
along with~\eqref{eq:2ps_3ps_A_solution_var_equation_oj-frame}.

In summary, the expression of the operator $e^{\tau \bm{L}_{A_V}}$ propagating $(q_l,p_l, \delta q_l, \delta  p_l)$ at time $t$ to $(q_l^\prime,p_l^\prime, \delta q_l^\prime, \delta  p_l^\prime)$ at time $t + \tau$ is given by
\begin{equation}
    e^{\tau \bm{L}_{A_V}}
    \colon 
    \begin{dcases}
        q_l ^\prime &= q_l \cos (\tau \alpha _l) +  p_l \sin (\tau \alpha _l) \\
        p_l ^\prime &= p_l \cos (\tau \alpha _l) -  q_l \sin (\tau \alpha _l) \\
        \delta q_l ^\prime &= \frac{q_l  \cos (\tau \alpha _l ) + p_l  \sin (\tau \alpha _l )}{ 2 J_l} \delta J_l  + \left(p_l  \cos (\tau \alpha _l ) - q_l  \sin (\tau \alpha _l )\right) \left(\beta \delta J_l  \tau + \delta \Theta  _l   \right) \\
        \delta p_l ^\prime  & = \frac{p_l  \cos (\tau \alpha _l ) - q_l \sin (\tau \alpha _l )}{2J_l } \delta J_l  - \left( q_l  \cos ( \tau \alpha _l ) + p_l \sin (\tau \alpha _l ) \right) \left(\beta \delta J_l  \tau + \delta \Theta  _l \right)
    \end{dcases},
    \label{eq:flow_elav_app_1}
\end{equation}
with
\[ 
J_l  = \frac{1}{2} (q_l  ^2 + p_l ^2), \quad \alpha _l = \epsilon _l + \beta  J_l,  
\]
\[
    \delta J_l   = q_l   \delta q_l   + p_l   \delta p_l, \quad \delta \Theta _l  = \frac{p_l  }{2J_l }\delta q_l  - \frac{q_l  }{2J_l } \delta p_l.
\]    

Let us now discuss in detail a particular problematic case appearing in~\eqref{eq:2ps_3ps_sol_var_eq_A1} and~\eqref{eq:flow_elav_app_1}, namely the situation where $J_l = (q_l^2 + p_l^2)/2 = 0$.
In that case, $q_l = p_l = 0$.
Then, from~\eqref{eq:2ps_3ps_A_var_equation_qp-frame}, we see that the variational equations take the form
\begin{equation}
  \begin{dcases}
    \dot{\delta q_l} = \epsilon_l \delta p_l & \\
    \dot{\delta p_l} = -\epsilon _l \delta q_l & \\
  \end{dcases}.
  \label{eq: 2ps 3ps var eq p=0 q=0_1}
\end{equation}
The solution of this set of ODEs is given by
\begin{equation} 
  \begin{dcases}
  \delta q_l (t) = \delta q_l(0) \cos \left(\epsilon _l t\right) + \delta p_l(0) \sin \left(\epsilon _l t\right) & \\
  \delta p_l (t) = \delta p_l (0) \cos \left(\epsilon _l t\right) - \delta q_l(0) \sin \left(\epsilon _l t\right)
  \end{dcases},
\label{eq: 2ps 3ps var eq p=0 q=0_2}
\end{equation}
assuming we started from $(\delta q_l (0), \delta p_l (0))$ at $t = 0$.
Consequently, the general form of the operator $e^{\tau \bm{L}_{A_V}}$ is given by~\eqref{eq:flow_elav_app_1} except for the particular case $J_l =0$ where we have 
\begin{equation}
  e^{L_{A_V}\tau} :=
    \begin{dcases}
      q_l^\prime = q_l = 0 \\
      p_l ^\prime = p_l = 0 \\
      \delta q_l ^\prime = \delta q_l \cos \left(\epsilon _l \tau\right) + \delta p_l \sin \left(\epsilon _l \tau\right) \\
      \delta p_l ^\prime = \delta p_l \cos \left(\epsilon _l \tau\right) - \delta q_l \sin \left(\epsilon _l \tau\right)    
    \end{dcases}.
    \label{eq:flow_elav_app_2}
\end{equation}

\subsection{\label{subsec:e_lbv_app}Analytic expression for the operator $e^{\tau {\bf L}_{B_V}}$}
Based on the analysis presented in the appendix of~\citep{gerlach2016symplectic}, let us now determine the solution of both the equations of motion and the variational equations of the $B$ Hamiltonian part~\eqref{eq:ham_2ps_gen}
\begin{equation}
  B = -\sum _l p_{l}p_{l+1} + q_{l}q_{l+ 1},
  \label{eq:b_twps_app}
\end{equation}
of the 1D DDNLS system~\eqref{eq:Hamilton_complex_dnls_1d_app}.
More specifically, we will determine the operator $e^{\tau \bm{L}_{B_V}}$ propagating an initial condition $\bm{Z}$ at time $t$ to $\bm{Z}^\prime$ at time $t + \tau$ according to
\begin{equation}
  e^{\tau L_{B_V}} \colon \bm{Z}^\prime = \bm{\Gamma} (\tau ) \cdot \bm{Z},
  \label{eq:evolution_2ps_general_final}
\end{equation}
with $\bm{Z} = \left(\bm{z}~\delta \bm{z}\right)^T$ such that $\bm{z} = \left(\bm{q}~\bm{p}\right)^T = \left(q_1~q_2~\ldots~q_N~p_1~p_2~\ldots~p_N\right)^T$, $\delta \bm{z} = \bm{w} = \left(\delta \bm{q}~\delta \bm{p}\right)^T = \left(\delta q_1~\delta q_2~\ldots~\delta q_N~\delta p_1~\delta p_2~\ldots~\delta p_N\right)^T$ and $\bm{\Gamma}$ being a $4N\times 4N$ matrix. 
Since SIs are used for fixed integration time steps, matrix $\bm{\Gamma} (\tau )$ in~\eqref{eq:evolution_2ps_general_final} is  constant (for fixed $\tau$ value). 
In addition, Hamiltonian $B$~\eqref{eq:b_twps_app} is a quadratic function in its variables $q_l$ and $p_l$ which in turn means that the equations of motion, and the variational equations are linear in $\bm{z}$, and $\bm{\delta z}$ and independent of each other (see~\eqref{eq:2ps_eqmotion_b_explicit} and~\eqref{eq:2ps_var_eq_b} below). 
Consequently, matrix $\bm{\Gamma}(\tau)$ takes the form 
\begin{equation}
    \bm{\Gamma} \left(\tau \right) = 
    \begin{bmatrix}
        \bm{C}\left(\tau \right) & \bm{0}_{2N} \\ 
        \bm{0}_{2N} & \bm{C}^\prime \left(\tau \right)
    \end{bmatrix},
    \label{eq:2ps_gamma_primary_form}
\end{equation}
with $\bm{C}(\tau)$ and $\bm{C}^\prime \left(\tau \right)$ being appropriate $2N\times 2N$ matrices respectively evolving the equations of motion and the variational equations according to 
\begin{equation}
    \begin{bmatrix}
        \bm{q}^\prime \\
        \bm{p}^\prime
    \end{bmatrix} 
    = \bm{C}\left(\tau \right) 
    \cdot 
    \begin{bmatrix}
        \bm{q} \\
        \bm{p}
    \end{bmatrix},
    \quad 
    \begin{bmatrix}
        \delta \bm{q}^\prime \\
        \delta \bm{p}^\prime
    \end{bmatrix} 
    = \bm{C}^\prime \left(\tau \right) 
    \cdot 
    \begin{bmatrix}
        \delta \bm{q} \\
        \delta \bm{p}
    \end{bmatrix}.
    \label{eq:2ps_gamma_primary_form_2}
\end{equation}

More specifically, the equations of motion of Hamiltonian $B$~\eqref{eq:b_twps_app} take the form 
\begin{equation}
  \frac{d \bm{z}}{dt} = \bm{L}_B \bm{z} \colon 
  \begin{dcases}
    \dot{q}_1 = - p_2 & \\
    \dot{q}_l = - p_{l - 1} - p_{l + 1} & \quad 2\leq l \leq N-1\\
    \dot{q}_N = - p_{N-1} &\\
    \dot{p} _1 = q_2 &\\
    \dot{p} _l = q_{l - 1} + q_{l + 1} & \quad 2\leq l \leq N-1\\
    \dot{p} _N = q_{N-1}
  \end{dcases}.
  \label{eq:2ps_eqmotion_b_explicit}
\end{equation}
Equation~\eqref{eq:2ps_eqmotion_b_explicit} can be written as
\begin{equation}
  \frac{d \bm{z}}{dt} =  \bm{L}_B \bm{z} = 
  \begin{pmatrix}
    \bm{0} & \bm{A} \\ 
    -\bm{A} & \bm{0}
  \end{pmatrix} 
  \cdot 
  \bm{z},
  \label{eq:2ps_eqmotion_b_contracted}
\end{equation}
with $\bm{A}$ being a tridiagonal $N\times N$ matrix whose main diagonal elements are equal to zero ($\bm{A}_{l, l} = 0, l = 1, 2, \ldots, N $), while all the elements of the first upper and lower diagonals are equal to $-1$ ($\bm{A}_{l, l + 1} = \bm{A}_{l - 1, l} = -1$) i.e.
\begin{equation}
  \bm{A} = 
  \begin{pmatrix}
      0 & -1      & 0   & \dots    & 0 & 0   \\
      -1 & 0       & -1   & \dots   & 0  & 0   \\
      \vdots  &      \vdots & \ddots         & \ddots    & \vdots & \vdots   \\
      0  &    0     &     0 & \dots     &     0      & -1   \\
      0  &    0     &     0 &  \dots     &      -1     & 0
  \end{pmatrix}.
  \label{eq:A_operator_form_app_1}
\end{equation}
The solution of~\eqref{eq:2ps_eqmotion_b_contracted} for an integration time step $\tau$ can be expressed in terms of a Lie series expansion~\eqref{eq:lie_series_method_gen} as
\begin{equation}
  \bm{z} (t + \tau ) = e^{\tau \bm{L}_B} \bm{z} (t) = \sum_{k = 0}^{\infty} \frac{\tau ^k}{k!} \bm{L}_B ^k \bm{z} (t) = \bm{C} (\tau) \bm{z} (t),
  \label{eq:2ps_formal_solution_of_b}
\end{equation}
with 
\begin{equation}
  \bm{C} \left(\tau \right) = \sum _{k = 0}^{\infty} \frac{\tau ^k}{k!} \bm{L}_B^k.
  \label{eq:2ps_formal_solution_of_b_2}
\end{equation}
As we mentioned in Sec.~\ref{subsec:taylor_series}, the next step to determine the expression of $\bm{C}(\tau)$ is to find the form of the operator $\bm{L}_B ^k$ in~\eqref{eq:2ps_formal_solution_of_b_2}.
Luckily, this task can be performed due to the rather simple form of operator $\bm{L}_B$~\eqref{eq:2ps_eqmotion_b_contracted}.
Let us initially note that 
\begin{equation}
    \bm{L}_B^{2k} = (-1)^k 
    \begin{pmatrix}
        \bm{A}^{2k} & \bm{0} \\
        \bm{0} & \bm{A}^{2k}
    \end{pmatrix}, 
    \quad
    \bm{L}_B^{2k + 1} = (-1)^k 
    \begin{pmatrix}
        \bm{0} & \bm{A}^{2k + 1}\\
        -\bm{A}^{2k + 1} & \bm{0}
    \end{pmatrix}.
    \label{eq:2ps_properties_of_lb}
\end{equation}
Then, using~\eqref{eq:2ps_formal_solution_of_b_2} and~\eqref{eq:2ps_properties_of_lb}, we get
\begin{align}
  \bm{C} (\tau) &= 
  \begin{pmatrix}
    \sum_{k = 0}^{\infty} \frac{(-1)^k }{(2k)!} \bm{A} ^{2k} \tau ^{2k} & \bm{0} \\
    \bm{0} &  \sum_{k = 0}^{\infty} \frac{(-1)^k }{(2k)!} \bm{A} ^{2k} \tau ^{2k}
  \end{pmatrix} \\
  &+  
  \begin{pmatrix}
    \bm{0} & \sum_{k = 0}^{\infty} \frac{(-1)^k }{(2k + 1)!} \bm{A} ^{2k + 1} \tau ^{2k + 1}  \\ 
    - \sum_{k = 0}^{\infty} \frac{(-1)^k }{(2k + 1)!} \bm{A}  ^{2k + 1} \tau ^{2k + 1} & \bm{0} 
  \end{pmatrix},
  \label{eq:C_operator_form_app_10}
\end{align}
which is equivalent to
\begin{equation}
  \bm{C} (\tau) = 
  \begin{pmatrix}
    \cos \left(\bm{A} \tau  \right) & \sin \left(\bm{A} \tau  \right) \\
    -\sin \left(\bm{A} \tau  \right) &  \cos \left(\bm{A} \tau  \right)
  \end{pmatrix}.
  \label{eq:expression_of_c_matrix_simple_form}
\end{equation}
The evaluation of the elements of matrices $\cos \left(\bm{A} \tau  \right)$ and $\sin \left(\bm{A} \tau  \right)$ can be obtained through the determination of the eigenvalues and eigenvectors of matrix $\bm{A}$~\eqref{eq:A_operator_form_app_1} itself~\citep{smith1985numerical}. 
In particular, the eigenvalues $\lambda _k$ of $\bm{A}$ are symmetric with respect to zero and are given by
\begin{equation}
    \lambda _k = -2\cos \left( \frac{k\pi}{N + 1} \right), \quad k = 1, 2, \dots, N, 
    \label{eq:eigenvalues_formula}
\end{equation}
while the normalized eigenvectors are orthogonal (since $\bm{A}$ is a symmetric matrix) and have the form
\begin{equation}
    \bm{v}_k = \frac{1}{V_k } \left(\sin \left( \frac{k\pi}{N + 1} \right), \sin \left( \frac{2k\pi}{N + 1} \right), \dots , \sin \left( \frac{jk\pi}{N + 1} \right), \dots, \sin \left( \frac{Nk\pi}{N + 1} \right) \right).  
    \label{eq:2ps_eigenvectors_b}
\end{equation}
The Euclidean norms $V_k$ of these eigenvectors are 
\begin{align}
  \label{eq:2ps_norm_coef_1}
  V_k  ^2 = \lVert \bm{v}_k \rVert^2 = \sum _{j = 1}^{N}\sin^2 \left( \frac{jk\pi}{N + 1} \right) & = \sum _{j = 1}^{N} \left[\frac{1}{2} - \frac{1}{2} \sum _{j = 1}^{N}\cos \left(\frac{2jk\pi}{N + 1} \right) \right], \\
  \label{eq:2ps_norm_coef_2}
  &= \frac{2N +  1}{4} - \frac{\sin \left( \frac{2N + 1}{N + 1} k\pi \right)}{4\sin \left( \frac{k\pi}{N + 1}  \right)}, \\
  \label{eq:2ps_norm_coef_3}
  &= \frac{N + 1}{2}.
\end{align} 
We note that going from~\eqref{eq:2ps_norm_coef_1} to~\eqref{eq:2ps_norm_coef_3}, we appropriately used the identities,
\begin{equation}
  \sum _{j = 0}^{N}\cos \left(jx \right) = \frac{1}{2}\left[1 + \frac{\sin \left[\left( N + \frac{1}{2} \right)x \right]}{\sin \left(\frac{x}{2} \right)} \right],
\end{equation}
which can be found for example at~\citep{jeffrey2007table} and 
\begin{equation}
  \sin \left( \frac{2N + 1}{N + 1} k\pi \right) = - \sin \left( \frac{k\pi}{N + 1}  \right). 
\end{equation}
The matrix $\bm{P}$ having as columns the eigenvectors~\eqref{eq:2ps_eigenvectors_b} i.e.~$\bm{P} = (\bm{v}_1^T~\bm{v}_2^T~\dots~\bm{v}_N^T)$ can be used to diagonalize $\bm{A}$~\eqref{eq:A_operator_form_app_1} so that
\begin{equation}
    \bm{A} = \bm{P}\cdot \bm{D}\cdot \bm{P}^{-1} = \bm{P}\cdot \bm{D}\cdot \bm{P}. 
    \label{eq:A_exp_diagonalize_app}
\end{equation}
We note that in~\eqref{eq:A_exp_diagonalize_app} $\bm{P}$ is a unitary $N\times N$ matrix, i.e. $\bm{P} = \bm{P}^{-1} = \bm{P}^T$, and $\bm{D}$ a diagonal $N\times N$ matrix having as non-zero elements the eigenvalues~\eqref{eq:eigenvalues_formula} i.e. $\bm{D} = diag\left(\lambda _1, \lambda _2, \dots, \lambda _N \right)$. 
Thus we can write the matrices $\cos \left( \bm{A}\tau \right)$ and $\sin \left( \bm{A}\tau \right)$ defining $\bm{C}(\tau)$ in~\eqref{eq:expression_of_c_matrix_simple_form} as
\[
\cos \left( \bm{A}\tau \right) = \bm{P}\cdot \bm{D}_c \cdot \bm{P}, \quad \sin \left( \bm{A}\tau \right) = \bm{P}\cdot \bm{D}_s \cdot \bm{P},
\]
with $\bm{D}_c = diag \left(\cos \left( \lambda _1 \tau \right), \cos \left( \lambda _2 \tau \right), \dots , \cos \left( \lambda _N \tau \right) \right)$, and $\bm{D}_s = diag \left(\sin \left( \lambda _1 \tau \right), \sin \left( \lambda _2 \tau \right), \dots , \sin \left( \lambda _N \tau \right) \right)$.

A similar treatment can be performed for the variational equations of the Hamiltonian $B$~\eqref{eq:b_twps_app} 
\begin{equation}
  \frac{d  \bm{w}}{dt} = \bm{L}_{B_V}\cdot \bm{w} \colon 
  \begin{dcases}
    \dot{\delta q}_1 = - \delta p_2 & \\
    \dot{\delta q}_l = - \delta p_{l - 1} - \delta p_{l + 1} & \quad 2\leq l \leq N-1\\
    \dot{\delta q}_N = - \delta p_{N-1} &\\
    \dot{\delta p} _1 = \delta q_2 &\\
    \dot{p} _l = \delta q_{l - 1} + \delta q_{l + 1} & \quad 2\leq l \leq N-1\\
    \dot{p} _N = \delta q_{N-1}
  \end{dcases},
  \label{eq:2ps_var_eq_b}
\end{equation}
which can be rewritten as 
\begin{equation}
  \frac{d \bm{w}}{dt} = \bm{L}_{B_V} \bm{w} = 
  \begin{pmatrix}
    \bm{0} & \bm{A} \\ 
    \bm{-A} & \bm{0}
  \end{pmatrix} \cdot \bm{w},
  \label{eq:2ps_var_eq_b contracted}
\end{equation}
is the system of ODEs~\eqref{eq:2ps_eqmotion_b_contracted}.
Thus,
\begin{equation}
  \bm{C}^\prime \left(\tau \right) = \bm{C} \left(\tau \right),
\end{equation}
in~\eqref{eq:2ps_gamma_primary_form} and~\eqref{eq:2ps_gamma_primary_form_2}.

\section{\label{sec:three-part_split_app}Three-part split symplectic integration of the 1D and 2D DDNLS models}
Following~\citep{skokos2014high,gerlach2016symplectic} and mainly~\citep{danieli2019computational}, we present here the analytic expressions of the operators involved in the symplectic integration of the 1D and 2D DDNLS systems when they are split in three integrable parts.

\subsection{\label{subsec:1D_DDNLS_thps_app}The 1D DDNLS system}

The Hamiltonian function $\mathcal{H}_{1D}$~\eqref{eq:ham_real_1d_ddnls_app} of the 1D DDNLS system can be split into three integrable parts as
\begin{equation}
    \mathcal{H}_{1D} = \mathcal{A} + \mathcal{B} + \mathcal{C},
    \label{eq:ham_3ps_1d_general}
\end{equation}
with
\begin{equation}
    \mathcal{A} = \sum_{l} \frac{\epsilon _l}{2}(q_l ^2 + p_l ^2) + \frac{\beta}{8}(q_l ^2 + p_l ^2)^2, \quad \mathcal{B}  = \sum_{l}- p_{l + 1}p_l, \quad \mathcal{C}  = \sum _l - q_{l + 1}q_l.
    \label{eq:3ps_1d_general_abc}
\end{equation}
Obviously, the $\mathcal{A}$ part in~\eqref{eq:3ps_1d_general_abc} coincides with the $A$ part in~\eqref{eq:ham_2ps_gen} so that the corresponding operator $e^{\tau \bm{L}_{\mathcal{A}_V}}$ is given by~\eqref{eq:flow_elav_app_1} and~\eqref{eq:flow_elav_app_2}.
The Hamiltonian functions $\mathcal{B}$ and $\mathcal{C}$ on the other hand possess $N$ cyclic coordinates as $\mathcal{B}$ depends only on momenta $p_l$ and $\mathcal{C}$ on the positions $q_l$.
Therefore it is easy to find the analytic expression of the operators $e^{\tau \bm{L}_{\mathcal{B}_V}}$ and $e^{\tau \bm{L}_{\mathcal{C}_V}}$ for the integration of their equations of motion and the variational equations.

More specifically, for the $\mathcal{B}$, we have the following system of equations of motion and variational equations
\begin{equation}
  \frac{d\bm{Z}}{dt} = \bm{L}_{\mathcal{B}_V}\bm{Z} \colon
  \begin{dcases}
    \dot{q}_1 = - p_{2} &\\
    \dot{q}_l = - p_{l - 1} - p_{l + 1} & l = 2, 3, \dots, N-1\\
    \dot{q}_N = - p_{N-1} &\\
    \dot{p}_l  = 0 & l = 1, 2, \dots, N \\
    \dot{\delta q}_1 = -\delta p_{2} &\\
    \dot{\delta q}_l = -\delta  p_{l - 1} - \delta  p_{l + 1} & l = 2, 3 \dots, N-1\\
    \dot{\delta q}_N = -\delta  p_{N-1} &\\
    \dot{\delta p}_l = 0 & l = 1, 2, \dots, N\\
  \end{dcases},
  \label{eq: 3ps eq motion b}
\end{equation}
which can be easily solved, leading to 
\begin{equation}
  e^{\tau \bm{L}_{\mathcal{B}_V}} \colon 
    \begin{dcases}
      q_1^\prime = q_1  - \tau p_2   &\\
      q_l^\prime = q_l  - \tau (p_{l - 1} + p_{l + 1}) & l = 2, 3, \dots, N-1 \\
      q_N^\prime = q_N - \tau p_{N-1}  &\\
      p_l^\prime = p_l & l = 1, 2, \dots, N \\
      \delta q_1^\prime = \delta  q_1  - \tau   \delta  p_{2} &\\
      \delta  q_l^\prime = \delta  q_l  - \tau (\delta  p_{l - 1} + \delta  p_{l + 1})  & l = 2, 3, \dots, N-1\\
      \delta  q_N^\prime = \delta  q_N  - \tau \delta  p_{N-1}  &\\
      \delta  p_l^\prime = \delta  p_l  & l = 1, 2, \dots, N\\
    \end{dcases}.
    \label{eq:3ps_var_eq_of_b}
\end{equation}
Similarly, the system of equations of motion and variational equations for the $\mathcal{C}$ Hamiltonian~\eqref{eq:3ps_1d_general_abc} is
\begin{equation}
  \frac{d\bm{Z}}{dt} = \bm{L}_{\mathcal{C}_V}\bm{Z} \colon
    \begin{dcases}
      \dot{p}_1 = \delta q_{2} &\\
      \dot{p}_l = \delta q_{l - 1} + \delta q_{l + 1} & l = 2, 3, \dots, N-1\\
      \dot{p}_N = \delta q_{N-1} &\\
      \dot{q}_l = 0 & l =1, 2, \dots, N \\
      \dot{\delta p}_1 = \delta q_{2} &\\
      \dot{\delta p}_l = \delta q_{l - 1} + \delta q_{l + 1} & l = 2, 3, \dots, N-1\\
      \dot{\delta p}_N = \delta q_{N-1} &\\
      \dot{\delta q}_l = 0 & l =1, 2, \dots, N \\
      \end{dcases}, 
  \label{eq:3ps_eq_motion_c}
\end{equation}
and its solution is given by
\begin{equation}
  e^{\tau \bm{L}_{\mathcal{C}_V}} \colon 
    \begin{dcases}
      q_l^\prime = q_l  & l =1, 2, \dots, N\\
      p_1^\prime = p_1 +   \tau  q_{2}  & \\
      p_l^\prime = p_l +   \tau (q_{l - 1} + q_{l + 1}) & l = 2, 3, \dots, N-1\\
      p_N^\prime = p_N +  \tau  q_{N-1}  & \\
      \delta q_l^\prime = \delta q_l   & l =1, 2, \dots, N \\
      \delta p_1^\prime = \delta p_1 + \tau \delta  q_{2} & \\
      \delta p_l^\prime = \delta p_l + \tau (\delta q_{l - 1} +\delta  q_{l + 1}) &  l =2, 3, \dots, N-1\\
      \delta p_N^\prime = \delta p_N + \tau \delta  q_{N-1}  & \\
    \end{dcases}.
  \label{EQ:e^LC_2_n-1}
\end{equation}

\subsection{\label{susec:3ps_2d_ddnsl_app}The 2D DDNLS system}
We can treat the 2D DDNLS Hamiltonian $\mathcal{H}_{2D}$~\eqref{eq:ham_real_2d_ddnls_app_oo} in a similar manner to the one implemented for the 1D DDNLS system in Sec.~\ref{subsec:1D_DDNLS_thps_app}.
In particular, the Hamiltonian can be split
\begin{equation}
  \mathcal{H}_{2D} = \mathsf{A} + \mathsf{B} + \mathsf{C},
  \label{eq:3ps_2d_ddnls_general}
\end{equation}
with 
\begin{equation}
  \mathsf{A} = \sum _{l,m} \frac{\epsilon _{l, m} }{2} \left(q_{l, m} ^2 + p_{l, m}^2\right) + \frac{\beta}{8} \left(q_{l, m} ^2 + p_{l, m}^2\right) ^2 ,
  \label{eq:a_3ps_2d_ddnls_app}
\end{equation}
\begin{equation}
  \mathsf{B} = \sum _{l, m} -  p _{l, m + 1} p _{l, m} - p _{l + 1, m}p _{l, m} , \qquad \mathsf{C} = \sum _{l, m} - q_{l, m + 1} q _{l, m} - q_{l + 1, m} q _{l, m}.
  \label{eq:bc_3ps_2d_ddnls_app}
\end{equation}

For the Hamiltonian $\mathsf{A}$~\eqref{eq:a_3ps_2d_ddnls_app}, following similar steps to the ones performed in Sec.~\ref{subsec:e_lav_app}, we obtain the following set of equations of motion and variational equations
\begin{equation}
  \begin{split}
    \frac{d\bm{Z}}{dt} =
    \bm{L}_{\mathsf{A}_V}\bm{Z} & : \left\{
    \begin{array}{rl}
      \dot{q}_{l, m} &= p_{l, m} \dot{\Theta}_{l, m}  \\
      \dot{p}_{l, m} & = -q_{l, m} \dot{\Theta}_{l, m} , \\
      \dot{\delta q}_{l, m} &= \left[ \dot{\Theta}_{l, m}  +  \beta p_{l, m} ^2\right] \delta p_{l, m} + \beta q_{l, m} p_{l, m} \delta q_{l, m} \\
      \dot{\delta p}_{l, m} &= - \left[ \dot{\Theta}_{l, m}  + \beta q_{l, m}^2\right] \delta q_{l, m} - \beta q_{l, m} p_{l, m} \delta p_{l, m}
    \end{array} \right., \\
  \end{split}
  \label{eq:DNLS_2D_3ps_A}
\end{equation}
where $\dot{\Theta}_{l, m} = \epsilon _{l, m} + \beta (q_{l, m} ^ 2 + p_{l, m} ^2 )/2$ are constants of motion.
Setting 
\[
    J_{l, m}  = \frac{1}{2} (q_{l, m}  ^2 + p_{l, m} ^2), \quad \alpha _{l, m} = \epsilon _{l, m} + \beta  J_{l, m},
\]
and  
\[ 
    \delta J_{l, m}   = q_{l, m}   \delta q_{l, m}   + p_{l, m}   \delta p_{l, m}, \quad \delta \Theta _{l, m}  = \frac{p_{l, m}  }{2J_{l, m} }\delta q_{l, m}  - \frac{q_{l, m}  }{2J_{l, m} } \delta p_{l, m},
\]
the solution of system~\eqref{eq:DNLS_2D_3ps_A} for $J_{l, m}\neq 0$ is obtained through the application of the operator
\begin{equation}
  \begin{split}
  e^{\tau \bm{L}_{\mathsf{A}_V}} \colon 
  \left\{
    \begin{array}{rl}
      q_{l, m} ^\prime &= q_{l, m} \cos (\tau \alpha _{l, m}) +  p_{l, m} \sin (\tau \alpha _{l, m}) \\
      p_{l, m} ^\prime &= p_{l, m} \cos (\tau \alpha _{l, m}) -  q_{l, m} \sin (\tau \alpha _{l, m}) \\
      \delta q_{l, m} ^\prime &= \frac{q_{l, m}  \cos (\tau \alpha _{l, m}) + p_{l, m}  \sin (\tau \alpha _{l, m} )}{ 2 J_{l, m}} \delta J_{l, m}  + \left(p_{l, m}  \cos (\tau \alpha _{l, m} ) - q_{l, m}  \sin (\tau \alpha _{l, m} )\right) \left(\beta \delta J_{l, m}  \tau + \delta \Theta  _{l, m}  \right) \\
      \delta p_{l, m} ^\prime  & = \frac{p_{l, m}  \cos (\tau \alpha _{l, m} ) - q_{l, m} \sin (\tau \alpha _{l, m} )}{2J_{l, m} } \delta J_{l, m}  - \left( q_{l, m}  \cos ( \tau \alpha _{l, m} ) + p_{l, m} \sin (\tau \alpha _{l, m} ) \right) \left(\beta \delta J_{l, m}  \tau + \delta \Theta  _{l, m} \right)
    \end{array} 
  \right.,
  \end{split}
\end{equation}
or when $J_{l,m} = 0$ through the operator 
\begin{equation}
  e^{\bm{L}_{\mathsf{A}_V}\tau} 
  \colon
  \begin{dcases}
      q_{l, m}^\prime = q_{l, m} = 0 \\
      p_{l, m} ^\prime = q_{l, m} = 0 \\
      \delta q_{l, m} ^\prime = \delta q_{l, m} \cos \left(\epsilon _{l, m} \tau\right) + \delta p_{l, m} \sin \left(\epsilon _{l, m} \tau\right) \\
      \delta p_{l, m} ^\prime = \delta p_{l, m} \cos \left(\epsilon _{l, m} \tau\right) - \delta q_{l, m} \sin \left(\epsilon _{l, m} \tau\right)    
  \end{dcases}.
  \label{eq:flow_elav_app_2_2d}
\end{equation}

For the Hamiltonian functions $\mathsf{B}$ and $\mathsf{C}$~\eqref{eq:bc_3ps_2d_ddnls_app} the set of equations of motion and variational equations are respectively
\begin{equation}
  \begin{split}
    \frac{d\bm{Z}}{dt}=\bm{L}_{\mathsf{B}_V}\bm{Z} & : 
    \left\{
      \begin{array}{lll}
        \dot{q}_{1, 1} &= -p_{1, 2} + p_{2, 1} & \\
        \dot{q}_{N, 1} & = - p_{N-1, 1} + p_{N, 2} & \\
        \dot{q}_{1, M} &= - p_{1, M-1} + p_{2, M} & \\
        \dot{q}_{N, M} &= - p_{N - 1, M} + p_{N, M-1} & \\
        \dot{q}_{l, 1} &= - p_{l-1, 1} + p_{l, 2} + p_{l+1, 1} & \mbox{$2\leq l \leq N - 1$}\\
        \dot{q}_{l, M} &= - p_{l - 1, M} + p_{l, M-1} + p_{l + 1, M} & \mbox{$2\leq l \leq N - 1$}\\
        \dot{q}_{1, m} &= - p_{1, m - 1} + p_{1, m + 1} + p_{2, m} & \mbox{$2\leq m \leq M - 1$}\\
        \dot{q}_{N, m} &= - p_{N - 1, m} + p_{N, m - 1} + p_{N, m + 1} & \mbox{$2\leq m \leq M - 1$} \\
        \dot{q}_{l, m} &= -  p_{l - 1, m} + p_{l, m - 1} + p_{l, m + 1} + p_{l + 1, m}  & \mbox{$2\leq l \leq N - 1$, $2\leq m\leq M-1$}\\
        \dot{p}_{l, m}  &= 0 & \mbox{$1\leq l \leq N$, $1\leq m\leq M$} \\
        \dot{\delta q}_{1, 1} &= - \delta p_{1, 2} + \delta p_{2, 1} &\\
        \dot{\delta q}_{N, 1} & =- \delta p_{N-1, 1} + \delta p_{N, 2} &\\
        \dot{\delta q}_{1, M} &= - \delta p_{1, M-1} + \delta p_{2, M} &\\
        \dot{\delta q}_{N, M} &= -\delta p_{N - 1, M} + \delta p_{N, M-1} &\\
        \dot{\delta q}_{l, 1} &= - \delta p_{l-1, 1} + \delta p_{l, 2} + \delta p_{l+1, 1} & \mbox{$2\leq l \leq N - 1$} \\
        \dot{\delta q}_{l, M} &= -\delta p_{l - 1, M} + \delta p_{l, M-1} + \delta p_{l + 1, M} & \mbox{$2\leq l \leq N - 1$} \\
        \dot{\delta q}_{1, m} &= -\delta p_{1, m - 1} + \delta p_{1, m + 1} + \delta p_{2, m} & \mbox{$2\leq m \leq M - 1$} \\
        \dot{\delta q}_{N, m} &= - \delta p_{N - 1, m} + \delta p_{N, m - 1} + \delta p_{N, m + 1} & \mbox{$2\leq m \leq M - 1$} \\
        \dot{\delta q}_{l, m} &= -  \delta p_{l - 1, m} + \delta p_{l, m - 1} + \delta p_{l, m + 1} + \delta p_{l + 1, m}  & \mbox{$2\leq l \leq N - 1$, $2\leq m\leq M-1$}\\
        \dot{\delta p}_{l, m} &= 0 & \mbox{$1\leq l \leq N$, $1\leq m\leq M$} \\
      \end{array} 
    \right.
  \end{split},
\end{equation}

and
\begin{equation}
  \begin{split}
  \frac{d\bm{Z}}{dt}=\bm{L}_{\mathsf{C}_V}\bm{Z} & : 
  \left\{
    \begin{array}{lll}
      \dot{q}_{l, m}  &= 0 & \mbox{$1\leq l \leq N$, $1\leq m\leq M$} \\
      \dot{p}_{1, 1} &= q_{1, 2} + q_{2, 1} & \\
      \dot{p}_{N, 1} & =  q_{N-1, 1} + q_{N, 2} & \\
      \dot{p}_{1, M} &= q_{1, M-1} + q_{2, M} & \\
      \dot{p}_{N, M} &=  q_{N - 1, M} + q_{N, M-1} & \\
      \dot{p}_{l, 1} &=q_{l-1, 1} + q_{l, 2} + q_{l+1, 1} & \mbox{$2\leq l \leq N - 1$}\\
      \dot{p}_{l, M} &= q_{l - 1, M} + q_{l, M-1} + q_{l + 1, M} & \mbox{$2\leq l \leq N - 1$}\\
      \dot{p}_{1, m} &=  q_{1, m - 1} + q_{1, m + 1} + q_{2, m} & \mbox{$2\leq m \leq M - 1$}\\
      \dot{p}_{N, m} &= q_{N - 1, m} + q_{N, m - 1} + q_{N, m + 1} & \mbox{$2\leq m \leq M - 1$} \\
      \dot{p}_{l, m} &=   q_{l - 1, m} + q_{l, m - 1} + q_{l, m + 1} + q_{l + 1, m}  & \mbox{$2\leq l \leq N - 1$, $2\leq m\leq M-1$}\\
      \dot{\delta q}_{l, m} &= 0 & \mbox{$1\leq l \leq N$, $1\leq m\leq M$} \\
      \dot{\delta p}_{1, 1} &= \delta q_{1, 2} + \delta q_{2, 1} &\\
      \dot{\delta p}_{N, 1} & = \delta q_{N-1, 1} + \delta q_{N, 2} &\\
      \dot{\delta p}_{1, M} &=  \delta q_{1, M-1} + \delta q_{2, M} &\\
      \dot{\delta p}_{N, M} &=  \delta q_{N - 1, M} + \delta q_{N, M-1} &\\
      \dot{\delta p}_{l, 1} &=  \delta q_{l-1, 1} + \delta q_{l, 2} + \delta q_{l+1, 1} & \mbox{$2\leq l \leq N - 1$} \\
      \dot{\delta p}_{l, M} &=  \delta q_{l - 1, M} + \delta q_{l, M-1} + \delta q_{l + 1, M} & \mbox{$2\leq l \leq N - 1$} \\
      \dot{\delta p}_{1, m} &=  \delta q_{1, m - 1} + \delta q_{1, m + 1} + \delta q_{2, m} & \mbox{$2\leq m \leq M - 1$} \\
      \dot{\delta p}_{N, m} &=  \delta q_{N - 1, m} + \delta q_{N, m - 1} + \delta q_{N, m + 1} & \mbox{$2\leq m \leq M - 1$} \\
      \dot{\delta p}_{l, m} &=  \delta q_{l - 1, m} + \delta q_{l, m - 1} + \delta q_{l, m + 1} + \delta q_{l + 1, m}  & \mbox{$2\leq l \leq N - 1$, $2\leq m\leq M-1$}\\
    \end{array} 
  \right..
  \end{split}
\end{equation}
Consequently, the operators $e^{\tau \bm{L}_{\mathsf{B}_V}}$ and $e^{\tau \bm{L}_{\mathsf{C}_V}}$ are given by 
\begin{equation}
  \begin{split}
    e^{\tau \bm{L}_{\mathsf{B}_V}} &:
    \left \{
      \begin{array}{lll}
        q_{1, 1}^\prime &= q_{1, 1} - \tau \left(p_{1, 2} + p_{2, 1}\right) & \\
        q_{N, 1}^\prime & = q_{N, 1} - \tau \left(p_{N-1, 1} + p_{N, 2}\right) & \\
        q_{1, M}^\prime &= q_{1, M} - \tau \left(p_{1, M-1} + p_{2, M}\right) & \\
        q_{N, M}^\prime &= q_{N, M} - \tau \left(p_{N - 1, M} + p_{N, M-1}\right) & \\
        q_{l, 1}^\prime &= q_{l, 1} - \tau \left(p_{l-1, 1} + p_{l, 2} + p_{l+1, 1}\right) & \mbox{$2\leq l \leq N - 1$}\\
        q_{l, M}^\prime &= q_{l, M} - \tau \left(p_{l - 1, M} + p_{l, M-1} + p_{l + 1, M}\right) & \mbox{$2\leq l \leq N - 1$}\\
        q_{1, m}^\prime &= q_{1, m} - \tau \left(p_{1, m - 1} + p_{1, m + 1} + p_{2, m}\right) & \mbox{$2\leq m \leq M - 1$}\\
        q_{N, m}^\prime &= q_{N, m} - \tau \left(p_{N - 1, m} + p_{N, m - 1} + p_{N, m + 1}\right) & \mbox{$2\leq m \leq M - 1$} \\
        q_{l, m}^\prime &= q_{l, m} - \tau \left( p_{l - 1, m} + p_{l, m - 1} + p_{l, m + 1} + p_{l + 1, m} \right) & \mbox{$2\leq l \leq N - 1$, $2\leq m\leq M-1$}\\
        p_{l, m}^\prime  &= p_{l, m} & \mbox{$1\leq l \leq N$, $1\leq m\leq M$} \\
        \delta q_{1, 1}^\prime &= \delta q_{1, 1} - \tau \left(\delta p_{1, 2} + \delta p_{2, 1}\right) &\\
        \delta q_{N, 1}^\prime & = \delta q_{N, 1} - \tau \left(\delta p_{N-1, 1} + \delta p_{N, 2}\right) &\\
        \delta q_{1, M}^\prime &= \delta q_{1, M} - \tau \left(\delta p_{1, M-1} + \delta p_{2, M}\right) &\\
        \delta q_{N, M}^\prime &= \delta q_{N, M} - \tau \left(\delta p_{N - 1, M} + \delta p_{N, M-1}\right) &\\
        \delta q_{l, 1}^\prime &= \delta q_{l, 1} - \tau \left(\delta p_{l-1, 1} + \delta p_{l, 2} + \delta p_{l+1, 1}\right) & \mbox{$2\leq l \leq N - 1$} \\
        \delta q_{l, M}^\prime &= \delta q_{l, M} - \tau \left(\delta p_{l - 1, M} + \delta p_{l, M-1} + \delta p_{l + 1, M}\right) & \mbox{$2\leq l \leq N - 1$} \\
        \delta q_{1, m}^\prime &= \delta q_{1, m} - \tau \left(\delta p_{1, m - 1} + \delta p_{1, m + 1} + \delta p_{2, m}\right) & \mbox{$2\leq m \leq M - 1$} \\
        \delta q_{N, m}^\prime &= \delta q_{N, m} - \tau \left(\delta p_{N - 1, m} + \delta p_{N, m - 1} + \delta p_{N, m + 1}\right) & \mbox{$2\leq m \leq M - 1$} \\
        \delta q_{l, m}^\prime &= \delta q_{l, m} - \tau \left( \delta p_{l - 1, m} + \delta p_{l, m - 1} + \delta p_{l, m + 1} + \delta p_{l + 1, m} \right) & \mbox{$2\leq l \leq N - 1$, $2\leq m\leq M-1$}\\
        \delta p_{l, m}^\prime &= \delta p_{l, m} & \mbox{$1\leq l \leq N$, $1\leq m\leq M$} \\
      \end{array} 
    \right. \\
  \end{split},
  \label{eq:e_LBZ_DNLS_2D_app}
\end{equation}
and
\begin{equation}
  \begin{split}
    e^{\tau \bm{L}_{\mathsf{C}_V}} & : 
    \left \{
      \begin{array}{lll}
        q_{l, m}^\prime  &= q_{l, m} & \mbox{$1\leq l \leq N$, $1\leq m\leq M$} \\
        p_{1, 1}^\prime &= p_{1, 1} + \tau \left(q_{1, 2} + q_{2, 1}\right) & \\
        p_{N, 1}^\prime & = p_{N, 1} + \tau \left(q_{N-1, 1} + q_{N, 2}\right) & \\
        p_{1, M}^\prime &= p_{1, M} + \tau \left(q_{1, M-1} + q_{2, M}\right) & \\
        p_{N, M}^\prime &= p_{N, M} + \tau \left(q_{N - 1, M} + q_{N, M-1}\right) & \\
        p_{l, 1}^\prime &= p_{l, 1} + \tau \left(q_{l-1, 1} + q_{l, 2} + q_{l+1, 1}\right) & \mbox{$2\leq l \leq N - 1$}\\
        p_{l, M}^\prime &= p_{l, M} + \tau \left(q_{l - 1, M} + q_{l, M-1} + q_{l + 1, M}\right) & \mbox{$2\leq l \leq N - 1$}\\
        p_{1, m}^\prime &= p_{1, m} + \tau \left(q_{1, m - 1} + q_{1, m + 1} + q_{2, m}\right) & \mbox{$2\leq m \leq M - 1$}\\
        p_{N, m}^\prime &= p_{N, m} + \tau \left(q_{N - 1, m} + q_{N, m - 1} + q_{N, m + 1}\right) & \mbox{$2\leq m \leq M - 1$} \\
        p_{l, m}^\prime &= p_{l, m} + \tau \left( q_{l - 1, m} + q_{l, m - 1} + q_{l, m + 1} + q_{l + 1, m} \right) & \mbox{$2\leq l \leq N - 1$, $2\leq m\leq M-1$}\\
        \delta q_{l, m}^\prime &= \delta q_{l, m} & \mbox{$1\leq l \leq N$, $1\leq m\leq M$} \\
        \delta p_{1, 1}^\prime &= \delta p_{1, 1} + \tau \left(\delta q_{1, 2} + \delta q_{2, 1}\right) &\\
        \delta p_{N, 1}^\prime & = \delta p_{N, 1} + \tau \left(\delta q_{N-1, 1} + \delta q_{N, 2}\right) &\\
        \delta p_{1, M}^\prime &= \delta p_{1, M} + \tau \left(\delta q_{1, M-1} + \delta q_{2, M}\right) &\\
        \delta p_{N, M}^\prime &= \delta p_{N, M} + \tau \left(\delta q_{N - 1, M} + \delta q_{N, M-1}\right) &\\
        \delta p_{l, 1}^\prime &= \delta p_{l, 1} + \tau \left(\delta q_{l-1, 1} + \delta q_{l, 2} + \delta q_{l+1, 1}\right) & \mbox{$2\leq l \leq N - 1$} \\
        \delta p_{l, M}^\prime &= \delta p_{l, M} + \tau \left(\delta q_{l - 1, M} + \delta q_{l, M-1} + \delta q_{l + 1, M}\right) & \mbox{$2\leq l \leq N - 1$} \\
        \delta p_{1, m}^\prime &= \delta p_{1, m} + \tau \left(\delta q_{1, m - 1} + \delta q_{1, m + 1} + \delta q_{2, m}\right) & \mbox{$2\leq m \leq M - 1$} \\
        \delta p_{N, m}^\prime &= \delta p_{N, m} + \tau \left(\delta q_{N - 1, m} + \delta q_{N, m - 1} + \delta q_{N, m + 1}\right) & \mbox{$2\leq m \leq M - 1$} \\
        \delta p_{l, m}^\prime &= \delta p_{l, m} + \tau \left( \delta q_{l - 1, m} + \delta q_{l, m - 1} + \delta q_{l, m + 1} + \delta q_{l + 1, m} \right) & \mbox{$2\leq l \leq N - 1$, $2\leq m\leq M-1$}\\
      \end{array} 
    \right..
  \end{split}
  \label{eq:e_LCZ_DNLS_2D_app}
\end{equation}

\end{appendix}



\bibliographystyle{abbrvnat}
\bibliography{references}


\printindex


\end{document}